\documentclass{aa}
\usepackage[varg]{txfonts}
\usepackage{latexsym}
\usepackage{graphicx}
\usepackage{natbib}
\usepackage{lscape}
\usepackage{afterpage}
\usepackage{booktabs}
\usepackage{xcolor}
 
\usepackage{cancel}
           
\usepackage[citecolor=blue]{hyperref}
\usepackage{color}
\usepackage{ulem}
\usepackage{arydshln}
\usepackage{multirow}

\begin{document}

\title{Compact molecular gas emission in local LIRGs among\\ low- and high-z galaxies}
\author{E. Bellocchi\inst{1}, M. Pereira-Santaella\inst{1}, L. Colina\inst{1}, A. Labiano\inst{1,2}, M. S\'anchez-Garc\'ia\inst{1,3}, A. Alonso-Herrero\inst{1}, S.~Arribas\inst{1}, S. Garc\'ia-Burillo\inst{3}, M. Villar-Mart\'in\inst{1}, D. Rigopoulou\inst{4}, F. Valentino\inst{5,6}, A. Puglisi\inst{7}, \\T.~D\'iaz-Santos\inst{8}, S.~Cazzoli\inst{9}, A. Usero\inst{3}
}
\institute{$^1$ Centro de Astrobiolog\'ia, (CSIC-INTA), Astrophysics Department, Madrid, Spain\\
$^2$ Telespazio UK, for the European Space Agency (ESA), ESAC, Villanueva de la Ca\~nada, Madrid, Spain\\ 
$^3$ Observatorio Astron\'omico Nacional (OAN-IGN)-Observatorio de Madrid, Alfonso XII, 3, 28014 Madrid, Spain\\
$^4$ Department of Physics, University of Oxford, Oxford OX1 3RH, UK\\
$^5$ Cosmic Dawn Center (DAWN), Copenhagen, Denmark\\
$^6$ Niels Bohr Institute, University of Copenhagen, Jagtvej 128, DK-2200, Copenhagen, Denmark\\
$^7$ Centre for Extragalactic Astronomy, Department of Physics, Durham University, South Road, Durham DH1 3LE, UK\\
$^8$ Institute of Astrophysics, Foundation for Research and Technology - Hellas (FORTH), Heraklion, 70013, Greece\\
$^9$ Instituto de Astrof\'isica de Andaluc\'ia (IAA-CSIC), Apdo. 3004, E-18008, Granada, Spain\\
\vskip0.5mm
\email{enrica.bellocchi@gmail.com, ebellocchi@cab.inta-csic.es}\\ 
}

\date{Received 1 December 2021 / Accepted 28 March 2022}

\abstract 
{
We present new CO(2-1) observations of a representative sample of 24 local (z$<$0.02) luminous infrared galaxies (LIRGs) at high spatial resolution ($<$100 pc) from the Atacama Large Millimeter/submillimeter Array (ALMA). Our LIRGs lie above the {\it Main-Sequence (MS)}, with typical stellar masses in the range 10$^{10}$-10$^{11}$ M$_\odot$ and SFR$\sim$30 M$_\odot$ yr$^{-1}$. We derive the effective radii of the CO(2-1) and the 1.3 mm continuum emissions using the curve-of-growth method.
LIRGs show an extremely compact cold molecular gas distribution (median R$_{CO}$ $\sim$0.7 kpc), which is a factor 2 smaller than the ionized gas (median R$_{H\alpha}$$\sim$1.4 kpc), and 3.5 times smaller than the stellar size (median R$_{star}$$\sim$2.4 kpc). The molecular size of LIRGs is similar to that of early-type galaxies (ETGs; R$_{CO}$$\sim$1 kpc) and about a factor of 6 more compact than local Spirals of similar stellar mass. Only the CO emission in low-z ULIRGs is more compact than these local LIRGs by a factor of 2.
Compared to high-z (1$<$z$<$6) systems, the stellar sizes and masses of local LIRGs are similar to those of high-z {\it MS} star-forming galaxies (SFG) and about a factor of 2-3 lower than sub-mm galaxies (SMG). The molecular sizes of high-z {\it MS} SFGs and SMGs are larger than those derived for LIRGs by a factor of $\sim$3 and $\sim$8, respectively.
Contrary to high-z SFGs and SMGs, which have comparable molecular and stellar sizes (median R$_{star}$/R$_{CO}$ = 1.8 and 1.2, respectively), the molecular gas in local LIRGs is more centrally concentrated (median R$_{star}$/R$_{CO}$ = 3.3).
A fraction of the low-z LIRGs and high-z galaxies share a similar range in the size of the ionized gas distribution, from 1 to 4 kpc. However, no LIRGs with very extended (above 4 kpc) radius are identified while for high-z galaxies no compact (less than 1 kpc) emission is detected.  
These results indicate that while low-z LIRGs and high-z {\it MS}-SFGs have similar stellar masses and sizes, the regions of current star formation (traced by the ionized gas) and of potential star-formation (traced by the molecular gas) are substantially smaller in LIRGs, and constrained to the central kpc region. High-z galaxies represent a wider population but their star-forming regions are more extended, even covering the overall size of the host galaxy. High-z galaxies have larger fractions of gas than low-z LIRGs, and therefore the formation of stars could be induced by interactions and mergers in extended disks or filaments with large enough molecular gas surface density involving physical mechanisms similar to those identified in the central kpc of LIRGs.  
}


\titlerunning{Compact molecular gas emission in local LIRGs}

\authorrunning{Bellocchi et al.}
\maketitle

\section{Introduction}

Luminous and ultraluminous infrared galaxies (i.e., LIRGs, L$_{IR}$ = L$_{[8-1000 \hskip1mm \mu m]}$ = 10$^{11}$-10$^{12}$ L$_\odot$, and ULIRGs, L$_{IR}$ > 10$^{12}$ L$_\odot$, respectively) host the most extreme star-forming events in the low-z Universe. 
They are characterized by extreme total infrared luminosity, since a large amount of star formation is hidden by dust, which reprocesses the UV photons that originate from hot young stars and/or active galactic nuclei (AGN; \citealt{U12} and references therein) and re-emit them at longer wavelengths, typically in the IR.
From the very early investigations, a large fraction of studies tried to determine what powers (U)LIRG systems. Both AGN and star forming activity (e.g., \citealt{Sanders96, Rigopoulou99, Veilleux99, Risaliti06, Valiante09, AAH12}) can co-exist in these systems. The fraction of galaxies dominated by the presence of an AGN increases with L$_{IR}$ (\citealt{Tran01, Nardini08, Veilleux09, AAH12}), while lower luminosity LIRGs are mainly powered by star-formation (see the review by \citealt{PT21}). (U)LIRGs show a large variety of morphologies, which suggest different dynamical phases: from isolated disks for low-luminosity LIRGs to a majority of merger remnants for ULIRGs (e.g., \citealt{Veilleux02, Arribas04, Kartaltepe10}). Their dynamical masses typically range between 10$^{10}$-10$^{11}$ M$_\odot$ (\citealt{Bellocchi13, Crespo21}).

These extreme populations are rare in the local Universe (e.g., \citealt{Lagache05}), but they are much more numerous at high-z and are supposed to be the dominant contributors to the star formation in the Universe at z $>$ 1 (e.g., \citealt{LF05, PG05, Caputi07, PG08, Magnelli13}). 
Local ULIRGs were initially assumed to be the local counterpart of high-z IR luminous galaxies discovered by {\it Spitzer} and the more luminous sub-mm galaxies (SMGs; e.g. \citealt{Blain02, Tacconi06}). 
Later on, several studies have suggested that high-z (z$\sim$2) ULIRGs and SMGs instead showed IR features more similar to those observed in local lower luminosity LIRGs (e.g., \citealt{Rujopakarn11}), rather than to local ULIRGs. 
In particular, the far-IR spectral energy distributions (SEDs) of ULIRGs and SMGs at z$\sim$2 differ from those of local galaxies of similar luminosity, since they appear cold as those of lower luminosity galaxies (LIRGs; \citealt{Muzzin10, Wuyts11a}). Such high-z systems are then assumed to be scaled-up version in size and star formation efficiency (SFE = SFR/M$_{star}$) of lower luminosity low-z (U)LIRGs, where low-z (U)LIRGs cover a similar SFR range than normal high-z star-forming galaxies (SFGs; $<$1000 M$_\odot$ yr$^{-1}$; \citealt{FS09, Rujopakarn11, Wuyts11, Arribas12}).

At different redshifts a tight correlation between the stellar mass and SFR has been observed in SFGs, called {\it Main-Sequence} ({\it MS}; \citealt{Elbaz07, Wuyts11}). 
Two main categories\footnote{At z$\sim$2 red and dead (passive) galaxies already exist at these cosmic epochs and form a separate sequence below the {\it MS} of SFGs.} can be identified: galaxies which lie on the {\it MS} and those lying above the {\it MS} (`outliers'). 
This tight correlation has thus invoked two distinct modes of star formation: the normal star-forming mode which describes galaxies lying on the SFR-M$_{star}$ relation ({\it MS} galaxies) which evolve through secular processes such as gas accretion (e.g., \citealt{Dekel09, Dave10, Genzel15}), while the starburst mode describes galaxies falling well above the {\it MS} (above-{\it MS} galaxies) which are likely driven by major mergers, representing a star-bursting period with respect to the galaxies on the {\it MS} (\citealt{Rodighiero11, Wuyts11, Cibinel19}).
Among the `outliers', local (U)LIRGs and SMGs at 1$<$z$<$4 show similar sSFR, possibly boosted by major merger events, although recent studies have found that some of SMGs can also be rotating disks (\citealt{Hodge16}) supporting the continuum gas accretion scenario through cold gas flow and minor mergers (\citealt{Dekel09}).

To shed more light on this scenario, we test for the first time the relation among the spatial extents (defined by the half-light radius) of different tracers in a sample of local LIRGs at high spatial resolution ($\sim$90 pc): in particular, the molecular emission traced by $^{12}$CO(2--1) (hereafter, CO(2--1)), the 1.3 mm (247 GHz) continuum, the stellar and the ionized gas emissions can be compared. The stellar and ionized gas sizes have been derived in previous works (\citealt{Arribas12, Bellocchi13}). 
With the advent of high resolution instrument as the Atacama Large Millimeter/Submillimeter Array (ALMA), we are now able to study the molecular emission in local galaxies at spatial resolution similar to that covered by typical giant molecular cloud (GMCs, $\sim$100 pc). 
Low-z LIRGs offer a unique opportunity to study, at high linear resolution and signal-to-noise ratio (S/N), extreme SF events and compare them with those observed locally (i.e., Spirals, early-type galaxies, ETGs, and ULIRGs) and at high-z.

Several works have tried to compare the effective size of different components in high-z systems. 
A large variety of objects covering a wide range in redshifts (1$\lesssim$z$\lesssim$6) and galaxy properties have been identified using different criteria (e.g., stellar mass, far-IR luminosity or optical compactness).
These systems can be mainly classified as (i) compact main-sequence ({\it MS}) SFGs (\citealt{Barro14}, \citealt{Barro16}), (ii) {\it MS} SFGs (\citealt{Straatman15}, \citealt{Tadaki17}, \citealt{FS18}, \citealt{Puglisi19}, \citealt{Kaasinen20}, \citealt{Fujimoto20}, \citealt{Wilman20}, \citealt{Cheng20}, \citealt{Valentino20}, \citealt{Puglisi21} and \citealt{Hogan21}) and the (iii) extreme SMGs, which mostly lie in the upper envelope of the {\it MS} plane (\citealt{Lang19}, \citealt{Chen17}, \citealt{Chen20}, \citealt{CR18}, \citealt{Hodge16}, \citealt{Gullberg18}). 
According to this classification, some of the so-called {\it MS} galaxies here mentioned can also include a few galaxies lying slightly off the {\it MS} (e.g., \citealt{Tadaki17, Puglisi19, Kaasinen20, Valentino20}): we should consider these systems to belong `around' the {\it MS} at their corresponding redshift.

The characterization of the distribution of all these tracers is key to understand how the different phases of the interstellar medium (ISM) evolve in size across the cosmic time studying several types of galaxies, at low- and at high-z.
In particular, we will be able to compare the size of the host galaxy (stellar component) with that derived for the ongoing star formation (ionized gas) as well as the size of the regions where stars are forming (molecular gas) in a local sample of LIRGs and compare them with those derived for local Spirals, ETGs and ULIRGs as well as high-z SFGs and SMGs.

The paper is structured as follows. In Sections 2 and 3 we describe, respectively, the sample as well as the observation and the data reduction. In Section 4, we describe the data analysis and the uncertainties related to the methods used. In Section 5 we present the results derived for the effective radii of the molecular and continuum components, comparing their values with those obtained for the stellar and ionized emissions. In Section 6 we discuss the results and we compare them with those derived for other local and high-z samples. In Section 7 we summarize our main findings and present our conclusions. In Appendix~\ref{app_classification} we present the morphological and kinematic classifications used to characterize our systems. In Appendix~\ref{app_maps} the CO(2--1), continuum at 1.3 mm and near-IR maps are shown for the whole sample. In Appendix~\ref{CO_sameFoV_Reff} the CO(2--1) and Two Micron All Sky Survey (2MASS; \citealt{Skrutskie06}) K-band images are compared for the whole sample, ordered according to their increasing molecular size, R$_{CO}$. Finally in Appendix~\ref{SEDs_sample}, we present the SEDs for the whole sample to estimate the continuum flux loss at 1.3 mm.

Throughout the paper, we consider the cosmology corrected quantities: H$_0$ = 67.8 km sec$^{-1}$ Mpc$^{-1}$, $\Omega_{M}$ = 0.308, $\Omega_{\Lambda}$ = 0.692. The redshift is corrected to the reference frame defined by the 3K~CMB.

\begin{table*}
\begin{tiny}
\caption{General properties of the LIRG sample. }
\label{Input_data}
\hskip-7mm
\begin{tabular}{ccccccccccc} 
\hline\hline\noalign{\smallskip}  
\multicolumn{2}{c}{Source}	 &		$\alpha$  & $\delta$		& z 	& D$_L$ & Scale & 	log L$_{IR}$ 		& SFR$_{IR}$	&	 Class	& AGN Ref \\
\cmidrule(lr){1-2}\cmidrule(lr){3-4}
IRAS & Other & {\tiny J2000.0} & {\tiny J2000.0} 	&&&&&&	 &Notes\\
 & & (h m s)  & ($^\circ$ $^{\prime}$ $^{\prime\prime}$) &  & (Mpc) & (pc/$\arcsec$)  	&	(L$_\odot$)		&  	(M$_\odot$ yr$^{-1}$) \\
(1) & (2) &  (3)& (4) &(5)  & (6)  &(7) &(8)	&	(9) 	& (10)	&	(11)\\
\hline\noalign{\smallskip} 	
{\tt F01341-3735 N}	 & {\tt ESO 297-G011}	&	 01 36 23.40 	& -37 19 17.6	&	0.0168 	& 75.2	&	353 & 10.84 (10.03)	&	10.5 (1.6)	&	{\tt 0 (P)}	&	(y) a, b \\
{\tt F01341-3735 S}	 & {\tt ESO 297-G012}	&	 01 36 24.17  	& -37 20 25.7	&	0.0172 	& 77.1	&	361 &  10.74 (9.92)		&	 8.3 (1.3)	&	{\tt 0 (R)}		\\
\hline\noalign{\smallskip} 
{\tt F04315-0840}	&{\tt NGC 1614}	&	04 33 59.85 	&  -08 34 44.0 & 	0.0159 	& 71.2	&	334	&  11.71 (9.74)	&  77.6 (0.8)	&	{\tt 2}	\\
\hline\noalign{\smallskip} 	
{\tt F06295-1735} & {\tt ESO 557-G002}	&	06 31 47.22  &	-17 37 17.3		&0.0208	&	93.4	&	435 	&  11.21 (9.86)	&	 24.5 (1.1)	& {\tt 0 (P)}		\\        
\hline\noalign{\smallskip} 	
{\tt F06592-6313 }	& --	&	06 59 40.25  &	-63 17 52.9	&	0.0224	& 100.7	&	467	&  11.26 (9.55)	&  27.5 (0.5)  &	{\tt 0 (P)}	\\
\hline\noalign{\smallskip} 	
{\tt F07160-6215}	&{\tt NGC 2369}		&07 16 37.73	& -62 20 37.4	& 0.0111	&	49.5	&	235	&  11.24 (9.24)   &	 26.3 (0.3)		&	{\tt 0 (P)}	&	(y) c\\
\hline\noalign{\smallskip} 	
{\tt  F10015-0614}	& {\tt NGC 3110} &  10 04 02.11  & -06  28  29.2 & 0.0163  &  73.0 & 343 &  11.27 (9.67)	&   28.2 (0.7) & {\tt 0 (P)} \\
\hline\noalign{\smallskip} 	
{\tt F10257-4339}	&	{\tt NGC 3256} & 10 27 51.27	&  -43 54 13.5  &0.0093 & 41.4	 &	197		&    11.69 (9.36) 		&	 74.1 (0.4)	&	{\tt 2}		\\
\hline\noalign{\smallskip} 	
{\tt F10409-4556} 	& {\tt ESO 264-G036}	&	10 43 07.67  & -46 12 44.6  & 0.0202	&  90.7 & 422 	&	 11.17 (10.29) 	&	 22.4 (3.0)   	&	{\tt 0 (R)}	\\
\hline\noalign{\smallskip} 	
{\tt F11255-4120} 	& 	{\tt ESO 319-G022} &	11 27 54.08  & -41 36 52.4	&   0.0161	&	72.1    &  338	&  11.00 (9.54) 	&   15.1 (0.5) 	&	{\tt 0 (P)}	\\     
\hline\noalign{\smallskip} 	
{\tt F11506-3851} 	& 	{\tt ESO 320-G030}	& 11 53 11.72 	&	-39 07 48.9  &  0.0102 &	45.5  &	216	&  11.27 (9.23)	&	 28.2 (0.3)		&	{\tt  0 (R)}\\
\hline\noalign{\smallskip} 	  
{\tt F12115-4546} 	& {\tt ESO 267-G030} &  12 14 12.88  & -47 13 42.3  & 0.0180 &   80.7  & 377  &  11.15 (9.98) 	&	 21.4 (1.5)		&	{\tt 0 (R)} 	& (y) d\\
\hline\noalign{\smallskip} 	
{\tt F12596-1529 (E+W)}		&	{\tt MCG-02-33-098}   & 13 02 20.37	& -15 45 59.7	&	0.0159	&	71.2	&	334	& E: 10.39 (8.91)	&	 3.7 (0.1) &	{\tt 1}	 & \\
&	  & 	& 	&	0.0156 &	69.8	&	328&  W: 10.84 (9.01)	&	 10.5 (0.2)		&	{\tt 1}	\\
\hline\noalign{\smallskip} 	
{\tt F13001-2339} 	& {\tt ESO 507-G070}	&	  13 02 52.35	& -23 55 17.7	& 0.0211 	&   94.8	&	441	& 11.54 (9.30)	   &	 52.5 (0.3)		&	{\tt 2}	\\
\hline\noalign{\smallskip} 	
{\tt F13229-2934}	&	{\tt NGC 5135} & 13 25 44.06 &-29 50 01.2 & 0.0136	& 60.8	& 287	&	 11.34 (9.46)	&  33.1 (0.4)	&	{\tt 0 (P)}	&	(y) a, b, c\\ 
\hline\noalign{\smallskip} 	
{\tt F14544-4255 (E+W)}	&	 {\tt IC 4518 (E+W)}	&	14 57 42.90	& -43 07 54.0 	&	0.0154	&	68.9	&	324	& E: 10.77 (9.44)	&  8.9 (0.4)	&	{\tt 1} & \\
 	&	 	&	 	&   &	 0.0160	&	 71.6	&	336 	&	W: 10.77 (9.47)  &	 8.9 (0.5) &	{\tt 1}	&	 (y) a, c, e\\
\hline\noalign{\smallskip} 	
{\tt F17138-1017} 	&	--	& 17 16 35.79	&	  -10 20 39.4	&	0.0172	&	77.1	&	361 & 11.45 (9.67)	&	 42.6 (0.7)    &     {\tt 2}		\\
\hline\noalign{\smallskip} 	
{\tt F18093-5744 N}		&	{\tt IC 4687}	       & 18 13 39.63 	&	-57 43 31.3 & 0.0163  & 73.0 & 343	&   11.26 (8.94)		&	 27.5 (0.13)		&	{\tt 0 (R)}	\\
\hline\noalign{\smallskip} 	
{\tt F18341-5732}		&	{\tt IC 4734} 	& 18 38 25.70  &  -57 29 25.6	& 0.0154	&	68.9	& 324	&	 11.37 (9.57)	&  35.5 (0.6)	&	{\tt 0 (R)}\\
\hline\noalign{\smallskip} 
{\tt F21453-3511}	&{\tt NGC 7130}	&	21 48 19.52 	&-34 57 04.5 	&	 0.0160 	&	71.6	&	 336 &  11.39 (9.65)	& 37.1 (0.7)		&    {\tt 2}	&		(y) a, b, c \\
\hline\noalign{\smallskip} 	
{\tt F22132-3705 }	&	{\tt IC 5179}	&	22 16 09.10  	&-36 50 37.4&	0.0112	&	49.9	& 	237 &  11.15 (9.41)	&	 21.4 (0.4) 	&	{\tt 0 (R)}	& \\
\hline\noalign{\smallskip} 	
{\tt F23007+0836}		& {\tt NGC 7469}	& 23 03 15.62	& +08 52 26.4	&0.0160	& 	71.6	&	336	&	11.59 (9.78)		&	58.9 (0.9)		&	{\tt 0 (R) }		& (y) b, c, e \\
\hline\hline\noalign{\smallskip} 	
\end{tabular}
\end{tiny}
\vskip2mm\hskip0mm\begin{minipage}{18.5cm}
\small
{{\bf Notes:} 
Columns:  (1) and (2): Object designation in the Infrared Astronomical Satellite (IRAS) Faint Source Catalog (FSC) and other identification; 
(3) and (4): Right ascension (hours, minutes and seconds) and declination (degrees, arcminutes and arcseconds) from the IRAS FSC;
(5): Redshift derived from the CO(2--1) emission line;
(6): Luminosity distance assuming a $\Lambda$DCM cosmology with H$_0$ = 67.8 km sec$^{-1}$ Mpc$^{-1}$, $\Omega_{M}$ = 0.308, $\Omega_{\Lambda}$ = 0.692, using the E. L. Wright Cosmology calculator, which is based on the prescription given by \cite{Wright06};
(7): Scale;
(8): Logarithmic infrared luminosity, L$_{IR}$, as derived in \cite{TDS17} assuming for the sample the luminosity distances, D$_L$, from \cite{Armus09}. We then scaled the L$_{IR}$ according to the $D_L$ used in this work. For individual galaxies in multiple systems, these are estimated dividing the total luminosity according to the relative fluxes in the MIPS images at 24 $\mu$m;
(9): SFR derived following \cite{Kennicutt12} relation with a Kroupa (\citealt{Kroupa01}) IMF (as in \citealt{Murphy11});
(10): `Composite' classification based on IRAC and {\it HST} morphological classifications complemented by the kinematic classification based on the ionized and molecular gas tracers (i.e., H$\alpha$ and CO). For further details see \S~\ref{sample_sect} and Appendix~\ref{app_classification}. {\tt 0 (R), 0 (P), 1} and {\tt 2} stand for isolated rotating disk, isolated perturbed disk, interacting and merger systems, respectively;
(11): (y) indicates whether the object shows evidence of an AGN. This information is taken from the references listed according to the following code: (a) \cite{Arribas14}, (b) \cite{Yuan10}, (c)~\cite{Pereira11}, (d)~\cite{JBailon07}, (e) \cite{AAH09}.
}
\end{minipage}
\end{table*}

\section{The sample}
\label{sample_sect}

\subsection{L$_{IR}$ range}

\begin{figure}
\centering
\includegraphics[width=0.45\textwidth,height=0.35\textwidth]{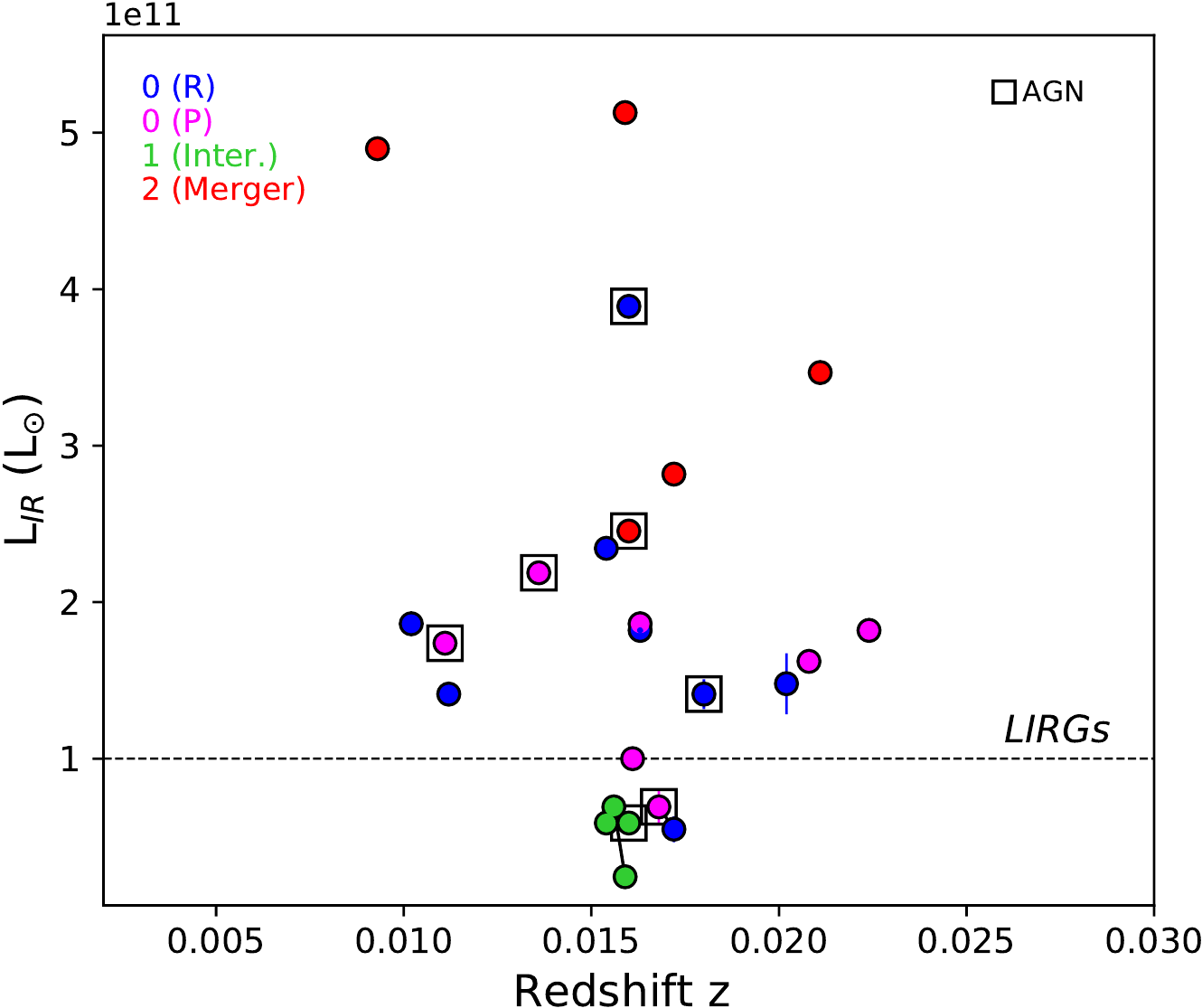}
\caption{Infrared luminosity {\it versus} redshift. The `composite' classification is also shown according to the following color-code: in blue are represented the rotating isolated disk galaxies {\tt (0 (R))}, in magenta the isolated perturbed disks {\tt (0 (P))}, in green the interacting systems and in red the post coalescence mergers. Galaxies which belong to the same system (ESO 297-G011/G012, F12596 E/W and IC 4518 E/W) are identified using a solid line. Galaxies containing an AGN are identified by a square.}
\label{z_LIR_sample}
\end{figure}

The volume-limited sample consists of 21 local LIRGs (24 individual galaxies) at z$\lesssim$0.02 for which we obtained ALMA data. The sample has been drawn from the IRAS Revised Bright Galaxy Sample (RBGS, \citealt{Sanders03}) with distance D$\lesssim$100~Mpc (a mean distance for the whole sample is D$\sim$75 Mpc, ranging from 41 Mpc to 101 Mpc). Our LIRGs have been previously observed in the optical band using VIMOS/VLT by \cite{Arribas08} and, 10 of them have been also analyzed in the near-IR using SINFONI/VLT data by \cite{Crespo21}. 

We derived the total infrared luminosity from \cite{TDS17}, which assumed the luminosity distance, D$_L$, considered in \cite{Armus09}. In particular, for each object they scaled the integrated IRAS IR and FIR (derived in the wavelength range 42.5-122.5 $\mu$m) luminosities with the ratio between the continuum flux density evaluated at 63 $\mu$m in the PACS spectrum (measured in the same aperture as the line) and the total IRAS 60 $\mu$m flux density. The normalization at 63 $\mu$m ensures that the emission can be associated to the dust continuum emission closely related to the star-formation. Finally, we scaled these values according to the D$_L$ used in this work. The derived L$_{IR}$ cover the range between 10$^{10.4}$-10$^{11.7}$ L$_\odot$ (see Fig.~\ref{z_LIR_sample}) with a uniform distribution. Therefore, it can be considered representative of the general properties of local LIRGs.
For individual galaxies in multiple systems (i.e., ESO 297-G011/-G012 (N/S), IC 4518 E/W and MCG-02-33-098 E/W), the individual contribution to the L$_{IR}$ of the system has been derived taking into account the relative fluxes of the individual galaxies in the MIPS images at 24 $\mu$m (see Tab.~\ref{Input_data}).

\subsection{Activity, morphology and dynamical phase classification}
\label{AGN_activity}

According to the nuclear optical spectroscopic classification (see \citealt{RZ11}) most of the sources are classified as HII galaxies, excluding NGC~7130, classified as LINER/Sy2, and IC~4518~W and NGC~5135, classified as Sy2 and NGC~7469 classified as Sy1, along with ESO~297-G011 which shows evidence of an AGN from the optical spectra (\citealt{Arribas14}).
Following complementary information obtained in the X-rays bands, 2 additional sources (2/24; NGC~2369 and ESO~267-G030) show evidence of an AGN. 
Mid-IR data are a good tool to search for obscured AGN. \cite{AAH12} searched for highly obscured AGN using mid-IR {\it Spitzer} data. Several sources are in common with those studied in this work: except for IC 4518~W, already classified as Sy2, we found that all these systems show a small AGN contribution at 24 $\mu$m (i.e., $\lesssim$8\%), confirming that the AGN contribution does not dominate the galaxy emission in our systems. Thus, we ended up with 7/24 galaxies in our sample which suggest the presence of an AGN (see Tab.~\ref{Input_data}).

The sample encompasses a wide variety of morphological types, suggesting different dynamical phases (isolated Spirals, interacting galaxies, and ongoing and post mergers). The majority of our LIRGs are isolated Spirals. 
In this work, the sources have been classified taking into account both the morphological information from {\it Spitzer}/IRAC and {\it Hubble Space Telescope (HST)} images and the kinematic information obtained from the ionized (H$\alpha$) and molecular (CO) gas phases. 
In the Appendix~\ref{app_classification} we give a detailed description of such classification (see also Tab.~\ref{Input_more_classifs}). Here, in brief, we highlight the characteristics of the final `composite' classification used in this work.
In particular, we defined four different classes as follows:

\begin{itemize}
\item {\tt 0 (R)}: single isolated objects, with relatively symmetric disk morphologies which show the typical kinematic maps of a {\tt rotating disk} ({\tt RD});
\vskip1mm
\item {\tt 0 (P)}: single isolated objects, with relatively symmetric disk morphologies but showing a somehow perturbed kinematics, hereafter, {\tt perturbed disk (PD)};
\vskip1mm
\item {\tt 1}: objects in a pre-coalescence phase with two well differentiated nuclei separated by a projected distance of at least 1.5 kpc up to a maximum distance of 15 kpc, showing a perturbed kinematics (hereafter, {\tt interacting});
\vskip1mm
\item {\tt 2}: objects with two nuclei separated by a projected distance $\leq$1.5 kpc or a single nucleus with a relatively asymmetric morphology, with perturbed or complex kinematics (hereafter, {\tt merger}).
\end{itemize}

According to our classification, the majority of our galaxies (2/3) show the presence of interaction or past merger activity in their morphology and/or kinematics (i.e., 7/24 are type {\tt 0 (P)}, 4/24 are type {\tt 1} and 5/24 are type {\tt 2}), while the remaining objects (1/3) are disky ({\tt 0 (R)}; see Tab.~\ref{Input_data} and App.~\ref{app_classification}).
In some specific cases the properties of individual galaxies in multiple systems could be inferred separately and were therefore treated individually.

\subsection{SFR -- M$_{star}$: location in the {\it Main Sequence} plane}

To estimate the stellar mass M$_{star}$ of our sample we used the integrated K-band near-IR magnitude obtained from 2MASS All-Sky Extended Source Catalog (XSC; \citealt{Jarrett00}). The stellar mass estimation in this band is considered a good tracer to the total stellar mass, since the bulk of the luminosity of a SED of simple stellar population (SSP) older than 10$^9$ yr is observed in the wavelength range 0.4 and 2.5 $\mu$m. 
As estimated in previous works (e.g., \citealt{AAH01, AAH06, PL12, Pereira15}), the visual extinction, A$_V$, in LIRG systems shows typical values of A$_V$~$<$~4 mag, being even smaller in the near-IR bands (e.g., A$_{Pa\alpha}$~$\sim$~0.4-1.0 mag). Besides this, the AGN contribution in our systems is negligible (see Sect.~\ref{AGN_activity}), and in the K-band it would only affect the nuclear K-band emission.
Furthermore, in the near-IR the contribution from young stars is usually negligible and the scatter in the mass-to-light (M/L) ratio for local LIRGs is relatively small ($\sim$0.4 dex; \citealt{Pereira11}). We then converted the magnitude to luminosity in the K band, L$_K$, and assuming a (mean) M$_{star}$/L$_K$ ratio of 0.4 as derived in \cite{Zibetti09}, we derived the stellar mass in the K band, M$_{star}$\footnote{The uncertainties associated to the stellar mass derivation are obtained taking into account the magnitude uncertainties derived in the K band from the 2MASS Extended Source Catalogue (XSC) along with the 0.4 dex uncertainty associated to the M/L ratio.}.

We derived the star-formation rate (SFR) from the L$_{IR}$ (see Tab.~\ref{Input_data}). 
This parameter can be considered a good tracer of star formation for all our systems because the AGN contribution in our sample to the total L$_{IR}$ is small ($\sim$5\% on average; see also \citealt{AAH12}). Such contribution is the result of the reprocessed emission originating in star-formation regions hidden by the large amount of dust. L$_{IR}$ has been then converted to SFR following \cite{Kennicutt12} relation for a Kroupa (\citealt{Kroupa01}) IMF (see \citealt{Murphy11}). 

\begin{figure*}
\centering
\includegraphics[width=0.9\textwidth, angle=180]{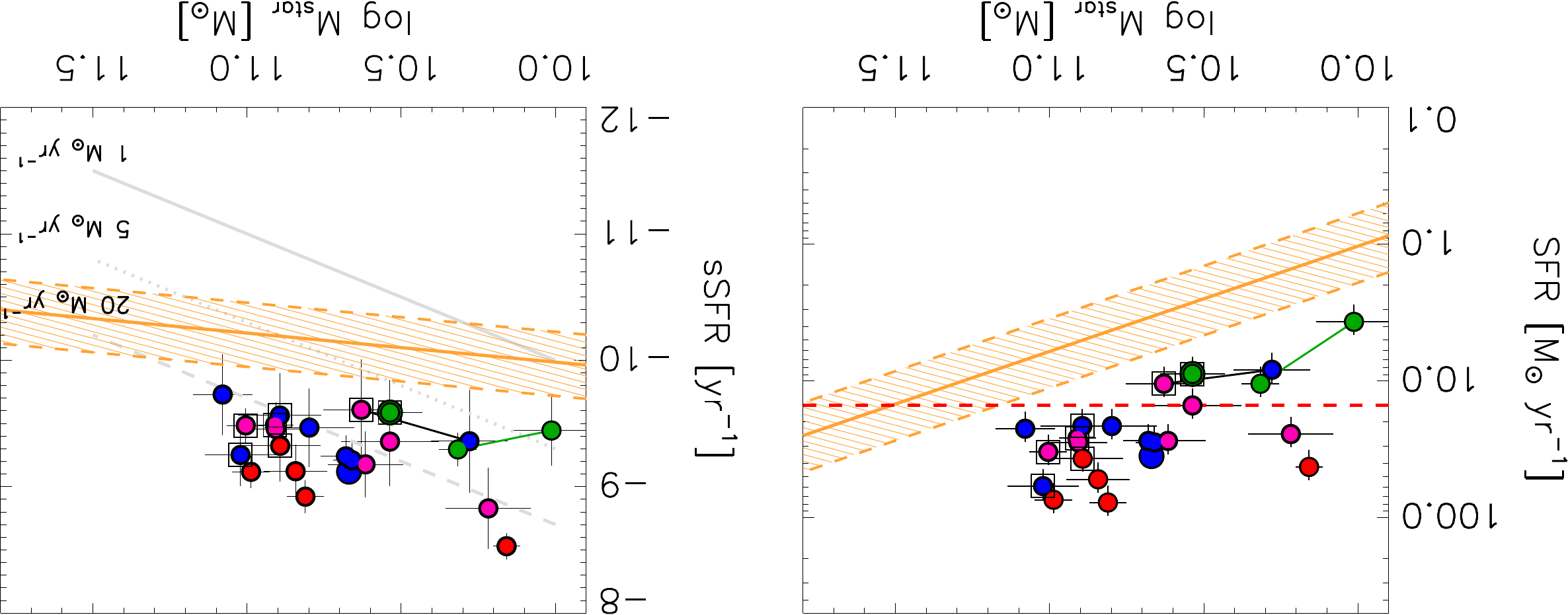}
\caption{{\it Left:} LIRG sample plotted in the SFR -- M$_{star}$ plane. 
The solid and dashed (orange) lines indicate the location of the local {\it MS} relation and the 1$\sigma$ scatter, respectively, obtained by \cite{Elbaz07} using SDSS galaxies at z$\sim$0 and converted to Kroupa IMF, in agreement with that derived by \cite{Whitaker12}.
Our LIRGs are shown according to the color code defined in Fig.~\ref{z_LIR_sample}: isolated disks ({\tt RD}), ({\tt PD}), interacting and merging systems are shown in blue, magenta, green and red, respectively. The solid lines between the points (green and black) link galaxies which belong to the same system. Galaxies containing an AGN are identified with an empty black square. The horizontal red dashed line represents the threshold SFR to reach the LIRG IR luminosity (log L$_{IR}$/L$_\odot$~$\geq$~11). 
{\it Right:} specific SFR as a function of the stellar mass M$_{star}$. The three (solid, dotted, and dashed) gray lines identify the different SFR regimes (i.e., 1, 5 and 20 M$_\odot$ yr$^{-1}$, respectively). The {\it MS} relation is also shown (in orange).
}
\label{MS_rels}
\end{figure*}

In Fig.~\ref{MS_rels} (left panel) the results obtained for the SFR {\it versus} stellar mass M$_{star}$ are shown for our sample.
The {\it MS} relation defined by \cite{Elbaz07} for SDSS galaxies at z$\sim$0 is shown with its uncertainties:

\begin{equation}
SFR_{Salp} \hskip1mm [M_\odot \hskip2mm yr^{-1}] = 8.7^{+7.4}_{-3.7} \times \frac{M_{star} [M_\odot]}{10^{11}}^{0.77}. 
\end{equation}

This relation used the Salpeter IMF, which has been converted to Kroupa according to the formula: SFR$_{Kroupa}$$\sim$0.7$\times$SFR$_{Salpeter}$ (see \citealt{Elbaz07}, \citealt{Madau14}). 
We derived the same {\it MS} relation when using the power law defined by \cite{Whitaker12} at z$\sim$0 using a Chabrier IMF. 
Our LIRGs lie a factor of 8 above the {\it MS} defined by \cite{Elbaz07}, and no clear trend is found among the different morphological classes in the {\it MS} plane (Tab.~\ref{MS_values_}).

Our LIRGs cover the stellar mass range between 10$^{10.0}$ and 10$^{11.1}$ M$_\odot$, where most of them (19/24) are in the range 10$^{10.5}$ and 10$^{11}$ M$_\odot$. 
All but one (i.e., F17138-1017) mergers lie in a smaller stellar mass range log M$_{star}$ between 10.8-11.0 M$_\odot$. The isolated galaxies (type {\tt 0}) cover a larger range of values, from log M$_{star}$~$\lesssim$ 10.2 to 11.1 M$_\odot$. The interacting (type {\tt 1}) systems show quite small values as a result of the disentanglement of the M$_{star}$ and SFR contributions of the individual galaxies.
Although the small number of AGN objects, our results suggest that LIRGs without an AGN show lower stellar mass ($\sim$40\%) with respect to those obtained for LIRGs with an AGN, although with similar SFR. 
If we distinguish among the different types of galaxies, we see that isolated disks (type {\tt 0}) and mergers (type {\tt 2}) share similar stellar masses but mergers are characterized by twice the SFR typical of isolated disks.

We then computed the specific SFR (sSFR = SFR/M$_{star}$) for our sample (right panel Fig.~\ref{MS_rels}), finding two extreme cases for which sSFR~$>$~1~Gyr$^{-1}$ with low stellar masses (log M$_{star}$ $\sim$10.2; ESO 557-G002 and F17138-1017).
According to the mean (and median) values, we found that LIRGs with an AGN show sSFR a factor of 2 lower than that derived when LIRGs without an AGN are considered, as a result of their larger (a factor of 2) M$_{star}$. Similar values are obtained for the type {\tt 0} and {\tt 1} galaxies while type {\tt 2} objects show the highest sSFR, although with larger scatter. 
The typical stellar masses and SFRs derived for our sample seem to be consistent the {\tt starburst} scenario (i.e., strong burst of recent, $<$100 Myr old, star-formation). The most extreme starbursts are those classified as mergers, followed by less extreme isolated disks.

\begin{table*}
\centering
\begin{small}
\caption{Mean (and median) stellar mass M$_{star}$, SFR and sSFR for the different LIRG subsamples.}
\label{MS_values_}
\begin{tabular}{ccccc} 
\hline\hline\noalign{\smallskip}  
Sample 	 &		\# &		M$_{star}$ &    	SFR		&	sSFR\\
	 	 &		 &		(10$^{10}$ M$_\odot$)	&    	(M$_\odot$ yr$^{-1}$)		&	 (Gyr$^{-1}$)	\\
(1)	&	(2)		&	(3)		&	(4)		&	(5)	\\
\hline\noalign{\smallskip}  
All LIRGs 			& 24 	& 5.6~$\pm$~3.2 (4.8) &  29.4~$\pm$~19.9 (26.9) 	& 0.63~$\pm$~0.58 (0.46)	\\
\hline\noalign{\smallskip}  
LIRGs w/o AGN  	& 17 	& 4.9~$\pm$~3.1 (4.6) &  29.9~$\pm$~21.4 (27.5) 	&	0.74~$\pm$~0.66 (0.57)	\\
LIRGs w AGN  	&   7   	& 7.4~$\pm$~2.7 (7.8) & 28.0~$\pm$~17.2 (26.3) 	&	0.35~$\pm$~0.12 (0.32) \\
\hline\noalign{\smallskip}  
 {\tt 0 RD} 	&	 8	& 	6.6~$\pm$~3.4 (5.5) 	&  28.0~$\pm$~14.7 (25.0)	&	0.47~$\pm$~0.19 (0.50)  \\
 {\tt 0 PD} 	&	 7	& 	5.7~$\pm$~3.1 (4.3) 	&  23.6~$\pm$~7.9 (26.3)	&	0.55~$\pm$~0.43 (0.35)	\\
  {\tt 1} 	&	4	&	2.5~$\pm$~1.2 (2.8)	&  8.0~$\pm$~3.0 (8.9)  	&     0.35~$\pm$~0.12 (0.31)  \\
  {\tt 2} 	&	5	&	6.5~$\pm$~3.1 (7.0) 	&  56.8~$\pm$~18.3 (52.5)	&	1.23~$\pm$~1.00 (0.76) \\
\hline\hline\noalign{\smallskip} 	
\end{tabular}
\end{small}
\vskip2mm\begin{minipage}{18cm}
\small
{{\bf Notes:} Columns: (1): Sample and subsample: class {\tt 0} identifies isolated disks, class {\tt 1} denotes interacting systems, and class {\tt 2} stands for mergers (see \S~\ref{sample_sect}). Subsamples of LIRGs with (w AGN) and without (w/o AGN) AGN are shown. 
(2): Number of objects in each sample;
(3): Stellar mass in units of 10$^{10}$ M$_\odot$;
(4): SFR in units of M$_\odot$~yr$^{-1}$; 
(5): Specific SFR (sSFR=SFR/M$_\star$) in units of Gyr$^{-1}$. 
}
\end{minipage}
\end{table*}

\section{Observations and data reduction}
\label{obs_data}

\subsection{ALMA data}

We obtained the CO(2--1) line and the continuum at 247~GHz ($\sim$1.3 mm) emissions of a sample of 18 individual local LIRGs observed with ALMA between August 2014 and August 2018, using the Band 6 from the project 2017.1.00255.S (PI: Miguel Pereira Santaella). This project was completed with 6 more galaxies, that belong to four different projects (see Tab.~\ref{BEAMS}).
The data for two of the LIRGs, IC4687 and ESO320-G030, were previously presented in \cite{PereiraS16}, \cite{Pereira16} and \cite{Pereira20}.

\begin{figure}
\centering
\includegraphics[width=0.44\textwidth]{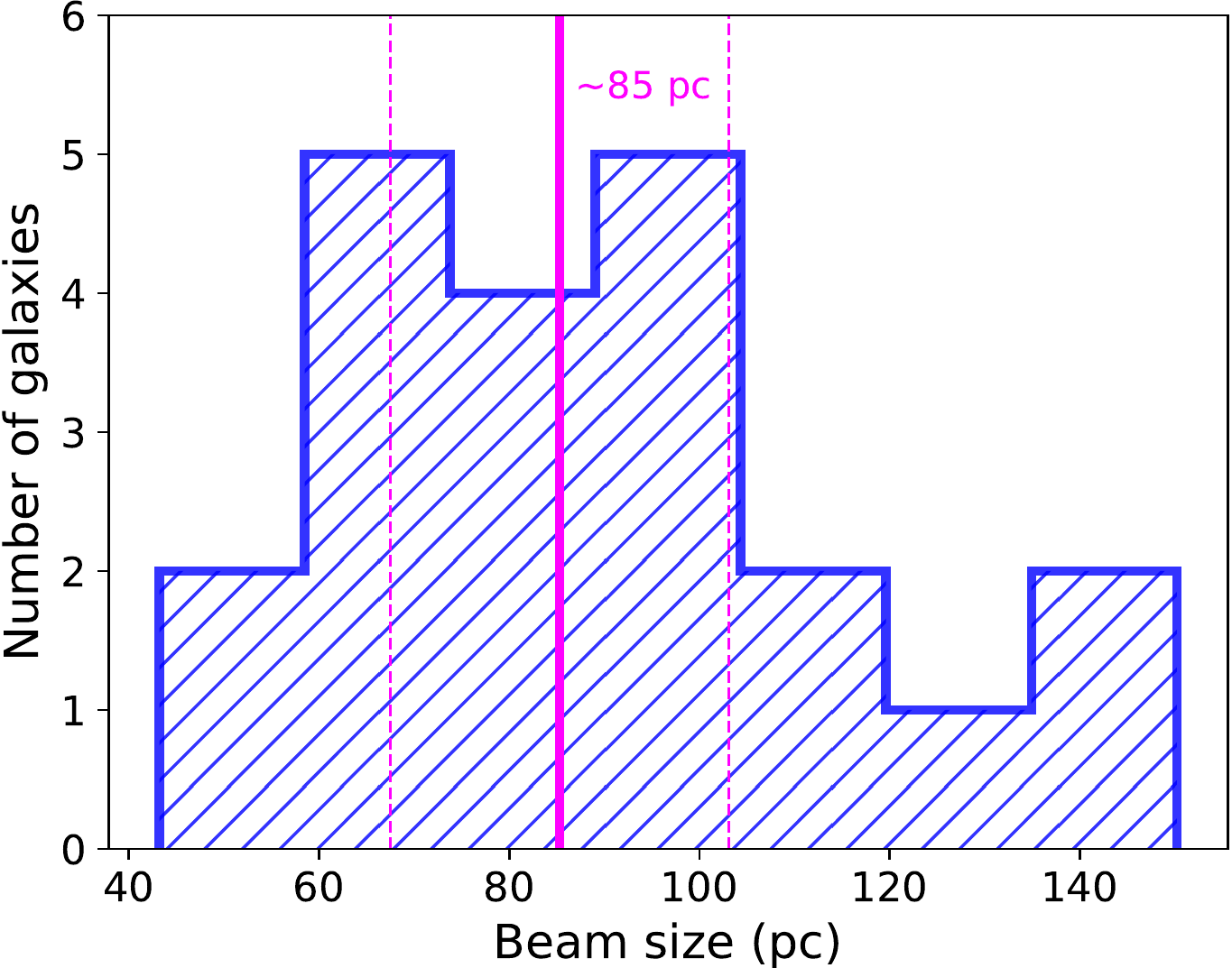}
\caption{Histogram showing the distribution in physical scales (parsecs) of the size of the beam for the galaxies of the sample. The median and MAD values (solid and dashed magenta lines, respectively) correspond to 85~$\pm$~18~pc.}
\label{LIRG_beams}
\end{figure}

The total integration time per source was $\sim$20-30~min in standard mode. The synthesized beam full width at half maximum (FWHM) of the sample ranges between $\sim$0.2\arcsec-0.4\arcsec: this corresponds to a median size of 85$\pm$18\footnote{Throughout the paper, the uncertainty associated to a median value are computed as the {\it median absolute deviation}, MAD. This uncertainty returns the median absolute deviation of a data set from the median, i.e., median(|data - median(data)|). It is a proxy for the standard deviation, but is more resistant against outliers.} pc at the distance of our LIRGs (see Fig.~\ref{LIRG_beams}). 
A combination of the extended and compact configurations was used to achieve the 0.2\arcsec\ angular resolution and a maximum recoverable scale of 10\arcsec-12\arcsec\ (3-5 kpc).
The field of view (FoV) imaged by a single pointing has a diameter ranging between $\sim$5-8 kpc, up to $\sim$10-17 kpc for the three mosaics (NGC 3256, NGC 7469 and MCG-02-33-098). 
Further details of the observations are listed in Tab.~\ref{BEAMS} for each source. 
Two spectral windows of 1.875 GHz (1.9 MHz $\sim$ 2.6 km s$^{-1}$ channel) were centered at the sky frequency $^{12}$CO(2-1) transition and 247 GHz continuum spectral window.

The data were calibrated using the standard ALMA reduction software Common Astronomy Software Applications ({\tt CASA}\footnote{\url{https://casa.nrao.edu/}} v5.1 \citealt{McMullin07}). In the CO(2--1) spectral window, the continuum emission was estimated using the line free channels and then this contribution was subtracted in the {\it uv}-plane. For the cleaning of CO(2--1) and continuum data, we used the natural weighting of the {\it uv}-plane, obtaining a spatial resolution in the range 50-150 pc (see Fig.~\ref{LIRG_beams} and Tab.~\ref{BEAMS}). The final CO(2--1) data cubes have channels of 4-23.4 MHz ($\sim$5-30 km~s$^{-1}$) and they were corrected for the primary beam. The pixel size is in the range 0.025\arcsec-0.06\arcsec. 
For the CO(2--1) data cubes the 1$\sigma$ sensitivity is $\sim$0.4-1.2~mJy beam$^{-1}$ per 10 km s$^{-1}$ channel while for the continuum is $\sim$0.02-0.1 mJy beam$^{-1}$ (see Tab.~\ref{BEAMS}).

\begin{table*}
\centering
\begin{tiny}
\caption{Beam sizes of the CO(2-1) and 1.3 mm continuum images for the whole sample.}
\label{BEAMS}
\begin{tabular}{cccc ccccc} 
\hline\hline\noalign{\smallskip}  
\multicolumn{2}{c}{Source}	    &  	\multicolumn{2}{c}{1.3 mm continuum}	&			 \multicolumn{2}{c}{CO(2--1)}			 	&$\theta^{max}_{circ}$, {\tt PA} 	&	Sensitivity &	Project \\ 
\cmidrule(lr){1-2}\cmidrule(lr){3-4}\cmidrule(lr){5-6}
 IRAS & Other 		&	$\theta_{maj}$ $\times$ $\theta_{min}$ 	& {\tt $\theta_{circ}$}	&	$\theta_{maj} $ $\times$ $\theta_{min}$  &  {\tt $\theta_{circ}$}		&	[pc, deg]	&	[$\mu$Jy beam$^{-1}$]		&  code	\\
 \noalign{\smallskip}  
(1) & (2) &  (3)& (4) &(5)  & (6) &(7)		&	(8)	&		(9)\\
\hline\noalign{\smallskip} 	
{\tt F01341-3735 N}	 &	{\tt ESO 297-G011}	& 	0.26\arcsec	 $\times$ 0.21\arcsec   &	  0.23\arcsec	&  0.27\arcsec $\times$ 0.21\arcsec	&     {\bf 0.24\arcsec}  &	 84, -76$^\circ$	 & 536/24	& {\tt MPS} \\
{\tt F01341-3735 S}	 & 	{\tt ESO 297-G012}	&	0.26\arcsec  $\times$ 0.21\arcsec  &     0.23\arcsec	& 0.27\arcsec  $\times$ 0.21\arcsec	&	{\bf 0.24\arcsec} &	&	567/24& \\
\hline\noalign{\smallskip} 	
{\tt F04315-0840} &	{\tt NGC 1614}	&	0.28\arcsec $\times$ 0.19\arcsec	& 0.23\arcsec	&  0.29\arcsec $\times$ 0.19\arcsec	&	{\bf 0.24\arcsec}& 	79, -67$^\circ$  	&		887/23& {\tt MPS}\\
\hline\noalign{\smallskip} 	
{\tt F06295-1735} &   {\tt ESO 557-G002} & 0.17\arcsec	$\times$ 0.12\arcsec	&     {\bf 0.14\arcsec}	&    0.16\arcsec $\times$ 0.12\arcsec	&     {\bf 0.14\arcsec} & 64, -82$^\circ$  	&	488/18	& {\tt MPS}	\\
\hline\noalign{\smallskip} 	
{\tt F06592-6313 }& 	-- & 	0.23\arcsec	$\times$ 0.21\arcsec		&{\bf 0.22\arcsec} 	& 0.22\arcsec $\times$ 0.20\arcsec	&	0.21\arcsec	& 107, -45$^\circ$ 	&	536/24	& {\tt MPS}	\\
\hline\noalign{\smallskip} 	
{\tt F07160-6215} &	{\tt NGC 2369} 	&0.31\arcsec $\times$ 0.27\arcsec	& {\bf 0.29\arcsec} & 0.31\arcsec	 $\times$ 0.26\arcsec	&     0.28\arcsec& 	69, 88$^\circ$  	& 1195/36	& {\tt MPS}\\
\hline\noalign{\smallskip} 	
{\tt F10015-0614} & 	{\tt NGC 3110} &  0.35\arcsec $\times$ 0.30\arcsec 	& {\bf 0.33\arcsec} & 0.33\arcsec $\times$ 0.28\arcsec 	&	0.30\arcsec& 125, -79$^\circ$   & 718/37	&{\tt MPS}\\
\hline\noalign{\smallskip} 	
{\tt F10257-4339} &	{\tt NGC 3256} &    0.23\arcsec $\times$ 0.20\arcsec	&	0.21\arcsec	&	0.23\arcsec $\times$ 0.21\arcsec& 	{\bf 0.22\arcsec}	&  \textcolor{blue}{48}, 57$^\circ$	&	1539/26	 & {\tt SK} \\
\hline\noalign{\smallskip} 	
{\tt F10409-4556} & {\tt ESO 264-G036}	& 0.20\arcsec $\times$ 0.18\arcsec 	& {\bf 0.19\arcsec} & 0.19\arcsec $\times$ 0.16\arcsec	&	0.18\arcsec  & 90, -81$^\circ$ 	&	413/13	& {\tt MPS}\\
\hline\noalign{\smallskip} 	
{\tt F11255-4120} & 	{\tt ESO 319-G022} &  0.17\arcsec $\times$ 0.14\arcsec	& {\bf 0.15\arcsec}	& 0.17\arcsec $\times$	0.14\arcsec 	&	{\bf 0.15\arcsec}&  55, +90$^\circ$  &	520/21	& {\tt MPS}	 \\
\hline\noalign{\smallskip} 	
{\tt F11506-3851}     & 	{\tt ESO 320-G030}	&	0.28\arcsec	 $\times$ 0.24\arcsec	&	0.26\arcsec		& 0.30\arcsec $\times$ 0.24\arcsec&	{\bf 0.27\arcsec}	&	68, 63$^\circ$ &	979/115	& {\tt LCa}	 \\
\hline\noalign{\smallskip} 	
{\tt F12115-4546} & {\tt ESO 267-G030} &  0.18\arcsec	$\times$ 0.14\arcsec 	& {\bf 0.16\arcsec} & 0.18\arcsec	 $\times$ 0.14\arcsec	&	{\bf 0.16\arcsec} & 65, -78$^\circ$	 	&	413/25	& {\tt MPS}\\
\hline\noalign{\smallskip} 	
{\tt F12596-1529}	&	{\tt MCG-02-33-098}   &     0.30\arcsec $\times$ 0.24\arcsec	&   {\bf 0.27\arcsec}	& 0.28\arcsec $\times$ 0.23\arcsec	&     0.25\arcsec     & 96, 86$^\circ$  	&	584/28	& {\tt MPS}\\
\hline\noalign{\smallskip} 	
{\tt F13001-2339} & {\tt ESO507-G070}	& 	0.20\arcsec $\times$ 0.17\arcsec	&     {\bf 0.18\arcsec} 	&     0.19\arcsec	 $\times$ 0.16\arcsec	&     0.17\arcsec & 85, -72$^\circ$ & 	655/27	& {\tt MPS}	  \\
\hline\noalign{\smallskip} 	
{\tt F13229-2934}     &	{\tt NGC 5135} &	0.35\arcsec	 $\times$ 0.28\arcsec	&	0.31\arcsec	&	0.38\arcsec $\times$ 0.30\arcsec	&	{\bf 0.34\arcsec}&  	105, 67$^\circ$	& 782/25	&	{\tt LCb} \\
\hline\noalign{\smallskip} 	
{\tt F14544-4255 E}  &	 {\tt IC 4518}	&	0.31\arcsec	 $\times$ 0.27\arcsec	& {\bf 0.29\arcsec}	& 0.30\arcsec $\times$ 0.26\arcsec	&  0.28\arcsec & 100, -89$^\circ$ 	& 507/36& {\tt MPS}\\
{\tt F14544-4255 W}  &	 {\tt IC 4518}	&	0.31\arcsec	 $\times$ 0.27\arcsec	& {\bf 0.29\arcsec}	& 0.30\arcsec $\times$ 0.26\arcsec	&  0.28\arcsec 	&& 704/26	& \\   
\hline\noalign{\smallskip} 	
{\tt F17138-1017} & 	--	&	0.27\arcsec	 $\times$ 0.25\arcsec	& 0.26\arcsec	& 0.28\arcsec $\times$ 0.26\arcsec	&	{\bf 0.27\arcsec} & 98, -66$^\circ$ 	&  852/17	& {\tt MPS}\\
\hline\noalign{\smallskip} 	
{\tt F18093-5744}    &	{\tt IC 4687}	&  	 0.38\arcsec $\times$ 0.29\arcsec	 	&	0.33\arcsec		& 0.44\arcsec $\times$	0.37\arcsec	 &	{\bf 0.40\arcsec} &	\textcolor{blue}{145}, -39$^\circ$ 	& 1086/48	 &{\tt LCa}\\
\hline\noalign{\smallskip} 	
{\tt F18341-5732}     &	{\tt IC 4734}	& 		0.25\arcsec $\times$ 0.20\arcsec	& 0.22\arcsec	& 0.26\arcsec $\times$ 0.21\arcsec 	&	{\bf 0.23\arcsec} & 75, -73$^\circ$	&	960/38	&	{\tt TDS}  \\
\hline\noalign{\smallskip} 	
{\tt F21453-3511} &{\tt NGC 7130}	&	0.44\arcsec $\times$ 0.36\arcsec	& 0.40\arcsec & 0.46\arcsec	 $\times$ 0.38\arcsec	&     {\bf 0.42\arcsec}	&  135, 72$^\circ$ &   	437/27	& {\tt MPS}\\
\hline\noalign{\smallskip} 	
{\tt F22132-3705 } &	{\tt IC 5179}	&	0.48\arcsec	 $\times$ 0.41\arcsec	& 0.44\arcsec 	& 0.49\arcsec $\times$ 0.42\arcsec	&	{\bf 0.46\arcsec} & 103, 52$^\circ$     & 	730/29	 & {\tt MPS}\\
\hline\noalign{\smallskip} 	
{\tt F23007+0836}   & {\tt NGC 7469}	& 0.23\arcsec $\times$ 0.17\arcsec	& {\bf 0.20\arcsec}	& 0.23\arcsec $\times$ 0.18\arcsec&	0.20\arcsec	& 63, -39$^\circ$  &  	546/17	&{\tt TDS}\\
\hline\hline\noalign{\smallskip} 	
\end{tabular}
\end{tiny}
\begin{minipage}{18cm}
\small
{{\bf Notes:} Columns: (1) and (2): IRAS Source name and alternative name;
(3) and (4): Major and minor FWHM ($\theta_{maj}$ and $\theta_{min}$) and circularized beam sizes of the 1.3 mm continuum map. The circularized beam, $\theta_{circ}$, has been derived as $\theta_{circularized}$ = $\sqrt{\theta_{min}\theta_{max}}$; 
(5) and (6): Major and minor FWHM and circularized beam sizes of the CO(2--1) map. For each source we used the maximum circularized beam derived for the CO(2--1) and continuum images, highlighted in boldface. 
(7): Largest circularized FWHM beam in parsec and its position angle {\tt (PA)} in degrees. The beams have been synthesized using the natural weighting. In blue are highlighted the smallest (NGC 3256) and largest (IC 4687) values;
(8): 1$\sigma$ line/continuum sensitivity of the CO(2--1) and 1.3 mm observations, respectively. We use the 7.8 MHz ($\sim$10 km s$^{-1}$) channels of the final data cube, with the exception of the galaxies ESO 320-G030 and NGC 5135 for which 5 km s$^{-1}$ and 30 km s$^{-1}$ channel are used; 
(9): Project code includes information on the principal investigator. The acronyms are listed using the following code: {\tt MPS}: Miguel Pereira-Santaella 2017.1.00255.S, {\tt SK}: Sliwa Kazimers 2015.1.00714, {\tt LCa}: Luis Colina 2013.1.00271.S, {\tt LCb}: Luis Colina 2013.1.00243.S, {\tt TDS}: Tanio D\'iaz-Santos 2017.1.00395.S.
}
\end{minipage}
\end{table*}

\subsection{Ancillary data}
\label{ancillary}

In order to compare the molecular size derived using ALMA data with the ionized and stellar emission in our LIRGs, we considered the results obtained in \cite{Arribas12} and \cite{Bellocchi13}.
In \cite{Arribas12} the authors derived the H$\alpha$ size of a local sample of (U)LIRGs, which includes all (but two\footnote{The galaxies NGC 7469 and IC 4734 are not included in their analysis.}) LIRGs of our sample. The analysis was based on integral field spectroscopic VIMOS/VLT data, which cover a FoV of $\sim$30\arcsec$\times$30\arcsec, at the angular resolution of $\sim$1.3\arcsec. The typical R$_{eff}$ derived for the ionized gas phase are shown in Tab.~2 in their work. The sizes obtained are intrinsic (i.e., deconvolved) sizes, that is, obtained by subtracting the contribution of the PSF in quadrature.
They computed the effective radii using the curve-of-growth {\tt (CoG)} and {\tt `A/2'} methods (see \S~\ref{sect_method} for further details). For extended objects, the limited VIMOS FoV only allowed a lower limit to the R$_{eff}$ estimation to be derived.

The bulk of the galaxy stellar component is well traced to first order by the rest frame near-IR continuum light. For the derivation of the effective radius of the stellar component we considered the 2MASS data in the K-band from \cite{Bellocchi13}.
2MASS images do not have any limitation by the FoV although they are characterized by a low angular resolution ($\sim$2\arcsec).
In their work the R$_{eff}$ have been derived using {\tt GALFIT} and the {\tt `A/2'} methods (see Tab.~B.1 in their work): these methods will be discussed in more detail in \S~\ref{sect_method}. 

Furthermore, we considered archival near-IR HST Pa$\alpha$ images\footnote{\url{https://hla.stsci.edu/hlaview.html}} to study how the extinction affects the derivation of the effective radius of the ionized component. Pa$\alpha$ images were obtained using the NICMOS2 camera on board the {\it HST} in combination with the narrow band F187N and F190N filters: the FoV of these images is $\sim$19.5\arcsec$\times$19.5\arcsec\ with an angular resolution FWHM$\sim$0.15\arcsec. Most of these observations were previously published by \cite{AAH01}, \cite{AAH02} and \cite{AAH06}.

\section{Data analysis}

\subsection{Molecular gas and 1.3 mm continuum distributions}

We generated CO(2--1) and 1.3 mm continuum maps for the whole sample, selecting the emission above 5$\sigma$ for both maps. These are all shown the Appendix~\ref{app_maps}. 
The galaxies of our sample show several morphologies: regular (ESO 320-G030, IC 5179), elongated (NGC 2369, IC 4518 E), and irregular objects (MCG 02-33-098 E-W, NGC 3256). Some of them are more compact (F06592-6313, IC 4734) while others more extended (IC 4687, IC~5179, NGC~3110). The 1.3 mm continuum emission is generally quite compact, and in some cases clumpier than CO(2--1). This result is discussed below.

\subsection{Effective radius determination and systematic effects}

\subsubsection{Selection of the methodology}
\label{sect_method}

The half-light (or effective) radius (R$_{eff}$) is defined as the radius which encloses half the total flux emission of the galaxy; it measures the light concentration and depends on the shape of the light profile (\citealt{Trujillo20}) as well as on the wavelength considered (e.g., \citealt{Graham08, Kennedy15}) and depth of the data.

To derive the size of a system, different methodologies can be applied: {\tt GALFIT}, the curve-of-growth {\tt (CoG)} and the so-called {\tt `A/2'} methods can be used according to the objects considered. {\tt GALFIT} is based on fitting the observed flux distribution to a galaxy model assuming standard surface brightness profiles (\citealt{Peng02, Peng10}). 
This method is accurate as long as the model is a good representation of the actual galaxy flux distribution. However, if the galaxies show irregular or clumpy emission (as in the case of interacting or merger systems) the half-light radii can be obtained from the {\tt CoG} of the flux at increasingly large apertures. 
The {\tt `A/2'} method computes the effective (circular) radius as R$_{eff}$~=~$\sqrt{(A/2)/\pi}$, where A/2 is the angular extent of the minimum number of pixels (or spaxels) encompassing 50\% of the total galaxy's light, A. This method is quite used for high-z systems (e.g., \citealt{Erb04}). 
The {\tt `A/2'} method does not require any knowledge of the galaxy center: indeed, it is not sensitive to the distribution of large and bright regions found within a galaxy but it is sensitive to the size of such bright regions. That is, this method depends on the number of pixels required to make up half of the galaxy light, but not on how those pixels are distributed within the FoV.
The R$_{eff}$ derived using these methods should be considered within the limitations imposed by the angular resolution, kind of tracer, sensitivity (e.g., \citealt{Lange15}), and FoV.

In our specific case, we consider {\tt CoG} method (i.e., the flux emission contained in circular apertures of increasing size are derived to compute the radius within which half of the total emission is contained) as a robust way to derive the R$_{eff}$ and it is commonly applied both at low- and high-z. 
We considered that an equivalent circular aperture as a good approximation of the (major) elliptical aperture. The effective radius in case of an elliptical aperture is defined as $\sqrt{ab}$, where {\tt a} and {\tt b} are the major and minor axes, respectively.
To check this point we considered ESO~320-G030, which shows in CO(2--1) a quite regular shape and surface brightness distribution and NGC 7130 which shows a more peculiar CO distribution. In the first case, the difference between the effective radius derived using a circular or elliptical aperture agrees within 5\% while in the latter the difference agrees within 12\%. 

We select the center of the (circular) aperture choosing the peak emission observed in the near-IR using {\it HST/NICMOS} F160W ($\lambda_c$ = 1.60 $\mu$m, FWHM = 0.34 $\mu$m) images. When this filter is not available, other near-IR {\it HST} filters like F110W, F190N are considered. When the {\it HST} images are not available, the peak emission in the continuum map at 1.3 mm is used. 
The {\it HST}/NICMOS astrometry has been corrected using stars within the NICMOS FoV in the F110W or F160W filters and the second {\it Gaia} data release (DR2) catalog as reference systems (further details in \citealt{SG22}).
 
In our sample we found very good agreement among the CO(2--1), dust-continuum and stellar peak emissions. In general, the dynamic center, as traced by the stellar emission, also well overlaps with the dust-continuum emission peak, even in very disturbed objects, as NGC~3256 and NGC~7130. Then, when no {\it HST} images are available, the dust-continuum peak position can be considered a good assumption of the center of the aperture. Furthermore, the CO kinematic center generally agrees well with the stellar peak emission position, although in more complex systems (NGC 3256 and NGC 7130) it is very hard to define.
If in disturbed galaxies, like NGC 7130, we modify the center of the aperture to the CO flux emission peak ($\sim$10 pixels), the new R$_{CO}$ and R$ _{cont}$ would be $\sim$0.72 kpc and 0.185 kpc, respectively, instead of $\sim$0.7 kpc and $\sim$0.16 kpc. Then, in such an extreme case, the variation of the effective radius would be of $\lesssim$4\% and $\lesssim$15\% for the CO and continuum distributions, respectively. The R$_{eff}$ estimation in the continuum seems to be more affected than in CO(2--1) when changing the center of the aperture as a result of its compact distribution.

The CO(2--1) and 1.3 mm continuum sizes presented in this work are observed, that is, not corrected for the beam. The intrinsic CO and 1.3 mm continuum sizes are also resolved (larger than the beam) and (on average) smaller than the observed size by $\lesssim$1\% and $\lesssim$4\%, respectively. Only for one object, ESO~557-G002, the intrinsic 1.3 mm continuum size would be $\sim$13\% smaller than the observed size (64 pc {\it vs.} 55 pc) while its intrinsic CO size would be only marginally affected ($<$1\% smaller).

\subsubsection{Correction factors and systemic effects associated to the choice of a method}
\label{CORR_FACTS}

\begin{figure}
\centering
\includegraphics[width=0.38\textwidth]{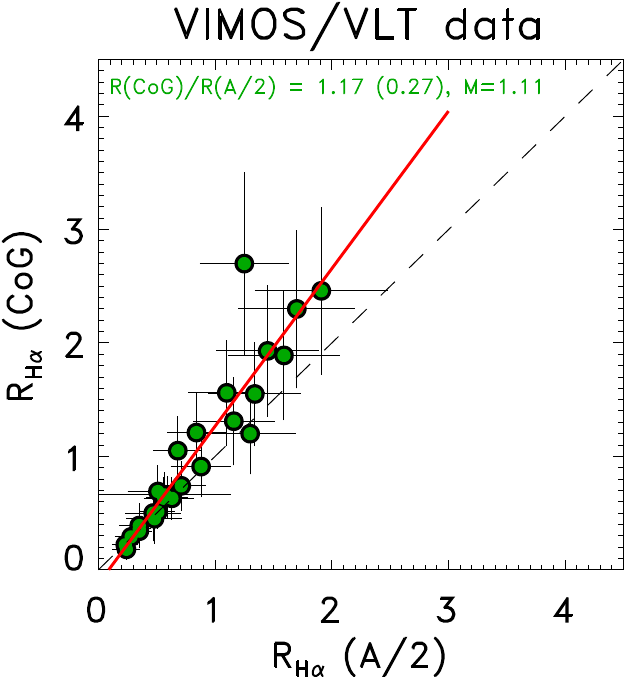}
\caption{Relation between the {\tt CoG} and {\tt `A/2'} results derived using VIMOS/VLT (H$\alpha$) data (\citealt{Arribas12}). To derive this relation close interacting pairs and the lower limits have been excluded (see text). The dashed black line represents the 1:1 relation between the radii while the red solid line is the best fit. In the plot the mean (standard deviation) and median (M) value of the R$_{H\alpha}${\tt (CoG)}/R$_{H\alpha}${\tt (A/2)} ratio are shown.}
\label{Conversions}
\end{figure}

In order to compare the CO(2-1) and 1.3 mm continuum sizes, obtained using the {\tt CoG} method with those obtained in \cite{Arribas12} and \cite{Bellocchi13} for the ionized and stellar components, respectively, we need to correct the effective radii of the stellar component, that were derived using the {\tt `A/2'} method. These radii can be indeed converted to {\tt CoG} measurements using the relation shown in Fig.~\ref{Conversions}. 
In particular, this relation has been derived considering the results shown in \cite{Arribas12} using the {\tt CoG} and {\tt `A/2'} methods to the H$\alpha$ emission for a local sample of 46 (U)LIRGs. To derive a robust trend able to relate the two methods, we excluded (15) lower limits (due to the limited VIMOS FoV) as well as the results obtained for systems in close\footnote{We excluded F06035-7102, F06206-6315, F08520-6850 and F23128-5919 (see \citealt{Bellocchi13}). The effective radius of the galaxy F12596-1529 was not available (\citealt{Arribas14}).} interaction phase, for which a reliable estimation of their size was not possible. 
For the remaining 27 galaxies, good agreement is found then between the results from the {\tt CoG} and {\tt `A/2'} methods and H$\alpha$ measurements for a given set of data. For this subsample, we found that 0.8~$\lesssim$~R$_{eff}$({\tt CoG})/R$_{eff}$({\tt A/2})~$\lesssim$ 2.2, with median ratio of 1.1. If we include the lower limits, we derive a similar median correction  factor (M=1.16) with larger dispersion.

To check the validity of the results achieved using the H$\alpha$ tracer, we also derived the R$_{eff}$  using the {\tt CoG} and {\tt `A/2'} methods to a subsample of nine\footnote{We considered NGC 3110 and NGC 7469 as isolated galaxies, IC 4687 N, IC 4518 E and W and ESO 297-G011 and ESO297-G012 as interacting galaxies, and NGC 1614 and IRAS F17138-1017 as merger systems.} LIRGs using the CO(2--1) maps. Among these sources we selected isolated, interacting and mergers systems deriving a mean (median) ratio {\tt CoG/`A/2'} $\sim$ 1.3 $\pm$ 0.2 (1.3), which is quite similar to that derived for the ionized component.

\subsubsection{Extinction impact: optical {\it versus} near-IR ionized gas tracers. Inclination effects.}
\label{EXTINC}

In order to study the extinction effect in our sample we compared the effective radii derived for the ionized gas component traced by the strongest hydrogen emission lines in the optical (H$\alpha$) and near-IR (Pa$\alpha$) for a subsample of 12\footnote{The following galaxies have been considered: ESO 320-G030, IC~4687, IC~5179, NGC~7130, NGC~1614, NGC~3256, NGC~3110, NGC~2369, NGC~5135, IRAS~F17138-1017, IC 4518 E and W.} sources from the ALMA sample for which Pa$\alpha$ images are available.

\begin{figure}
\centering
\includegraphics[width=0.45\textwidth]{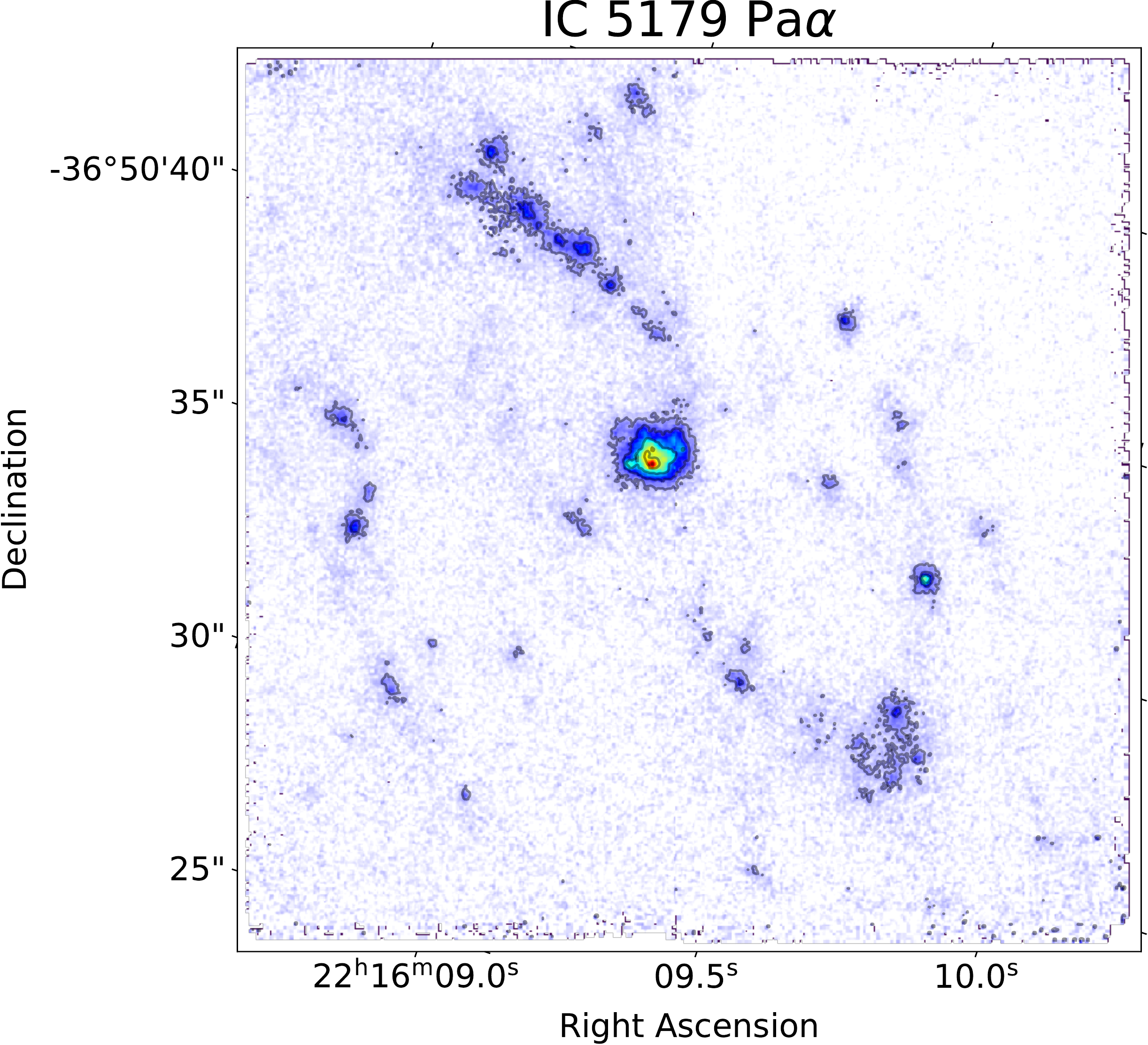}
\caption{Pa$\alpha$ image of the most extended galaxy of the sample, IC 5179, obtained from {\it HST}/NIC2 data (filters F187N and F190N). Gray contours highlight the regions in which the star-formation is taking place. The FoV is about 19.5\arcsec$\times$19.5\arcsec\ with a FWHM~$\sim$~0.15\arcsec. North is up and east to the left. }
\label{IC5179_Pa_example}
\end{figure}

The H$\alpha$ hydrogen recombination line is a direct probe of the current unobscured star formation activity in a galaxy: in presence of large amounts of dust, as in (U)LIRG systems, its emission could be strongly attenuated. On the other hand, the near-IR Pa$\alpha$ tracer is less affected by dust extinction. The combination of the two datasets can give us hints on the distribution of the extinction in our sample. 
The higher angular resolution Pa$\alpha$ images reveal in great detail the morphology of the high surface brightness HII regions, with typical physical scales of the order of a few tens of parsecs and it is thus able to resolve the ongoing star forming regions found in our LIRG systems (see \citealt{AAH09}). 
However, {\it NICMOS/HST} images may be insensitive to diffuse Pa$\alpha$ emission (e.g., \citealt{AAH06, Calzetti07}), which, in turn, can suffer much less extinction than that affecting the HII regions (e.g., \citealt{Rieke09}). 
On the other hand, VIMOS H$\alpha$ imaging has a lower angular resolution (physical scale involved is a few hundreds of parsec), which implies that the observations are more sensitive to the diffuse emission of lower surface brightness. 
Although the NICMOS data are characterized by a smaller FoV than VIMOS, the bulk of Pa$\alpha$ emission for this subsample falls well within the NICMOS FoV.
In Fig.~\ref{IC5179_Pa_example} we show the Pa$\alpha$ emission observed in IC~5179, one of the most (intrinsically) extended objects in our sample, that is fully covered by the {\it HST}/NICMOS imaging, showing a number of lower surface brightness star-forming regions in addition to the bright nuclear emission.

\begin{figure*}
\centering
\includegraphics[width=0.65\textwidth]{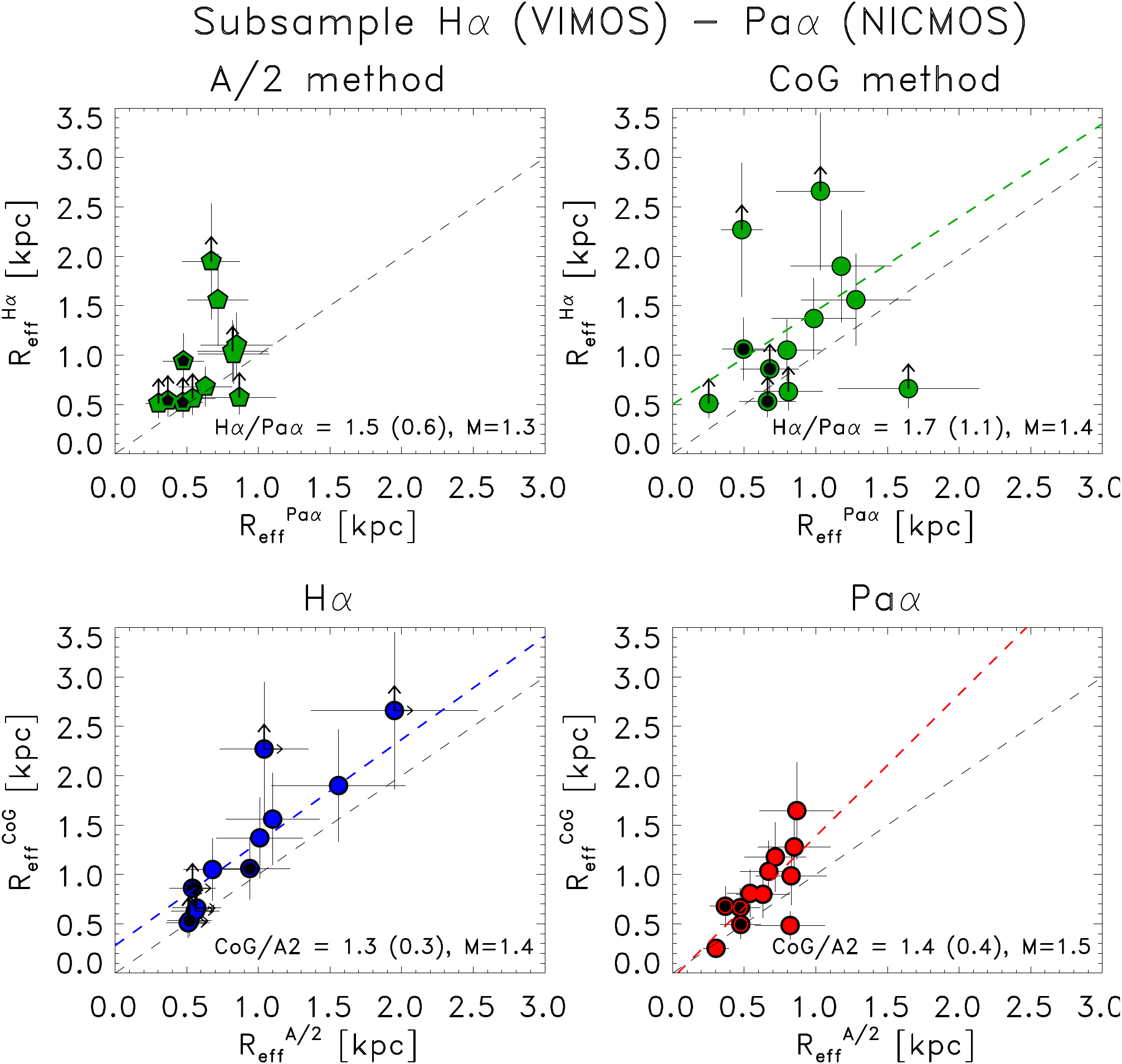}
\caption{Relations between the observed (not corrected for the extinction) H$\alpha$ and Pa$\alpha$ tracers when using {\tt `A/2'} and {\tt CoG} methods applied to a subsample of 12 galaxies. Lower limits in H$\alpha$ measurements are due to the VIMOS limited FoV. {\it Top panels:} R$_{eff}^{H\alpha}$ as a function of the R$_{eff}^{Pa\alpha}$ when applying the {\tt `A/2'} ({\it left}) and {\tt CoG} ({\it right}) methods. {\it Bottom panels:} Comparison of the R$_{eff}$ obtained using the {\tt CoG} and the {\tt `A/2'} methods derived for the H$\alpha$ ({\it left}) and Pa$\alpha$ ({\it right}). 
For each plot the mean (and standard deviation) as well as the median (M) values are shown. Galaxies containing an AGN are identified using an additional small black symbol.
The black dashed line represents the 1:1 relation between the parameters considered. The colored dashed lines identify the trend obtained using a linear least square fit for each kind of data, derived, in this case, excluding the lower limits.} 
\label{Ha_Pa_rel}
\end{figure*}

In Fig.~\ref{Ha_Pa_rel} we compare the effective radius estimations derived when using the {\tt `A/2'} and {\tt CoG} methods for both tracers without applying the extinction correction to these fluxes. In the top panels, we show the relation between the two tracers when applying the {\tt `A/2'} method (left) and the {\tt CoG} (right). 
Taking those systems compact enough to be fully covered by the VIMOS FoV, we derived a median ratio R$_{H\alpha}$/R$_{P\alpha}$~$\sim$1.4. This result should be confirmed with a larger sample covered by a larger (imaging or IFS) FoV.
In the bottom panels, the relations between the {\tt `A/2'} and {\tt CoG} for H$\alpha$ (left) and Pa$\alpha$ (right) are shown. The observed H$\alpha$ radii derived using the {\tt CoG} method cover a large range of values (0.5-3 kpc) than Pa$\alpha$ (0.3-1.5 kpc). For this subsample we found typical mean (median) values of R$_{H\alpha}$ = 1.26$\pm$0.71 (1.08) kpc and R$_{Pa\alpha}$ = 0.86$\pm$0.39 (0.81) kpc. The distribution of the optical tracer is 1.5 (mean value) higher than that derived for the near-IR tracer.

\cite{AAH09} compared the morphology of the high surface-brightness HII regions in the H$\alpha$ and NICMOS Pa$\alpha$ emissions for a sample of LIRGs at z~$<$~0.02. Their systems share similar morphology of high surface-brightness HII regions in the H$\alpha$ and Pa$\alpha$ emissions, suggesting that the extinction effects on H$\alpha$ are not severe, except in the very nuclear regions. 
Due to the central obscuration, the outer regions have a relatively larger fraction of flux in the H$\alpha$ maps, leading to larger R$_{eff}$.
This would result in a higher estimation of the R$_{eff}$ with respect to the one derived if the emission would be corrected for the extinction. 
\cite{Arribas12} corrected their images with a simple model extinction in order to understand how the extinction affects the determination of R$_{eff}$. They found that the corrected maps show a mean reduction in size of 25-30\% with respect to the uncorrected values. If we correct our H$\alpha$ radii for this factor we end up with an intrinsic mean (median) R$_{eff}^{H\alpha}\sim$~0.9 (0.8) kpc, closer to the value obtained with the Pa$\alpha$ tracer.

Furthermore, inclination effects could play an important role when deriving the size of a galaxy. In particular, \cite{Graham08} found that the effects of the inclination on the derivation of the R$_{eff}$ between the optical and near-IR bands is very low. 
Indeed, they found that the intrinsic scale length of a galaxy in the B- and K-bands would be, respectively, 1.35 and 1.05 times lower than the observed scale length (i.e., h$_{obs}$/h$_{intr}$~$\sim$1.35 and $\sim$1.05, assuming a mean inclination of 52$^\circ$, as that derived for our LIRGs, see \citealt{Bellocchi13} and \citealt{Law09}). 
A smaller factor ($\sim$1.2) is derived in the R band. This means that the intrinsic R$_{eff}$ is expected to be slightly lower than the observed radius in the optical and near-IR bands. 
Under this assumption, we also expect a small conversion at longer (mm) wavelengths: for this reason, no inclination correction has been applied to our sample.

\subsection{Instrumental bias: limitation in the FoV, angular resolution and sensitivity}

We now focus our attention on the limitations due to the instrumental setup such as the angular resolution of the observations, their sensitivity and a limited FoV.
We investigate these effects in three galaxies (NGC~1614, NGC~7130 and NGC~3256) classified as post-coalescence systems, that show a quite complex structure (e.g., tidal tails). 
These galaxies are good targets to understand how the flux distribution at different bands and angular resolutions affects the R$_{eff}$ estimation in extreme systems. 
We took into account near-IR continuum images from {\it HST}/NIC2 (or {\it HST}/WFPC3), 2MASS and {\it Spitzer}/IRAC1 images at 1.6, 2.2 and 3.6 $\mu$m, with angular resolutions of $\sim$0.15\arcsec (0.26\arcsec), 2\arcsec\ and 2.4\arcsec, respectively. 
The comparison among the different instrumental setups with that characterizing 2MASS observations allows us to derive the following results:

\begin{enumerate}
 
\item under the same angular resolution, sensitivity (\citealt{Benjamin03}) and FoV (without limitation), as those achieved by {\it Spitzer}/IRAC1 data, we derived similar (or $\sim$30\% smaller) R$_{eff}$ between these data sets: {\tt R$_{IRAC1}$$\sim$(30\%)$\times$R$_{2MASS}$}; 
\vskip2mm
\item at higher angular resolution ($\times$10) and sensitivity than 2MASS data, as in the case of {\it HST}/WFPC3 data characterized by a large FoV ($\sim$2\arcmin$\times$2\arcmin), we derived similar R$_{eff}$: {\tt R$_{HST/WFPC3}$~$\sim$~R$_{2MASS}$} ;
\vskip2mm
\item when the highest angular resolution among the instruments is considered (0.15\arcsec) along with very good sensitivity (i.e., slightly lower than WFPC3), but a smaller FoV is involved, as in the case of {\it HST}/NIC2, we derived lower R$_{eff}$ for this data set: {\tt R$_{NIC2}$ $\lesssim$ 20\%$\times$R$_{2MASS}$}. 
\end{enumerate}

Thus, under similar conditions ({\it Spitzer}/IRAC1 and 2MASS data) slightly smaller (or similar) sizes are derived at 3.6 $\mu$m than at smaller wavelength ($\sim$2 $\mu$m, i.e., 2MASS H and K bands).
High angular resolution and sensitivity data, as those achieved by the {\it HST}/WFPC3 images, allow to derive similar size as those obtained using 2MASS data in the same band, unless a limited FoV is involved (i.e., {\it HST}/NIC2) which then implies smaller effective radii.

\section{Results}
\label{results_par}

In this section we present the R$_{eff}$ results obtained for the CO(2--1) and 1.3 mm emissions using the {\tt CoG} method of our sample. We compare these results with those previously obtained using the same method for the stellar and ionized (observed H$\alpha$) gas emissions.
Further comparison between the molecular CO(2--1) and ionized (unobscured) Pa$\alpha$ emissions is also discussed.

In Appendix~\ref{CO_sameFoV_Reff} we present the CO emission (above 5$\sigma$) and K-band maps for the whole sample, ordered according to their increasing molecular effective radii, R$_{CO}$. The same FoV (14$\times$14 kpc$^2$) is considered for all the galaxies for a direct comparison of their size. R$_{eff}$ are given for both the CO(2--1) and K-band maps.

\subsection{Molecular and 1.3 mm continuum emissions }
\label{molecular_size}

The molecular gas distribution derived for our LIRG sample is very compact. It is characterized by R$_{CO}$ sizes spanning a range of values in between a few hundreds of parsecs up to $\sim$1.3~kpc size (see Tab.~\ref{sizes_LIRGs}). On the other hand, the 1.3 mm continuum emission is even more compact than the molecular gas.
We derived that the CO(2--1) emission is about twice the size of the 1.3 mm continuum emission, with typical mean (median) sizes of R$_{CO}$ = 0.66 $\pm$ 0.33 (0.7) kpc and R$_{cont}$ = 0.37 $\pm$ 0.31 (0.31) kpc.
Such compactness could be explained as due to a sensitivity effect, rather than to a limited FoV. In particular, the high angular resolution of our ALMA data prevent us from detecting the contribution of the faintest and most extended emission. As a result, the brightest dust emission then might appear more compact. However, the maximum recoverable scale of these observations is 3-5 kpc so that the extended diffuse and faint emission is unlikely filtered-out. A more detailed analysis of the molecular continuum emission is needed to confirm this point, and it will be presented in a future work.
To investigate further on this point, we presented in the Appendix \ref{SEDs_sample} the SED fitting results to quantify the flux loss at 1.3 mm with our ALMA observations. 
The 1.3 mm continuum fluxes derived with ALMA at $>$5$\sigma$ are well below the SED emission (Flux(SED)/Flux(ALMA)$\sim$2-15) as a result of the sensitivity effects, for which the faint emission at larger radii is not observed. 
With that in mind, we assume that the effective radii estimation of the 1.3 mm continuum emission is affected by the flux loss of the outer regions at this frequency: for this reason, we will not discuss the results derived for this tracer.

On the other hand, we can compare the CO(2--1) fluxes derived with ALMA with those obtained using single dish observations only for a couple of galaxies (IC~4687 and ESO~320-G030; see \citealt{PereiraS16, Pereira16}). For these systems, good agreement is found between the integrated ALMA fluxes and single dish observations, within a factor of $<$15\%.

\begin{table*}
\centering
\begin{small}
\caption{Molecular and 1.3 mm continuum effective radii determinations of the LIRG sample.}
\label{sizes_LIRGs}
\begin{tabular}{cccccc} 
\hline\hline\noalign{\smallskip}  
\multicolumn{1}{c}{Source}	 &		\multicolumn{2}{c}{R$_{eff}$}	& \multicolumn{2}{c}{Flux (R$_{eff}$)}	& AGN	 \\
IRAS/Other 	&	$^{12}$CO	& 1.3 mm &	 $^{12}$CO	& 1.3 mm \\
\cmidrule(lr){2-3}  \cmidrule(lr){4-5}
		&	[pc]	& [pc]	& [Jy] & [mJy] 	\\
(1) & (2) &  (3)& (4)  &(5)	 \\
\hline\noalign{\smallskip} 	
{\tt ESO 297-G011 (N)}	&   	598$\pm$69	&  499$\pm$54   & 161$\pm$32	& 0.7$\pm$0.2	&	(y)  \\
{\tt ESO 297-G012 (S)}	&  	369$\pm$65 & 152 $\pm$ 31 	& 103$\pm$21	& 1.2$\pm$0.3	&	\\
\hline\noalign{\smallskip} 	
{\tt NGC 1614}	&	634$\pm$110	&	264$\pm$29	& 340$\pm$68	& 8.0$\pm$1.6	&	\\
\hline\noalign{\smallskip} 	
{\tt ESO 557-G002}	&	300$\pm$57	&	64$\pm$9	& 68$\pm$14	& 1.3$\pm$0.3	&	\\        
\hline\noalign{\smallskip} 	
{\tt F06592-6313 }	& 310 $\pm$ 57	&	139$\pm$24	& 67$\pm$14	& 1.9$\pm$0.4	&	\\
\hline\noalign{\smallskip} 	
{\tt NGC 2369}		& 777$\pm$189	&	404$\pm$190	  & 648$\pm$129	& 11.5$\pm$2.3	&(y)  \\
\hline\noalign{\smallskip} 	
{\tt NGC 3110} & 	1134$\pm$173 &	312$\pm$50	& 236$\pm$47	& 1.9$\pm$0.4	\\
\hline\noalign{\smallskip} 	
{\tt NGC 3256} & 1279$\pm$160		&	517$\pm${356}	& 1354$\pm$269	& 10.7$\pm$2.1	 \\
\hline\noalign{\smallskip} 	
{\tt ESO 264-G036}	&	755$\pm$337	& 365$\pm$15	& 755$\pm$35	& 0.9$\pm$0.2	\\
\hline\noalign{\smallskip} 	
{\tt ESO 319-G022} &	206$\pm$36	&	68$\pm$8	& 45$\pm$9	& 1.9$\pm$0.4	 \\     
\hline\noalign{\smallskip} 	
{\tt ESO 320-G030}	& 381$\pm$164 &	83$\pm$10 	& 401$\pm$80	& 14.4$\pm$2.9\\
\hline\noalign{\smallskip} 	
{\tt ESO 267-G030} &  	894$\pm$182	&  374$\pm$302	& 133$\pm$27	& 0.21$\pm$0.04		&(y)    \\
\hline\noalign{\smallskip} 	
{\tt MCG-02-33-098 E}    & 360$\pm$58 	&	210$\pm$57	& 12$\pm$2	& 0.3$\pm$0.1		\\
{\tt MCG-02-33-098 W}   & 246$\pm$58	&	126$\pm$16	& 54$\pm$11	& 1.1$\pm$0.2	\\
\hline\noalign{\smallskip} 	
 {\tt ESO 507-G070}	&	325$\pm$76	&	150$\pm$25 & 180$\pm$36	& 8.3$\pm$1.7	 \\
\hline\noalign{\smallskip} 	
{\tt NGC 5135} & 	797$\pm$104	&	666$\pm$71	& 408$\pm$81	& 9.4$\pm$1.9	&	(y) \\ 
\hline\noalign{\smallskip} 	
{\tt IC 4518 E}	&	930$\pm$196	& 	865$\pm$234	& 121$\pm$23	& 0.6$\pm$0.1	 \\
{\tt IC 4518 W}	& 497$\pm$95  &	147$\pm$45	& 125$\pm$25	& 1.4$\pm$0.3		&	(y) \\
\hline\noalign{\smallskip} 	
{\tt F17138-1017} 	&	897$\pm$146	& 738$\pm$ 25		& 253$\pm$51	& 4.9$\pm$1.0	  \\
\hline\noalign{\smallskip} 	
{\tt IC 4687}	&  1154$\pm$191	&	1256$\pm$169 	& 214$\pm$43	& 5.2$\pm$1.1\\
\hline\noalign{\smallskip} 	
{\tt IC 4734}	&	366$\pm$89 	&	164$\pm$19	& 192$\pm$38	& 4.7$\pm$0.9	\\
\hline\noalign{\smallskip} 	
{\tt NGC 7130}	& 		699$\pm$219	&	160$\pm$29	&332$\pm$66	& 5.2$\pm$1.0	&	(y) \\ 
\hline\noalign{\smallskip} 	
{\tt IC 5179}	& 	{1182$\pm$198}	&	{844$\pm$349}	& 394$\pm$78	& 3.0$\pm$0.6	\\
\hline\noalign{\smallskip} 	
{\tt NGC 7469}	& {701$\pm$89}	& {405$\pm$63}	&344$\pm$69	& 4.5$\pm$0.9	 & (y)  \\
\hline\hline\noalign{\smallskip} 	
\end{tabular}
\end{small}
\vskip5mm\begin{minipage}{18cm}
\small
{{\bf Notes:} Column (1): Source name (IRAS and other). Columns (2, 3): Effective radius derived for the CO(2--1) line and 1.3 mm continuum emissions using the {\tt CoG} method. The uncertainties associated to the derivation of the R$_{eff}$  have been derived as half the difference between the R$_{eff}$ derived at 4$\sigma$ and 6$\sigma$, considered as the major source of uncertainty for the derivation of this parameter (i.e., $\Delta$R$_{eff}$ = (R$_{eff,4\sigma}$ - R$_{eff,6\sigma}$)/2). Columns (4, 5): Flux within the effective radius derived for the CO(2--1) line and 1.3 mm continuum emissions. The uncertainties have been derived as half the difference between the flux within the R$_{eff}$ derived at 4$\sigma$ and 6$\sigma$ (i.e., $\Delta$F~(R$_{eff}$)~=~$\frac{Flux(R_{eff,4\sigma}) - Flux(R_{eff,6\sigma})}{2}$. 
Column (6): Presence of an AGN in the galaxy identified by (y) (see Tab.~\ref{Input_data} for further details).}
\end{minipage}
\end{table*}

\begin{table*}
\centering
\begin{small}
\caption{Mean (and median) half-light radius of the different tracers analyzed in the LIRG sample. }
\label{Mix_radii_sample}
\begin{tabular}{cccccc} 
\hline\hline\noalign{\smallskip}  
Sample 	 &		\# &		R$_{CO}$ [kpc] &    	R$_{cont}$ [kpc]	&  R$_{star}$ [kpc] 	& R$_{H\alpha}$ [kpc] \\
(1)		&	(2)		&(3)		&	(4)		&	(5)		&	(6)\\
\hline\noalign{\smallskip}  
All LIRGs 	& 24 & 0.66$\pm$0.33 (0.67$\pm$0.29) & 0.37$\pm$0.31 (0.29) 	& 2.21$\pm$0.81 (2.41$\pm$0.72)		& 1.42$\pm$0.89 (1.22$\pm$0.68)	\\
LIRGs w/o AGN  	& 17 & 0.64$\pm$0.38 (0.38$\pm$0.18) & 0.37$\pm$0.34 (0.24) 	& 2.12$\pm$0.88 (1.95$\pm$0.75)	& 1.48$\pm$0.97 (1.21$\pm$0.69)	 \\
LIRGs w AGN  	& 7  & 0.72$\pm$0.14 (0.70$\pm$0.10) & 0.38$\pm$0.20 (0.40) 	& 2.44$\pm$0.62 (2.50$\pm$0.12)		& 1.29$\pm$0.71 (1.20$\pm$0.60)		\\
\hline\noalign{\smallskip} 	
{\tt 0 RD}		&	8	&	0.73$\pm$0.34 (0.73$\pm$0.35)	&	0.46$\pm$0.40 (0.38)	&	2.00$\pm$0.83 (2.08$\pm$0.50)	&	1.59$\pm$1.00 (1.56$\pm$0.43)			\\
{\tt 0 PD}		&	7	&	0.59$\pm$0.34 (0.60$\pm$0.29)	&	0.31$\pm$0.23 (0.31)	&	2.58$\pm$0.65 (2.65$\pm$0.21)	&	1.64$\pm$1.03 (1.89$\pm$0.81)			\\
{\tt 1}			&	4	&	0.51$\pm$0.30 (0.43$\pm$0.13)	&	0.34$\pm$0.35 (0.18)	&	1.96$\pm$0.89 (1.85$\pm$0.65)	&	0.86$\pm$0.28 (0.86$\pm$0.20)			\\
{\tt 2}			&	5	&	0.77$\pm$0.35 (0.70$\pm$0.20)	&	0.37$\pm$0.26 (0.26)	&	2.22$\pm$0.99 (1.95$\pm$0.91)	&	1.14$\pm$0.71 (0.91$\pm$0.40)			\\
\hline\hline\noalign{\smallskip} 	
\end{tabular}
\end{small}
\vskip3mm\begin{minipage}{18cm}
\small
{{\bf Notes:} Column (1): Sample; Column (2): Number of galaxies in each sample; Columns (3, 4, 5, 6): Effective radius derived for the CO(2--1), 1.3 mm continuum, stellar and ionized H$\alpha$ emissions. All values are in units of kpc. See text for further details. }
\end{minipage}
\end{table*}

\subsection{No relation between the molecular size and the morphological type}
\label{rco_morpho}

\begin{figure*}
\centering
\includegraphics[width=0.28\textwidth, height=0.24\textwidth]{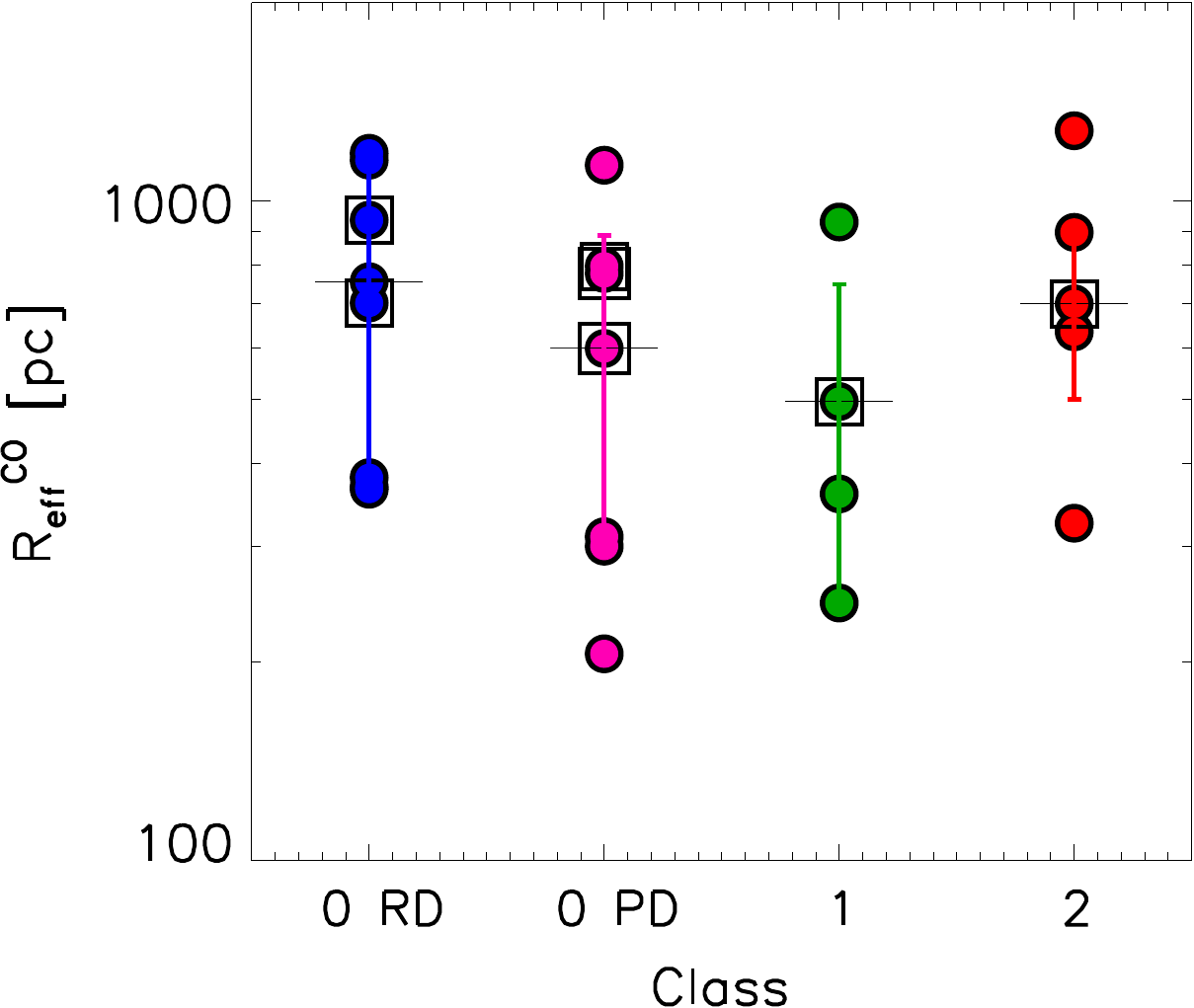}
\hskip10mm
\includegraphics[width=0.3\textwidth]{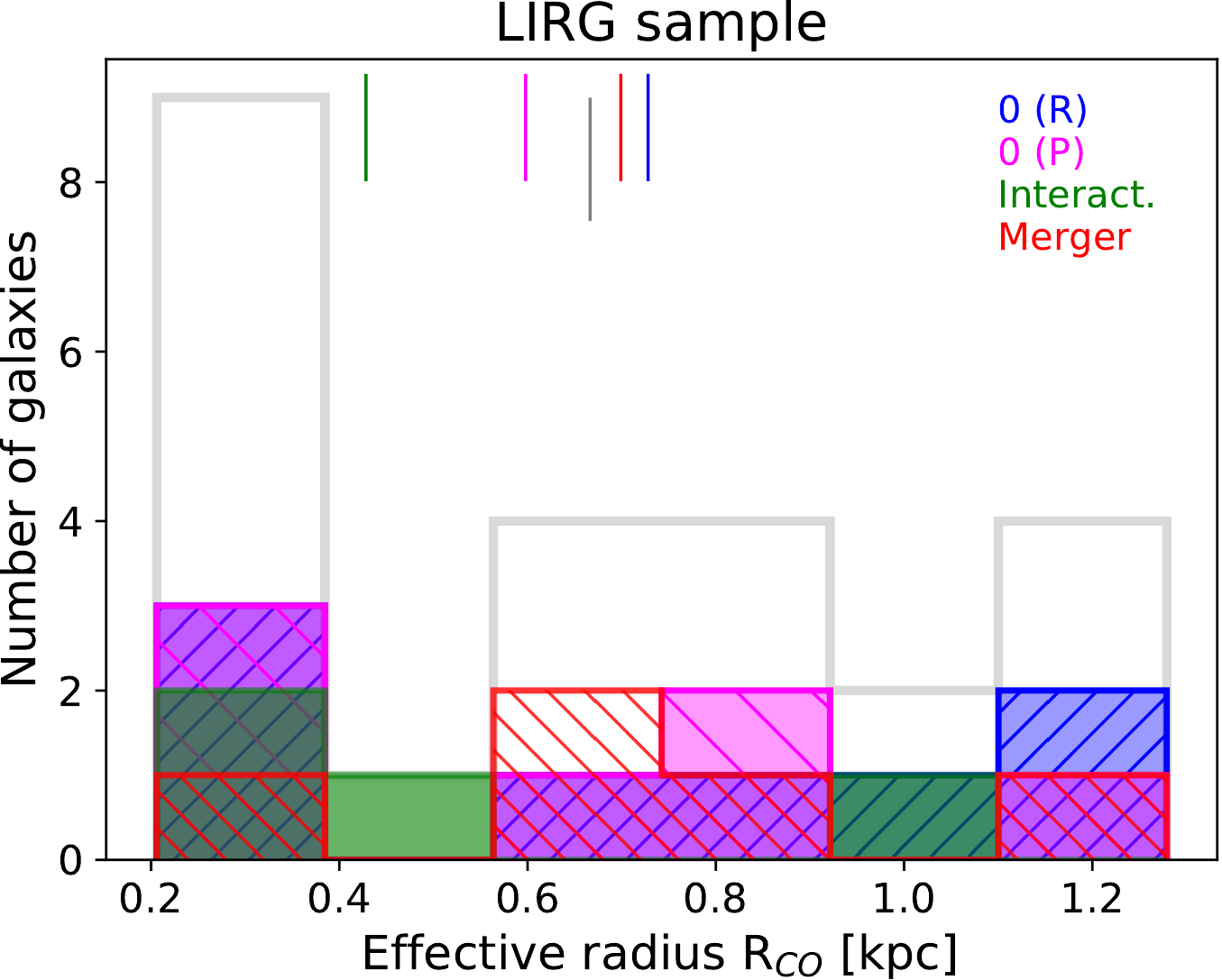}
\hskip3mm\includegraphics[width=0.3\textwidth]{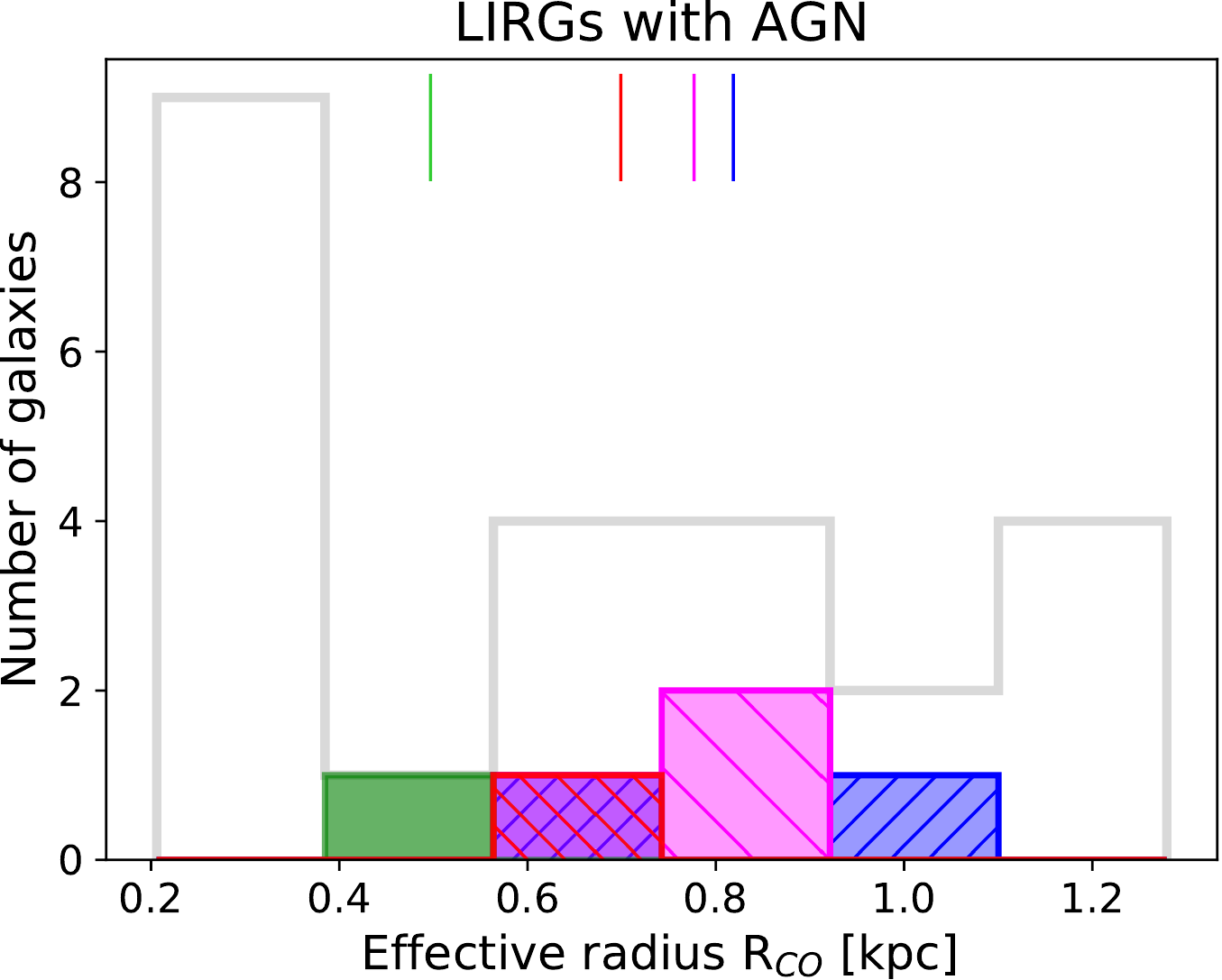}
\caption{{\it Left:} CO(2--1) effective radius as a function of the class. Type {\tt 0 (R), 0 (P)}, {\tt 1} and {\tt 2} are shown in blue, magenta, green and red, respectively. The median R$_{CO}$ values (horizontal lines) are shown for each class. Empty squares identify galaxies containing an AGN. {\it Middle:} R$_{CO}$ distribution for the whole sample (gray solid line) and for the individual subgroups (same color code as in the left panel). The colored vertical lines in the upper part of the panel represent the median values of each distribution, following the same color code. {\it Right:} R$_{CO}$ distribution for the whole sample (gray solid line) and for LIRGs with an AGN. 
}
\label{RCO_morpho}
\end{figure*}

Our sample shows a large variety of morphology, characterized by compact (R$_{eff}<$~0.4 kpc), elongated and complex\footnote{Compact systems include IC~4734, ESO~297-G012 and ESO~320-G030 as type {\tt 0 (R)} galaxies, ESO~319-G022, ESO~557-G002 and F06592-6313 as type {\tt 0 (P)} and ESO~507-G070 as type {\tt 2} objects, elongated systems include ESO~507-G070 and NGC~2369 and complex systems include NGC3256.} CO(2--1) distributions.
The distribution of the sample according to the size of the CO emission (Fig.~\ref{RCO_morpho} left) suggests a bimodal behavior, with compact (R$_{eff}$~$<$~0.4 kpc) and more extended (R$_{eff}$~$>$~0.6 kpc) galaxies. Smaller (median) radii are found for interacting systems (type {\tt 1}), while mergers (type {\tt 2}) share similar size with isolated (type {\tt 0}) objects.

We derived the following results (see Fig.~\ref{RCO_morpho}, right): 
(a) some disks (type {\tt 0}; 6/14, 43\%) have effective radii in the range 0.2 and 0.4 kpc, and between $\sim$0.6 and 0.8 kpc (5/14, 36\%), while 3/14 (21\%) sources show radii $\sim$1.2 kpc; (b) the interacting systems size peaks around 0.3 kpc; (c) finally, mergers are more commonly distributed around 0.7~kpc, with some outliers found at 0.3 and 1.2 kpc. 
We can thus claim that no relation has been found between the molecular size, R$_{CO}$, and the galaxy types.

\subsection{The impact of AGN on the molecular gas}
\label{AGN_discuss}

As described in Sect.~\ref{sample_sect}, a small number of objects in our sample are classified as Seyfert galaxies or show signs of the presence of an AGN, the contribution of which does not dominate the galaxy emission. 
When an AGN is present, both the continuum from the active nucleus and young stars can contribute to the ionization of the gas.
We distinguished our sample in LIRGs with (w AGN, 7/24) and without (w/o AGN, 17/24) an AGN to look for correlation between them. 
An extra flux produced by a bright AGN in the nuclear regions of a galaxy is expected to lead to a smaller effective radius estimation (as for the ionized gas emission; see \citealt{Arribas12}).

In our particular case, larger (median) molecular radii by a factor of 2 are derived when considering galaxies with an AGN: the majority of the sources without an AGN show R$_{CO}$ sizes within 0.2-0.4 kpc, while LIRGs with an AGN are characterized by lager molecular size (0.4-1.1 kpc; Fig.~\ref{RCO_morpho} middle and right panels). 
A similar trend is observed for the stellar emission, for which we derived larger stellar radii (by a factor of $\sim$1.3) for those sources containing an AGN. 
On the other hand, the average R$_{H\alpha}$ of the ionized component is $\times$0.9 more compact in presence of an AGN, similar to what derived in \cite{Arribas12} (Tab.~\ref{Mix_radii_sample}), without any variation in their median values. 
 
Applying the Kolmogorov-Smirnov (KS) test to the estimation of the effective radii derived for the several tracers with or without an AGN we can see whether the two subsamples are drawn from the same distribution. For the molecular emission the test suggests that the two subsamples (AGN {\it versus} non-AGN) are not drawn from the same parent sample (p-value=0.08) while for the stellar and ionized (H$\alpha$) components we found a better correlation between the two subsamples (p = 0.2 and 0.8, respectively). 

The presence of a high surface brightness CO emission located in the extra-nuclear regions (e.g., spiral arms, off-nuclear structures) in LIRGs with an AGN could give us some hints to explain the larger molecular size, R$_{CO}$, derived in these systems.

\subsection{A possible relation between R$_{CO}$ and the {\it Main Sequence} parameters?}

\begin{figure*}
\centering
\includegraphics[width=0.9\textwidth, angle=180]{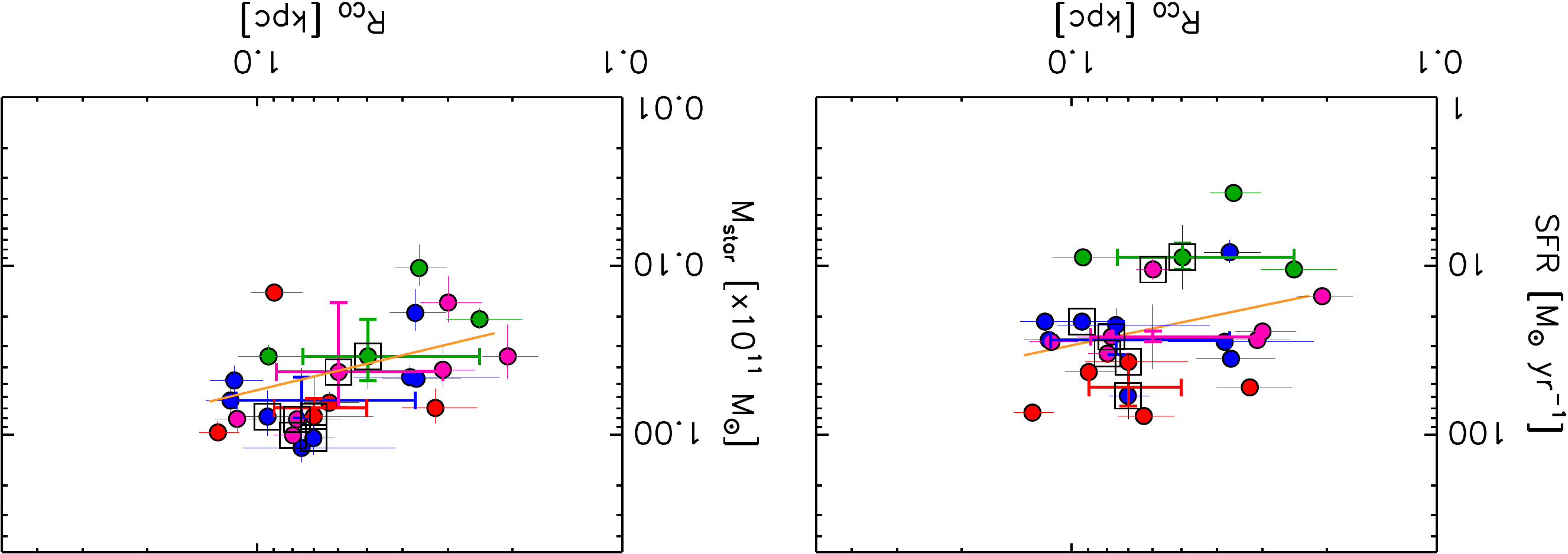}
\caption{{\it Left:} SFR as a function of the effective radius R$_{CO}$. Galaxies containing an AGN are surrounded by a square. A Spearman's rank correlation coefficient $\rho$ = 0.32 is derived (see text). {\it Right:} Stellar mass M$_{star}$ {\it versus} R$_{CO}$ ($\rho$ = 0.53). The different types of galaxies are identified with the same color code used in Fig.~\ref{RCO_morpho}. The orange solid line represents the best fit (linear least square fit).  
}
\label{SFR_Mstar_Rco}
\end{figure*}

There is a slight tendency (Fig.~\ref{SFR_Mstar_Rco}, left) for the galaxies with higher SFR to have larger CO sizes (Spearman's rank correlation coefficient $\rho$ = 0.32, with probability of no correlation p = 0.12). This tendency appears to be stronger when the stellar mass is considered (Fig.~\ref{SFR_Mstar_Rco}, right), i.e., more massive galaxies tend to have large CO sizes ($\rho$ = 0.53, p = 8$\times$10$^{-3}$). However, despite the stellar mass and SFR ranges covered by the LIRG sample span factor of ten and more (10$^{10}$-10$^{11}$ M$_\odot$ and 10-100 M$_\odot$ yr$^{-1}$, respectively), the CO regions still are compact, with sizes of less than 1 kpc.

\subsection{Molecular {\it versus} ionized (Pa$\alpha$) emissions }
\label{Pa_tracers_sect}

\begin{figure*}
\centering
\includegraphics[width=0.95\textwidth]{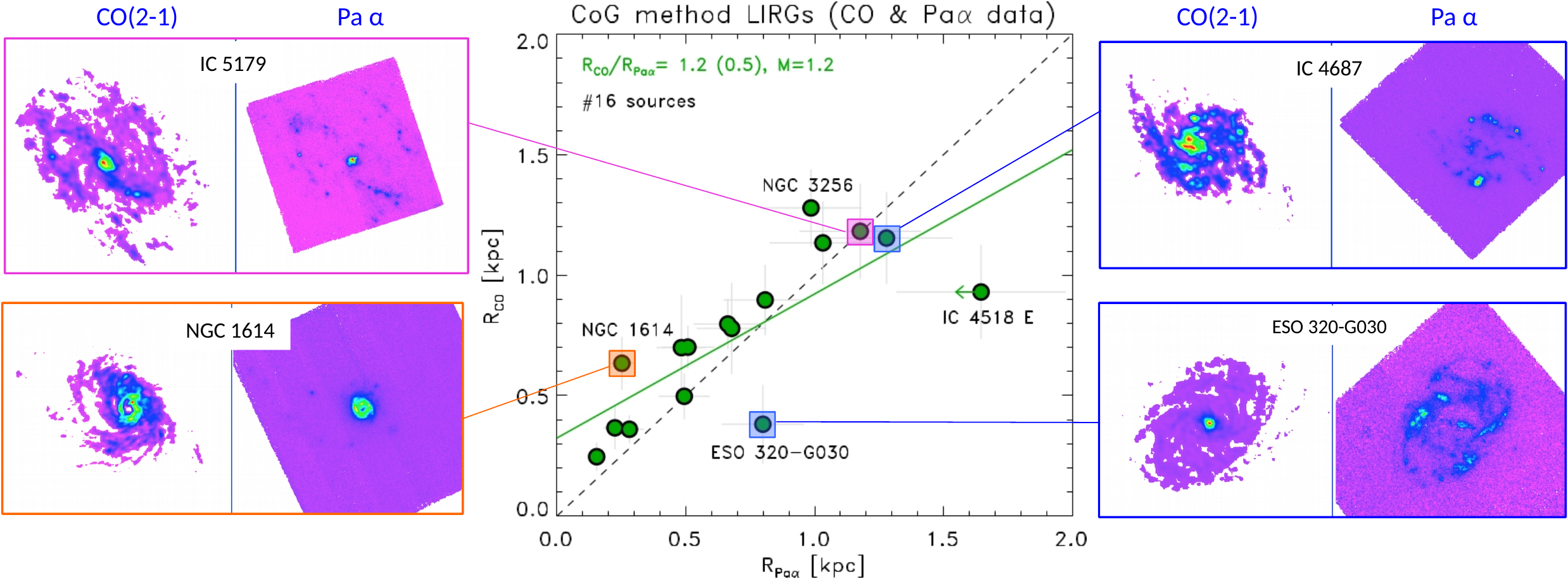}
\caption{Effective radii derived for the CO(2--1) and Pa$\alpha$ tracers using the {\tt CoG} method. The black dashed line represents the 1:1 relation. The solid green line identifies the derived best fit to the data, derived using a linear least square fit. Two outliers (IC 4687 and ESO 320-G030), for which the R$_{Pa\alpha}$ $\gtrsim$ R$_{CO}$, are identified while NGC 1614 is characterized by R$_{CO}$ $\gtrsim$ R$_{Pa\alpha}$. 
The most extreme R$_{Pa\alpha}$ is derived for IC~4518~E, and considered as an upper limit (see text for details). IC 5179 is a galaxy which shows similar effective radii for both gas tracers (R$_{Pa\alpha}$ $\sim$ R$_{CO}$).
}
\label{CO_Pa_rel}
\end{figure*}

Under the assumption discussed in \S~\ref{EXTINC}, we now compare the Pa$\alpha$ results of the ionized component with those derived for the molecular CO(2--1) line (Fig.~\ref{CO_Pa_rel}).
In this case we found a tight correlation between the two tracers: indeed, for this subsample (see \S~\ref{EXTINC}) the effective radius derived for CO(2--1) emission are very similar to that derived for the Pa$\alpha$ emission.
We derived typical mean (median) effective radii of R$_{CO}$ = 0.75$\pm$0.33 (0.79) kpc and R$_{Pa\alpha}$~=~0.72$\pm$0.42 (0.68)~kpc. 
This result highlights the good agreement between the effective radii indicating that on average the molecular and ionized (Pa$\alpha$) gas distributions are very compact. However, different morphologies can be appreciated within the sample (Fig.~\ref{CO_Pa_rel}). 
There are systems where the molecular gas is centrally concentrated while the Pa$\alpha$ shows a more extended distribution with star-forming regions in the spiral arms (ESO 320-G030 and IC 4687, probably due to the high extinction of the nuclear regions, A$_V$$\sim$3 mag for both galaxies; see \citealt{AAH06}), others show equally compact nuclear distribution (NGC 1614), while others show similar distributions with both nuclear as well as extended (IC 5179). 
As discussed in \cite{PereiraS16, Pereira16}, a different distribution of the HII regions traced by Pa$\alpha$ emission with the molecular CO(2--1) regions within the 100-200 pc scales can suggest that the SF law breaks down on sub-kpc scales (e.g., \citealt{SG22}).

\subsection{Ionized (H$\alpha$) and stellar emissions in LIRGs: relations with the molecular distribution}
\label{mol_ion_star}

\begin{figure*}
\centering
\includegraphics[width=0.95\textwidth]{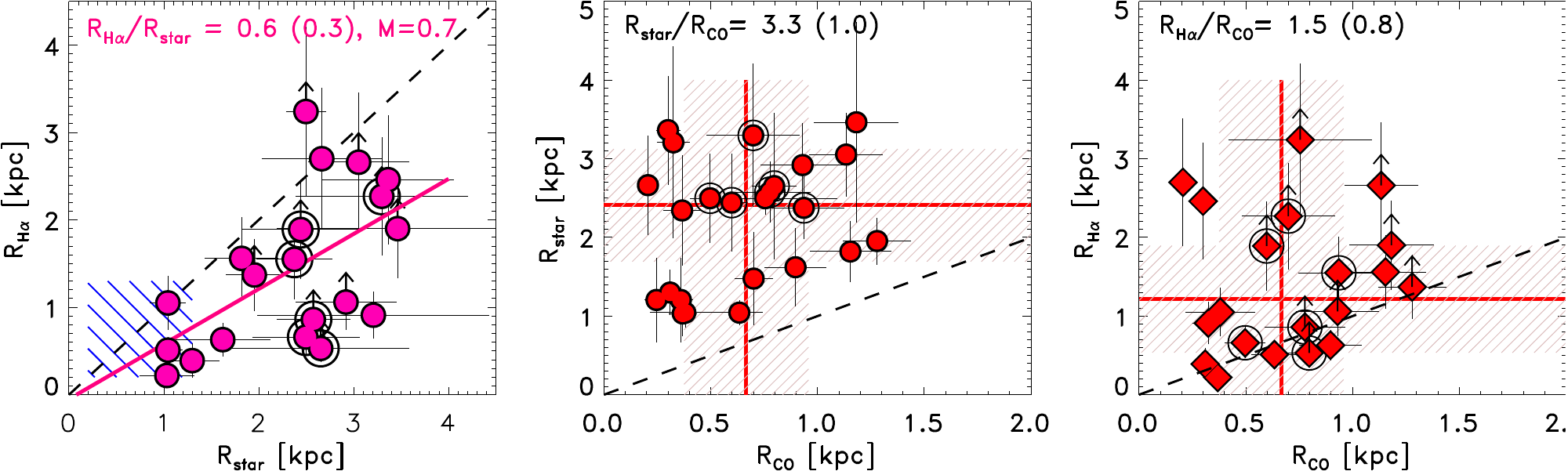}
\caption{{\it Left:} Ionized gas size traced by the H$\alpha$ emission, R$_{H\alpha}$, as a function of the stellar continuum size in the K band, R$_{star}$. The 1:1 relation is shown using the dashed black line while the best fit is shown in magenta using a solid line, derived using a linear least square fit. In the plot the mean (standard deviation) and median (M) values are shown. Double circles represent galaxies containing an AGN. The dashed blue area represents the size covered by the CO emission for a direct comparison. {\it Middle, Right:} Molecular size {\it versus} stellar (middle) and ionized gas (right) distributions. Double circle identifies galaxies with an AGN. The dashed black line represents the 1:1 ratio. The solid lines and the shaded areas represent the median values (MAD). The mode values of these ratios are similar to the median values when considering the stellar results, while we derive a mode of $\sim$1.3 when considering the H$\alpha$ results.
}
\label{star_Ha_4_plots}
\end{figure*}

We star by comparing the effective radii derived for the the (observed, not corrected for the extinction) ionized (H$\alpha$) gas and stellar continuum emissions for all galaxies of our sample (Fig.~\ref{star_Ha_4_plots} left panel; see \S~\ref{ancillary}). We excluded from this analysis two galaxies (NGC 7469 and IC 4734) for which no VIMOS (H$\alpha$) data are available. As discussed in \S~\ref{sect_method}, we considered the results obtained using the {\tt CoG} method. 
It is apparent that the stellar emission is characterized by larger size than the ionized component, by a factor $\sim$2. The stellar extension is in the range 1~$\lesssim$~R$_{eff}$~$\lesssim$~4~kpc, with median value of 2.4 kpc while the typical median H$\alpha$ size is $\sim$1.4 kpc. Since several H$\alpha$ measurements are lower limits, the ratio R$_{H\alpha}$/R$_{star}$=0.7 can be considered a lower limit as well.

We next compare the stellar and ionized gas distributions to the CO size (Fig.~\ref{star_Ha_4_plots} middle and right panels).  
The ratio between the stellar and the molecular CO sizes give a (median) value of R$_{star}$/R$_{CO}$ = 3.3$\pm$1.0, while the ionized and molecular gas shows more similar values (R$_{H\alpha}$/R$_{CO}$~=~1.5$\pm$0.8). 
According to the results derived for all the different tracers we found that in our local LIRGs the stellar emission is the most extended component, for which we derived a typical mean (median) value of 2.2$\pm$0.8 (2.4) kpc. The molecular gas is the most compact tracer (if we exclude the 1.3 mm continuum distribution), characterized by typical size of $\sim$0.7$\pm$0.3 (0.7) kpc. According to these results the molecular gas is a factor of 3 more compact than the stellar emission and a factor of 2 more compact than the ionized (H$\alpha$) gas. 
As discussed in \S~\ref{EXTINC}, larger H$\alpha$ size with respect to Pa$\alpha$ tracer are derived as a result of the extinction. Indeed, when comparing the molecular distribution with that traced by the ionized Pa$\alpha$ emission, the difference decreases, deriving similar effective radii for both tracers.

\begin{figure*}
\centering
\includegraphics[width=0.88\textwidth]{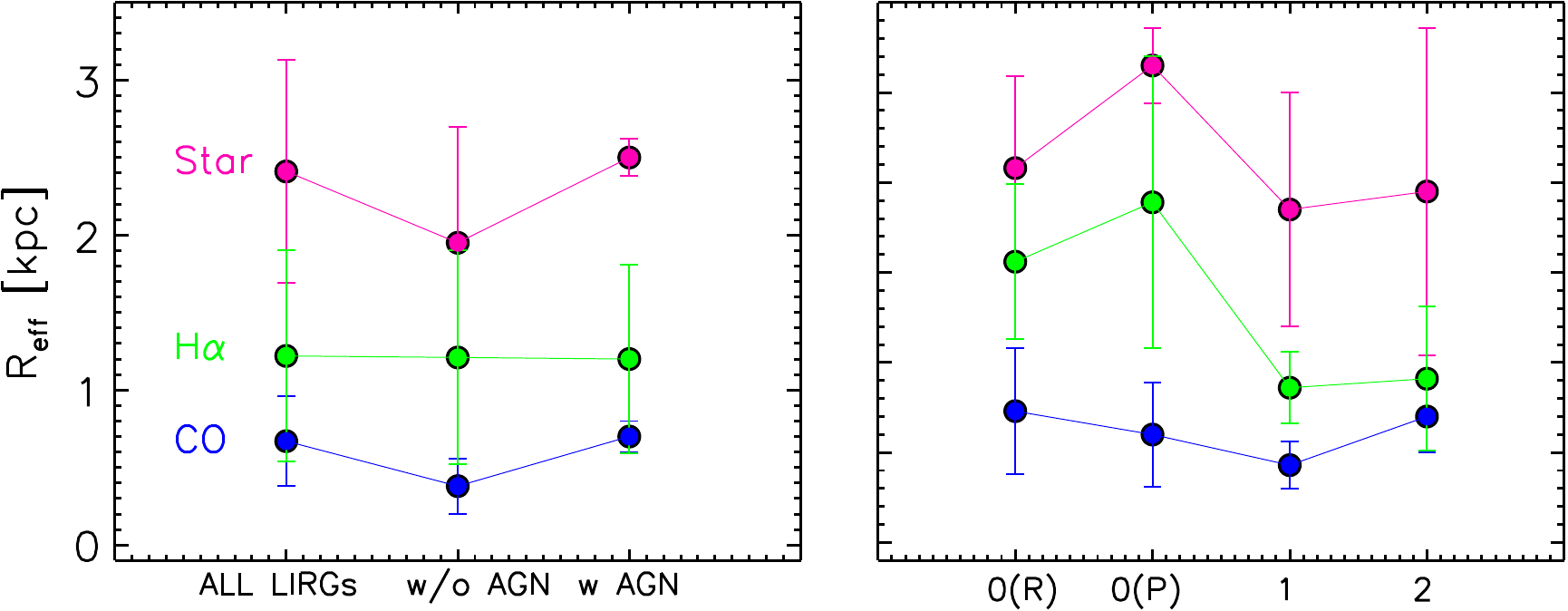}
\caption{Median effective radii (and MAD) derived for the different tracers and subsamples. The stellar, ionized and CO(2--1) R$_{eff}$ are shown in magenta, green and blue, respectively. {\it Left:} The median values for the whole sample are shown (i.e., `ALL LIRGs'), along with the median values derived for the LIRG subsample with (w AGN) and without (w/o AGN) an AGN. {\it Right:} The median R$_{eff}$ for the four subsamples defined in Tab.~\ref{Input_data} ({\tt 0(R) 0(P), 1, 2}) are shown.}
\label{Different_reff_samples} 
\end{figure*}
 
Furthermore, isolated (type {\tt 0}) objects seem to show less compact stellar and ionized (H$\alpha$) gas distributions ($\lesssim$2 kpc), while interacting and mergers are more compact (1-1.5 kpc; right panel in Fig.~\ref{Different_reff_samples}).  
On the other hand, the molecular gas size does not seem to depend on the specific type of galaxy, remaining constant ($\sim$0.7 kpc) among the different subgroups (see \S~\ref{rco_morpho}). Typical mean (and median) values of the different components are summarized in Tab.~\ref{Mix_radii_sample}.

\section{Discussion}
 
\subsection{LIRGs {\it versus} low-z ETGs, Spirals and ULIRGs}

The molecular and stellar sizes have been studied at low-z for Spirals (\citealt{Bolatto17}\footnote{In \cite{Bolatto17} interferometric CO(1--0) observations made with the Combined Array for Millimeter-wave Astronomy (CARMA) interferometer are presented, with galaxies taken from the Extragalactic Database for Galaxy Evolution survey (EDGE) at the angular resolution $\sim$4\arcsec-5\arcsec. A total of 126 galaxies (hereafter, Spirals EDGE-CALIFA) have been selected at a distance of $<$120 Mpc, deriving the molecular size R$_{CO}$ as the radius enclosing half the CO(1-0)~flux.} and \citealt{Leroy21}\footnote{\cite{Leroy21} analyzed a representative sample of 90 nearby {\it MS} SFGs at a distance of $<$90 Mpc observed by the Physics at High Angular resolution in Nearby Galaxies (PHANGS)-ALMA survey at the spatial resolution comparable to that of our LIRGs ($\sim$100 pc) observing the CO(2--1) emission, also deriving R$_{star}$ and M$_{star}$ among other parameters. Their galaxies are mostly Spirals, including both early- and late-type spirals, but not irregular galaxies. We maintain this nomenclature to distinguish the two Spiral samples.}) and ETGs (i.e., ellipticals and lenticulars; \citealt{Cappellari13, Davis13, Davis14}\footnote{ETG have been taken from the ATLAS$^{3D}$ {\it parent} sample, which consists of 871 galaxies: 68 ellipticals, 192 lenticular (a total of 260 ETGs) and 611 spiral and irregular galaxies galaxies. In \cite{Davis13} the R$_{CO}$ has been derived using CARMA data, which provide a spatial resolution of 4\arcsec-5\arcsec, to observe the CO(1--0) emission in a sample of $\sim$40 ETGs selected from the {\it parent} sample.}). Recently \cite{Pereira21} analyzed a local sample of ULIRGs using ALMA data, for which they derived the R$_{CO}$ and R$_{cont}$ sizes.
Here we have studied, for the first time, the corresponding molecular and stellar sizes for luminous SFGs located above the {\it MS} of local SFGs. Their relation with the SFR and stellar mass parameters is also investigated and compared within the aforementioned local systems.

\begin{figure*}
\centering
\includegraphics[width=0.98\textwidth]{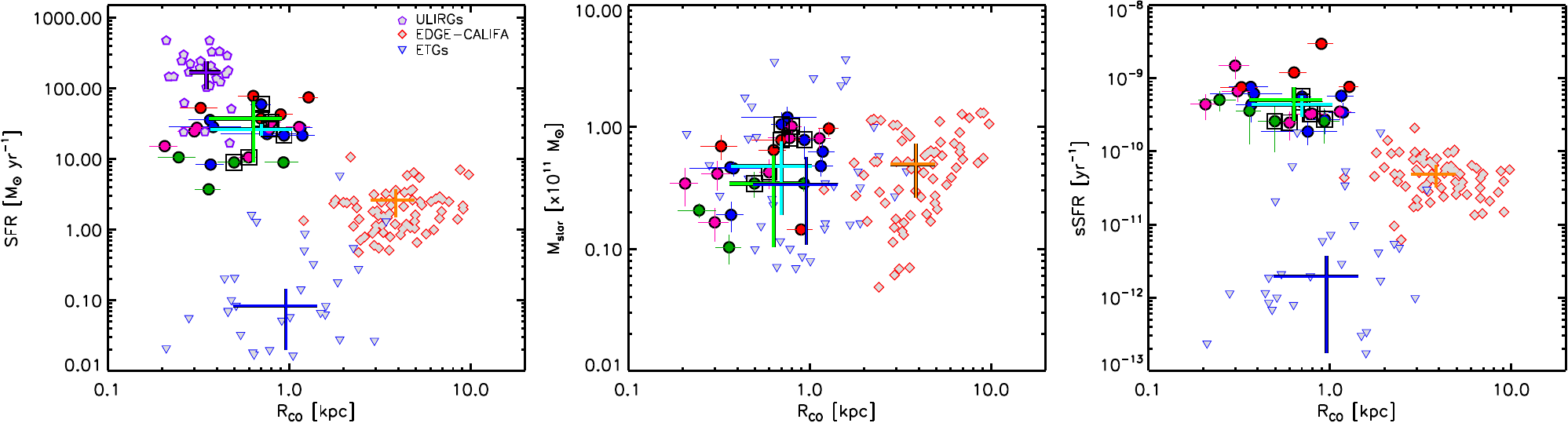}
\caption{Distribution of LIRGs and other low-z galaxy samples in the SFR-R$_{CO}$ (left) and M$_{star}$-R$_{CO}$ (middle) and sSFR-R$_{CO}$ (right) planes. Our LIRGs are shown according to the color-code presented in Fig.~\ref{z_LIR_sample}.
Local Spiral (gray red-contour diamonds) values are taken from the {\tt EDGE-CALIFA} survey (\citealt{Bolatto17}) while ETGs (blue contour down pointing triangles) are taken from the ATLAS$^{3D}$ sample (\citealt{Cappellari13}). ULIRGs (purple pentagons) values are derived from ALMA data (\citealt{Pereira21}).
The median values of each sample is shown according to the following color-code: purple for ULIRGs, light green for interacting and merger LIRGs, light blue for disky (i.e., {\tt RD} and {\tt PD}) LIRGs, orange for Spirals and dark blue for ETGs. In ULIRGs the AGN contribution has been removed to estimate their SFRs.
}
\label{low_lirg_spirlas}
\end{figure*}

\subsubsection{SFR {\it versus} R$_{CO}$}

LIRGs are completely decoupled from Spirals and ETGs in the SFR\footnote{For Spirals taken from the CALIFA sample the SFR has been derived using the H$\alpha$ emission corrected for the Balmer-decrement inferred extinction using a Salpeter IMF (see \citealt{Bolatto17} for further details). In normal Spirals the SFR is not severely affected by obscured star formation, that would invisible in the optical. To properly compare our results with those derived in their work for the stellar mass and SFR, we transformed their values to the Kroupa IMF.
The SFR (and stellar masses) of {\it MS} SFGs taken from the PHANGS-ALMA sample have been derived using a combination of {\it Galaxy Evolution Explorer (GALEX)} and {\it Wide-field Infrared Survey Explorer (WISE)} photometric bands and a Chabrier IMF (see also \citealt{Leroy19}). 
Finally, the SFR for ETGs has been derived using a combination of {\it WISE} 22 $\mu$m and {\it GALEX} far-ultraviolet emissions, using the Kroupa IMF (\citealt{Davis14}). The stellar mass has been obtained as described in \cite{Cap13a}.} 
to R$_{CO}$ diagram (see Fig.~\ref{low_lirg_spirlas} left panel). While Spirals\footnote{\cite{Leroy21} assumed that the stellar size in {\it MS} SFGs is similar to the molecular size, as expected for Spirals (\citealt{Wuyts11a}).} are extended systems with CO(2--1) radius of about 2 to 10 kpc, LIRGs have radii about a factor $\lesssim$6 smaller. Moreover, the size of LIRGs cover a range similar to that covered by ETGs, while their SFR is a factor 60 higher than for ETGs. 
The large discrepancy in the molecular size between Spirals and local LIRGs seems to be in agreement with what discussed in \cite{Bolatto17}: they found that galaxies that are more compact in the molecular gas than in stars tend to show signs of interaction (the presence of bar can also affect the emission distribution). The majority of our galaxy show the presence of interactions or past merger activity, for which type {\tt 0 PD} and {\tt 1} galaxies show the most compact molecular size. On the other hand, merger (type {\tt 2}) LIRGs show similar molecular size than rotating disks: thus interactions may have an important role in the compaction of the molecular size, although this requires further investigation. 
The work by \cite{Pereira21} seems to support the aforementioned result: they derived still more compact molecular size for their sources with respect to our LIRGs. Their sources are more extreme in terms of SFR ($\sim$340 M$_\odot$ yr$^{-1}$) and even show a more disturbed morphology. Their molecular size is a factor of 2 more compact than our LIRGs (see Tab.~\ref{Mix_radii_sample_3}).

The derived sSFR of the different samples follows a similar trend than that shown for the SFR parameter. Local Spirals show sSFR of $\sim$10$^{-10}$ yr$^{-1}$ while ETGs are characterized by a sSFR a factor of 100 lower than local Spirals.

\subsubsection{Stellar mass {\it versus} R$_{CO}$}

In the stellar mass to R$_{CO}$ diagram (Fig.~\ref{low_lirg_spirlas} right panel), LIRGs cover a mass range similar to that of low-z samples of Spirals and ETGs, with values in the 10$^{10}$-10$^{11}$~M$_\odot$ range.
Then, compared with different types of low-z galaxies, LIRGs are characterized by been located in galaxy hosts with intermediate stellar mass ($\sim$5$\times$10$^{10}$ M$_\odot$; Tab.~\ref{Mix_radii_sample_3}), forming stars at rates a factor $\gtrsim$10 above the Spirals, and with compact CO sizes of R$_{CO}$$\sim$0.7~kpc, similar to  that of ETGs (R$_{CO}$$\sim$1 kpc).

\subsubsection{Stellar mass - size plane}

\begin{figure}
\centering
\includegraphics[width=0.44\textwidth]{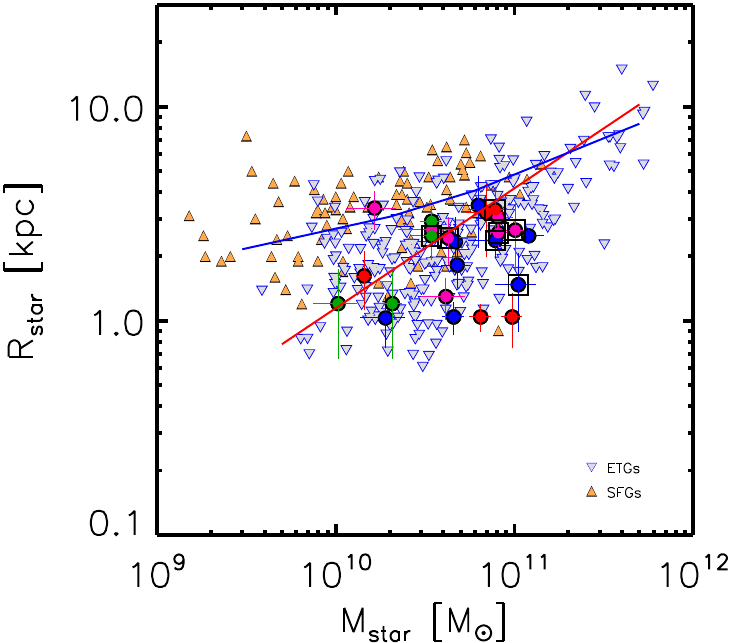}
\caption{
Mass-size distribution for our LIRGs including ETGs and local Spirals (i.e., {\it MS} SFGs; \citealt{Leroy21}). ETGs are identified using (blue contour) down pointing triangles and {\it MS} SFGs with up pointing orange triangles. LIRGs are identified following the color-code used in previous figures. The blue and red slopes represent the mass-size relations derived in \cite{Shen03} (see also \citealt{FL13}) for late- and early-type local galaxies, respectively.
}
\label{ETG_Sp_Mstar_Reff}
\end{figure}

A different trend than that derived for the stellar mass and molecular size component is derived for our sample when the stellar size is considered (Fig.~\ref{ETG_Sp_Mstar_Reff}). 
As before, local Spirals (i.e., {\it MS} SFGs from \citealt{Leroy21}) and ETGs are included in the diagram. 
LIRGs share similar stellar mass and stellar size with ETGs, while they are more compact (by a factor of $\sim$5) and more massive (by a factor of $\lesssim$2) than local {\it MS} SFGs (see Tab.~\ref{Size_summary_all}).
From \cite{Leroy13} we know that the distributions of molecular gas and stellar disk in galaxies follow each other closely in nearby disk galaxies (R$_{star}$$\sim$R$_{CO}$), while the stellar size is larger than the molecular size for ETGs and for our LIRGs, by a factor of $\lesssim$3.

According to evolutionary scenarios, many works support the idea that (U)LIRGs can transform gas rich Spirals into intermediate (stellar) mass (10$^{10}$-10$^{11}$ M$_\odot$) Ellipticals through merger events (\citealt{Genzel01, Tacconi02, Dasyra06, Dasyra_2_06, Kawakatu06, Cappellari13}). The kinematic study of local (U)LIRGs (\citealt{Bellocchi13}) highlights that these systems fill the gap between rotation-dominated Spirals and dispersion-dominated ETGs in the v/$\sigma$--$\sigma$ plane. 
Following our present results, most of the LIRGs share similar properties with ETGs while only a few overlap with the region covered by Spirals.

\subsection{LIRGs {\it versus} high-z SFGs}

\begin{figure*}
\centering
\includegraphics[width=0.93\textwidth]{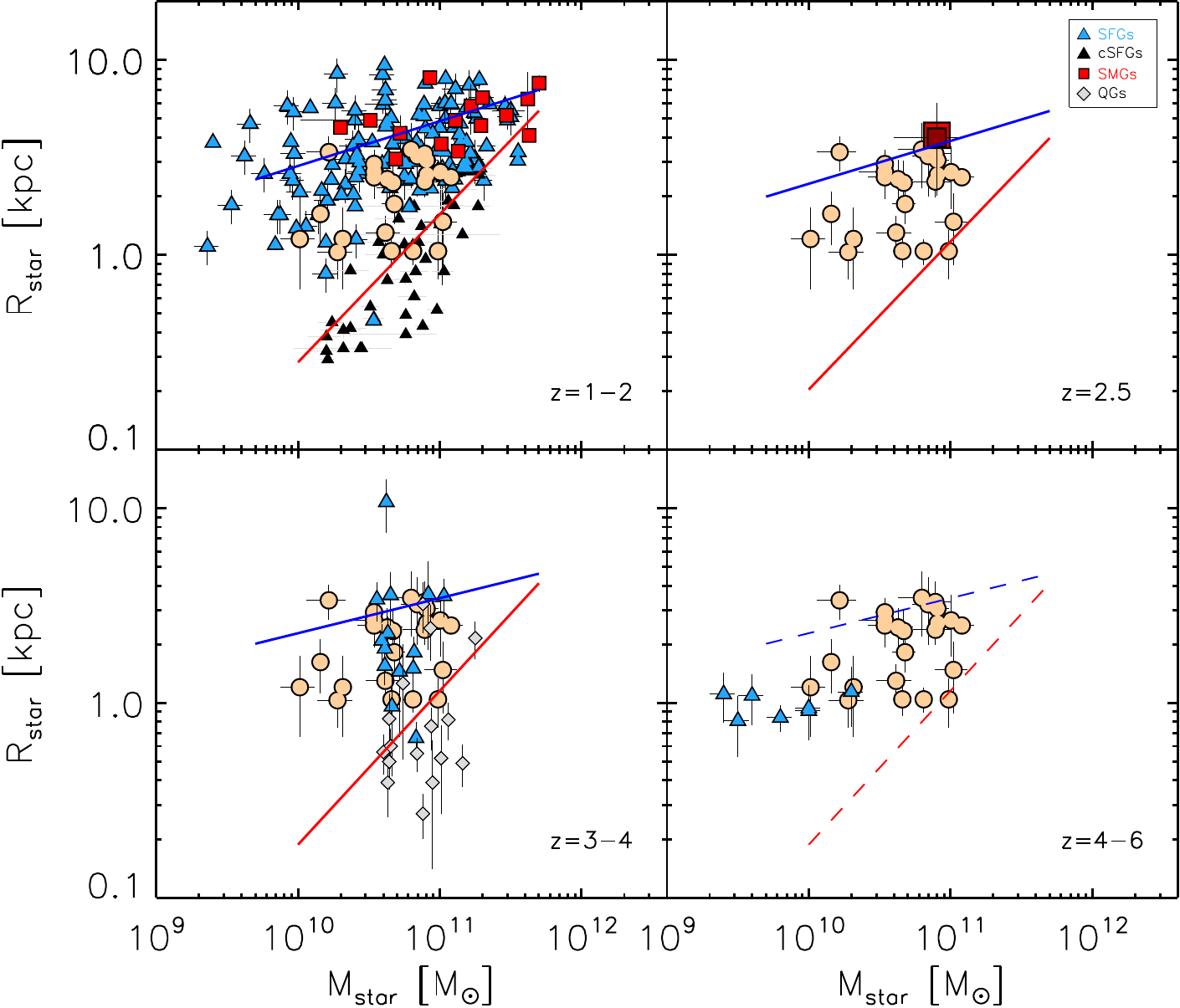}
\caption{Mass-size distribution for our LIRGs and high-z galaxies (Tab.~\ref{Mix_radii_sample_3}). 
LIRGs from this work are shown using light orange circles. High-z SFGs and SMGs data are taken from the following works: SMGs from \cite{Hodge16}, \cite{Chen17}, \cite{CR18}, \cite{Lang19} while SFGs are from \cite{Barro14}, \cite{Straatman15}, \cite{Tadaki17}, \cite{FS18}, \cite{Cheng20}, \cite{Fujimoto20}, \cite{Kaasinen20}, \cite{Valentino20} and \cite{Puglisi21}. The high-z data are shown following the color and symbol code shown in the legend. The blue and red lines in each redshift range represent the mass to size relations for late- and early-type galaxies, respectively, at {\it z}$\sim$~1.75, 2.25, 2.75 derived from the {\it 3D-HST}+{\it CANDELS} surveys (\citealt{vdWel14}). Since at {\it z}=4-6 the mass to size relation for late- and early-type galaxies has not been derived, the dashed blue and red lines still show the behavior considered at {\it z}=3-4.
}
\label{R_M_star_highz}
\end{figure*}

Deep cosmological imaging surveys have identified a wide variety of high-z galaxies ranging from quiescent systems (QGs; \citealt{Straatman15}) to compact SFGs (cSFGs; \citealt{Barro14, Barro16}), {\it MS} SFGs (\citealt{Straatman15}, \citealt{Tadaki17}, \citealt{FS18}, \citealt{Puglisi19}, \citealt{Cheng20}, \citealt{Kaasinen20}, \citealt{Valentino20} and \citealt{Puglisi21}\footnote{The sources analyzed in \cite{Valentino20} and \cite{Puglisi21} are far-IR selected encompassing the upper envelope of the main sequence and off {\it MS} galaxies.}), and more extreme starbursts which generally lie above the {\it MS}\footnote{\cite{Dudzeviciute20} found that at z$\sim$1 SMGs lie a factor of 6 above the {\it MS} while at z$\sim$4 they lie a factor of 2 above the {\it MS}, as a result of the strong evolution of the sSFR.}, as in the case of SMGs (\citealt{Hodge16}, \citealt{Chen17}, \citealt{CR18}, \citealt{Gullberg18}, \citealt{Lang19}, \citealt{Chen20}; see Tab.~\ref{Mix_radii_sample_3}).

All these systems have been selected using different criteria (i.e., stellar mass, optical radius or IR luminosity) and observational techniques also covering a broad range of redshifts (1~$\lesssim$~z~$\lesssim$~6) and galaxy properties. 
Therefore they represent the rich diversity of galaxies during the first half in the history of the Universe. The comparison between the results derived for our LIRGs and a compilation of measurements for different samples of high-z systems  will help to better understand how galaxies form and evolve at different cosmic times. In the following, such a comparison in the stellar mass-size plane (M$_{star}$-R$_{star}$), and in the sizes of the (molecular and ionized) ISM {\it versus} their stellar host is presented and discussed.

\begin{table*}
\begin{tiny}
\caption{Mean (and median) half-light radius of the different tracers along with their stellar mass for local and high-z systems.}
\label{Mix_radii_sample_3}
\hskip-0mm
\begin{tabular}{ccc|cccc|cc} 
\hline\hline\noalign{\smallskip}  
Sample & z 	 &	\# &   R$_{CO}$  &    R$_{cont}$ &  R$_{star}$ 	& R$_{H\alpha}$ & M$_{star}$ 	&	Ref.\\
& & &	[kpc] & [kpc]	&   [kpc] &  [kpc]  &  [$\times$10$^{10}$] M$_\odot$	&	\\
(1)		&	(2)		&(3)		&	(4)		&	(5)		&	(6)	& (7)	&(8)	&(9)\\
\hline\noalign{\smallskip}  
{\tt LIRG} & $\lesssim$0.02 & 24 & 0.66$\pm$0.33 (0.67)	&	0.37$\pm$0.31 (0.29)	&  2.21$\pm$0.81 (2.41) 	& 1.42$\pm$0.89 (1.37)	&	5.6$\pm$3.2 (4.8)	&	{\tt this work}	\\
{\tt E-S0} & 0.02 & 258 (49$^a$)	& 1.15$\pm$0.77 (1.01)	&	\dots	&	2.74$\pm$1.98 (2.20) & \dots & 7.3$\pm$9.6 (3.8)	&	{\tt C13, D13, D14}\\
{\tt Spiral} & $\lesssim$0.03 & 68 &		4.37$\pm$1.97 (3.86)	& \dots & \dots & \dots  &	7.6$\pm$5.0 (7.1) & {\tt Bo17}	\\
{\tt MS SFG} & $<$0.004  & 90 & \dots & \dots & 3.45$\pm$1.41 (3.40)  & -  & 2.9$\pm$2.8 (2.3)   & {\tt L21}\\
{\tt ULIRG}	& $<$0.17 & 30 (7, 23$^b$)& 0.33$\pm$0.09 (0.36) &  0.12$\pm$0.10 (0.11) &  3.45$\pm$1.79 (3.80) &  2.02$\pm$1.55 (1.58)	& \dots & {\tt P21, A12, B13}\\
\hline\noalign{\smallskip}  
\textcolor{cyan}{\tt SFG}	& 1-1.7 & 82 (72$^c$) & 1.95$\pm$1.23 (1.67)$^*$ &  \dots &  3.43$\pm$1.49 (3.10) &  \dots & 7.9$\pm$7.9 (5.7)	& {\tt V20, Pu21}\\
{\tt cSFG} & 2  & 45 & \dots & \dots&	1.26$\pm$0.88 (1.00) & \dots   &	8.1$\pm$5.5 (10.8) & {\tt Ba14} \\
\textcolor{cyan}{\tt SFG} & 2  & 3 & 4.90$\pm$1.31 (5.50) & 4.83$\pm$4.18 (3.90)  & 6.93$\pm$1.76 (7.90)  & \dots	& 20.0$\pm$9.5 (19.0)	& {\tt K20}\\
\textcolor{cyan}{\tt SFG} & 2  & 11 & \dots & 1.5$\pm$1.2 (1.2) &	 3.7$\pm$1.5 (3.1) & \dots &	15.6$\pm$6.2 (11.7) & {\tt T17} \\
\textcolor{cyan}{\tt SFG} & 2  & 38 & \dots & \dots&	3.82$\pm$2.20 (3.20) &\dots & 4.3$\pm$6.1 (2.2) & {\tt FS18} \\
\textcolor{cyan}{\tt SFG} & 2  & 4 & \dots & 4.63$\pm$0.65 (4.80) 	&	5.20$\pm$0.62 (5.20) & 3.43$\pm$1.46 (3.90)&	12.7$\pm$13.3 (8.3) & {\tt Cheng20} \\
\textcolor{cyan}{\tt SFG} & 0.7-2.7  & 280 & \dots & \dots 	&	3.39$\pm$1.63 (3.14)  & 4.26$\pm$2.74 (3.43) &	 \dots  & {\tt W20} \\
\hdashline\noalign{\smallskip}  
\textcolor{red}{\tt SMG} & 2  & 14 & \dots & 1.74$\pm$0.51 (1.95)	&	5.03$\pm$1.48 (4.75) & \dots    &	18.6$\pm$16.0 (13.2)& {\tt L19} \\
\textcolor{red}{\tt SMG} & 2  & 1 & 6.6$\pm$0.9  &  1.2$\pm$0.1 & 6.4$\pm$0.5&	6.6$\pm$0.9 & 20.0	& {\tt Chen17} \\
\textcolor{red}{\tt SMG}& 2  & (6) m & \dots   & 1.82$\pm$0.31 (1.85) & \dots  & 3.80$\pm$1.40 (3.95) &   16.0 & {\tt Chen20} \\
\hline\noalign{\smallskip}  
{\tt cSFG} & 2.5  & 6 & \dots &0.90$\pm$0.30 (0.81) &	1.8$\pm$0.93 (1.57) & \dots    &	\dots & {\tt Ba16} \\
\textcolor{cyan}{\tt SFG}   & 2-2.5 & 6&  \dots & \dots  &  \dots &  4.18$\pm$1.58 (3.85)	& 3.7$\pm$2.5 (2.5) & {\tt H21}\\
\textcolor{red}{\tt SMG} & 2.5  & (4) m & 3.8$\pm$0.1 & 1.7$\pm$0.1 &	4.0$\pm$2.0 & \dots &		8.0 & {\tt CR18} \\
\textcolor{red}{\tt SMG} & 2.5 & (16) m & \dots& 1.8$\pm$0.2 &	4.1$\pm$0.8 & \dots  &	8.0$\pm$1.0  & {\tt H16} \\
\hline\noalign{\smallskip}  
\textcolor{gray}{\tt QG} & 3-4  & 16 & \dots & \dots &	0.98$\pm$0.86 (0.58) & \dots&	8.2$\pm$3.9 (7.6) & {\tt S15} \\
\textcolor{cyan}{\tt SFG} & 3-4  & 14 &\dots & \dots  &	2.79$\pm$2.49 (2.09) &\dots &	5.5$\pm$2.0 (4.6) & {\tt S15} \\
\hline\noalign{\smallskip}
\textcolor{red}{\tt SMG} & 4--5  & 4 & - & 1.13$\pm$0.18 (1.05)	& \dots& \dots & \dots  & {\tt G18} \\    
\textcolor{cyan}{\tt SFG} & 4--6  & 18 (7$^c$) & \dots & \dots    &	0.98$\pm$0.14 (0.94) & \dots & 0.9$\pm$0.5 (0.8)  & {\tt F20, Fa20} \\
\hline\hline\noalign{\smallskip} 	
\end{tabular}
\end{tiny}
\vskip1mm 
\begin{minipage}{18.5cm}
\small
{{\bf Notes:} 
  Column~(1): Sample. The ones highlighted using different colors are the samples shown in Fig.~\ref{R_M_star_highz};
  Column~(2): Redshift;
  Column~(3): Number of galaxies in each sample. When `m' is shown it means that the {\it median value} is considered; (a) number of sources considered in \cite{Davis14} to derive R$_{CO}$; (b) number of ULIRGs considered in \cite{Bellocchi13} and \cite{Arribas12} to derive R$_{star}$ and R$_{H\alpha}$, respectively, from the VIMOS/VLT sample and combining VIMOS/VLT with the INTEGRAL/WHT samples; (c) number of sources for which the R$_{star}$ can be derived; 
  Column (4, 5, 6, 7): Effective radius derived for the CO(2--1), dust, stellar and ionized gas (i.e., H$\alpha$) emissions, respectively. In column 4 the ${`*'}$ symbol means that the molecular size has been derived as a combination of the CO, [CI] and dust tracers, as a robust average size of the cold ISM phase in their systems.  
  Column (8): Stellar mass.  
  Column (9): References with the following code: A12 \cite{Arribas12}, B13 \cite{Bellocchi13}, C13 \cite{Cappellari13}, D13 \cite{Davis13}, D14 \cite{Davis14}, Bo17 \cite{Bolatto17}, K20 \cite{Kaasinen20}, L19 \cite{Lang19}, Chen17 \cite{Chen17}, Chen20 \cite{Chen20}, Cheng20 \cite{Cheng20}, T17 \cite{Tadaki17}, Ba14 \cite{Barro14}, Ba16 \cite{Barro16}, CR18 \cite{CR18}, H16 \cite{Hodge16}, S15 \cite{Straatman15}, FS18 \citealt{FS18}, G18 \cite{Gullberg18}, F20 \cite{Fujimoto20}, Fa20 \cite{Faisst20}, V20 \cite{Valentino20}, W20 \cite{Wilman20}, L21 \cite{Leroy21}, P21 \cite{Pereira21}, H21 \cite{Hogan21}, Pu21 \cite{Puglisi21}.
}
\end{minipage}
\end{table*}

\begin{table*}
\centering
\begin{small}
\caption{Ratios of the molecular, ionized and stellar ISM distributions of different types of galaxies relative to those of LIRGs.}
\label{Size_summary_all}
\begin{tabular}{ccccccccc} 
\hline\hline\noalign{\smallskip} 	  
{\tt Parameter} & \multicolumn{8}{c}{\tt Sample} \\\cline{1-1}\cline{2-9} 	\noalign{\smallskip}
 & \multicolumn{5}{c}{\tt Low-z}  & \multicolumn{3}{c}{\tt High-z} \\\noalign{\smallskip} 	
\cmidrule(lr){2-6} \cmidrule(lr){7-9} 
  &\textcolor{blue}{\tt LIRGs} & {\tt Spirals}  &  {\tt MS SFGs} & {\tt ETGs}&  {\tt ULIRGs} & {\tt SMGs} & {\tt SFGs} & {\tt cSFGs} \\\noalign{\smallskip} 	
 (1)	& (2) & (3) & (4) & (5) & (6) & (7)  & (8) & (9)\\\hline\hline\noalign{\smallskip} 
{\tt R$_{mol}$} & \textcolor{blue}{\tt $\sim$0.7 kpc} & $\times$6   &	$\times$5 &$\sim$1 &	$\times$0.5&	$\times$7.8& $\times$2.6  &	\dots     \\
\hline\noalign{\smallskip} 	
{\tt R$_{ion}$ }  & \textcolor{blue}{\tt $\sim$1.4 kpc} &\dots & \dots & \dots & $\sim$1 	& $\times$2.7 & $\times$2.4	& \dots \\
\hline\noalign{\smallskip} 	
{\tt R$_{star}$}& \textcolor{blue}{\tt $\sim$2.4 kpc} & $\times$1.6 &$\times$1.4  &$\sim$1  &  $\times$1.6	&  $\times$1.9 	&$\times$1.3	&	$\times$0.5 \\
\hline\noalign{\smallskip} 	
{\tt M$_{star}$} & \textcolor{blue}{\tt $\sim$5$\times$10$^{10}$ M$_\odot$}  &$\times$1.5  & $\times$0.5 & $\sim$1	&  \dots    &$\times$2.7 	& $\times$0.9 	& $\times$2 	\\
\hline\hline\noalign{\smallskip} 	
\end{tabular}
\end{small}
\vskip2mm\begin{minipage}{18cm}
\small
{{\bf Notes:} Column (1): Parameters considered in the comparison; Column (2): (Median) size of the different tracers as well as the (median) stellar mass of our LIRG sample;
Columns~(3-9): Scaling factors derived when comparing the size of the different tracers and stellar mass of low- (Columns 3-6) and high-z (Columns 7-9) samples with those derived for our LIRGs. The samples involved in the comparison are: Spirals from the {\tt EDGE-CALIFA} (Column 3) and {\tt PHANGS-ALMA} samples (Column 4), ETGs (Column 5), local ULIRGs (Column 6), high-z SMGs (Column 7), high-z SFGs (Column 8) and high-z compact SFGs (Column 9; see text for details). Ellipsis dots (\dots) indicate missing values.
}
\end{minipage}
\end{table*}

\subsubsection{LIRGs {\it versus} high-z SFGs. Stellar hosts}

The vast majority of high-z galaxies have masses in the $\sim$10$^{10}$-2$\times$10$^{11}$ M$_\odot$ mass range, but their sizes cover a much wider range, which reflects the large variety of systems considered at high-z (Fig.~\ref{R_M_star_highz}, Tab.~\ref{Mix_radii_sample_3}). 
Indeed, the stellar hosts can be divided in three broad classes according their effective radii: (i) compact hosts with radii of less than 1 kpc, (ii) star-forming galaxies with intermediate sizes between 1 and 4 kpc, and (iii) extended hosts with radii above 4 kpc and up to 10 kpc. The stellar mass and size of the low-z LIRGs are consistent with that of the intermediate size SFGs and well differentiated from both the compact and the extended hosts. Among high-z systems, SMG hosts are significantly more massive than local LIRGs by a (median) factor of $\sim$3 and also more extended (i.e., stellar size) up to a factor of $\sim$2 (Tab.~\ref{Size_summary_all}). On the other hand, low-z LIRGs appear a factor about two larger than the massive, H-band selected compact SFGs at redshift 2 (\citealt{Barro14}). 

The comparison with {\it MS} SFGs indicates that low-z LIRGs have stellar sizes and masses similar to those of K-band selected SFGs at redshifts 3-4 (\citealt{Straatman15}). However, LIRGs  appear (on average) slightly smaller in size (factor 1.3) than the {\it MS} SFGs at redshifts $\sim$1-2 (\citealt{FS18}, \citealt{Valentino20}, \citealt{Puglisi21}) while their masses (average of 5.6$\times$ 10$^{10}$ M$_\odot$) are within the wide range of stellar masses (from 4.3 to 20.0 $\times$ 10$^{10}$ M$_\odot$) covered by the hosts of the {\it MS} SFGs. These results indicate that LIRGs, that represent the bursty above-{\it MS} SFGs at low-z, appear to be similar to the bulk population of {\it MS} SFGs at intermediate redshifts (z$\sim$1-4) when it comes to the stellar mass and size of their hosts.

\cite{Wang18} proposed a two-step scenario to explain how local galaxies evolve from extended star-forming galaxies (eSFGs), through compact star-forming galaxies (cSFGs), to a quenched galaxies (QGs). According to this scenario, eSFGs are transformed into cSFGs through {\tt compaction} mechanisms like minor mergers or interactions with close companions. These mechanisms play a role in enhancing star formation and contributing to the build up of the stellar cores or bulges. At low-z the compaction processes are more gentle and have longer timescales than those found in the high-z (z$\sim$1) Universe (e.g., \citealt{Zolotov15}), mainly triggered by major merger events.
Then, the {\tt quenching} mechanism is needed to consume or lose the cold gas in these systems, although they are still able to sustain their star formation activity and the central stellar mass assembly. 
Finally, from cSFGs to QGs, they proposed that a strong dissipational processes are invoked in which the systems consume all their cold gas and quench their SFR as well as the build up of their bulges or stellar cores.

At high-z, a similar scenario has been suggested by \cite{Barro14}, who found that compact SFGs (cSFGs) at z$\sim$2-3 could be the natural progenitors of compact QGs (cQGs) at z$\sim$2. In this respect, it is interesting to note that all the low-z LIRGs as well as a large fraction of the high-z galaxies appear in the stellar mass to size diagram (Fig.~\ref{R_M_star_highz}) in between the relations of early- and late-type galaxies at {\it z} $\sim$ 1.75 (typical redshift value for the majority of the high-z systems considered in this work), as derived from the {\it 3D-HST} and {\it CANDELS} surveys (\citealt{vdWel14} and references therein). 
As observed in local LIRGs, distant SMG populations also show a mixture of dynamical phases, hosting merger-driven starbursts (\citealt{Aguirre13}) as well as ordered rotating disks (\citealt{Hodge16}). 
This could indicate that many of the high-z galaxies could be in a transitory phase related to tidally perturbed disks or galaxies involved in interactions and mergers, as low-z LIRGs. Whether this transitory phase could represent the evolution from extended disks to compact spheroids requires also spatially resolved kinematical information traced by the molecular and ionized ISM.

\subsubsection{LIRGs {\it versus} high-z SFGs. Sizes of the (un)obscured star formation traced by molecular and ionized ISM}

\begin{figure*}
\vskip5mm
\centering
\includegraphics[width=1\textwidth, angle=180]{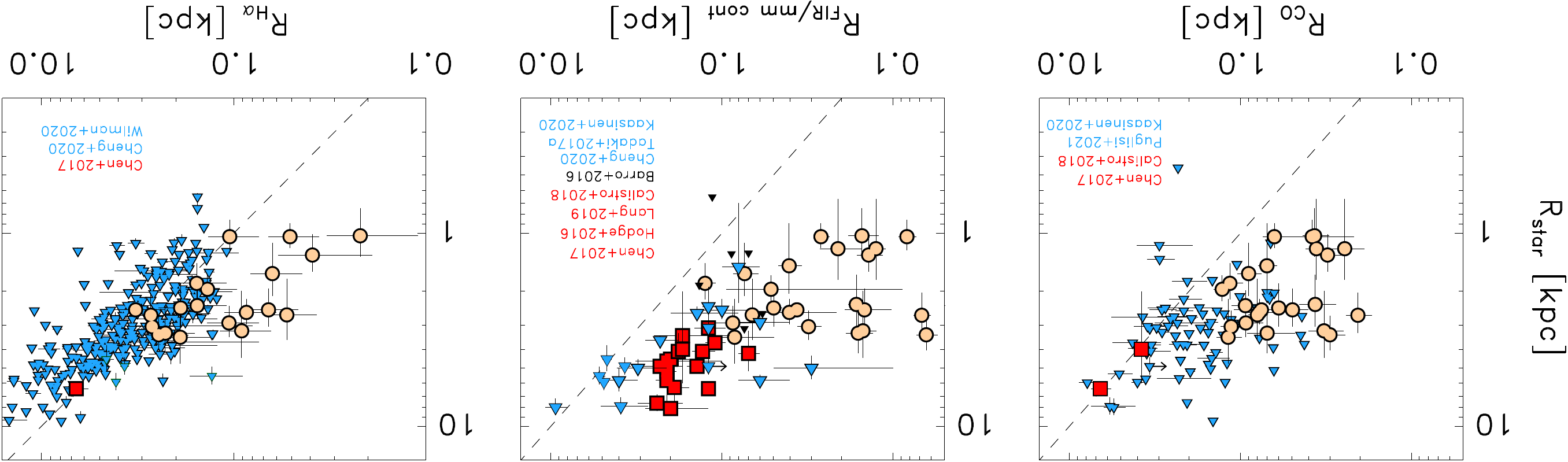}
\caption{Ratios of different tracers in high-z (z=2-2.5) galaxies samples relative to the low-z LIRG sample (Tab.~\ref{Mix_radii_sample_3}). {\it Left:} stellar {\it versus} molecular sizes. {\it Middle:} stellar {\it versus} mm/sub-mm (dust-)continuum sizes. {\it Right:} stellar {\it versus} ionized gas sizes. Our sample is shown using light orange circles, while the high-z galaxies are shown following the same symbol and color code shown in Fig.~\ref{R_M_star_highz}. In each panel we report the references from which the high-z data are taken. R$_{mm\hskip1mm cont}$ derived for our LIRGs were derived from our 1.3 mm ALMA observations, that we considered as lower limits due to the limited sensitivity of our observations.}
\label{R_mix_highz}
\end{figure*}

To further explore these evolutionary scenarios empirically, a direct comparison between the tracers of the ISM and the host galaxy is required both at low- and high-z. While the properties of the stellar hosts are traced by the optical and near-infrared continuum, the raw molecular material that will be transformed into stars is traced by the CO emitting gas, while the active regions of obscured and unobscured star-formation are traced by the continuum dust emission and hydrogen emission lines, respectively. Thus, the effective radius of the dust, molecular and ionized gas, relative to the stellar light distribution in these galaxies will provide a key information about how galaxies build their stellar mass, and how   young, massive starbursts could impact the stellar host through stellar winds, and affect its evolution (see Tab.~\ref{Mix_radii_sample_3} and Fig.~\ref{R_mix_highz}). However, the simultaneous information of all these tracers is still very limited for high-z samples, in particular when it comes to the ionized component of the ISM traced by the hydrogen lines. This will change in the near future with the advent of spectroscopy in the near- and mid-infrared spectral range with the James Webb Space Telescope (JWST).
As the measurements are very limited for high-z samples, we grouped the various samples into three main categories, independent of their redshift: i) the compact SFG, selected as bright H-band selected galaxies, iii) the regular SFGs, that are mostly galaxies classified as {\it MS} SFGs at their respective redshift range, and iii) the SMGs, that are extreme starbursts above the main sequence of SFGs (see Tabs.~\ref{Mix_radii_sample_3} and ~\ref{Size_summary_all} for details).

LIRGs appear as more compact than high-z SFGs in their molecular gas distribution (factor 2.6 smaller) while having a slightly smaller size (factor 1.3 smaller) in their stellar host (Fig.~\ref{R_mix_highz}, left). The difference is even larger when comparing with the few SMGs and extended SFGs with available data. SMGs and extended SFGs are far larger (factors $\sim$8) than LIRGs in their molecular gas while only two times larger in their hosts. In summary, the distribution of the molecular gas (i.e. the raw material for the formation of new stars) in high-z galaxies is more extended than in low-z LIRGs by factors 2.6 to 7.8 in extremely extended galaxies. Moreover, while all LIRGs appear to be located far away from the 1:1 stellar-to-molecular radius relation, the high-z SFGs and SMGs tend to be closer, with similar CO and stellar sizes in several systems. 
Up to now, the number of high-z galaxies with available measurements of the molecular, ionized and stellar light distributions is quite low, such that the relations found here should be investigated further with larger samples to draw firm conclusions. 

If the molecular size (CO) traces the regions of future {\it in-situ} star formation in galaxies, this result could indicate a key difference in the process of star formation and evolution of high-z galaxies with respect to that of starburst galaxies in the nearby Universe: while the star formation in LIRGs is concentrated in the central regions of the host, in high-z systems, the star-formation can proceed over the entire size of the host. However to validate this scenario, the size of the active regions of star-formation, both obscured and unobscured, relative to the stellar host needs to be established. This comparison can be obtained by measuring the size of the far-infrared continuum emitting region (Fig.~\ref{R_mix_highz}, middle) and of the hydrogen recombination lines H{$\alpha$} (Fig.~\ref{R_mix_highz}, right), that are considered tracers of the dust and ionized gas emission associated with dust-enshrouded and unobscured star-formation, respectively.

Several works have studied the dust-continuum emission at 870 $\mu$m in high-z SFGs as a proxy for dust-obscured star formation (e.g., \citealt{Simpson15}, \citealt{Lang19}). This component is found to be compact and centrally concentrated, being a common feature among massive ($\sim$10$^{11}$ M$_\odot$) high-z SFGs and SMGs (e.g., \citealt{Simpson15}, \citealt{Hodge16, Chen17, Tadaki17, CR18, Kaasinen20}; see Tab.~\ref{Mix_radii_sample_3}).  
Unlike local Spirals (e.g. KINGFISH sample; \citealt{Hunt15}), for which the dust shares similar scales as the stellar disks (R$_{dust}$$\sim$R$_{star}$), these high-z SF systems are characterized by centrally enhanced FIR continuum emission. 
The compactness of the dust emission with respect to the stellar host in high-z SFGs might be due to the high gas fractions observed in the high-z systems compared to z$\sim$0 objects (\citealt{Lang19}). This will involve the presence of very intense and highly obscured star-formation with the subsequent growth of stellar mass, mostly in the central regions in these galaxies.
However, high-z SFGs and SMGs do not appear in general as compact as low-z LIRGs (see Fig.~\ref{R_mix_highz} where R$_{cont}$ values at 1.3 mm for the sample of LIRGs is also presented). Even if the sizes measured in LIRGs (average R$_{cont}$ of 0.37 kpc) could be considered as a lower limit to the real size due to sensitivity effects, the size would not likely be larger than a factor $\sim$2, if the dust distribution follows the molecular gas emission (average R$_{CO}$ of 0.66 kpc). Moreover, the sizes of the FIR continuum emission in LIRGs are similar to those derived in a sample of most luminous LIRGs and ULIRGs using the 33 GHz radio emission (\cite{BM17}, GOALS survey). This could therefore represent an intrinsically structural distinction between LIRGs and high-z SFG in that most high-z systems appear as to far more extended than LIRGs in their dust emission relative to the size of their host. However, the number of high-z galaxies with high-angular resolution ($\leq$0.1\arcsec-0.2\arcsec) ALMA observations is still scarce, and data on larger samples are required before drawing firm conclusions.

The size of the ongoing (unobscured) star formation in high-z SFGs as well as in low-z LIRGs is traditionally traced by the H$\alpha$ emission line in the optical (Fig.~\ref{R_mix_highz}, right panel; \citealt{Chen17, Cheng20, Wilman20}). 
Detection of other FIR lines like the [CII]158$\mu$m emission line with high angular resolution has been also studied in systems in the redshift range 4 to 6 as part of the ALPINE survey (\citealt{Fujimoto20}), for which a typical size of R$_{[CII]}$$\sim$2.2 kpc has been derived. However, [CII] not only traces ionized gas but also the PDR interface between the atomic and molecular gas phases (\citealt{Lagache18}, \citealt{Zanella18} \citealt{Heintz21} and references therein).

The sizes of the ionized gas relative to that of FIR continuum emission in low-z LIRGs and high-z SFGs show some relevant differences. On the one hand, the size of the ionized gas is in general larger than that of the FIR emission (e.g., \citealt{Chen17, Chen20}), and closer to the size of the host galaxy (i.e., closer to the 1:1 relation represented in Fig.~\ref{R_mix_highz} right; see \citealt{Chen17, Cheng20, Wilman20}). 
As presented by \cite{Popping21} in their simulations, the smaller observed-frame 850 $\mu$m half-light radius compared to the observed-frame 1.6 $\mu$m half-size can be explained as the result of obscuration of the stellar emission at 1.6 $\mu$m.
Indeed, the dust plays an important role in attenuating the 1.6 $\mu$m emission in the inner regions of the galaxies (flatter profile), deriving larger stellar size with respect to the case where no dust attenuation is considered. 
The compactness increases with the redshift because the observed H-band emission traces bluer rest-frame wavelengths, which are more affected by the dust attenuation at increasing redshift.

On the other hand, while a fraction of the LIRGs and high-z SFGs share similar sizes (1 to 4 kpc effective radius) in their ionized gas distribution, the two samples are well differentiated in their overall distribution.
A substantial fraction of high-z SFGs have very extended H$\alpha$ emitting regions with effective radii of up to 10 kpc, and even larger. 
This is in agreement with the results found by \cite{Nelson16}. In their work they analyzed a large sample of SFGs observed at z$\sim$1 and found that their ionized H$\alpha$ emission is more extended than the stellar continuum emission, demonstrating that galaxies at this epoch are growing in size due to star formation. Their result is consistent with inside-out assembly of galactic disks.

On the contrary, half of the low-z LIRGs have sizes of less than 1 kpc, not present in high-z SFGs. If the ionized gas represents the size of the ongoing unobscured star formation regions, and associated extended nebulae due to outflows and stellar winds, these results would indicate that in general high-z systems are forming stars in regions distributed over the entire host galaxy while low-z LIRGs is concentrated in smaller central regions. However, the size of the ionized gas is larger than that of the stellar host in an important fraction of the high-z systems (\citealt{Wilman20}; see Fig.~\ref{R_mix_highz}, right panel). In these systems the size of the ionized gas could not necessarily trace the real distribution of the star-forming clumps but rather the extended ionized nebulae in the gas rich environment as expected in high-z galaxies. Similar to the extended ionized nebulae around quasars, ionizing photons could reach further out and ionize regions well outside the star-forming clumps, even beyond the size of the stellar host.

\section{Conclusions}

We present the first study which provides detailed measurements of the size (i.e., effective radius) of the molecular gas traced by CO(2--1) in a sample of 21 LIRG systems at low-z. To this aim, we used high resolution ALMA data, which allow us to reach sub-kpc spatial resolution scale ($<$100 pc, i.e. size of GMCs). The sample encompasses a wide variety of morphological types, suggesting different dynamical phases (isolated Spirals, interacting galaxies, and ongoing and post-coalescence mergers). All LIRGs are characterized by SFR and stellar masses that place them `above' MS plane at low-z. The sample also represents the closest analogs to the intermediate- and high-z IR luminous star-forming galaxies.

We performed a comprehensive study of the molecular, stellar (\citealt{Bellocchi13}) and ionized (\citealt{Arribas12}) gas distributions and their relative sizes. A comparison samples of local galaxies and high-z systems have also been included, in order to place the low-z LIRGs in a general context. 
The main results of the present study can be summarized as follows:

 \vskip5mm
 \begin{itemize}
 \item[]{\tt Low-z LIRGs:} 
\vskip2mm
 \item[$\bullet$]  
  The molecular gas distribution as traced by the CO(2--1) line is compact in local LIRGs and concentrated in the central regions, with typical (median) R$_{CO}$~$\sim$~0.7 kpc. The stellar host and the ionized gas distribution in these systems are factors of $\sim$3.5 and $\sim$2 larger than the molecular distribution. The continuum size at 1.3 mm is half the molecular size, but this should be considered as a lower limit because it is affected by sensitivity effects;
\vskip5mm

\item[]{\tt Comparison with low-z Spirals, ETGs and ULIRGs:}
\vskip2mm
\item[$\bullet$] LIRGs are indistinguishable from ETGs in the M$_{star}$-R$_{star}$ plane, sharing a similar range in their stellar mass and size. However, LIRGs have a stellar host more compact than that of local Spirals, by a factor 1.6. This seems to support the evolutionary scenario in which LIRGs can transform Spirals to Ellipticals (\citealt{Genzel01, Tacconi02, Dasyra06}); 
\vskip2mm
\item[$\bullet$] LIRGs are well separated from low-z Spirals and ETGs in the SFR {\it versus} CO size plane as a consequence of their SFR and size: the molecular size of LIRG is $\sim$6 times more compact than local Spirals of similar stellar mass, while similar to local ETGs, although characterized by higher SFR.
\vskip2mm
\item[]{\tt Comparison with high-z SFGs:}
\vskip2mm
\item[$\bullet$] LIRGs, representing the population of bursty above-{\it MS} SFGs at low-z, appear to be similar to the bulk population of {\it MS} SFGs at intermediate redshfit (z$\sim$1-4) in terms of the stellar mass and the size of their host galaxies;
\vskip2mm
\item[$\bullet$] LIRGs appear to be more compact than high-z SFGs in their molecular gas distribution (factor 2.6 smaller) while having a slightly smaller size (factor 1.3 smaller) in their stellar host. Also, while all LIRGs appear to lie a factor of $\sim$3 above the 1:1 stellar-to-CO radius relation, the high-z SFGs tend to be closer, that is with similar CO and stellar sizes in several galaxies (i.e., \citealt{Chen17}, \citealt{CR18}, \citealt{Kaasinen20} and \citealt{Puglisi21});
\vskip2mm
\item[$\bullet$] The size of the ionized gas distribution in LIRGs and high-z SFGs is larger than the distribution of the FIR emission, and is closer to the size of the host galaxy (i.e., closer to the 1:1 stellar to ionized gas radius relation). 
Excluding a small fraction of the LIRGs which share similar ionized gas distribution (1 to 4 kpc effective radius) with the high-z SFGs analyzed in \cite{Wilman20}, these two populations show different distributions, with high-z systems being found on the 1:1 relation. 
A substantial fraction of high-z SFGs have very extended H$\alpha$ emitting regions with effective radii of up to 10 kpc (\citealt{Nelson16}), and even larger, with no LIRGs with sizes above 4 kpc. On the other hand, half of the low-z LIRGs are less than 1 kpc in size, while no high-z SFGs of this size are found in our sample.

\end{itemize}

\begin{acknowledgements}

We thank the anonymous referee for useful comments and suggestions which helped us to improve the quality and presentation of the manuscript.
EB and AL acknowledge the support from Comunidad de Madrid through the Atracci\'on de Talento grant 2017-T1/TIC-5213. This research has been partially funded by the Spanish State Research Agency (AEI) Project MDM-2017-0737 Unidad de Excelencia `Mar\'ia de Maeztu'- Centro de Astrobiolog\'ia (INTA-CSIC). 
MPS acknowledges support from the Comunidad de Madrid through the Atracci\'on de Talento Investigador Grant 2018-T1/TIC-11035 and PID2019-105423GA-I00 (MCIU/AEI FEDER, UE). 
AAH, MVM and SGB acknowledge support from grant PGC2018-094671-B-I00 (MCIU/AEI/FEDER,UE). SGB acknowledges support from the research project PID2019-106027GA-C44 of the Spanish Ministerio de Ciencia e Innovaci\'on.
DR acknowledges support from STFC through grant ST/S000488/1.
AL, LC and SA acknowledge the support from grant PID2019-106280GB-I00 (MCIU/AEI/FEDER,UE).
SC acknowledges financial support from the State Agency for Research of the Spanish MCIU through the Center of Excellence Severo Ochoa award to the Instituto de Astrof\'isica de Andaluc\'ia (SEV-2017-0709).
MSG acknowledges support from the Spanish Ministerio de Econom\'ia y Competitividad through the grants BES-2016-078922 and ESP2017-83197-P.
AP gratefully acknowledges financial support from STFC through grant ST/T000244/1. 
FV acknowledges support from the Carlsberg Foundation Research Grant CF18-0388 `Galaxies: Rise and Death'. The Cosmic Dawn Center (DAWN) is funded by the Danish National Research Foundation under grant No. 140. 
AU acknowledges support from the Spanish grants PGC2018-094671-B-I00, funded by MCIN/AEI/10.13039/501100011033 and by ``ERDF A way of making Europe'', and PID2019-108765GB-I00, funded by MCIN/AEI/10.13039/501100011033.

\end{acknowledgements}

\bibliographystyle{aa} 
\bibliography{biblio}

\begin{thebibliography}{129}
\expandafter\ifx\csname natexlab\endcsname\relax\def\natexlab#1{#1}\fi

\bibitem[{{Aguirre} {et~al.}(2013){Aguirre}, {Baker}, {Menanteau}, {Lutz}, \&
  {Tacconi}}]{Aguirre13}
{Aguirre}, P., {Baker}, A.~J., {Menanteau}, F., {Lutz}, D., \& {Tacconi}, L.~J.
  2013, \apj, 768, 164

\bibitem[{{Alonso-Herrero} {et~al.}(2009){Alonso-Herrero},
  {Garc{\'\i}a-Mar{\'\i}n}, {Monreal-Ibero}, {Colina}, {Arribas},
  {Alfonso-Garz{\'o}n}, \& {Labiano}}]{AAH09}
{Alonso-Herrero}, A., {Garc{\'\i}a-Mar{\'\i}n}, M., {Monreal-Ibero}, A.,
  {et~al.} 2009, \aap, 506, 1541

\bibitem[{{Alonso-Herrero} \& {Knapen}(2001)}]{AAH01}
{Alonso-Herrero}, A. \& {Knapen}, J.~H. 2001, \aj, 122, 1350

\bibitem[{{Alonso-Herrero} {et~al.}(2012){Alonso-Herrero}, {Pereira-Santaella},
  {Rieke}, \& {Rigopoulou}}]{AAH12}
{Alonso-Herrero}, A., {Pereira-Santaella}, M., {Rieke}, G.~H., \& {Rigopoulou},
  D. 2012, \apj, 744, 2

\bibitem[{{Alonso-Herrero} {et~al.}(2006){Alonso-Herrero}, {Rieke}, {Rieke},
  {Colina}, {P{\'e}rez-Gonz{\'a}lez}, \& {Ryder}}]{AAH06}
{Alonso-Herrero}, A., {Rieke}, G.~H., {Rieke}, M.~J., {et~al.} 2006, \apj, 650,
  835

\bibitem[{{Alonso-Herrero} {et~al.}(2002){Alonso-Herrero}, {Rieke}, {Rieke}, \&
  {Scoville}}]{AAH02}
{Alonso-Herrero}, A., {Rieke}, G.~H., {Rieke}, M.~J., \& {Scoville}, N.~Z.
  2002, \aj, 124, 166

\bibitem[{{Armus} {et~al.}(2009){Armus}, {Mazzarella}, {Evans}, {Surace},
  {Sanders}, {Iwasawa}, {Frayer}, {Howell}, {Chan}, {Petric}, {Vavilkin},
  {Kim}, {Haan}, {Inami}, {Murphy}, {Appleton}, {Barnes}, {Bothun}, {Bridge},
  {Charmandaris}, {Jensen}, {Kewley}, {Lord}, {Madore}, {Marshall},
  {Melbourne}, {Rich}, {Satyapal}, {Schulz}, {Spoon}, {Sturm}, {U}, {Veilleux},
  \& {Xu}}]{Armus09}
{Armus}, L., {Mazzarella}, J.~M., {Evans}, A.~S., {et~al.} 2009, \pasp, 121,
  559

\bibitem[{{Arribas} {et~al.}(2004){Arribas}, {Bushouse}, {Lucas}, {Colina}, \&
  {Borne}}]{Arribas04}
{Arribas}, S., {Bushouse}, H., {Lucas}, R.~A., {Colina}, L., \& {Borne}, K.~D.
  2004, \aj, 127, 2522

\bibitem[{{Arribas} {et~al.}(2012){Arribas}, {Colina}, {Alonso-Herrero},
  {Rosales-Ortega}, {Monreal-Ibero}, {Garc{\'\i}a-Mar{\'\i}n},
  {Garc{\'\i}a-Burillo}, \& {Rodr{\'\i}guez-Zaur{\'\i}n}}]{Arribas12}
{Arribas}, S., {Colina}, L., {Alonso-Herrero}, A., {et~al.} 2012, \aap, 541,
  A20

\bibitem[{{Arribas} {et~al.}(2014){Arribas}, {Colina}, {Bellocchi}, {Maiolino},
  \& {Villar-Mart{\'\i}n}}]{Arribas14}
{Arribas}, S., {Colina}, L., {Bellocchi}, E., {Maiolino}, R., \&
  {Villar-Mart{\'\i}n}, M. 2014, \aap, 568, A14

\bibitem[{{Arribas} {et~al.}(2008){Arribas}, {Colina}, {Monreal-Ibero},
  {Alfonso}, {Garc{\'\i}a-Mar{\'\i}n}, \& {Alonso-Herrero}}]{Arribas08}
{Arribas}, S., {Colina}, L., {Monreal-Ibero}, A., {et~al.} 2008, \aap, 479, 687

\bibitem[{{Barcos-Mu{\~n}oz} {et~al.}(2017){Barcos-Mu{\~n}oz}, {Leroy},
  {Evans}, {Condon}, {Privon}, {Thompson}, {Armus}, {D{\'\i}az-Santos},
  {Mazzarella}, {Meier}, {Momjian}, {Murphy}, {Ott}, {Sanders}, {Schinnerer},
  {Stierwalt}, {Surace}, \& {Walter}}]{BM17}
{Barcos-Mu{\~n}oz}, L., {Leroy}, A.~K., {Evans}, A.~S., {et~al.} 2017, \apj,
  843, 117

\bibitem[{{Barro} {et~al.}(2014){Barro}, {Faber}, {P{\'e}rez-Gonz{\'a}lez},
  {Pacifici}, {Trump}, {Koo}, {Wuyts}, {Guo}, {Bell}, {Dekel}, {Porter},
  {Primack}, {Ferguson}, {Ashby}, {Caputi}, {Ceverino}, {Croton}, {Fazio},
  {Giavalisco}, {Hsu}, {Kocevski}, {Koekemoer}, {Kurczynski}, {Kollipara},
  {Lee}, {McIntosh}, {McGrath}, {Moody}, {Somerville}, {Papovich}, {Salvato},
  {Santini}, {Tal}, {van der Wel}, {Williams}, {Willner}, \&
  {Zolotov}}]{Barro14}
{Barro}, G., {Faber}, S.~M., {P{\'e}rez-Gonz{\'a}lez}, P.~G., {et~al.} 2014,
  \apj, 791, 52

\bibitem[{{Barro} {et~al.}(2016){Barro}, {Kriek}, {P{\'e}rez-Gonz{\'a}lez},
  {Trump}, {Koo}, {Faber}, {Dekel}, {Primack}, {Guo}, {Kocevski},
  {Mu{\~n}oz-Mateos}, {Rujopakarn}, \& {Seth}}]{Barro16}
{Barro}, G., {Kriek}, M., {P{\'e}rez-Gonz{\'a}lez}, P.~G., {et~al.} 2016,
  \apjl, 827, L32

\bibitem[{{Bellocchi} {et~al.}(2016){Bellocchi}, {Arribas}, \&
  {Colina}}]{Bellocchi16}
{Bellocchi}, E., {Arribas}, S., \& {Colina}, L. 2016, \aap, 591, A85

\bibitem[{{Bellocchi} {et~al.}(2013){Bellocchi}, {Arribas}, {Colina}, \&
  {Miralles-Caballero}}]{Bellocchi13}
{Bellocchi}, E., {Arribas}, S., {Colina}, L., \& {Miralles-Caballero}, D. 2013,
  \aap, 557, A59

\bibitem[{{Benjamin} {et~al.}(2003){Benjamin}, {Churchwell}, {Babler}, {Bania},
  {Clemens}, {Cohen}, {Dickey}, {Indebetouw}, {Jackson}, {Kobulnicky},
  {Lazarian}, {Marston}, {Mathis}, {Meade}, {Seager}, {Stolovy}, {Watson},
  {Whitney}, {Wolff}, \& {Wolfire}}]{Benjamin03}
{Benjamin}, R.~A., {Churchwell}, E., {Babler}, B.~L., {et~al.} 2003, \pasp,
  115, 953

\bibitem[{{Blain} \& {Phillips}(2002)}]{Blain02}
{Blain}, A.~W. \& {Phillips}, T.~G. 2002, \mnras, 333, 222

\bibitem[{{Bolatto} {et~al.}(2017){Bolatto}, {Wong}, {Utomo}, {Blitz}, {Vogel},
  {S{\'a}nchez}, {Barrera-Ballesteros}, {Cao}, {Colombo}, {Dannerbauer},
  {Garc{\'\i}a-Benito}, {Herrera-Camus}, {Husemann}, {Kalinova}, {Leroy},
  {Leung}, {Levy}, {Mast}, {Ostriker}, {Rosolowsky}, {Sandstrom}, {Teuben},
  {van de Ven}, \& {Walter}}]{Bolatto17}
{Bolatto}, A.~D., {Wong}, T., {Utomo}, D., {et~al.} 2017, \apj, 846, 159

\bibitem[{{Calistro Rivera} {et~al.}(2018){Calistro Rivera}, {Hodge}, {Smail},
  {Swinbank}, {Weiss}, {Wardlow}, {Walter}, {Rybak}, {Chen}, {Brandt},
  {Coppin}, {da Cunha}, {Dannerbauer}, {Greve}, {Karim}, {Knudsen},
  {Schinnerer}, {Simpson}, {Venemans}, \& {van der Werf}}]{CR18}
{Calistro Rivera}, G., {Hodge}, J.~A., {Smail}, I., {et~al.} 2018, \apj, 863,
  56

\bibitem[{{Calzetti} {et~al.}(2007){Calzetti}, {Kennicutt}, {Engelbracht},
  {Leitherer}, {Draine}, {Kewley}, {Moustakas}, {Sosey}, {Dale}, {Gordon},
  {Helou}, {Hollenbach}, {Armus}, {Bendo}, {Bot}, {Buckalew}, {Jarrett}, {Li},
  {Meyer}, {Murphy}, {Prescott}, {Regan}, {Rieke}, {Roussel}, {Sheth}, {Smith},
  {Thornley}, \& {Walter}}]{Calzetti07}
{Calzetti}, D., {Kennicutt}, R.~C., {Engelbracht}, C.~W., {et~al.} 2007, \apj,
  666, 870

\bibitem[{{Cappellari} {et~al.}(2013{\natexlab{a}}){Cappellari}, {McDermid},
  {Alatalo}, {Blitz}, {Bois}, {Bournaud}, {Bureau}, {Crocker}, {Davies},
  {Davis}, {de Zeeuw}, {Duc}, {Emsellem}, {Khochfar}, {Krajnovi{\'c}},
  {Kuntschner}, {Morganti}, {Naab}, {Oosterloo}, {Sarzi}, {Scott}, {Serra},
  {Weijmans}, \& {Young}}]{Cappellari13}
{Cappellari}, M., {McDermid}, R.~M., {Alatalo}, K., {et~al.}
  2013{\natexlab{a}}, \mnras, 432, 1862

\bibitem[{{Cappellari} {et~al.}(2013{\natexlab{b}}){Cappellari}, {Scott},
  {Alatalo}, {Blitz}, {Bois}, {Bournaud}, {Bureau}, {Crocker}, {Davies},
  {Davis}, {de Zeeuw}, {Duc}, {Emsellem}, {Khochfar}, {Krajnovi{\'c}},
  {Kuntschner}, {McDermid}, {Morganti}, {Naab}, {Oosterloo}, {Sarzi}, {Serra},
  {Weijmans}, \& {Young}}]{Cap13a}
{Cappellari}, M., {Scott}, N., {Alatalo}, K., {et~al.} 2013{\natexlab{b}},
  \mnras, 432, 1709

\bibitem[{{Caputi} {et~al.}(2007){Caputi}, {Lagache}, {Yan}, {Dole},
  {Bavouzet}, {Le Floc'h}, {Choi}, {Helou}, \& {Reddy}}]{Caputi07}
{Caputi}, K.~I., {Lagache}, G., {Yan}, L., {et~al.} 2007, \apj, 660, 97

\bibitem[{{Chen} {et~al.}(2020){Chen}, {Harrison}, {Smail}, {Swinbank},
  {Turner}, {Wardlow}, {Brandt}, {Calistro Rivera}, {Chapman}, {Cooke},
  {Dannerbauer}, {Dunlop}, {Farrah}, {Micha{\l}owski}, {Schinnerer}, {Simpson},
  {Thomson}, \& {van der Werf}}]{Chen20}
{Chen}, C.-C., {Harrison}, C.~M., {Smail}, I., {et~al.} 2020, \aap, 635, A119

\bibitem[{{Chen} {et~al.}(2017){Chen}, {Hodge}, {Smail}, {Swinbank}, {Walter},
  {Simpson}, {Calistro Rivera}, {Bertoldi}, {Brandt}, {Chapman}, {da Cunha},
  {Dannerbauer}, {De Breuck}, {Harrison}, {Ivison}, {Karim}, {Knudsen},
  {Wardlow}, {Wei{\ss}}, \& {van der Werf}}]{Chen17}
{Chen}, C.-C., {Hodge}, J.~A., {Smail}, I., {et~al.} 2017, \apj, 846, 108

\bibitem[{{Cheng} {et~al.}(2020){Cheng}, {Ibar}, {Smail}, {Molina}, {Sobral},
  {Escala}, {Best}, {Cochrane}, {Gillman}, {Swinbank}, {Ivison}, {Huang},
  {Hughes}, {Villard}, \& {Cirasuolo}}]{Cheng20}
{Cheng}, C., {Ibar}, E., {Smail}, I., {et~al.} 2020, \mnras, 499, 5241

\bibitem[{{Chu} {et~al.}(2017){Chu}, {Sanders}, {Larson}, {Mazzarella},
  {Howell}, {D{\'\i}az-Santos}, {Xu}, {Paladini}, {Schulz}, {Shupe},
  {Appleton}, {Armus}, {Billot}, {Chan}, {Evans}, {Fadda}, {Frayer}, {Haan},
  {Ishida}, {Iwasawa}, {Kim}, {Lord}, {Murphy}, {Petric}, {Privon}, {Surace},
  \& {Treister}}]{Chu17}
{Chu}, J.~K., {Sanders}, D.~B., {Larson}, K.~L., {et~al.} 2017, \apjs, 229, 25

\bibitem[{{Cibinel} {et~al.}(2019){Cibinel}, {Daddi}, {Sargent}, {Le Floc'h},
  {Liu}, {Bournaud}, {Oesch}, {Amram}, {Calabr{\`o}}, {Duc}, {Pannella},
  {Puglisi}, {Perret}, {Elbaz}, \& {Kokorev}}]{Cibinel19}
{Cibinel}, A., {Daddi}, E., {Sargent}, M.~T., {et~al.} 2019, \mnras, 485, 5631

\bibitem[{{Condon} \& {Ransom}(2016)}]{Condon16}
{Condon}, J.~J. \& {Ransom}, S.~M. 2016, {Essential Radio Astronomy}

\bibitem[{{Crespo G{\'o}mez} {et~al.}(2021){Crespo G{\'o}mez}, {Piqueras
  L{\'o}pez}, {Arribas}, {Pereira-Santaella}, {Colina}, \& {Rodr{\'\i}guez del
  Pino}}]{Crespo21}
{Crespo G{\'o}mez}, A., {Piqueras L{\'o}pez}, J., {Arribas}, S., {et~al.} 2021,
  \aap, 650, A149

\bibitem[{{Dasyra} {et~al.}(2006{\natexlab{a}}){Dasyra}, {Tacconi}, {Davies},
  {Genzel}, {Lutz}, {Naab}, {Sanders}, {Veilleux}, \& {Baker}}]{Dasyra_2_06}
{Dasyra}, K.~M., {Tacconi}, L.~J., {Davies}, R.~I., {et~al.}
  2006{\natexlab{a}}, \nar, 50, 720

\bibitem[{{Dasyra} {et~al.}(2006{\natexlab{b}}){Dasyra}, {Tacconi}, {Davies},
  {Naab}, {Genzel}, {Lutz}, {Sturm}, {Baker}, {Veilleux}, {Sanders}, \&
  {Burkert}}]{Dasyra06}
{Dasyra}, K.~M., {Tacconi}, L.~J., {Davies}, R.~I., {et~al.}
  2006{\natexlab{b}}, \apj, 651, 835

\bibitem[{{Dav{\'e}} {et~al.}(2010){Dav{\'e}}, {Finlator}, {Oppenheimer},
  {Fardal}, {Katz}, {Kere{\v{s}}}, \& {Weinberg}}]{Dave10}
{Dav{\'e}}, R., {Finlator}, K., {Oppenheimer}, B.~D., {et~al.} 2010, \mnras,
  404, 1355

\bibitem[{{Davis} {et~al.}(2013){Davis}, {Alatalo}, {Bureau}, {Cappellari},
  {Scott}, {Young}, {Blitz}, {Crocker}, {Bayet}, {Bois}, {Bournaud}, {Davies},
  {de Zeeuw}, {Duc}, {Emsellem}, {Khochfar}, {Krajnovi{\'c}}, {Kuntschner},
  {Lablanche}, {McDermid}, {Morganti}, {Naab}, {Oosterloo}, {Sarzi}, {Serra},
  \& {Weijmans}}]{Davis13}
{Davis}, T.~A., {Alatalo}, K., {Bureau}, M., {et~al.} 2013, \mnras, 429, 534

\bibitem[{{Davis} {et~al.}(2014){Davis}, {Young}, {Crocker}, {Bureau}, {Blitz},
  {Alatalo}, {Emsellem}, {Naab}, {Bayet}, {Bois}, {Bournaud}, {Cappellari},
  {Davies}, {de Zeeuw}, {Duc}, {Khochfar}, {Krajnovi{\'c}}, {Kuntschner},
  {McDermid}, {Morganti}, {Oosterloo}, {Sarzi}, {Scott}, {Serra}, \&
  {Weijmans}}]{Davis14}
{Davis}, T.~A., {Young}, L.~M., {Crocker}, A.~F., {et~al.} 2014, \mnras, 444,
  3427

\bibitem[{{Dekel} {et~al.}(2009){Dekel}, {Sari}, \& {Ceverino}}]{Dekel09}
{Dekel}, A., {Sari}, R., \& {Ceverino}, D. 2009, \apj, 703, 785

\bibitem[{{D{\'\i}az-Santos} {et~al.}(2017){D{\'\i}az-Santos}, {Armus},
  {Charmandaris}, {Lu}, {Stierwalt}, {Stacey}, {Malhotra}, {van der Werf},
  {Howell}, {Privon}, {Mazzarella}, {Goldsmith}, {Murphy}, {Barcos-Mu{\~n}oz},
  {Linden}, {Inami}, {Larson}, {Evans}, {Appleton}, {Iwasawa}, {Lord},
  {Sanders}, \& {Surace}}]{TDS17}
{D{\'\i}az-Santos}, T., {Armus}, L., {Charmandaris}, V., {et~al.} 2017, \apj,
  846, 32

\bibitem[{{Dudzevi{\v{c}}i{\={u}}t{\.{e}}}
  {et~al.}(2020){Dudzevi{\v{c}}i{\={u}}t{\.{e}}}, {Smail}, {Swinbank}, {Stach},
  {Almaini}, {da Cunha}, {An}, {Arumugam}, {Birkin}, {Blain}, {Chapman},
  {Chen}, {Conselice}, {Coppin}, {Dunlop}, {Farrah}, {Geach}, {Gullberg},
  {Hartley}, {Hodge}, {Ivison}, {Maltby}, {Scott}, {Simpson}, {Simpson},
  {Thomson}, {Walter}, {Wardlow}, {Weiss}, \& {van der Werf}}]{Dudzeviciute20}
{Dudzevi{\v{c}}i{\={u}}t{\.{e}}}, U., {Smail}, I., {Swinbank}, A.~M., {et~al.}
  2020, \mnras, 494, 3828

\bibitem[{{Elbaz} {et~al.}(2007){Elbaz}, {Daddi}, {Le Borgne}, {Dickinson},
  {Alexander}, {Chary}, {Starck}, {Brandt}, {Kitzbichler}, {MacDonald},
  {Nonino}, {Popesso}, {Stern}, \& {Vanzella}}]{Elbaz07}
{Elbaz}, D., {Daddi}, E., {Le Borgne}, D., {et~al.} 2007, \aap, 468, 33

\bibitem[{{Erb} {et~al.}(2004){Erb}, {Steidel}, {Shapley}, {Pettini}, \&
  {Adelberger}}]{Erb04}
{Erb}, D.~K., {Steidel}, C.~C., {Shapley}, A.~E., {Pettini}, M., \&
  {Adelberger}, K.~L. 2004, \apj, 612, 122

\bibitem[{{Faisst} {et~al.}(2020){Faisst}, {Schaerer}, {Lemaux}, {Oesch},
  {Fudamoto}, {Cassata}, {B{\'e}thermin}, {Capak}, {Le F{\`e}vre}, {Silverman},
  {Yan}, {Ginolfi}, {Koekemoer}, {Morselli}, {Amor{\'\i}n}, {Bardelli},
  {Boquien}, {Brammer}, {Cimatti}, {Dessauges-Zavadsky}, {Fujimoto},
  {Gruppioni}, {Hathi}, {Hemmati}, {Ibar}, {Jones}, {Khusanova}, {Loiacono},
  {Pozzi}, {Talia}, {Tasca}, {Riechers}, {Rodighiero}, {Romano}, {Scoville},
  {Toft}, {Vallini}, {Vergani}, {Zamorani}, \& {Zucca}}]{Faisst20}
{Faisst}, A.~L., {Schaerer}, D., {Lemaux}, B.~C., {et~al.} 2020, \apjs, 247, 61

\bibitem[{{Fern{\'a}ndez Lorenzo} {et~al.}(2013){Fern{\'a}ndez Lorenzo},
  {Sulentic}, {Verdes-Montenegro}, \& {Argudo-Fern{\'a}ndez}}]{FL13}
{Fern{\'a}ndez Lorenzo}, M., {Sulentic}, J., {Verdes-Montenegro}, L., \&
  {Argudo-Fern{\'a}ndez}, M. 2013, \mnras, 434, 325

\bibitem[{{F{\"o}rster Schreiber} {et~al.}(2009){F{\"o}rster Schreiber},
  {Genzel}, {Bouch{\'e}}, {Cresci}, {Davies}, {Buschkamp}, {Shapiro},
  {Tacconi}, {Hicks}, {Genel}, {Shapley}, {Erb}, {Steidel}, {Lutz},
  {Eisenhauer}, {Gillessen}, {Sternberg}, {Renzini}, {Cimatti}, {Daddi},
  {Kurk}, {Lilly}, {Kong}, {Lehnert}, {Nesvadba}, {Verma}, {McCracken},
  {Arimoto}, {Mignoli}, \& {Onodera}}]{FS09}
{F{\"o}rster Schreiber}, N.~M., {Genzel}, R., {Bouch{\'e}}, N., {et~al.} 2009,
  \apj, 706, 1364

\bibitem[{{F{\"o}rster Schreiber} {et~al.}(2018){F{\"o}rster Schreiber},
  {Renzini}, {Mancini}, {Genzel}, {Bouch{\'e}}, {Cresci}, {Hicks}, {Lilly},
  {Peng}, {Burkert}, {Carollo}, {Cimatti}, {Daddi}, {Davies}, {Genel}, {Kurk},
  {Lang}, {Lutz}, {Mainieri}, {McCracken}, {Mignoli}, {Naab}, {Oesch},
  {Pozzetti}, {Scodeggio}, {Shapiro Griffin}, {Shapley}, {Sternberg},
  {Tacchella}, {Tacconi}, {Wuyts}, \& {Zamorani}}]{FS18}
{F{\"o}rster Schreiber}, N.~M., {Renzini}, A., {Mancini}, C., {et~al.} 2018,
  \apjs, 238, 21

\bibitem[{{Fujimoto} {et~al.}(2020){Fujimoto}, {Silverman}, {Bethermin},
  {Ginolfi}, {Jones}, {Le F{\`e}vre}, {Dessauges-Zavadsky}, {Rujopakarn},
  {Faisst}, {Fudamoto}, {Cassata}, {Morselli}, {Maiolino}, {Schaerer}, {Capak},
  {Yan}, {Vallini}, {Toft}, {Loiacono}, {Zamorani}, {Talia}, {Narayanan},
  {Hathi}, {Lemaux}, {Boquien}, {Amorin}, {Ibar}, {Koekemoer},
  {M{\'e}ndez-Hern{\'a}ndez}, {Bardelli}, {Vergani}, {Zucca}, {Romano}, \&
  {Cimatti}}]{Fujimoto20}
{Fujimoto}, S., {Silverman}, J.~D., {Bethermin}, M., {et~al.} 2020, \apj, 900,
  1

\bibitem[{{Genzel} {et~al.}(2015){Genzel}, {Tacconi}, {Lutz}, {Saintonge},
  {Berta}, {Magnelli}, {Combes}, {Garc{\'\i}a-Burillo}, {Neri}, {Bolatto},
  {Contini}, {Lilly}, {Boissier}, {Boone}, {Bouch{\'e}}, {Bournaud}, {Burkert},
  {Carollo}, {Colina}, {Cooper}, {Cox}, {Feruglio}, {F{\"o}rster Schreiber},
  {Freundlich}, {Gracia-Carpio}, {Juneau}, {Kovac}, {Lippa}, {Naab}, {Salome},
  {Renzini}, {Sternberg}, {Walter}, {Weiner}, {Weiss}, \& {Wuyts}}]{Genzel15}
{Genzel}, R., {Tacconi}, L.~J., {Lutz}, D., {et~al.} 2015, \apj, 800, 20

\bibitem[{{Genzel} {et~al.}(2001){Genzel}, {Tacconi}, {Rigopoulou}, {Lutz}, \&
  {Tecza}}]{Genzel01}
{Genzel}, R., {Tacconi}, L.~J., {Rigopoulou}, D., {Lutz}, D., \& {Tecza}, M.
  2001, \apj, 563, 527

\bibitem[{{Graham} \& {Worley}(2008)}]{Graham08}
{Graham}, A.~W. \& {Worley}, C.~C. 2008, \mnras, 388, 1708

\bibitem[{{Gullberg} {et~al.}(2018){Gullberg}, {Swinbank}, {Smail}, {Biggs},
  {Bertoldi}, {De Breuck}, {Chapman}, {Chen}, {Cooke}, {Coppin}, {Cox},
  {Dannerbauer}, {Dunlop}, {Edge}, {Farrah}, {Geach}, {Greve}, {Hodge}, {Ibar},
  {Ivison}, {Karim}, {Schinnerer}, {Scott}, {Simpson}, {Stach}, {Thomson}, {van
  der Werf}, {Walter}, {Wardlow}, \& {Weiss}}]{Gullberg18}
{Gullberg}, B., {Swinbank}, A.~M., {Smail}, I., {et~al.} 2018, \apj, 859, 12

\bibitem[{{Heintz} {et~al.}(2021){Heintz}, {Watson}, {Oesch}, {Narayanan}, \&
  {Madden}}]{Heintz21}
{Heintz}, K.~E., {Watson}, D., {Oesch}, P., {Narayanan}, D., \& {Madden}, S.~C.
  2021, arXiv e-prints, arXiv:2108.13442

\bibitem[{{Hodge} {et~al.}(2016){Hodge}, {Swinbank}, {Simpson}, {Smail},
  {Walter}, {Alexander}, {Bertoldi}, {Biggs}, {Brandt}, {Chapman}, {Chen},
  {Coppin}, {Cox}, {Dannerbauer}, {Edge}, {Greve}, {Ivison}, {Karim},
  {Knudsen}, {Menten}, {Rix}, {Schinnerer}, {Wardlow}, {Weiss}, \& {van der
  Werf}}]{Hodge16}
{Hodge}, J.~A., {Swinbank}, A.~M., {Simpson}, J.~M., {et~al.} 2016, \apj, 833,
  103

\bibitem[{{Hogan} {et~al.}(2021){Hogan}, {Rigopoulou}, {Magdis},
  {Pereira-Santaella}, {Garc{\'\i}a-Bernete}, {Thatte}, {Grisdale}, \&
  {Huang}}]{Hogan21}
{Hogan}, L., {Rigopoulou}, D., {Magdis}, G.~E., {et~al.} 2021, \mnras, 503,
  5329

\bibitem[{{Hunt} {et~al.}(2015){Hunt}, {Draine}, {Bianchi}, {Gordon}, {Aniano},
  {Calzetti}, {Dale}, {Helou}, {Hinz}, {Kennicutt}, {Roussel}, {Wilson},
  {Bolatto}, {Boquien}, {Croxall}, {Galametz}, {Gil de Paz}, {Koda},
  {Mu{\~n}oz-Mateos}, {Sandstrom}, {Sauvage}, {Vigroux}, \& {Zibetti}}]{Hunt15}
{Hunt}, L.~K., {Draine}, B.~T., {Bianchi}, S., {et~al.} 2015, \aap, 576, A33

\bibitem[{{Jarrett} {et~al.}(2000){Jarrett}, {Chester}, {Cutri}, {Schneider},
  {Skrutskie}, \& {Huchra}}]{Jarrett00}
{Jarrett}, T.~H., {Chester}, T., {Cutri}, R., {et~al.} 2000, \aj, 119, 2498

\bibitem[{{Jim{\'e}nez-Bail{\'o}n} {et~al.}(2007){Jim{\'e}nez-Bail{\'o}n},
  {Loiseau}, {Guainazzi}, {Matt}, {Rosa-Gonz{\'a}lez}, {Piconcelli}, \&
  {Santos-Lle{\'o}}}]{JBailon07}
{Jim{\'e}nez-Bail{\'o}n}, E., {Loiseau}, N., {Guainazzi}, M., {et~al.} 2007,
  \aap, 469, 881

\bibitem[{{Kaasinen} {et~al.}(2020){Kaasinen}, {Walter}, {Novak}, {Neeleman},
  {Smail}, {Boogaard}, {Cunha}, {Weiss}, {Liu}, {Decarli}, {Popping},
  {Diaz-Santos}, {Cort{\'e}s}, {Aravena}, {Werf}, {Riechers}, {Inami}, {Hodge},
  {Rix}, \& {Cox}}]{Kaasinen20}
{Kaasinen}, M., {Walter}, F., {Novak}, M., {et~al.} 2020, \apj, 899, 37

\bibitem[{{Kartaltepe} {et~al.}(2010){Kartaltepe}, {Sanders}, {Le Floc'h},
  {Frayer}, {Aussel}, {Arnouts}, {Ilbert}, {Salvato}, {Scoville}, {Surace},
  {Yan}, {Capak}, {Caputi}, {Carollo}, {Cassata}, {Civano}, {Hasinger},
  {Koekemoer}, {Le F{\`e}vre}, {Lilly}, {Liu}, {McCracken}, {Schinnerer},
  {Smol{\v{c}}i{\'c}}, {Taniguchi}, {Thompson}, {Trump}, {Baldassare}, \&
  {Fiorenza}}]{Kartaltepe10}
{Kartaltepe}, J.~S., {Sanders}, D.~B., {Le Floc'h}, E., {et~al.} 2010, \apj,
  721, 98

\bibitem[{{Kawakatu} {et~al.}(2006){Kawakatu}, {Anabuki}, {Nagao}, {Umemura},
  \& {Nakagawa}}]{Kawakatu06}
{Kawakatu}, N., {Anabuki}, N., {Nagao}, T., {Umemura}, M., \& {Nakagawa}, T.
  2006, \apj, 637, 104

\bibitem[{{Kennedy} {et~al.}(2015){Kennedy}, {Bamford}, {Baldry},
  {H{\"a}u{\ss}ler}, {Holwerda}, {Hopkins}, {Kelvin}, {Lange}, {Moffett},
  {Popescu}, {Taylor}, {Tuffs}, {Vika}, \& {Vulcani}}]{Kennedy15}
{Kennedy}, R., {Bamford}, S.~P., {Baldry}, I., {et~al.} 2015, \mnras, 454, 806

\bibitem[{{Kennicutt} \& {Evans}(2012)}]{Kennicutt12}
{Kennicutt}, R.~C. \& {Evans}, N.~J. 2012, \araa, 50, 531

\bibitem[{{Krajnovi{\'c}} {et~al.}(2006){Krajnovi{\'c}}, {Cappellari}, {de
  Zeeuw}, \& {Copin}}]{Krajnovic06}
{Krajnovi{\'c}}, D., {Cappellari}, M., {de Zeeuw}, P.~T., \& {Copin}, Y. 2006,
  \mnras, 366, 787

\bibitem[{{Kroupa}(2001)}]{Kroupa01}
{Kroupa}, P. 2001, \mnras, 322, 231

\bibitem[{{Lagache} {et~al.}(2018){Lagache}, {Cousin}, \&
  {Chatzikos}}]{Lagache18}
{Lagache}, G., {Cousin}, M., \& {Chatzikos}, M. 2018, \aap, 609, A130

\bibitem[{{Lagache} {et~al.}(2005){Lagache}, {Puget}, \& {Dole}}]{Lagache05}
{Lagache}, G., {Puget}, J.-L., \& {Dole}, H. 2005, \araa, 43, 727

\bibitem[{{Lang} {et~al.}(2019){Lang}, {Schinnerer}, {Smail},
  {Dudzevi{\v{c}}i{\={u}}t{\.{e}}}, {Swinbank}, {Liu}, {Leslie}, {Almaini},
  {An}, {Bertoldi}, {Blain}, {Chapman}, {Chen}, {Conselice}, {Cooke}, {Coppin},
  {Dunlop}, {Farrah}, {Fudamoto}, {Geach}, {Gullberg}, {Harrington}, {Hodge},
  {Ivison}, {Jim{\'e}nez-Andrade}, {Magnelli}, {Micha{\l}owski}, {Oesch},
  {Scott}, {Simpson}, {Smol{\v{c}}i{\'c}}, {Stach}, {Thomson}, {Toft},
  {Vardoulaki}, {Wardlow}, {Weiss}, \& {van der Werf}}]{Lang19}
{Lang}, P., {Schinnerer}, E., {Smail}, I., {et~al.} 2019, \apj, 879, 54

\bibitem[{{Lange} {et~al.}(2015){Lange}, {Driver}, {Robotham}, {Kelvin},
  {Graham}, {Alpaslan}, {Andrews}, {Baldry}, {Bamford}, {Bland-Hawthorn},
  {Brough}, {Cluver}, {Conselice}, {Davies}, {Haeussler}, {Konstantopoulos},
  {Loveday}, {Moffett}, {Norberg}, {Phillipps}, {Taylor},
  {L{\'o}pez-S{\'a}nchez}, \& {Wilkins}}]{Lange15}
{Lange}, R., {Driver}, S.~P., {Robotham}, A. S.~G., {et~al.} 2015, \mnras, 447,
  2603

\bibitem[{{Law} {et~al.}(2009){Law}, {Steidel}, {Erb}, {Larkin}, {Pettini},
  {Shapley}, \& {Wright}}]{Law09}
{Law}, D.~R., {Steidel}, C.~C., {Erb}, D.~K., {et~al.} 2009, \apj, 697, 2057

\bibitem[{{Le Floc'h} {et~al.}(2005){Le Floc'h}, {Papovich}, {Dole}, {Bell},
  {Lagache}, {Rieke}, {Egami}, {P{\'e}rez-Gonz{\'a}lez}, {Alonso-Herrero},
  {Rieke}, {Blaylock}, {Engelbracht}, {Gordon}, {Hines}, {Misselt}, {Morrison},
  \& {Mould}}]{LF05}
{Le Floc'h}, E., {Papovich}, C., {Dole}, H., {et~al.} 2005, \apj, 632, 169

\bibitem[{{Leroy} {et~al.}(2019){Leroy}, {Sandstrom}, {Lang}, {Lewis}, {Salim},
  {Behrens}, {Chastenet}, {Chiang}, {Gallagher}, {Kessler}, \&
  {Utomo}}]{Leroy19}
{Leroy}, A.~K., {Sandstrom}, K.~M., {Lang}, D., {et~al.} 2019, \apjs, 244, 24

\bibitem[{{Leroy} {et~al.}(2021){Leroy}, {Schinnerer}, {Hughes}, {Rosolowsky},
  {Pety}, {Schruba}, {Usero}, {Blanc}, {Chevance}, {Emsellem}, {Faesi},
  {Herrera}, {Liu}, {Meidt}, {Querejeta}, {Saito}, {Sandstrom}, {Sun},
  {Williams}, {Anand}, {Barnes}, {Behrens}, {Belfiore}, {Benincasa},
  {Be{\v{s}}li{\'c}}, {Bigiel}, {Bolatto}, {den Brok}, {Cao}, {Chandar},
  {Chastenet}, {Chiang}, {Congiu}, {Dale}, {Deger}, {Eibensteiner}, {Egorov},
  {Garc{\'\i}a-Rodr{\'\i}guez}, {Glover}, {Grasha}, {Henshaw}, {Ho}, {Kepley},
  {Kim}, {Klessen}, {Kreckel}, {Koch}, {Kruijssen}, {Larson}, {Lee}, {Lopez},
  {Machado}, {Mayker}, {McElroy}, {Murphy}, {Ostriker}, {Pan}, {Pessa},
  {Puschnig}, {Razza}, {S{\'a}nchez-Bl{\'a}zquez}, {Santoro}, {Sardone},
  {Scheuermann}, {Sliwa}, {Sormani}, {Stuber}, {Thilker}, {Turner}, {Utomo},
  {Watkins}, \& {Whitmore}}]{Leroy21}
{Leroy}, A.~K., {Schinnerer}, E., {Hughes}, A., {et~al.} 2021, arXiv e-prints,
  arXiv:2104.07739

\bibitem[{{Leroy} {et~al.}(2013){Leroy}, {Walter}, {Sandstrom}, {Schruba},
  {Munoz-Mateos}, {Bigiel}, {Bolatto}, {Brinks}, {de Blok}, {Meidt}, {Rix},
  {Rosolowsky}, {Schinnerer}, {Schuster}, \& {Usero}}]{Leroy13}
{Leroy}, A.~K., {Walter}, F., {Sandstrom}, K., {et~al.} 2013, \aj, 146, 19

\bibitem[{{Madau} \& {Dickinson}(2014)}]{Madau14}
{Madau}, P. \& {Dickinson}, M. 2014, \araa, 52, 415

\bibitem[{{Magnelli} {et~al.}(2013){Magnelli}, {Popesso}, {Berta}, {Pozzi},
  {Elbaz}, {Lutz}, {Dickinson}, {Altieri}, {Andreani}, {Aussel},
  {B{\'e}thermin}, {Bongiovanni}, {Cepa}, {Charmandaris}, {Chary}, {Cimatti},
  {Daddi}, {F{\"o}rster Schreiber}, {Genzel}, {Gruppioni}, {Harwit}, {Hwang},
  {Ivison}, {Magdis}, {Maiolino}, {Murphy}, {Nordon}, {Pannella}, {P{\'e}rez
  Garc{\'\i}a}, {Poglitsch}, {Rosario}, {Sanchez-Portal}, {Santini}, {Scott},
  {Sturm}, {Tacconi}, \& {Valtchanov}}]{Magnelli13}
{Magnelli}, B., {Popesso}, P., {Berta}, S., {et~al.} 2013, \aap, 553, A132

\bibitem[{{McMullin} {et~al.}(2007){McMullin}, {Waters}, {Schiebel}, {Young},
  \& {Golap}}]{McMullin07}
{McMullin}, J.~P., {Waters}, B., {Schiebel}, D., {Young}, W., \& {Golap}, K.
  2007, in Astronomical Society of the Pacific Conference Series, Vol. 376,
  Astronomical Data Analysis Software and Systems XVI, ed. R.~A. {Shaw},
  F.~{Hill}, \& D.~J. {Bell}, 127

\bibitem[{{Murphy} {et~al.}(2011){Murphy}, {Condon}, {Schinnerer}, {Kennicutt},
  {Calzetti}, {Armus}, {Helou}, {Turner}, {Aniano}, {Beir{\~a}o}, {Bolatto},
  {Brandl}, {Croxall}, {Dale}, {Donovan Meyer}, {Draine}, {Engelbracht},
  {Hunt}, {Hao}, {Koda}, {Roussel}, {Skibba}, \& {Smith}}]{Murphy11}
{Murphy}, E.~J., {Condon}, J.~J., {Schinnerer}, E., {et~al.} 2011, \apj, 737,
  67

\bibitem[{{Muzzin} {et~al.}(2010){Muzzin}, {van Dokkum}, {Kriek}, {Labb{\'e}},
  {Cury}, {Marchesini}, \& {Franx}}]{Muzzin10}
{Muzzin}, A., {van Dokkum}, P., {Kriek}, M., {et~al.} 2010, \apj, 725, 742

\bibitem[{{Nardini} {et~al.}(2008){Nardini}, {Risaliti}, {Salvati}, {Sani},
  {Imanishi}, {Marconi}, \& {Maiolino}}]{Nardini08}
{Nardini}, E., {Risaliti}, G., {Salvati}, M., {et~al.} 2008, \mnras, 385, L130

\bibitem[{{Nelson} {et~al.}(2016){Nelson}, {van Dokkum}, {F{\"o}rster
  Schreiber}, {Franx}, {Brammer}, {Momcheva}, {Wuyts}, {Whitaker}, {Skelton},
  {Fumagalli}, {Hayward}, {Kriek}, {Labb{\'e}}, {Leja}, {Rix}, {Tacconi}, {van
  der Wel}, {van den Bosch}, {Oesch}, {Dickey}, \& {Ulf Lange}}]{Nelson16}
{Nelson}, E.~J., {van Dokkum}, P.~G., {F{\"o}rster Schreiber}, N.~M., {et~al.}
  2016, \apj, 828, 27

\bibitem[{{Peng} {et~al.}(2002){Peng}, {Ho}, {Impey}, \& {Rix}}]{Peng02}
{Peng}, C.~Y., {Ho}, L.~C., {Impey}, C.~D., \& {Rix}, H.-W. 2002, \aj, 124, 266

\bibitem[{{Peng} {et~al.}(2010){Peng}, {Ho}, {Impey}, \& {Rix}}]{Peng10}
{Peng}, C.~Y., {Ho}, L.~C., {Impey}, C.~D., \& {Rix}, H.-W. 2010, \aj, 139,
  2097

\bibitem[{{Pereira-Santaella} {et~al.}(2015){Pereira-Santaella},
  {Alonso-Herrero}, {Colina}, {Miralles-Caballero}, {P{\'e}rez-Gonz{\'a}lez},
  {Arribas}, {Bellocchi}, {Cazzoli}, {D{\'\i}az-Santos}, \& {Piqueras
  L{\'o}pez}}]{Pereira15}
{Pereira-Santaella}, M., {Alonso-Herrero}, A., {Colina}, L., {et~al.} 2015,
  \aap, 577, A78

\bibitem[{{Pereira-Santaella} {et~al.}(2011){Pereira-Santaella},
  {Alonso-Herrero}, {Santos-Lleo}, {Colina}, {Jim{\'e}nez-Bail{\'o}n},
  {Longinotti}, {Rieke}, {Ward}, \& {Esquej}}]{Pereira11}
{Pereira-Santaella}, M., {Alonso-Herrero}, A., {Santos-Lleo}, M., {et~al.}
  2011, \aap, 535, A93

\bibitem[{{Pereira-Santaella} {et~al.}(2016{\natexlab{a}}){Pereira-Santaella},
  {Colina}, {Garc{\'\i}a-Burillo}, {Alonso-Herrero}, {Arribas}, {Cazzoli},
  {Emonts}, {Piqueras L{\'o}pez}, {Planesas}, {Storchi Bergmann}, {Usero}, \&
  {Villar-Mart{\'\i}n}}]{PereiraS16}
{Pereira-Santaella}, M., {Colina}, L., {Garc{\'\i}a-Burillo}, S., {et~al.}
  2016{\natexlab{a}}, \aap, 594, A81

\bibitem[{{Pereira-Santaella} {et~al.}(2020){Pereira-Santaella}, {Colina},
  {Garc{\'\i}a-Burillo}, {Gonz{\'a}lez-Alfonso}, {Alonso-Herrero}, {Arribas},
  {Cazzoli}, {Piqueras-L{\'o}pez}, {Rigopoulou}, \& {Usero}}]{Pereira20}
{Pereira-Santaella}, M., {Colina}, L., {Garc{\'\i}a-Burillo}, S., {et~al.}
  2020, \aap, 643, A89

\bibitem[{{Pereira-Santaella} {et~al.}(2021){Pereira-Santaella}, {Colina},
  {Garc{\'\i}a-Burillo}, {Lamperti}, {Gonz{\'a}lez-Alfonso}, {Perna},
  {Arribas}, {Alonso-Herrero}, {Aalto}, {Combes}, {Labiano},
  {Piqueras-L{\'o}pez}, {Rigopoulou}, \& {van der Werf}}]{Pereira21}
{Pereira-Santaella}, M., {Colina}, L., {Garc{\'\i}a-Burillo}, S., {et~al.}
  2021, arXiv e-prints, arXiv:2104.08238

\bibitem[{{Pereira-Santaella} {et~al.}(2016{\natexlab{b}}){Pereira-Santaella},
  {Colina}, {Garc{\'\i}a-Burillo}, {Planesas}, {Usero}, {Alonso-Herrero},
  {Arribas}, {Cazzoli}, {Emonts}, {Piqueras L{\'o}pez}, \&
  {Villar-Mart{\'\i}n}}]{Pereira16}
{Pereira-Santaella}, M., {Colina}, L., {Garc{\'\i}a-Burillo}, S., {et~al.}
  2016{\natexlab{b}}, \aap, 587, A44

\bibitem[{{P{\'e}rez-Gonz{\'a}lez} {et~al.}(2005){P{\'e}rez-Gonz{\'a}lez},
  {Rieke}, {Egami}, {Alonso-Herrero}, {Dole}, {Papovich}, {Blaylock}, {Jones},
  {Rieke}, {Rigby}, {Barmby}, {Fazio}, {Huang}, \& {Martin}}]{PG05}
{P{\'e}rez-Gonz{\'a}lez}, P.~G., {Rieke}, G.~H., {Egami}, E., {et~al.} 2005,
  \apj, 630, 82

\bibitem[{{P{\'e}rez-Gonz{\'a}lez} {et~al.}(2008){P{\'e}rez-Gonz{\'a}lez},
  {Rieke}, {Villar}, {Barro}, {Blaylock}, {Egami}, {Gallego}, {Gil de Paz},
  {Pascual}, {Zamorano}, \& {Donley}}]{PG08}
{P{\'e}rez-Gonz{\'a}lez}, P.~G., {Rieke}, G.~H., {Villar}, V., {et~al.} 2008,
  \apj, 675, 234

\bibitem[{{P{\'e}rez-Torres} {et~al.}(2021){P{\'e}rez-Torres}, {Mattila},
  {Alonso-Herrero}, {Aalto}, \& {Efstathiou}}]{PT21}
{P{\'e}rez-Torres}, M., {Mattila}, S., {Alonso-Herrero}, A., {Aalto}, S., \&
  {Efstathiou}, A. 2021, \aapr, 29, 2

\bibitem[{{Piqueras L{\'o}pez} {et~al.}(2012){Piqueras L{\'o}pez}, {Colina},
  {Arribas}, {Alonso-Herrero}, \& {Bedregal}}]{PL12}
{Piqueras L{\'o}pez}, J., {Colina}, L., {Arribas}, S., {Alonso-Herrero}, A., \&
  {Bedregal}, A.~G. 2012, \aap, 546, A64

\bibitem[{{Popping} {et~al.}(2021){Popping}, {Pillepich}, {Calistro Rivera},
  {Schulz}, {Hernquist}, {Kaasinen}, {Marinacci}, {Nelson}, \&
  {Vogelsberger}}]{Popping21}
{Popping}, G., {Pillepich}, A., {Calistro Rivera}, G., {et~al.} 2021, \mnras

\bibitem[{{Puglisi} {et~al.}(2019){Puglisi}, {Daddi}, {Liu}, {Bournaud},
  {Silverman}, {Circosta}, {Calabr{\`o}}, {Aravena}, {Cibinel}, {Dannerbauer},
  {Delvecchio}, {Elbaz}, {Gao}, {Gobat}, {Jin}, {Le Floc'h}, {Magdis},
  {Mancini}, {Riechers}, {Rodighiero}, {Sargent}, {Valentino}, \&
  {Zanisi}}]{Puglisi19}
{Puglisi}, A., {Daddi}, E., {Liu}, D., {et~al.} 2019, \apjl, 877, L23

\bibitem[{{Puglisi} {et~al.}(2021){Puglisi}, {Daddi}, {Valentino}, {Magdis},
  {Liu}, {Kokorev}, {Circosta}, {Elbaz}, {Bournaud}, {Gomez-Guijarro}, {Jin},
  {Madden}, {Sargent}, \& {Swinbank}}]{Puglisi21}
{Puglisi}, A., {Daddi}, E., {Valentino}, F., {et~al.} 2021, arXiv e-prints,
  arXiv:2103.12035

\bibitem[{{Rieke} {et~al.}(2009){Rieke}, {Alonso-Herrero}, {Weiner},
  {P{\'e}rez-Gonz{\'a}lez}, {Blaylock}, {Donley}, \& {Marcillac}}]{Rieke09}
{Rieke}, G.~H., {Alonso-Herrero}, A., {Weiner}, B.~J., {et~al.} 2009, \apj,
  692, 556

\bibitem[{{Rigopoulou} {et~al.}(1999){Rigopoulou}, {Spoon}, {Genzel}, {Lutz},
  {Moorwood}, \& {Tran}}]{Rigopoulou99}
{Rigopoulou}, D., {Spoon}, H.~W.~W., {Genzel}, R., {et~al.} 1999, \aj, 118,
  2625

\bibitem[{{Risaliti} {et~al.}(2006){Risaliti}, {Maiolino}, {Marconi}, {Sani},
  {Berta}, {Braito}, {Della Ceca}, {Franceschini}, \& {Salvati}}]{Risaliti06}
{Risaliti}, G., {Maiolino}, R., {Marconi}, A., {et~al.} 2006, \mnras, 365, 303

\bibitem[{{Rodighiero} {et~al.}(2011){Rodighiero}, {Daddi}, {Baronchelli},
  {Cimatti}, {Renzini}, {Aussel}, {Popesso}, {Lutz}, {Andreani}, {Berta},
  {Cava}, {Elbaz}, {Feltre}, {Fontana}, {F{\"o}rster Schreiber},
  {Franceschini}, {Genzel}, {Grazian}, {Gruppioni}, {Ilbert}, {Le Floch},
  {Magdis}, {Magliocchetti}, {Magnelli}, {Maiolino}, {McCracken}, {Nordon},
  {Poglitsch}, {Santini}, {Pozzi}, {Riguccini}, {Tacconi}, {Wuyts}, \&
  {Zamorani}}]{Rodighiero11}
{Rodighiero}, G., {Daddi}, E., {Baronchelli}, I., {et~al.} 2011, \apjl, 739,
  L40

\bibitem[{{Rodr{\'\i}guez-Zaur{\'\i}n}
  {et~al.}(2011){Rodr{\'\i}guez-Zaur{\'\i}n}, {Arribas}, {Monreal-Ibero},
  {Colina}, {Alonso-Herrero}, \& {Alfonso-Garz{\'o}n}}]{RZ11}
{Rodr{\'\i}guez-Zaur{\'\i}n}, J., {Arribas}, S., {Monreal-Ibero}, A., {et~al.}
  2011, \aap, 527, A60

\bibitem[{{Rujopakarn} {et~al.}(2011){Rujopakarn}, {Rieke}, {Eisenstein}, \&
  {Juneau}}]{Rujopakarn11}
{Rujopakarn}, W., {Rieke}, G.~H., {Eisenstein}, D.~J., \& {Juneau}, S. 2011,
  \apj, 726, 93

\bibitem[{{S{\'a}nchez-Garc{\'\i}a} {et~al.}(2022){S{\'a}nchez-Garc{\'\i}a},
  {Pereira-Santaella}, {Garc{\'\i}a-Burillo}, {Colina}, {Alonso-Herrero},
  {Villar-Mart{\'\i}n}, {Saito}, {D{\'\i}az-Santos}, {Piqueras L{\'o}pez},
  {Arribas}, {Bellocchi}, {Cazzoli}, \& {Labiano}}]{SG22}
{S{\'a}nchez-Garc{\'\i}a}, M., {Pereira-Santaella}, M., {Garc{\'\i}a-Burillo},
  S., {et~al.} 2022, \aap, 659, A102

\bibitem[{{Sanders} {et~al.}(2003){Sanders}, {Mazzarella}, {Kim}, {Surace}, \&
  {Soifer}}]{Sanders03}
{Sanders}, D.~B., {Mazzarella}, J.~M., {Kim}, D.~C., {Surace}, J.~A., \&
  {Soifer}, B.~T. 2003, \aj, 126, 1607

\bibitem[{{Sanders} \& {Mirabel}(1996)}]{Sanders96}
{Sanders}, D.~B. \& {Mirabel}, I.~F. 1996, \araa, 34, 749

\bibitem[{{Shen} {et~al.}(2003){Shen}, {Mo}, {White}, {Blanton}, {Kauffmann},
  {Voges}, {Brinkmann}, \& {Csabai}}]{Shen03}
{Shen}, S., {Mo}, H.~J., {White}, S. D.~M., {et~al.} 2003, \mnras, 343, 978

\bibitem[{{Simpson} {et~al.}(2015){Simpson}, {Smail}, {Swinbank}, {Almaini},
  {Blain}, {Bremer}, {Chapman}, {Chen}, {Conselice}, {Coppin}, {Danielson},
  {Dunlop}, {Edge}, {Farrah}, {Geach}, {Hartley}, {Ivison}, {Karim}, {Lani},
  {Ma}, {Meijerink}, {Micha{\l}owski}, {Mortlock}, {Scott}, {Simpson},
  {Spaans}, {Thomson}, {van Kampen}, \& {van der Werf}}]{Simpson15}
{Simpson}, J.~M., {Smail}, I., {Swinbank}, A.~M., {et~al.} 2015, \apj, 799, 81

\bibitem[{{Skrutskie} {et~al.}(2006){Skrutskie}, {Cutri}, {Stiening},
  {Weinberg}, {Schneider}, {Carpenter}, {Beichman}, {Capps}, {Chester},
  {Elias}, {Huchra}, {Liebert}, {Lonsdale}, {Monet}, {Price}, {Seitzer},
  {Jarrett}, {Kirkpatrick}, {Gizis}, {Howard}, {Evans}, {Fowler}, {Fullmer},
  {Hurt}, {Light}, {Kopan}, {Marsh}, {McCallon}, {Tam}, {Van Dyk}, \&
  {Wheelock}}]{Skrutskie06}
{Skrutskie}, M.~F., {Cutri}, R.~M., {Stiening}, R., {et~al.} 2006, \aj, 131,
  1163

\bibitem[{{Straatman} {et~al.}(2015){Straatman}, {Labb{\'e}}, {Spitler},
  {Glazebrook}, {Tomczak}, {Allen}, {Brammer}, {Cowley}, {van Dokkum},
  {Kacprzak}, {Kawinwanichakij}, {Mehrtens}, {Nanayakkara}, {Papovich},
  {Persson}, {Quadri}, {Rees}, {Tilvi}, {Tran}, \& {Whitaker}}]{Straatman15}
{Straatman}, C. M.~S., {Labb{\'e}}, I., {Spitler}, L.~R., {et~al.} 2015, \apjl,
  808, L29

\bibitem[{{Tacconi} {et~al.}(2002){Tacconi}, {Genzel}, {Lutz}, {Rigopoulou},
  {Baker}, {Iserlohe}, \& {Tecza}}]{Tacconi02}
{Tacconi}, L.~J., {Genzel}, R., {Lutz}, D., {et~al.} 2002, \apj, 580, 73

\bibitem[{{Tacconi} {et~al.}(2006){Tacconi}, {Neri}, {Chapman}, {Genzel},
  {Smail}, {Ivison}, {Bertoldi}, {Blain}, {Cox}, {Greve}, \&
  {Omont}}]{Tacconi06}
{Tacconi}, L.~J., {Neri}, R., {Chapman}, S.~C., {et~al.} 2006, \apj, 640, 228

\bibitem[{{Tadaki} {et~al.}(2017){Tadaki}, {Genzel}, {Kodama}, {Wuyts},
  {Wisnioski}, {F{\"o}rster Schreiber}, {Burkert}, {Lang}, {Tacconi}, {Lutz},
  {Belli}, {Davies}, {Hatsukade}, {Hayashi}, {Herrera-Camus}, {Ikarashi},
  {Inoue}, {Kohno}, {Koyama}, {Mendel}, {Nakanishi}, {Shimakawa}, {Suzuki},
  {Tamura}, {Tanaka}, {{\"U}bler}, \& {Wilman}}]{Tadaki17}
{Tadaki}, K.-i., {Genzel}, R., {Kodama}, T., {et~al.} 2017, \apj, 834, 135

\bibitem[{{Tran} {et~al.}(2001){Tran}, {Lutz}, {Genzel}, {Rigopoulou}, {Spoon},
  {Sturm}, {Gerin}, {Hines}, {Moorwood}, {Sanders}, {Scoville}, {Taniguchi}, \&
  {Ward}}]{Tran01}
{Tran}, Q.~D., {Lutz}, D., {Genzel}, R., {et~al.} 2001, \apj, 552, 527

\bibitem[{{Trujillo} {et~al.}(2020){Trujillo}, {Chamba}, \&
  {Knapen}}]{Trujillo20}
{Trujillo}, I., {Chamba}, N., \& {Knapen}, J.~H. 2020, \mnras, 493, 87

\bibitem[{{U} {et~al.}(2012){U}, {Sanders}, {Mazzarella}, {Evans}, {Howell},
  {Surace}, {Armus}, {Iwasawa}, {Kim}, {Casey}, {Vavilkin}, {Dufault},
  {Larson}, {Barnes}, {Chan}, {Frayer}, {Haan}, {Inami}, {Ishida},
  {Kartaltepe}, {Melbourne}, \& {Petric}}]{U12}
{U}, V., {Sanders}, D.~B., {Mazzarella}, J.~M., {et~al.} 2012, \apjs, 203, 9

\bibitem[{{Valentino} {et~al.}(2020){Valentino}, {Daddi}, {Puglisi}, {Magdis},
  {Liu}, {Kokorev}, {Cortzen}, {Madden}, {Aravena}, {G{\'o}mez-Guijarro},
  {Lee}, {Le Floc'h}, {Gao}, {Gobat}, {Bournaud}, {Dannerbauer}, {Jin},
  {Dickinson}, {Kartaltepe}, \& {Sanders}}]{Valentino20}
{Valentino}, F., {Daddi}, E., {Puglisi}, A., {et~al.} 2020, \aap, 641, A155

\bibitem[{{Valiante} {et~al.}(2009){Valiante}, {Lutz}, {Sturm}, {Genzel}, \&
  {Chapin}}]{Valiante09}
{Valiante}, E., {Lutz}, D., {Sturm}, E., {Genzel}, R., \& {Chapin}, E.~L. 2009,
  \apj, 701, 1814

\bibitem[{{van der Wel} {et~al.}(2014){van der Wel}, {Franx}, {van Dokkum},
  {Skelton}, {Momcheva}, {Whitaker}, {Brammer}, {Bell}, {Rix}, {Wuyts},
  {Ferguson}, {Holden}, {Barro}, {Koekemoer}, {Chang}, {McGrath},
  {H{\"a}ussler}, {Dekel}, {Behroozi}, {Fumagalli}, {Leja}, {Lundgren},
  {Maseda}, {Nelson}, {Wake}, {Patel}, {Labb{\'e}}, {Faber}, {Grogin}, \&
  {Kocevski}}]{vdWel14}
{van der Wel}, A., {Franx}, M., {van Dokkum}, P.~G., {et~al.} 2014, \apj, 788,
  28

\bibitem[{{Veilleux}(1999)}]{Veilleux99}
{Veilleux}, S. 1999, \apss, 266, 67

\bibitem[{{Veilleux} {et~al.}(2002){Veilleux}, {Kim}, \&
  {Sanders}}]{Veilleux02}
{Veilleux}, S., {Kim}, D.~C., \& {Sanders}, D.~B. 2002, \apjs, 143, 315

\bibitem[{{Veilleux} {et~al.}(2009){Veilleux}, {Rupke}, {Kim}, {Genzel},
  {Sturm}, {Lutz}, {Contursi}, {Schweitzer}, {Tacconi}, {Netzer}, {Sternberg},
  {Mihos}, {Baker}, {Mazzarella}, {Lord}, {Sanders}, {Stockton}, {Joseph}, \&
  {Barnes}}]{Veilleux09}
{Veilleux}, S., {Rupke}, D.~S.~N., {Kim}, D.~C., {et~al.} 2009, \apjs, 182, 628

\bibitem[{{Wang} {et~al.}(2018){Wang}, {Kong}, \& {Pan}}]{Wang18}
{Wang}, E., {Kong}, X., \& {Pan}, Z. 2018, \apj, 865, 49

\bibitem[{{Whitaker} {et~al.}(2012){Whitaker}, {van Dokkum}, {Brammer}, \&
  {Franx}}]{Whitaker12}
{Whitaker}, K.~E., {van Dokkum}, P.~G., {Brammer}, G., \& {Franx}, M. 2012,
  \apjl, 754, L29

\bibitem[{{Wilman} {et~al.}(2020){Wilman}, {Fossati}, {Mendel}, {Saglia},
  {Wisnioski}, {Wuyts}, {F{\"o}rster Schreiber}, {Beifiori}, {Bender}, {Belli},
  {{\"U}bler}, {Lang}, {Chan}, {Davies}, {Nelson}, {Genzel}, {Tacconi},
  {Galametz}, {Davies}, {Lutz}, {Price}, {Burkert}, {Tadaki}, {Herrera-Camus},
  {Brammer}, {Momcheva}, \& {van Dokkum}}]{Wilman20}
{Wilman}, D.~J., {Fossati}, M., {Mendel}, J.~T., {et~al.} 2020, \apj, 892, 1

\bibitem[{{Wright}(2006)}]{Wright06}
{Wright}, E.~L. 2006, \pasp, 118, 1711

\bibitem[{{Wuyts} {et~al.}(2011{\natexlab{a}}){Wuyts}, {F{\"o}rster Schreiber},
  {Lutz}, {Nordon}, {Berta}, {Altieri}, {Andreani}, {Aussel}, {Bongiovanni},
  {Cepa}, {Cimatti}, {Daddi}, {Elbaz}, {Genzel}, {Koekemoer}, {Magnelli},
  {Maiolino}, {McGrath}, {P{\'e}rez Garc{\'\i}a}, {Poglitsch}, {Popesso},
  {Pozzi}, {Sanchez-Portal}, {Sturm}, {Tacconi}, \& {Valtchanov}}]{Wuyts11a}
{Wuyts}, S., {F{\"o}rster Schreiber}, N.~M., {Lutz}, D., {et~al.}
  2011{\natexlab{a}}, \apj, 738, 106

\bibitem[{{Wuyts} {et~al.}(2011{\natexlab{b}}){Wuyts}, {F{\"o}rster Schreiber},
  {van der Wel}, {Magnelli}, {Guo}, {Genzel}, {Lutz}, {Aussel}, {Barro},
  {Berta}, {Cava}, {Graci{\'a}-Carpio}, {Hathi}, {Huang}, {Kocevski},
  {Koekemoer}, {Lee}, {Le Floc'h}, {McGrath}, {Nordon}, {Popesso}, {Pozzi},
  {Riguccini}, {Rodighiero}, {Saintonge}, \& {Tacconi}}]{Wuyts11}
{Wuyts}, S., {F{\"o}rster Schreiber}, N.~M., {van der Wel}, A., {et~al.}
  2011{\natexlab{b}}, \apj, 742, 96

\bibitem[{{Yuan} {et~al.}(2010){Yuan}, {Kewley}, \& {Sanders}}]{Yuan10}
{Yuan}, T.~T., {Kewley}, L.~J., \& {Sanders}, D.~B. 2010, \apj, 709, 884

\bibitem[{{Zanella} {et~al.}(2018){Zanella}, {Daddi}, {Magdis}, {Diaz Santos},
  {Cormier}, {Liu}, {Cibinel}, {Gobat}, {Dickinson}, {Sargent}, {Popping},
  {Madden}, {Bethermin}, {Hughes}, {Valentino}, {Rujopakarn}, {Pannella},
  {Bournaud}, {Walter}, {Wang}, {Elbaz}, \& {Coogan}}]{Zanella18}
{Zanella}, A., {Daddi}, E., {Magdis}, G., {et~al.} 2018, \mnras, 481, 1976

\bibitem[{{Zibetti} {et~al.}(2009){Zibetti}, {Charlot}, \& {Rix}}]{Zibetti09}
{Zibetti}, S., {Charlot}, S., \& {Rix}, H.-W. 2009, \mnras, 400, 1181

\bibitem[{{Zolotov} {et~al.}(2015){Zolotov}, {Dekel}, {Mandelker}, {Tweed},
  {Inoue}, {DeGraf}, {Ceverino}, {Primack}, {Barro}, \& {Faber}}]{Zolotov15}
{Zolotov}, A., {Dekel}, A., {Mandelker}, N., {et~al.} 2015, \mnras, 450, 2327

\end{thebibliography}

\clearpage

\begin{appendix}

\section{Classifying the LIRG sample using morphological and kinematic information}
\label{app_classification}

\begin{figure*}
\centering
\includegraphics[width=0.6\textwidth]{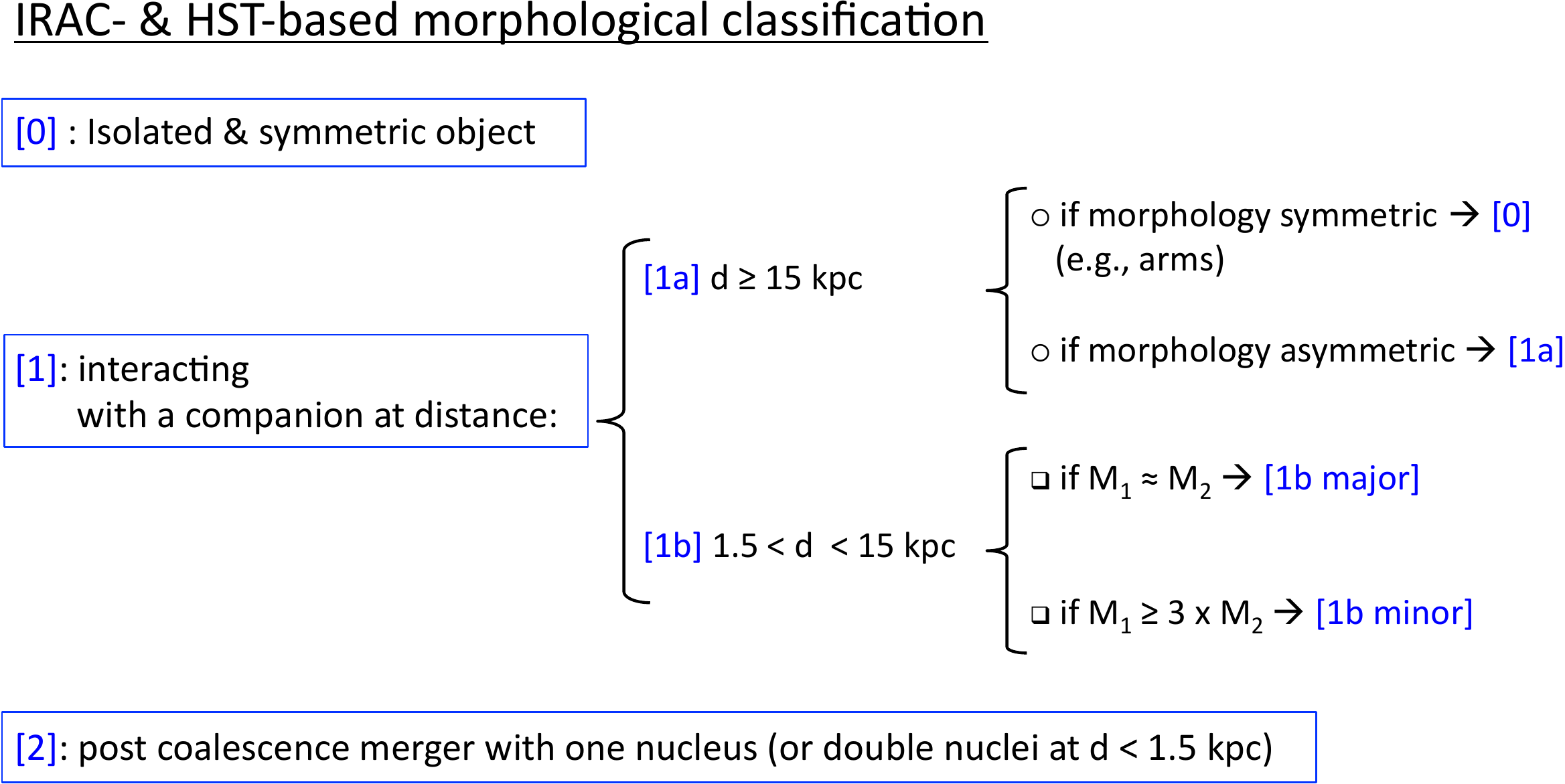}
\caption{Schematic view of the morphological classification based on {\it Spitzer}/IRAC and {\it HST} images for the whole sample. }
\label{IRAC_class_HST}
\end{figure*}

In this appendix we describe the criteria used to classified our sample. 
As shown in Tab.~\ref{Input_more_classifs}, the final classification is the result of the combination between the morphological information obtained using {\it Spitzer} and {\it HST} images and the kinematic information derived from the ionized and molecular gas velocity and velocity dispersion maps. We refer to this as `composite' classification.
In particular, the morphological classification is based on archival near-IR {\it Spitzer}/IRAC band at 3.6 $\mu$m and {\it HST}/WFPC3 images when available. Briefly, the morphological classes are defined following the simplified version of the \cite{Veilleux02} classification. As shown in Fig.~\ref{IRAC_class_HST}, we define three main classes:

\begin{itemize}
\item {\tt Class 0}: objects that appear to be single isolated objects, with relatively symmetric disk morphologies and without evidence for strong past or ongoing interaction;
\vskip1mm
\item {\tt Class 1}: objects in a pre-coalescence phase with two well differentiated nuclei separated by a projected distance of at least 1.5 kpc. For these objects, it is still possible to identify (in some cases using {\it HST} imaging) the individual merging galaxies and their corresponding tidal structures due to the interaction;
\vskip1mm
\item {\tt Class 2}: objects with two nuclei separated by a projected distance $\leq$1.5 kpc or a single nucleus with a relatively asymmetric morphology, suggesting a post-coalescence merging phase (hereafter, {\tt merger}).
\end{itemize}

Within the type 1 objects, we also defined other sub-classes according to the distance involved in the system: 

\begin{itemize}
\item {\tt 1a}: interacting objects with a projected separation larger than 15 kpc; in case of showing a symmetric morphology (i.e., symmetric spiral arms), the galaxy is finally classified as {\tt `0'}, otherwise, in case of asymmetric morphology (e.g., asymmetric spiral arms), we confirm the morphological class as {\tt `1a'};  
\vskip1mm
\item {\tt 1b}: interacting objects with a projected separation in between 1.5 and 15 kpc; if the interacting galaxies are characterized by similar mass we classify them as {\tt `1b major'}, otherwise, if the masses are different we classify them as {\tt `1b minor'}.
\end{itemize}

In some cases the morphological classification alone could be misleading, thus we complemented this information using the kinematic information available for the different gas  (i.e., ionized and molecular) phases. 
As mentioned in the Introduction, the ionized emission (as traced by the H$\alpha$ line) in our LIRG sample has been analyzed in previous works with the aim of studying the kinematic asymmetries in the H$\alpha$ maps, through the visual inspection (\citealt{Bellocchi13}) and through the {\tt kinemetry} method (\citealt{Bellocchi16}). We reported these results in Tab.~\ref{Input_more_classifs} (Columns 4 and 5). In brief, according to the visual classification we classify the objects as {\tt rotating disk (RD)}, {\tt perturbed disk (PD)} and {\tt complex kinematics (CK)} (see \citealt{Bellocchi13} for further details). According to the { \tt kinemetry} method, we classify the galaxies as {\tt disk (D)} or {\tt merger (M)} or, when the galaxies belong to the `transition region', as {\tt disk* (D*)} or {\tt merger* (M*)} (see \citealt{Bellocchi16} for details). The `transition region' defines the area where disks and mergers coexist, making more uncertain their classification. 
Furthermore, following the same scheme used for the H$\alpha$ emission, we visually classify the kinematic maps produced for the molecular CO(2--1) line. These maps are not shown in this work, because it goes beyond the scope of this paper and they will be presented in future works. However, these maps allowed us to better characterize the kinematics of our systems.

Combining both the morphological and kinematic information, we finally proposed the `composite' classification shown in Tab.~\ref{Input_more_classifs} (Column 7), defining the following classes:

\begin{itemize}
\item {\tt 0 (RD)}: single isolated objects, with relatively symmetric disk morphologies and without evidence for strong past or ongoing interaction. Their kinematic (i.e., velocity field and velocity dispersion) maps show the typical `rotating disk' pattern: i.e., {\it point-antisymmetric} velocity field and {\it point-symmetric} velocity dispersion (see \citealt{Krajnovic06} for a detailed description; hereafter, {\tt rotating disk (RD)});
\vskip1mm
\item {\tt 0 (PD)}: single isolated objects, with relatively symmetric disk morphologies but showing a somehow perturbed kinematics (perturbed velocity field and/or dispersion maps), hinting some past interactions (hereafter, {\tt perturbed disk (PD)});
\vskip1mm
\item {\tt 1}: 
objects in a pre-coalescence phase with two well differentiated nuclei separated by a projected distance of at least 1.5 kpc up to a maximum distance of 15 kpc (hereafter, {\tt interacting}) showing perturbed or complex kinematics. Beyond this separation ($>$15 kpc) the galaxies are also classified as {\tt `RD'} or {\tt `PD'}, like type {\tt 0} ones (e.g., ESO 297-G011/G012); 
\vskip1mm
\item {\tt 2}: objects with two nuclei separated by a projected distance $\leq$1.5 kpc or a single nucleus with a relatively asymmetric morphology (e.g., tidal tail, asymmetric spiral arms), which suggests a post-coalescence merging phase (hereafter, {\tt merger}). These systems show perturbed or complex kinematics. 
\end{itemize}

In the specific case of NGC 2369, this is an isolated galaxy observed edge-on, showing a perturbed kinematics in H$\alpha$ and CO. However, using the {\tt kinemetry} method it is classified as {\tt D*}, then lying in the transition region: we proposed the {\tt `0 (PD)'} classification.
NGC 5135 is an isolated face-on galaxy which shows quite complex kinematics in H$\alpha$ and CO and classified as {\tt D*} with {\tt kinemetry}. Furthermore, in this case, the H$\alpha$ emission is perpendicular to the continuum emission. As in NGC 2369, we propose the classification {\tt `0 (PD)'}. 
These are two extreme cases for which the perturbed and distorted kinematics might be due to the inclination effect of the sources: although the galaxies are isolated, in the face-on galaxy the rotational motion of the gas does not seem to clearly dominate on the dispersion (v~$\gtrsim$~$\sigma$) while in the edge-one galaxy projection effects could affect the observed velocity pattern.\\ 
Finally, IC 4687 belongs a triple system and it is the northern galaxy of the system. This source is separated by $\sim$10~kpc from the central galaxy (IC 4686) and by $\sim$30~kpc from the southern companion (IC 4689). The ratio of the dynamical masses between the northern (N), central (C) and southern (S) companion is 1~:~$\frac{1}{9}$~:~1 (N:C:S), allowing us to classify the NC system as {\tt `1b~minor'}. According to its kinematics, IC~4687 follows a quite regular rotational pattern both in the ionized and molecular gas phases, thus we finally propose the {\tt `0~(RD)'} classification.

The visual kinematic classification proposed for our galaxies using the ionized and molecular gas tracers allowed a direct comparison between the two. Furthermore, the visual classification could be also useful to check the results obtained with {\tt kinemetry}. Indeed, although this method allows to quantify the asymmetries in the kinematic maps, the frontier which separates disks from mergers is not universal neither clearly defined. This method thus involves certain degree of uncertainty in its classification.

For this reason, the final `composite' classification has been achieved taking into account the morphological and kinematic information, in which the latter includes both the visual and the {\tt kinemetry} classifications: this allowed us to analyze the sample under several aspects and finally draw more robust conclusions on the real nature of the systems.

\begin{table*}
\centering
\begin{small}
  \caption{Morphological and kinematic properties of the LIRG sample. The morphological classification is based on {\it Spitzer}/IRAC and {\it HST}/WFPC3 images. The kinematic classification is based using the ionized (H$\alpha$) and molecular (CO) gas kinematic maps according to the visual and {\tt kinemetry} classifications. The final `composite' classification is also shown as a result of the combination of the two classifications.}
\label{Input_more_classifs}
\begin{tabular}{ccccccc} 
\hline\hline\noalign{\smallskip}  
\multicolumn{2}{c}{Source}	 &		{\it Spitzer}/{\it HST} &\multicolumn{2}{c}{H$\alpha$ kinematics} & \multicolumn{1}{c}{$^{12}$CO(2--1) kinematics}	& {\tt Composite}	 \\
IRAS & Other 	&	Morphology	&	Visual class.	&	{\tt kinemetry}	&	Visual class. &{\tt classification}	\\
\cmidrule(lr){1-2}\cmidrule(lr){3-3}\cmidrule(lr){4-5}\cmidrule(lr){6-6}\cmidrule(lr){7-7}
(1) & (2) &  (3)& (4) &(5)  & (6)  & (7)	 \\
\hline\hline\noalign{\smallskip} 	
{\tt F01341-3735 N}	 & {\tt ESO 297-G011}	& 1a  	& PD	&D 	& PD	&	 {\tt 0 (PD)}\\
{\tt F01341-3735 S}	 & {\tt ESO 297-G012}	& 0 	& PD	& D		&	RD	&	{\tt 0 (RD)}	\\
\hline\noalign{\smallskip} 	
{\tt F04315-0840}	&{\tt NGC 1614}	&	2	&	CK		&	D*	& 	PD		&	{\tt 2} \\
\hline\noalign{\smallskip} 	
{\tt F06295-1735} & {\tt ESO 557-G002}	&	1a	&	PD		& D*		&	PD	&	{\tt 0 (PD)}	\\        
\hline\noalign{\smallskip} 	
{\tt F06592-6313 }	& --	& 0	&	PD	&	D	&	PD 	& 	{\tt 0 (PD)}\\
\hline\noalign{\smallskip} 	
{\tt F07160-6215}	&{\tt NGC 2369}		&	0	&	PD (CK)	&	D* & PD (CK)			&	{\tt 0 (PD)} \\
\hline\noalign{\smallskip} 	
{\tt  F10015-0614}	& {\tt NGC 3110} &1a	&	PD	&	D&	PD	&	{\tt 0 (PD)}	 \\
\hline\noalign{\smallskip} 	
{\tt F10257-4339}		&	{\tt NGC 3256} & 	2	&	PD	 & M  &	CK	&	{\tt 2}\\
\hline\noalign{\smallskip} 	
{\tt F10409-4556} 	& {\tt ESO 264-G036}	&	0	&	RD	& D & RD	&	{\tt 0 (RD)}	\\
\hline\noalign{\smallskip} 	
{\tt F11255-4120} 	& 	{\tt ESO 319-G022} &	0	&	PD	& D 	 & PD	&	{\tt 0 (PD)}\\     
\hline\noalign{\smallskip} 	
{\tt F11506-3851} 	& 	{\tt ESO 320-G030}	& 0	&	RD		& D & RD  	&	{\tt 0 (RD)}	 \\
\hline\noalign{\smallskip} 	
{\tt F12115-4546} 	& {\tt ESO 267-G030} &  1a 	&	RD	& D	&  RD	&	{\tt 0 (RD)}	 \\
\hline\noalign{\smallskip} 	
{\tt F12596-1529 E}		&	{\tt MCG-02-33-098}   & 1b major	&	(...)	&	M	&	PD	&	 {\tt 1}\\
{\tt F12596-1529 W}		&		 	& 1b major	&	(...)	&	M	&	PD	&	 {\tt 1}\\
\hline\noalign{\smallskip} 	
{\tt F13001-2339} 	& {\tt ESO 507-G070}	&	1b minor  	&	CK & M*		&		PD	&	{\tt 2}	\\
\hline\noalign{\smallskip} 	
{\tt F13229-2934}		&	{\tt NGC 5135} & 	0	&	CK	& D*  	&	CK	&	{\tt 0 (PD)}	\\
\hline\noalign{\smallskip} 	
{\tt F14544-4255 E}		&	 {\tt IC 4518}	&1b major 	&	PD & D		& PD	&	{\tt 1}	\\
{\tt F14544-4255 W}	&	 {\tt IC 4518}	&1b major	&	CK (PD)			& D* &	CK	&	{\tt 1}\\
\hline\noalign{\smallskip} 	
{\tt F17138-1017} 	&		--	&	2	&	PD	  & D*	&	PD	&	{\tt 2} \\
\hline\noalign{\smallskip} 	
{\tt F18093-5744 N}		&	{\tt IC 4687}	& 1b minor 	& RD	& D	&	RD	&	{\tt 0 (RD)}	\\
\hline\noalign{\smallskip} 	
{\tt F18341-5732}		&	{\tt IC 4734}	&	0	&	(...)	&	(...)		&RD 		&	{\tt 0 (RD)}	\\
\hline\noalign{\smallskip} 	
{\tt F21453-3511}	&{\tt NGC 7130}	& 	1a  	&	PD	& D*  	& PD (CK)	&	{\tt 2}	\\
\hline\noalign{\smallskip} 	
{\tt F22132-3705 }	&	{\tt IC 5179}	& 	0	&	RD	& D  &	RD	&	{\tt 0 (RD)}	\\
\hline\noalign{\smallskip} 	 
{\tt F23007+0836}		& {\tt NGC 7469}	&	 0	&	(...)	&	(...)	 & RD	& {\tt 0 (RD)}	 \\
\hline\hline\noalign{\smallskip} 	
\end{tabular}
\end{small}
\vskip1mm\hskip0mm\begin{minipage}{18cm}
\small
{{\bf Notes:} 
Column (1): object designation in the Infrared Astronomical Satellite (IRAS) Faint Source Catalog (FSC). Column (2): other identification.
Column (3): morphological classification based on {\it Spitzer} and {\it HST} images. The galaxies are classified using the scheme shown in Fig.~\ref{IRAC_class_HST}.
Column (4): H$\alpha$ visual kinematic classification from \cite{Bellocchi13}. The sources are classified as rotating disk {\tt (RD)}, perturbed disk {\tt (PD)} or complex kinematics {\tt (CK)}. 
Column (5): H$\alpha$ {\tt kinemetry} classification from \cite{Bellocchi16}. 
The galaxies are classified as {\tt disk (D)}, {\tt merger (M)}, {\tt disk* (D*)} or {\tt merger* (M*)}. The star {\tt (*)} symbol identifies the galaxies which belong to the `transition region' (i.e., the area where disks and mergers coexist, making their classification more uncertain; see \citealt{Bellocchi16} for details).
Column (6): CO(2--1) visual kinematic classification (this work). We used the same criteria used for the H$\alpha$ line from \cite{Bellocchi13}.
Column (7): final {\tt composite} classification: the morphological and kinematic information are merged together. Four different classes are defined: isolated rotating disks {\tt `0~(RD)'}, isolated perturbed disks {\tt `0 (PD)'}, interacting systems {\tt `1'} and and post-coalescence mergers {\tt `2'} (see text for details).
}
\end{minipage}
\end{table*}

\section{CO(2--1) and 1.3~mm continuum ALMA maps and {\it HST}/NICMOS image}
\label{app_maps}

In this appendix we present the CO(2--1), 1.3 mm continuum flux density maps along with the stellar emission, observed using the near-IR {\it HST}/NICMOS F160W filter, for the whole sample. 

We centered the CO(2--1) and 1.3 mm continuum images using the near-IR peak stellar emission.  When the F160W filter is not available other near-IR {\it HST} filters (e.g., F190W, F110W) are considered. When {\it HST} images are not available, the continuum peak emission at 1.3 mm is considered as reference.

\begin{figure*}
\centering
\vskip5mm
\includegraphics[width=0.98\textwidth]{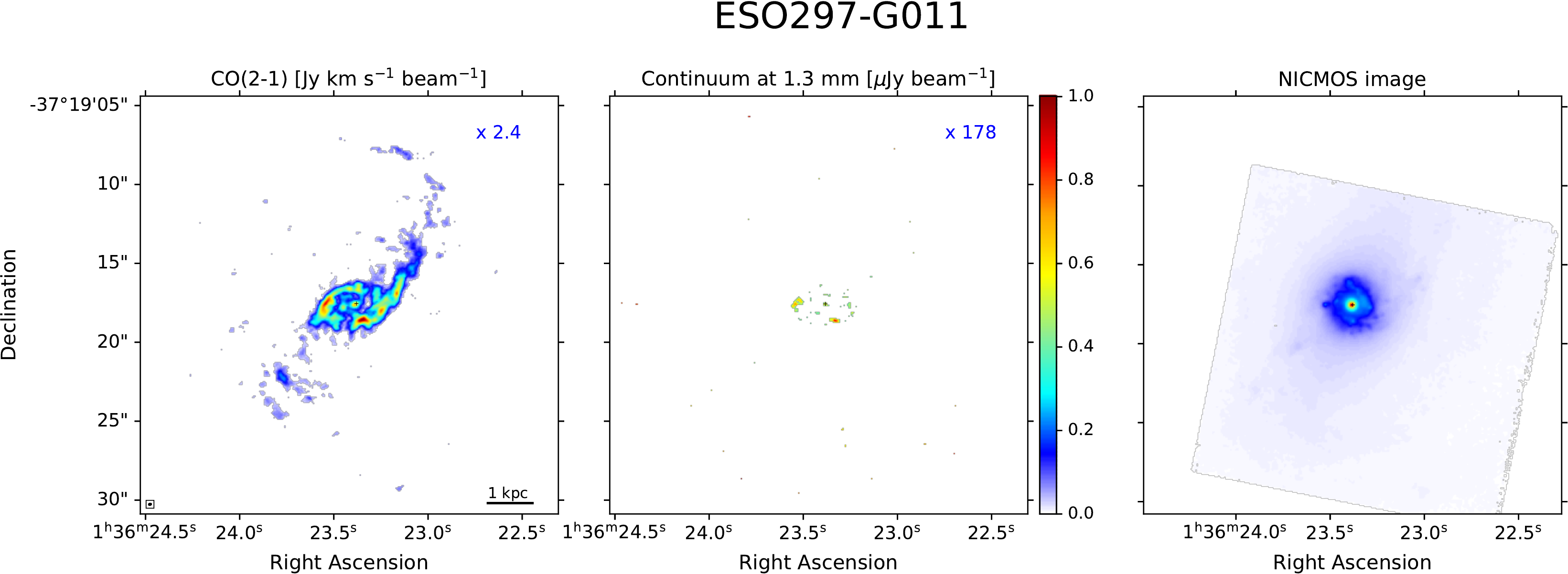}
\vskip5mm
\includegraphics[width=0.98\textwidth]{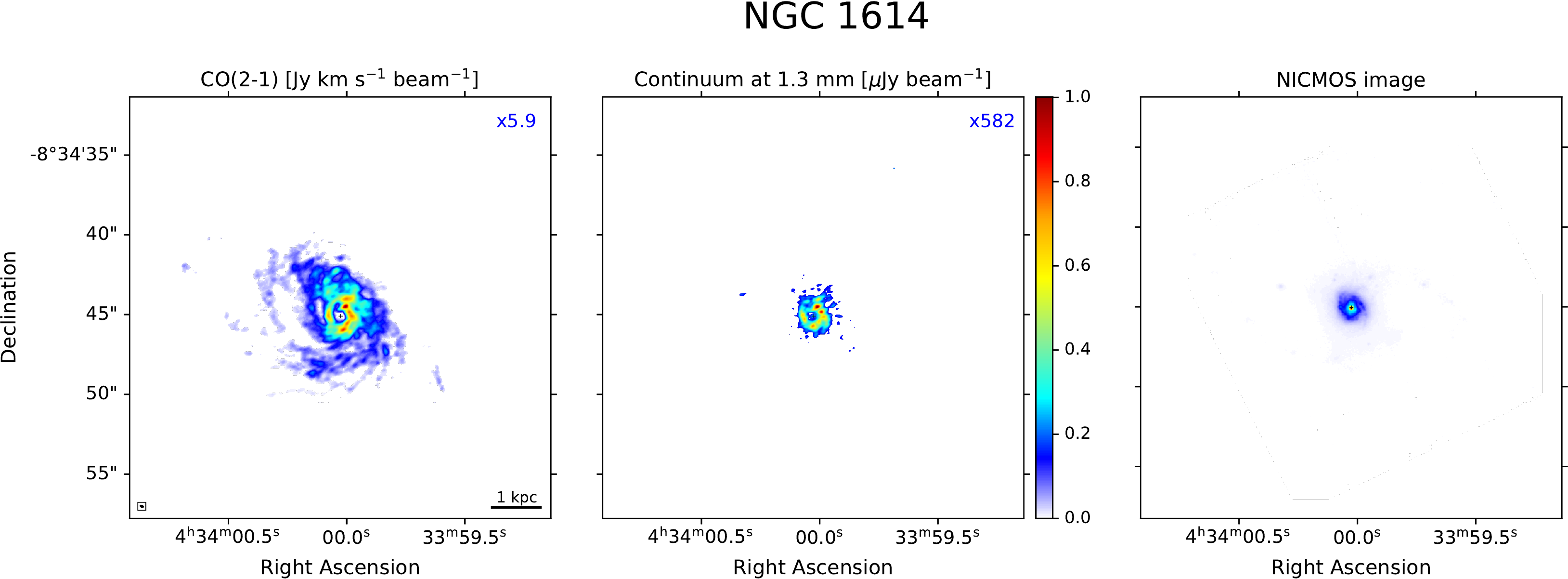}
\vskip5mm
\includegraphics[width=0.98\textwidth]{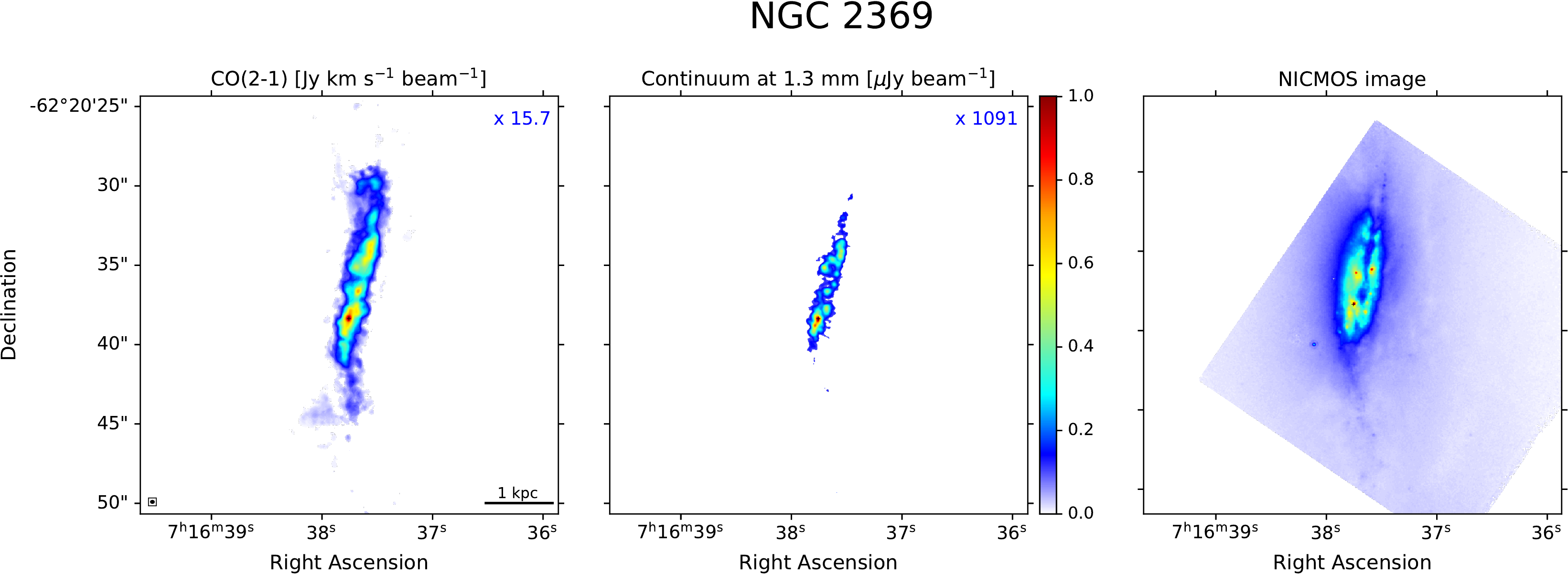}
\caption{{\it From left to right:} CO(2--1) and 1.3 mm continuum maps obtained with ALMA, complemented by {\it HST}/NICMOS image when available. The CO(2--1) and 1.3 mm emission is $>$5$\sigma$. The color bar is normalized to the maximum value of each ALMA maps: in order to derive the real maximum value, a factor scale, shown in the top-right of each panel, has to be applied. The cross in the three panels represents the stellar peak emission identified using {\it HST}/NICMOS image.  The beam size and the physical scale in kpc are also shown in the left panel. In all panel, north is to the top and east to the left.}
\label{mix_panels_ALMA_HST_0}
\end{figure*}

\begin{figure*}
\centering
\vskip5mm
\includegraphics[width=0.98\textwidth]{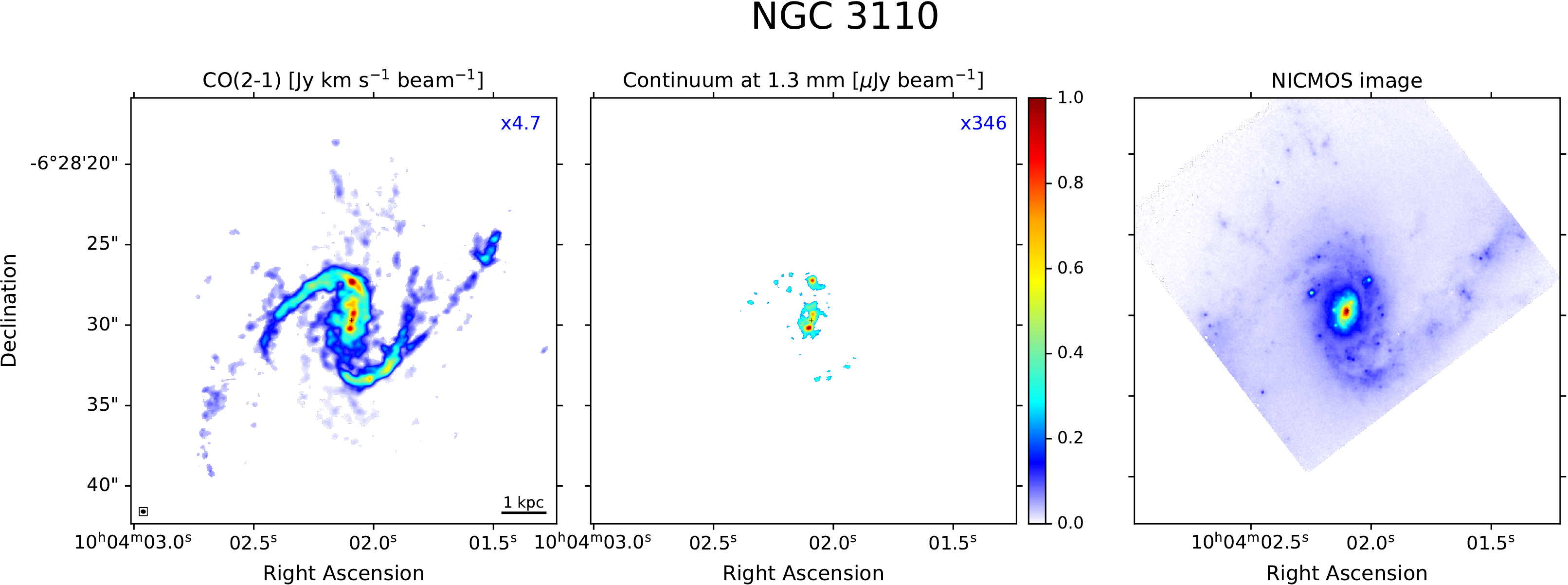}
\vskip7mm
\includegraphics[width=0.98\textwidth]{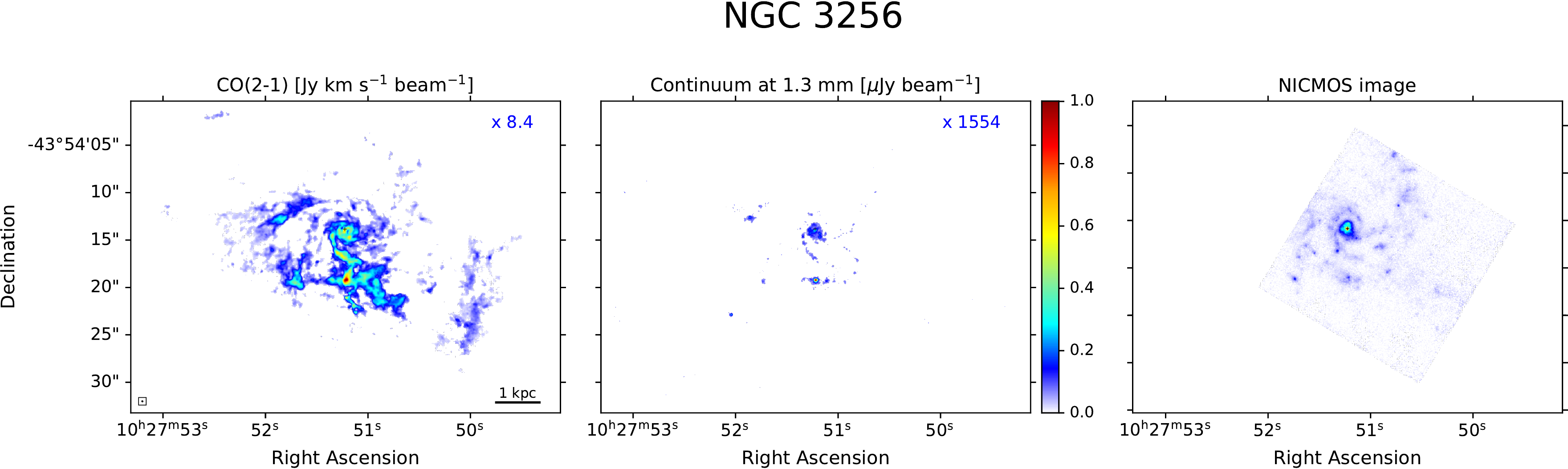}
\vskip7mm
\includegraphics[width=0.98\textwidth]{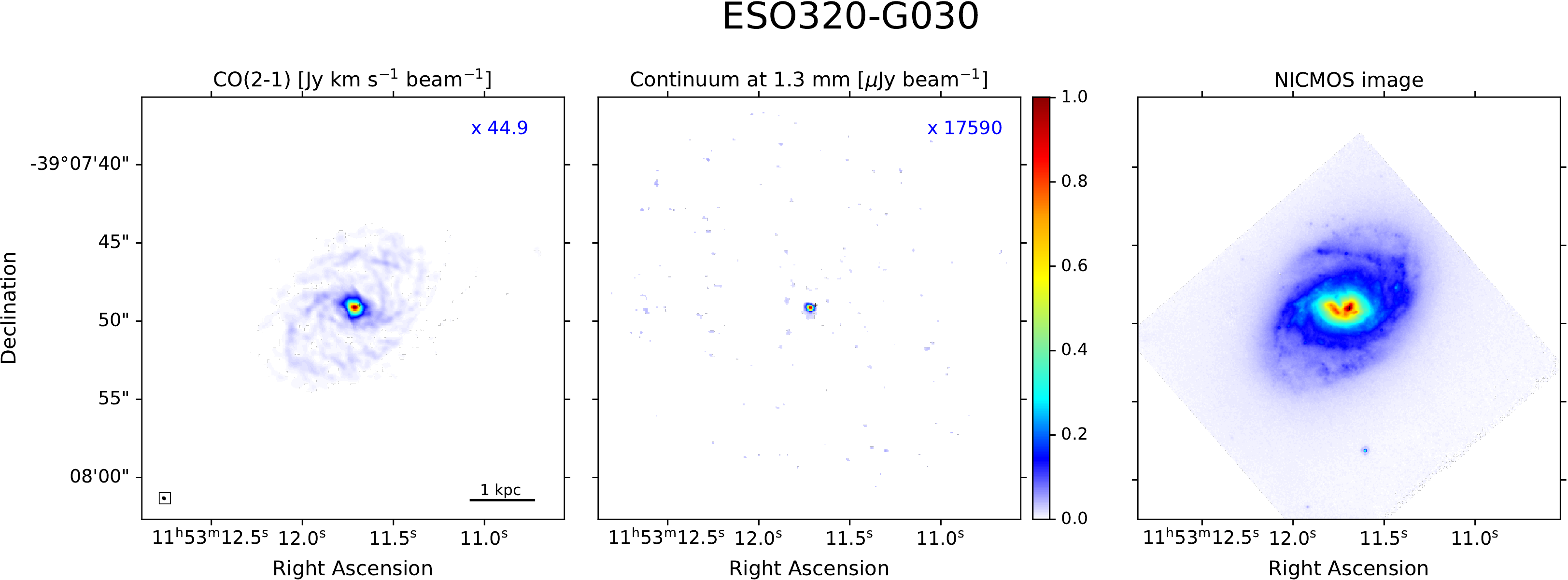}
\caption{Same figure caption as in Fig.~\ref{mix_panels_ALMA_HST_0}}
\label{mix_panels_ALMA_HST}
\end{figure*}

\begin{figure*}
\centering
\vskip7mm
\includegraphics[width=0.98\textwidth]{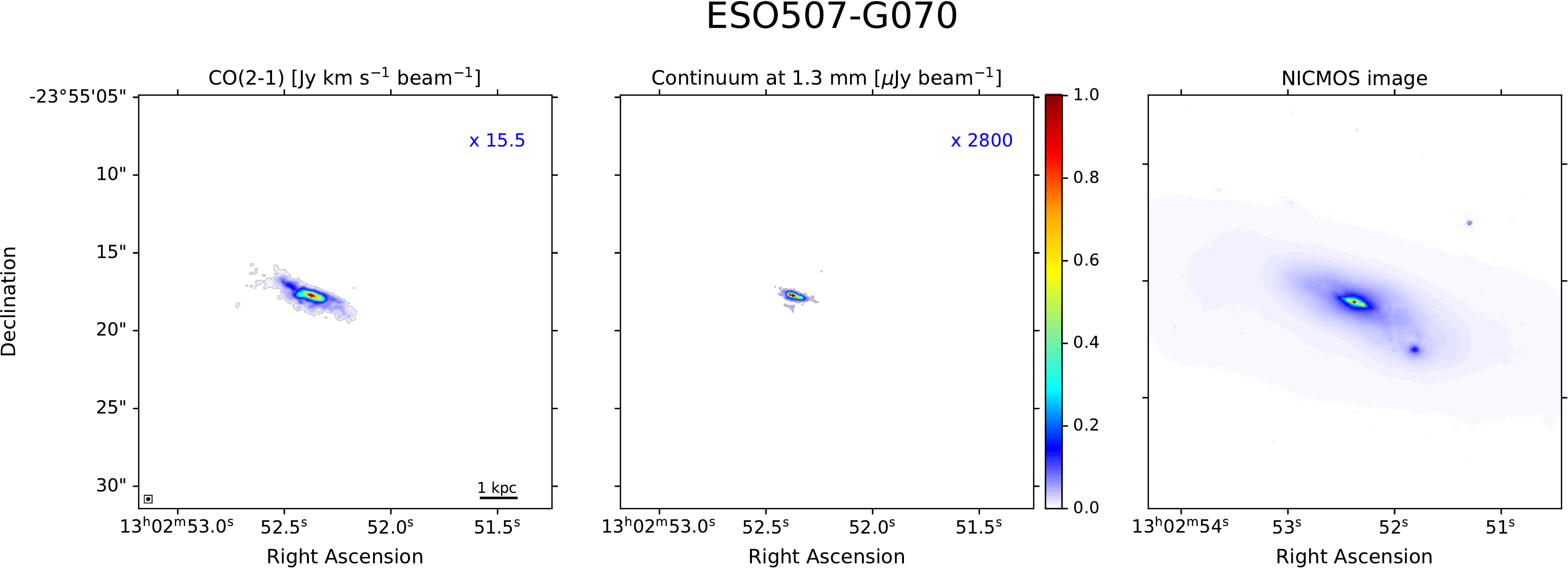}
\vskip7mm
\includegraphics[width=0.98\textwidth]{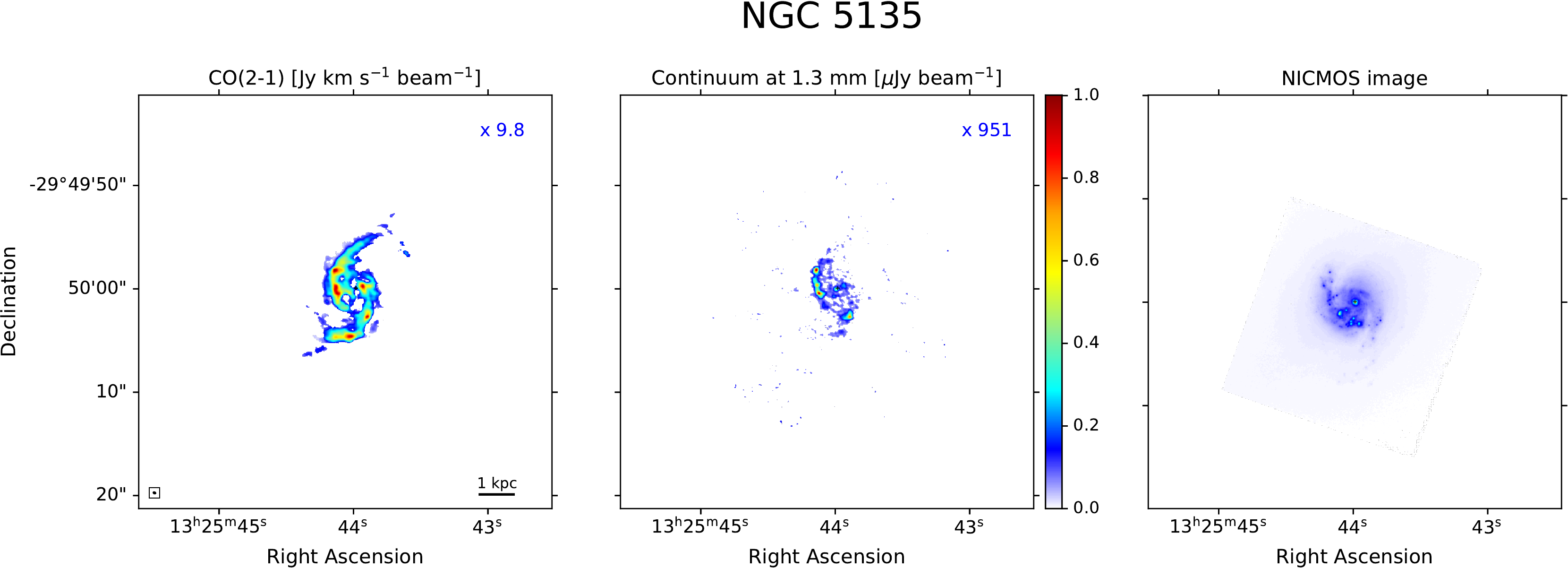}
\vskip7mm
\includegraphics[width=0.98\textwidth]{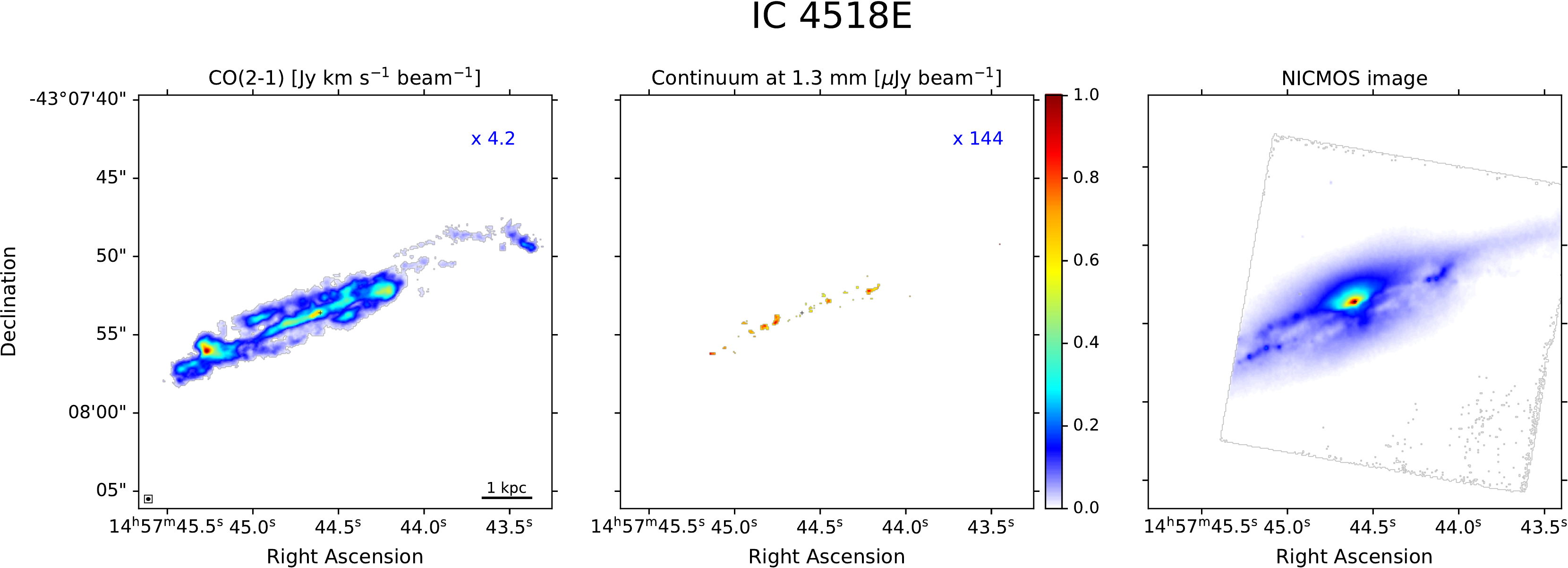}
\caption{Same figure caption as in Fig.~\ref{mix_panels_ALMA_HST_0}}
\label{mix_panels_ALMA_HST}
\end{figure*}

\begin{figure*}
\centering
\vskip7mm
\includegraphics[width=0.98\textwidth]{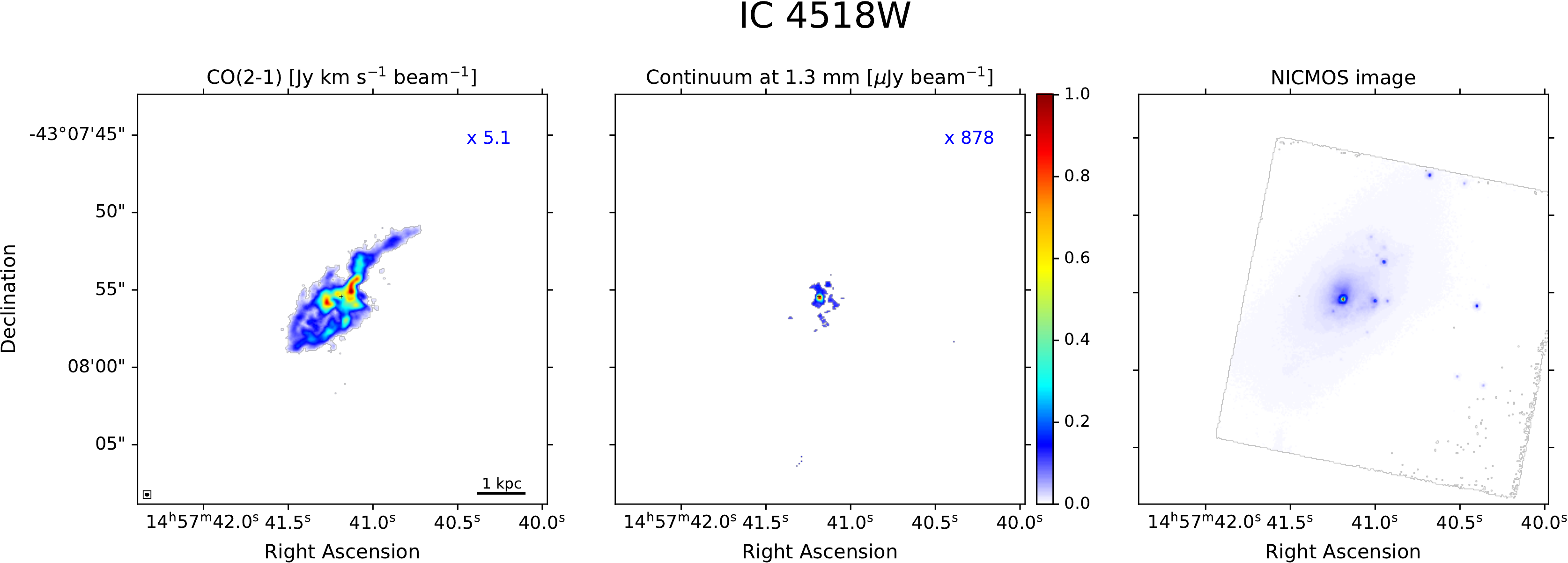}
\vskip7mm
\includegraphics[width=0.98\textwidth]{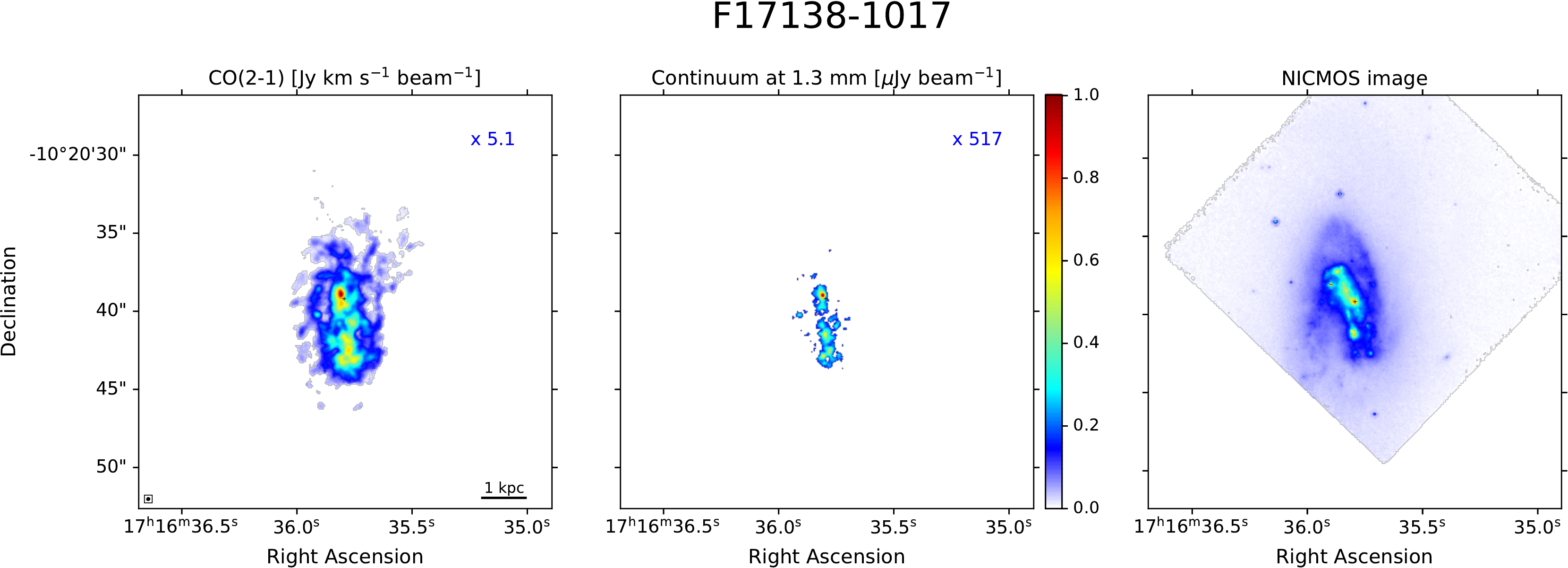}
\vskip7mm
\includegraphics[width=0.98\textwidth]{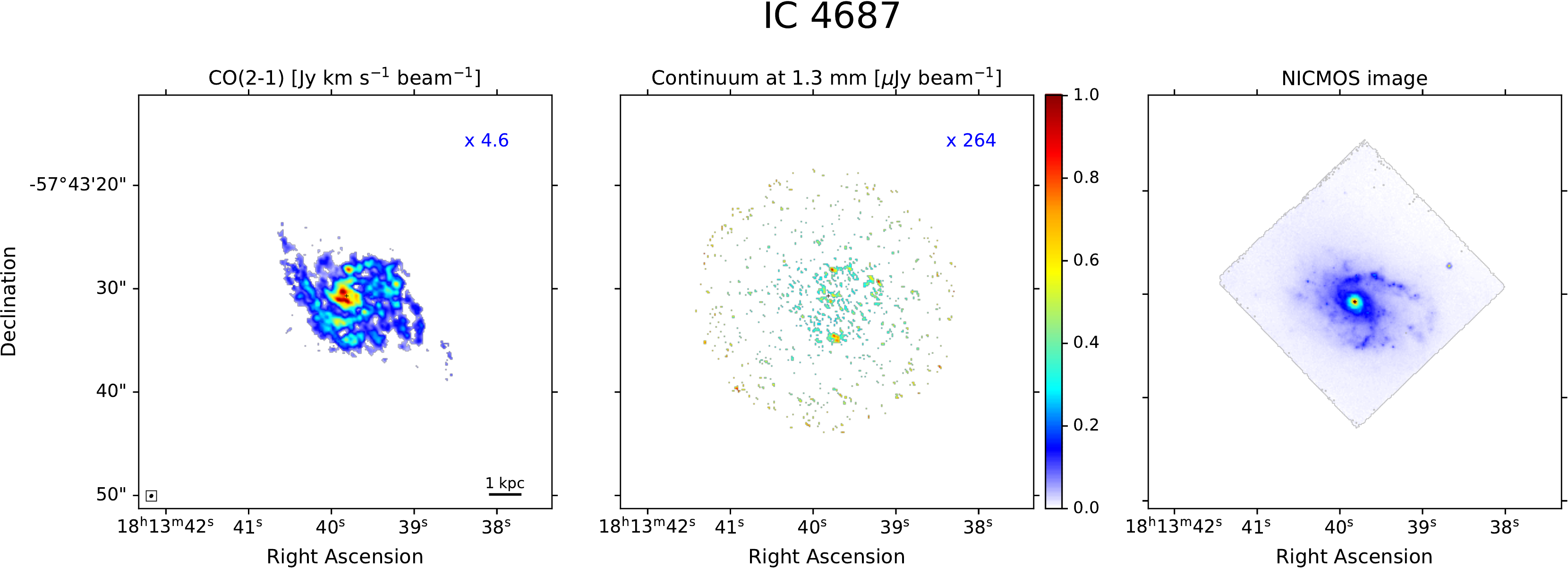}
\caption{Same figure caption as in Fig.~\ref{mix_panels_ALMA_HST_0}}
\label{mix_panels_ALMA_HST}
\end{figure*}

\begin{figure*}
\centering
\vskip7mm
\hskip4mm\includegraphics[width=1\textwidth]{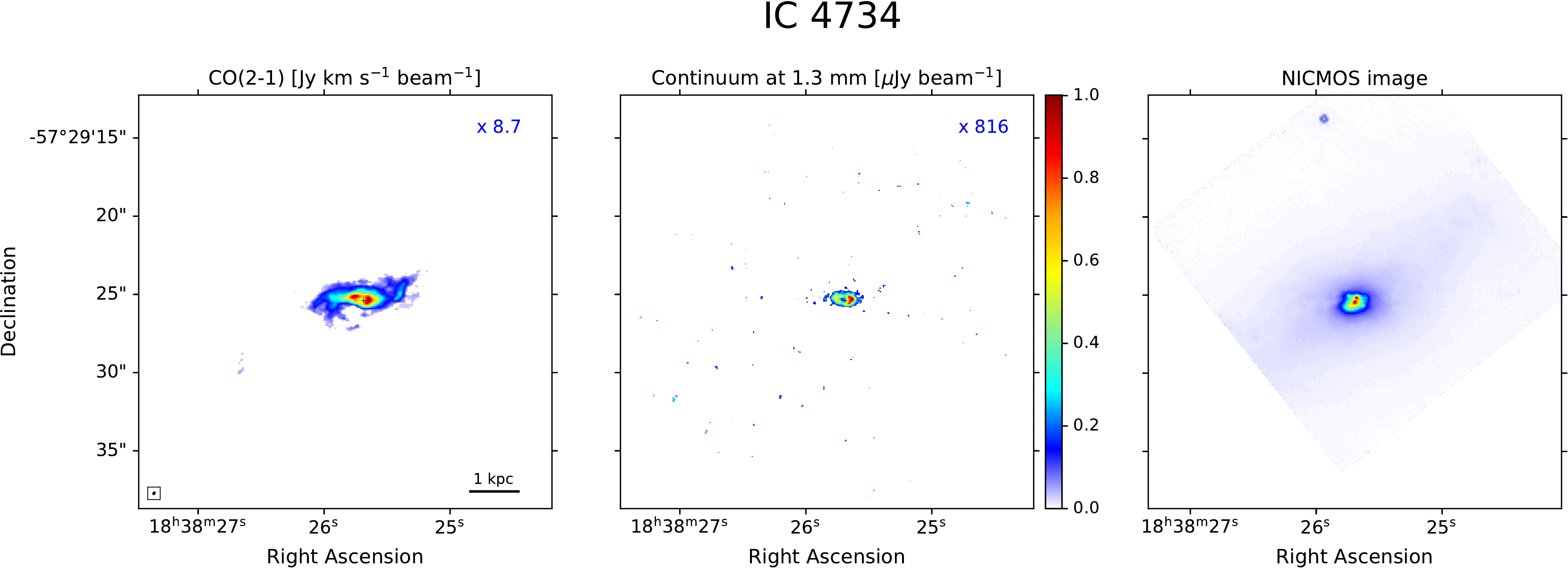}
\vskip7mm
\includegraphics[width=0.99\textwidth]{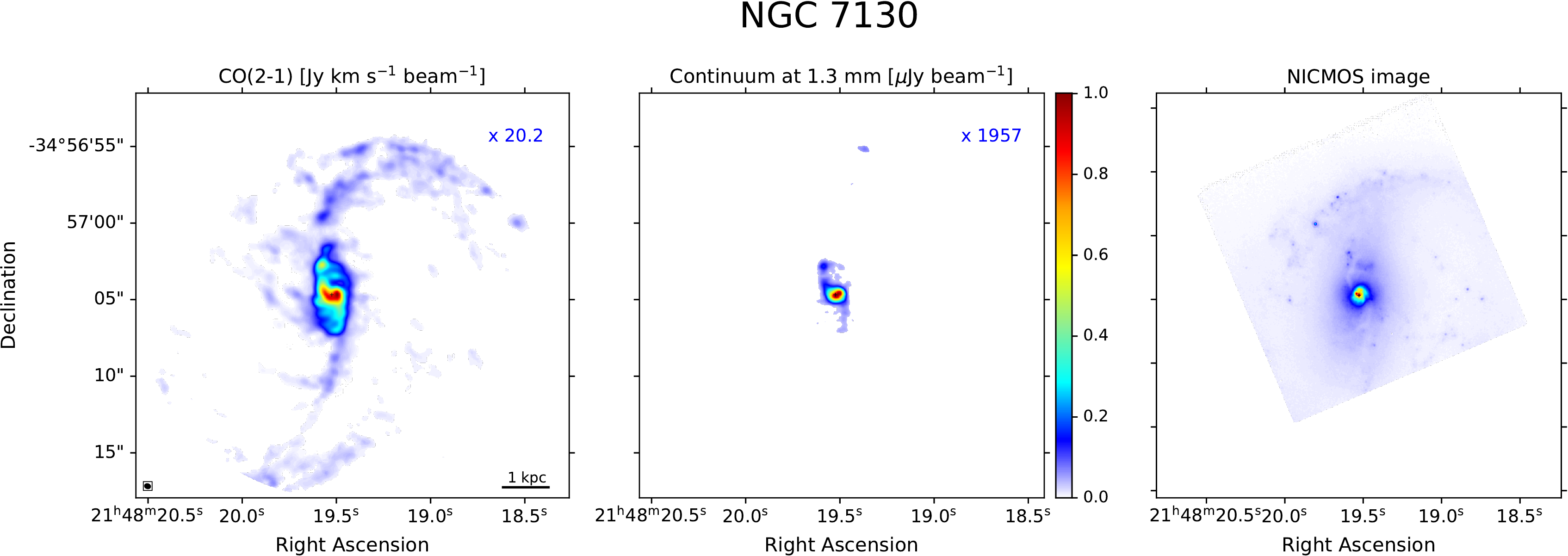}
\vskip7mm
\includegraphics[width=0.99\textwidth]{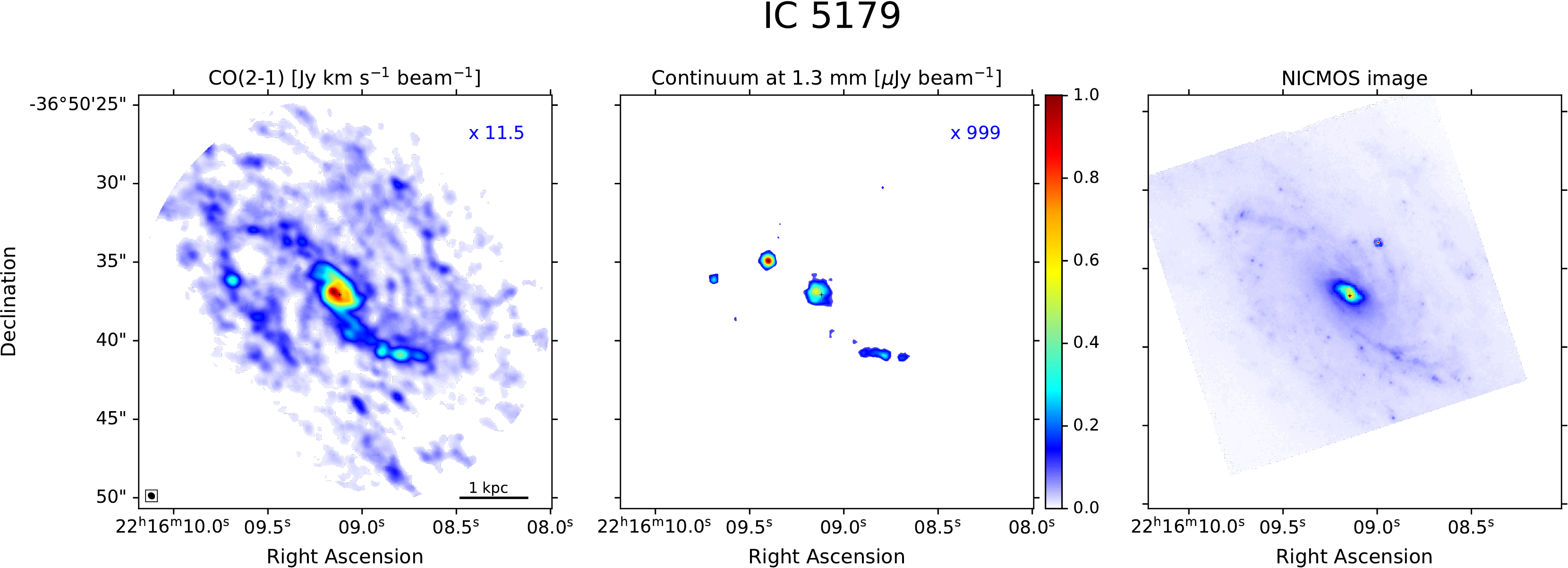}
\caption{Same figure caption as in Fig.~\ref{mix_panels_ALMA_HST_0}}
\label{mix_panels_ALMA_HST}
\end{figure*}

\begin{figure*}
\centering
\includegraphics[width=0.98\textwidth]{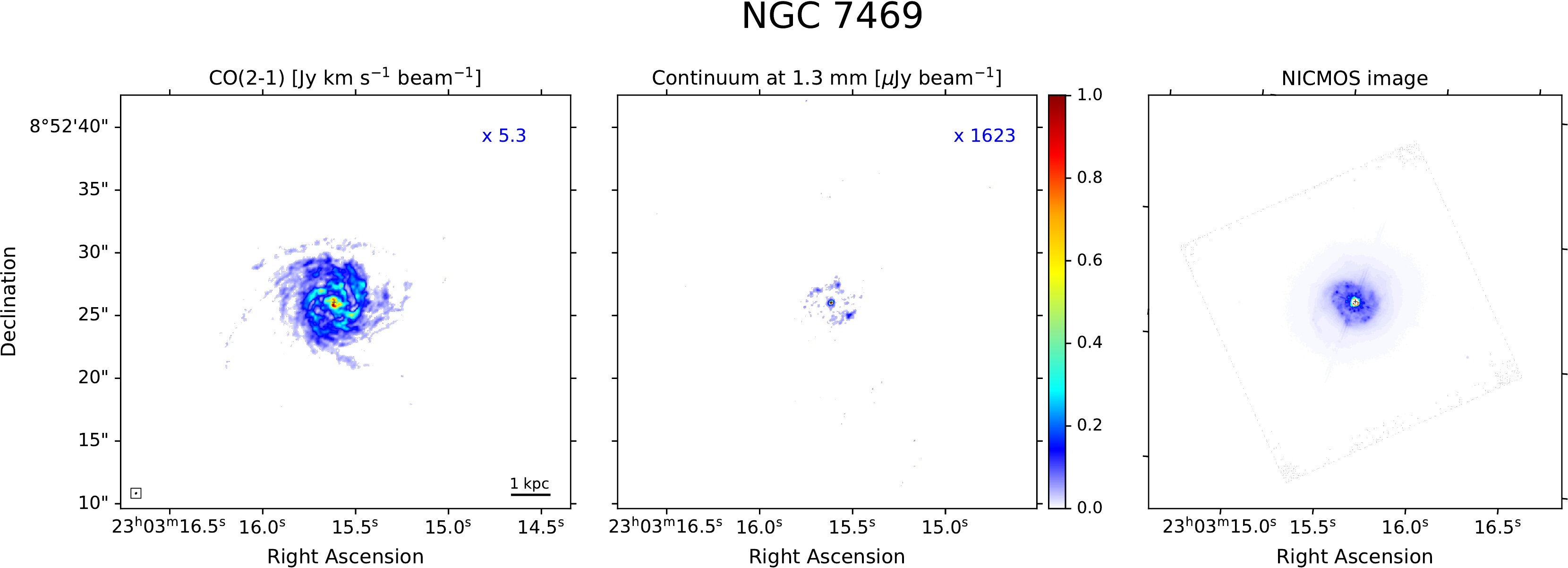}
\caption{Same figure caption as in Fig.~\ref{mix_panels_ALMA_HST_0}}
\vskip5mm
\includegraphics[width=0.75\textwidth]{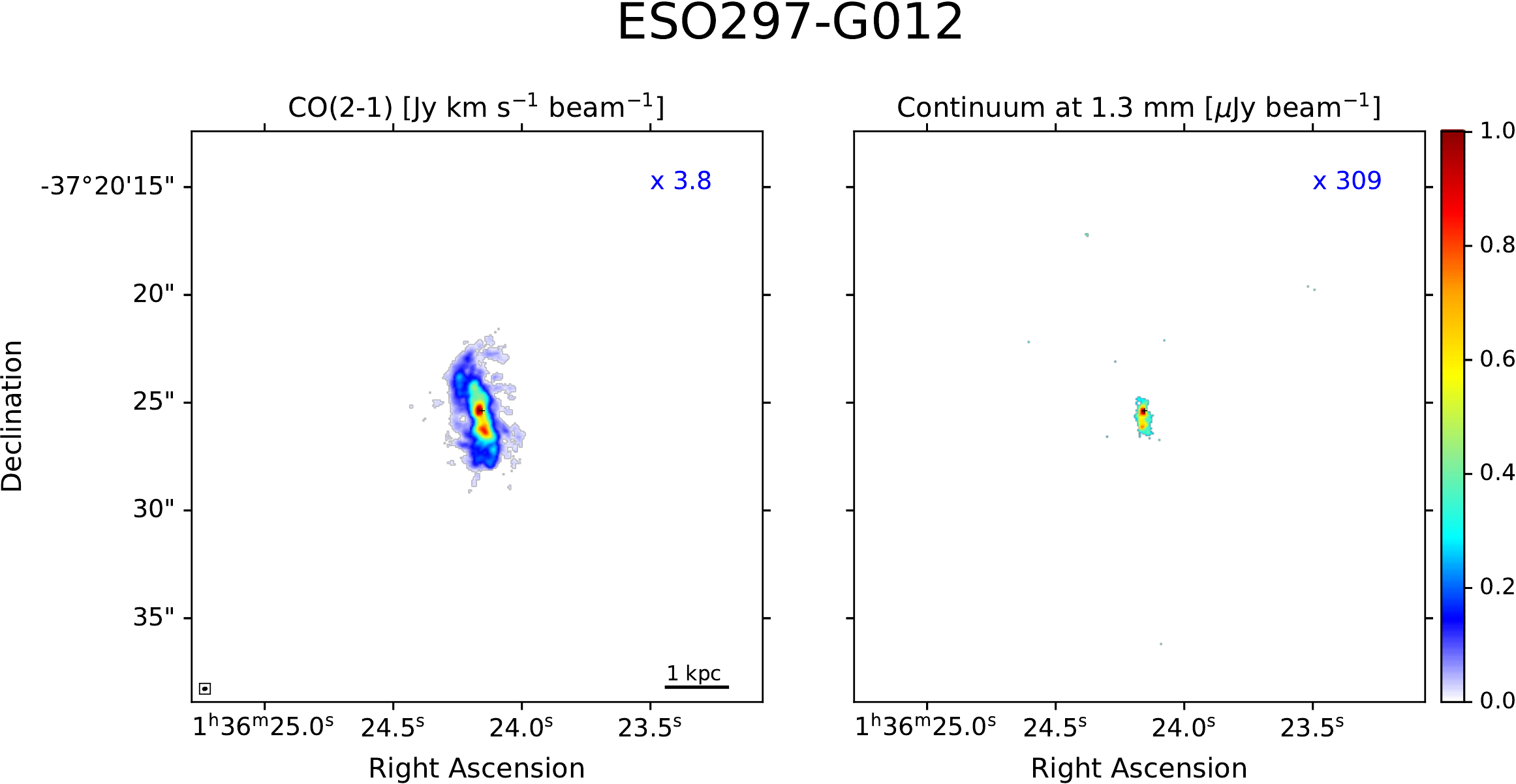}
\vskip5mm
\includegraphics[width=0.75\textwidth]{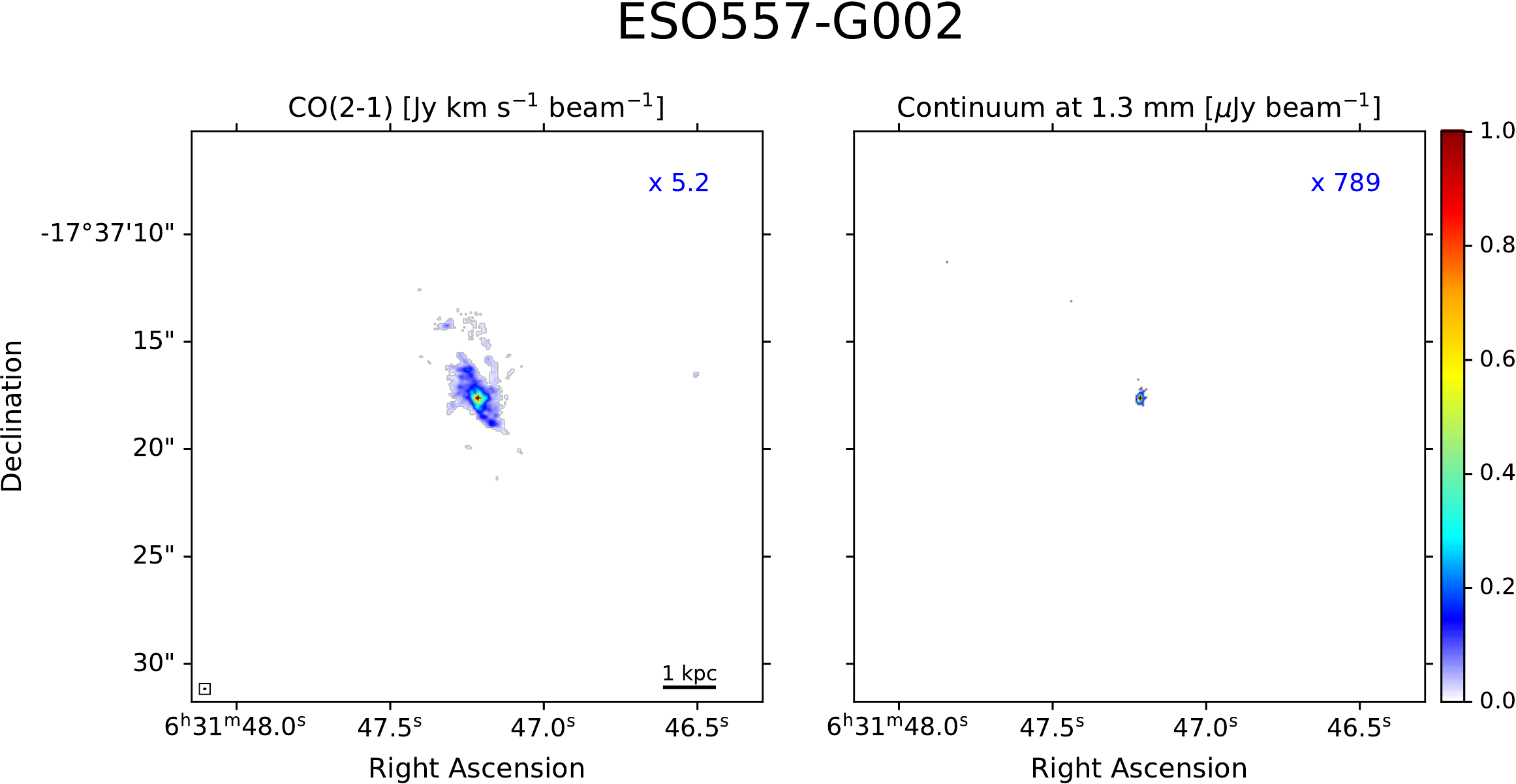}
\caption{Same figure caption as in Fig.~\ref{mix_panels_ALMA_HST_0} for the CO(2--1) and 1.3 mm continuum images, for which no {\it HST}/NICMOS data are available. The cross identifies the continuum peak emission at 1.3 mm. }
\label{mix_panels_ALMA_HST_2}
\end{figure*}

\begin{figure*}
\centering
\includegraphics[width=0.8\textwidth]{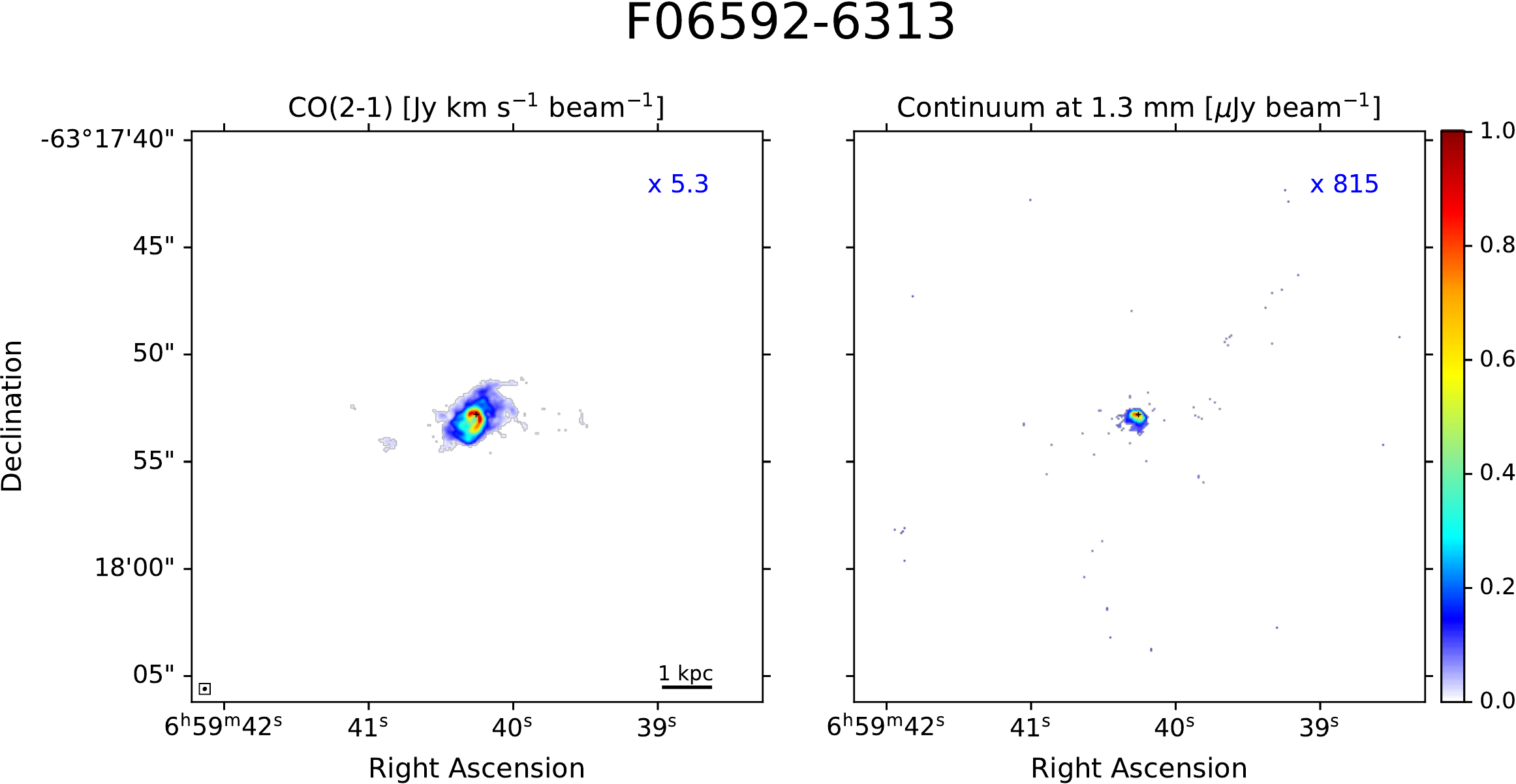}
\vskip5mm
\includegraphics[width=0.8\textwidth]{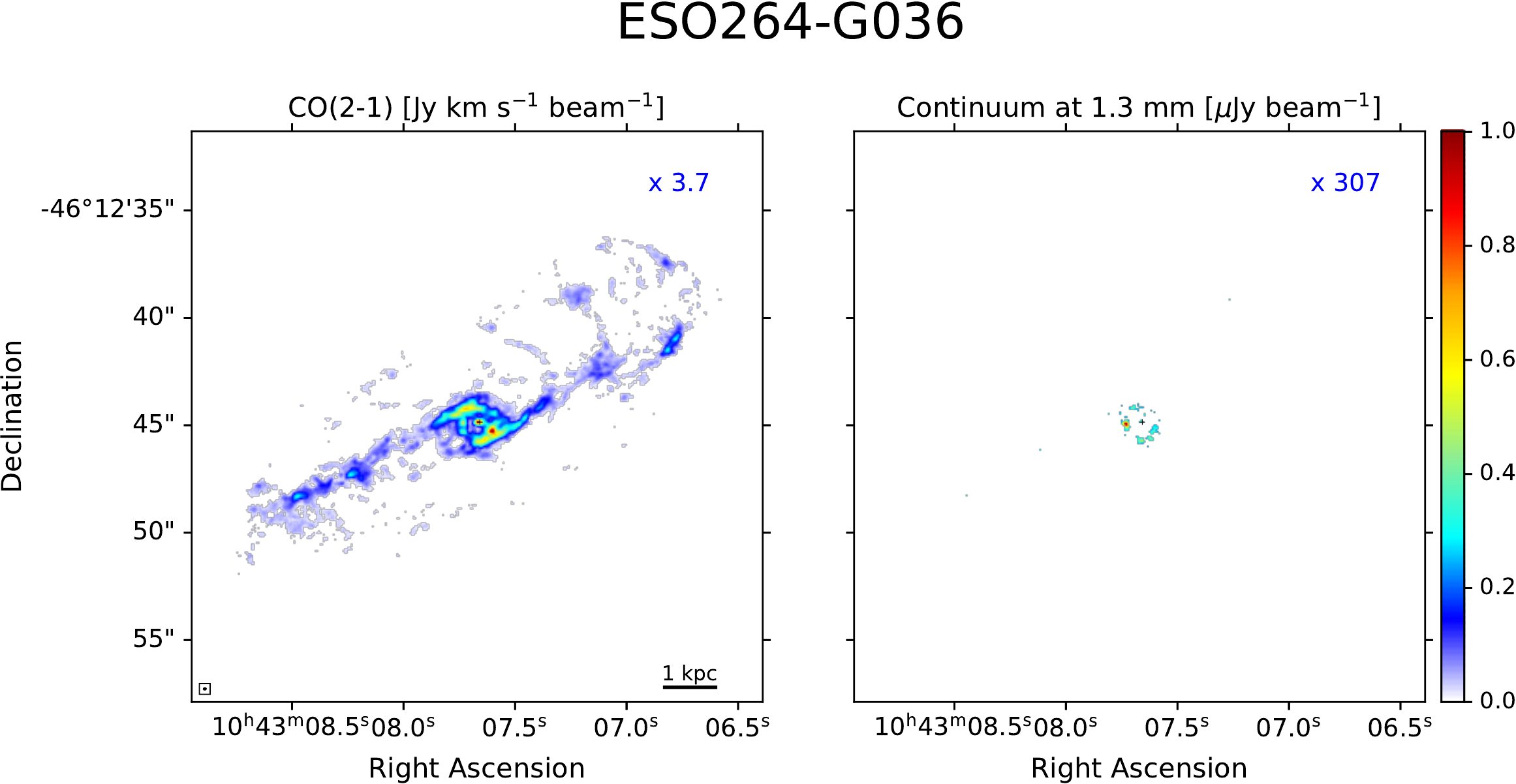}
\vskip5mm
\includegraphics[width=0.8\textwidth]{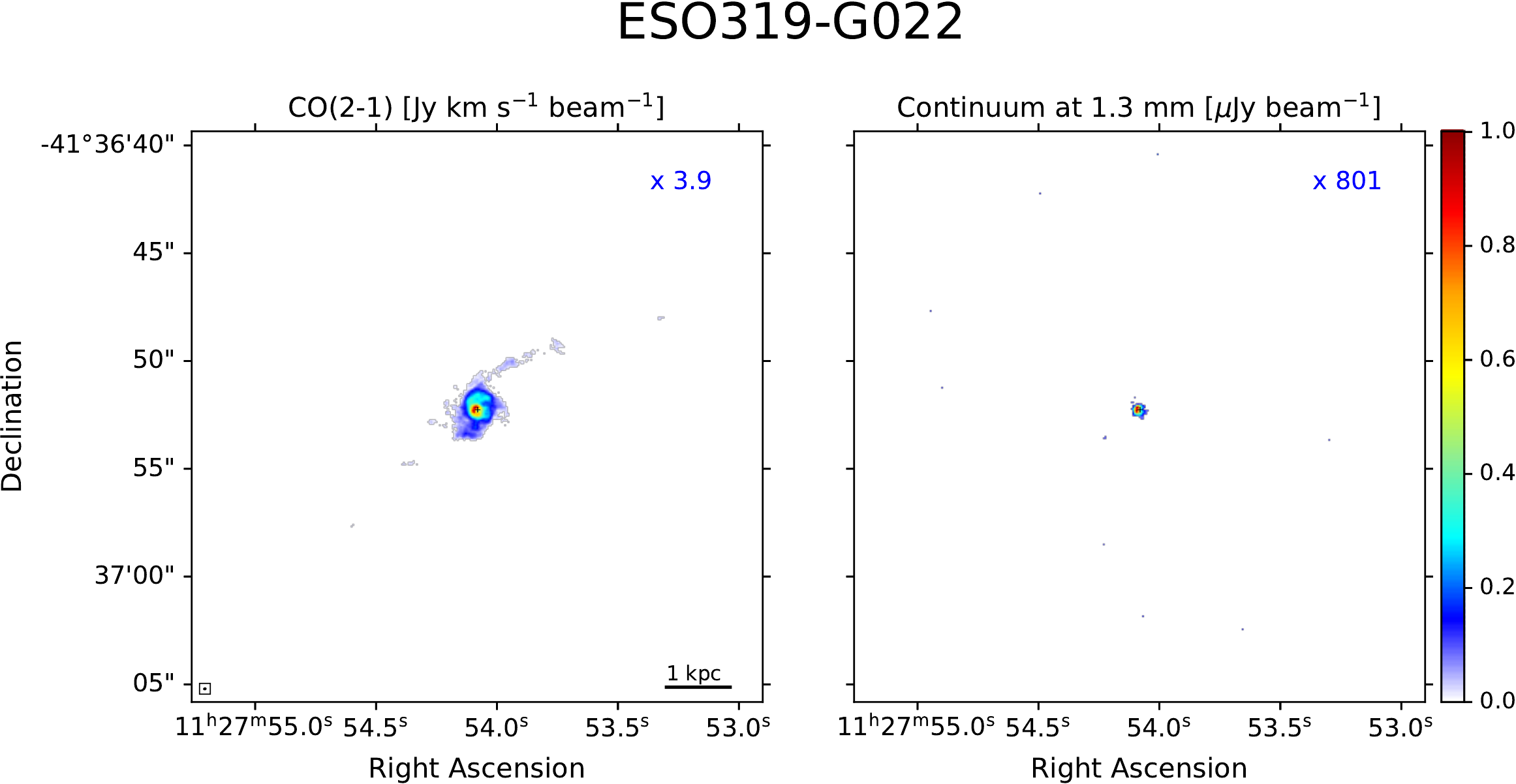}
\caption{Same figure caption as in Fig.~\ref{mix_panels_ALMA_HST_2}}
\label{mix_panels_ALMA_HST_3}
\end{figure*}

\begin{figure*}
\centering
\includegraphics[width=0.8\textwidth]{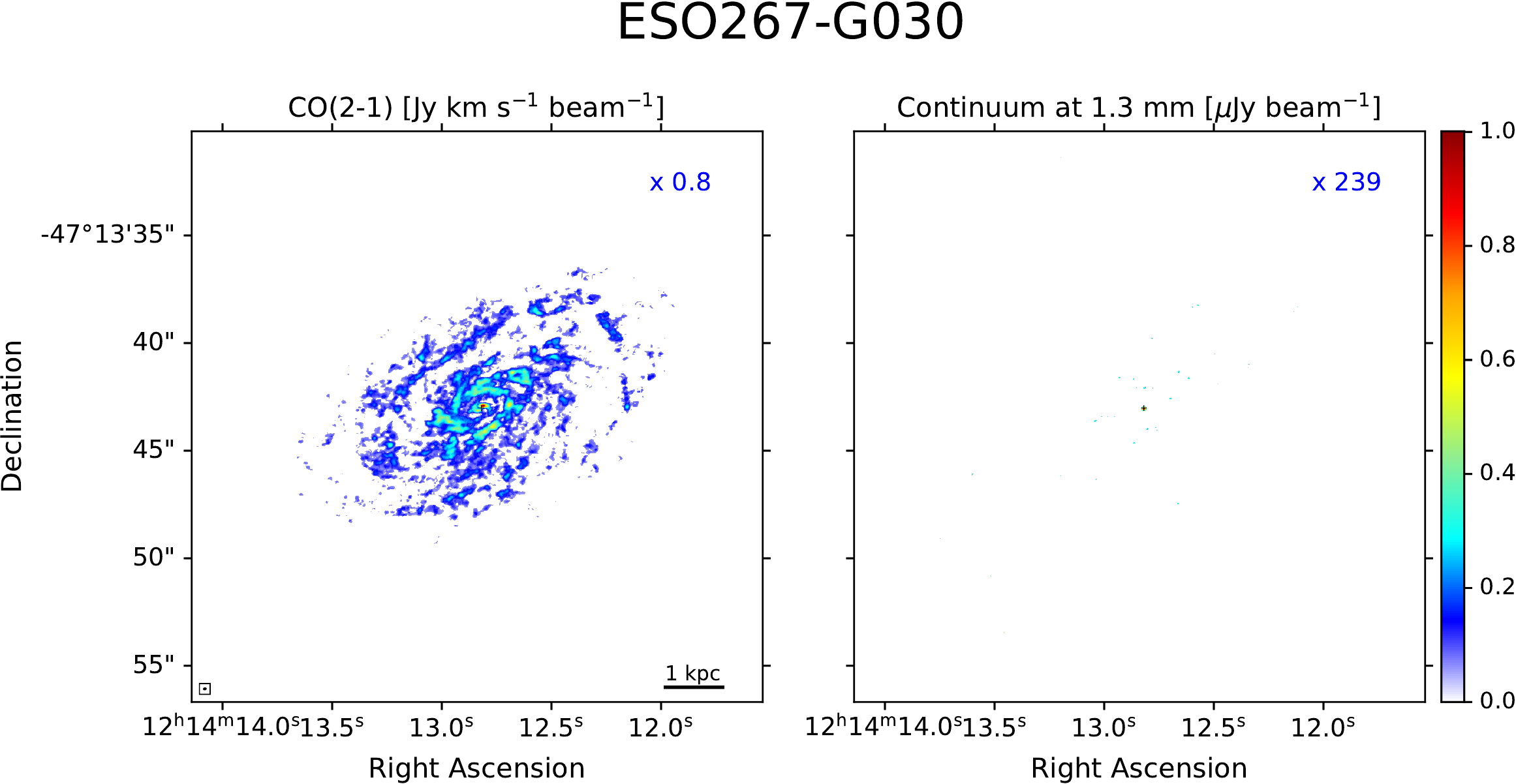}
\vskip5mm
\includegraphics[width=0.8\textwidth]{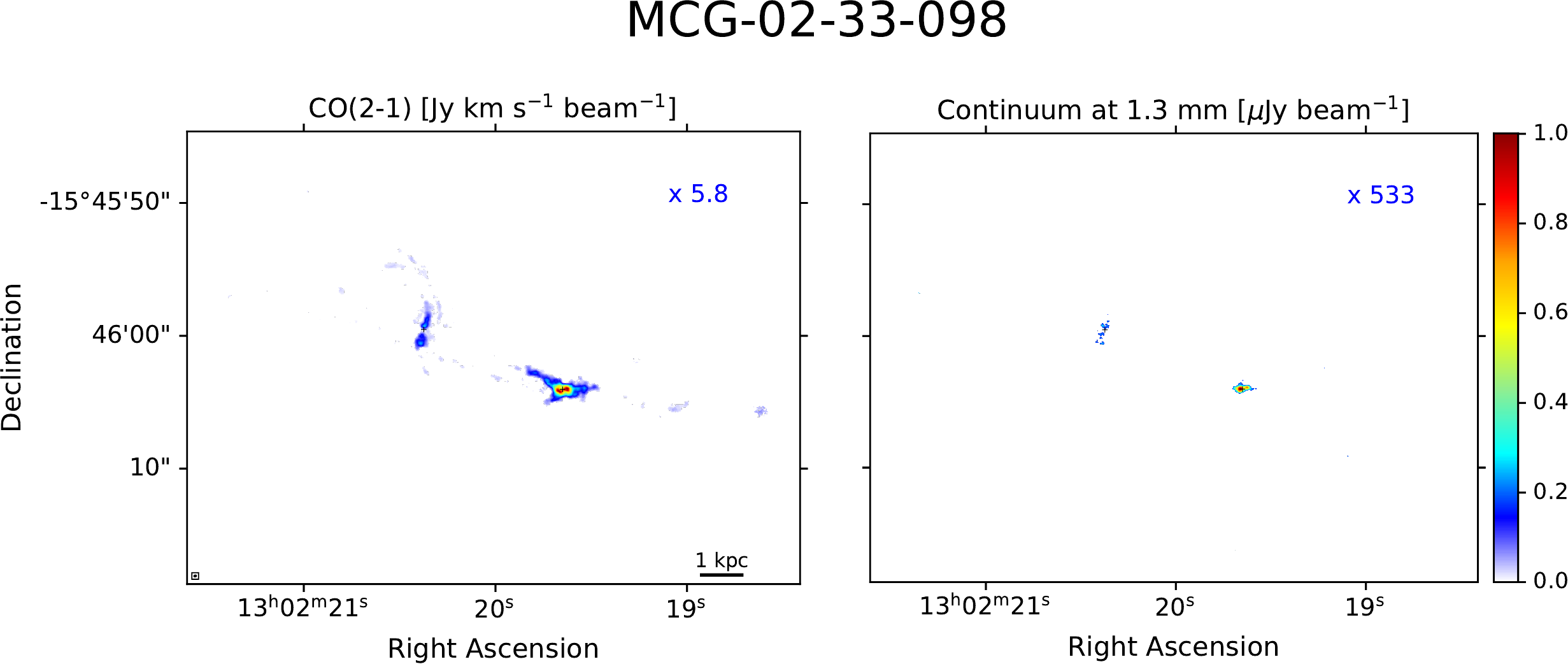}
\caption{Same figure caption as in Fig.~\ref{mix_panels_ALMA_HST_2}}
\label{mix_panels_ALMA_HST_3}
\end{figure*}

\section{CO(2--1) flux intensity maps ordered according to their increasing R$_{CO}$}
\label{CO_sameFoV_Reff}

\begin{figure*}
\centering
\includegraphics[width=0.45\textwidth]{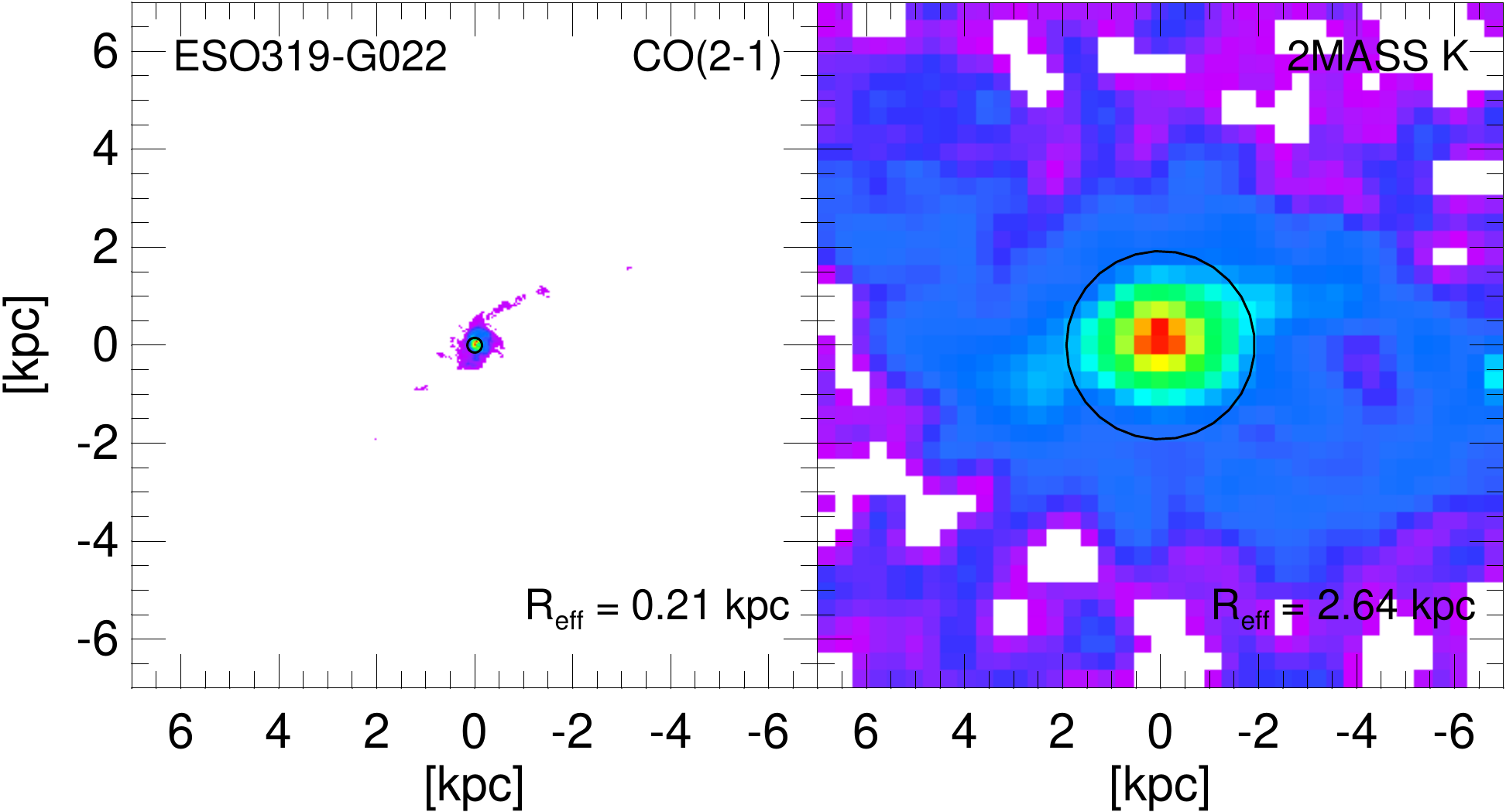}
\includegraphics[width=0.45\textwidth]{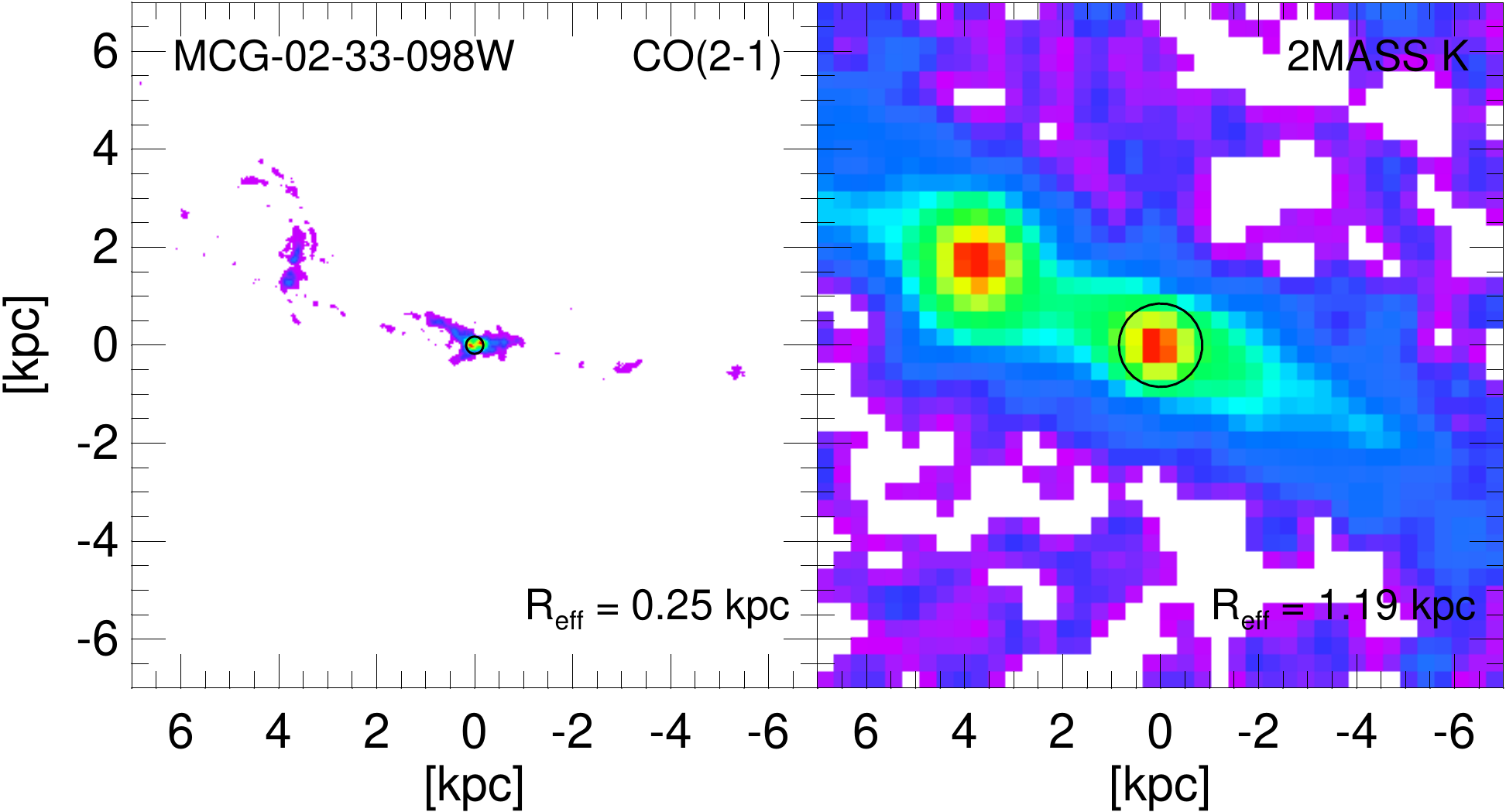}
\vskip3mm
\includegraphics[width=0.45\textwidth]{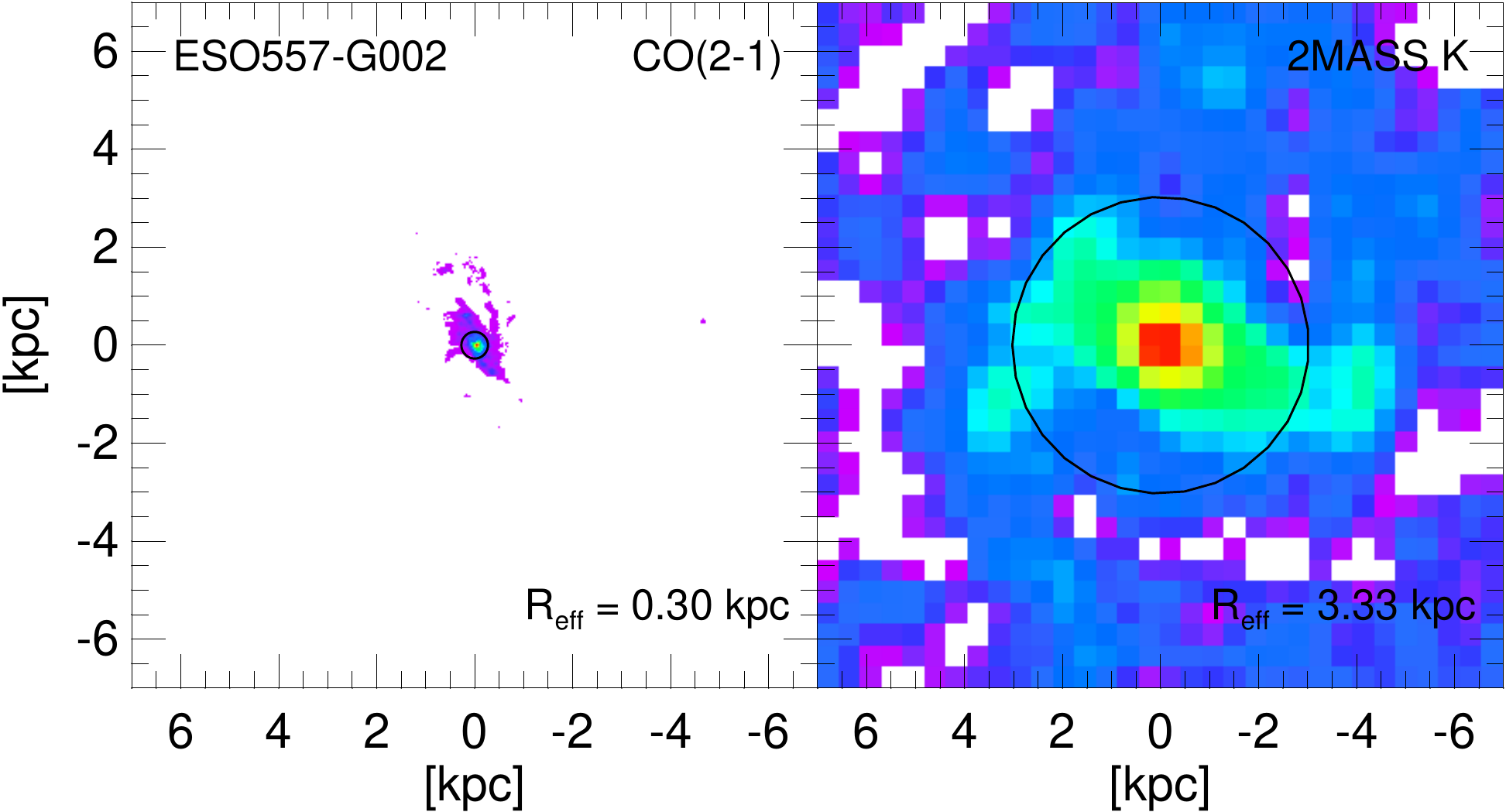}
\includegraphics[width=0.45\textwidth]{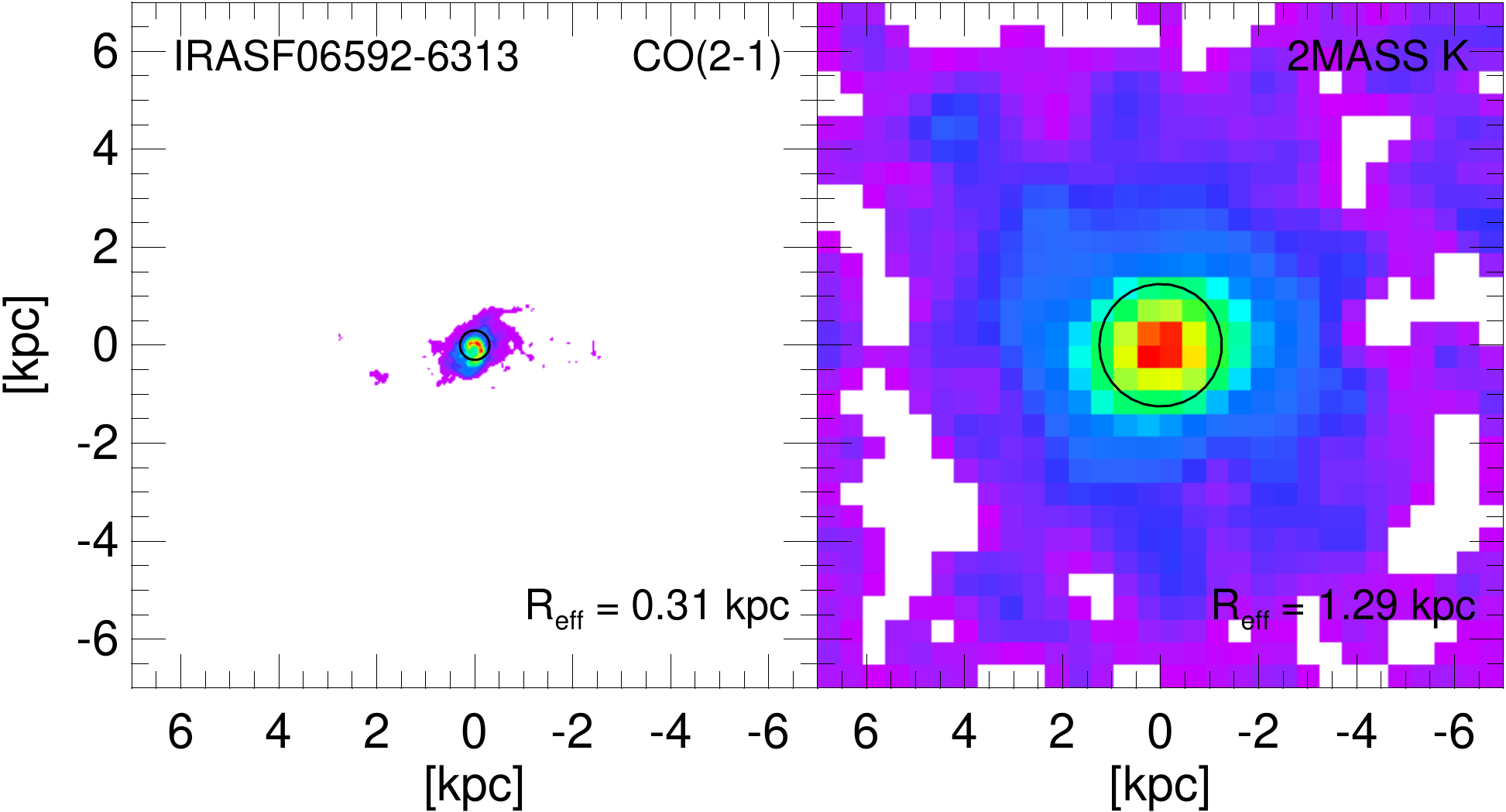}
\vskip3mm
\includegraphics[width=0.45\textwidth]{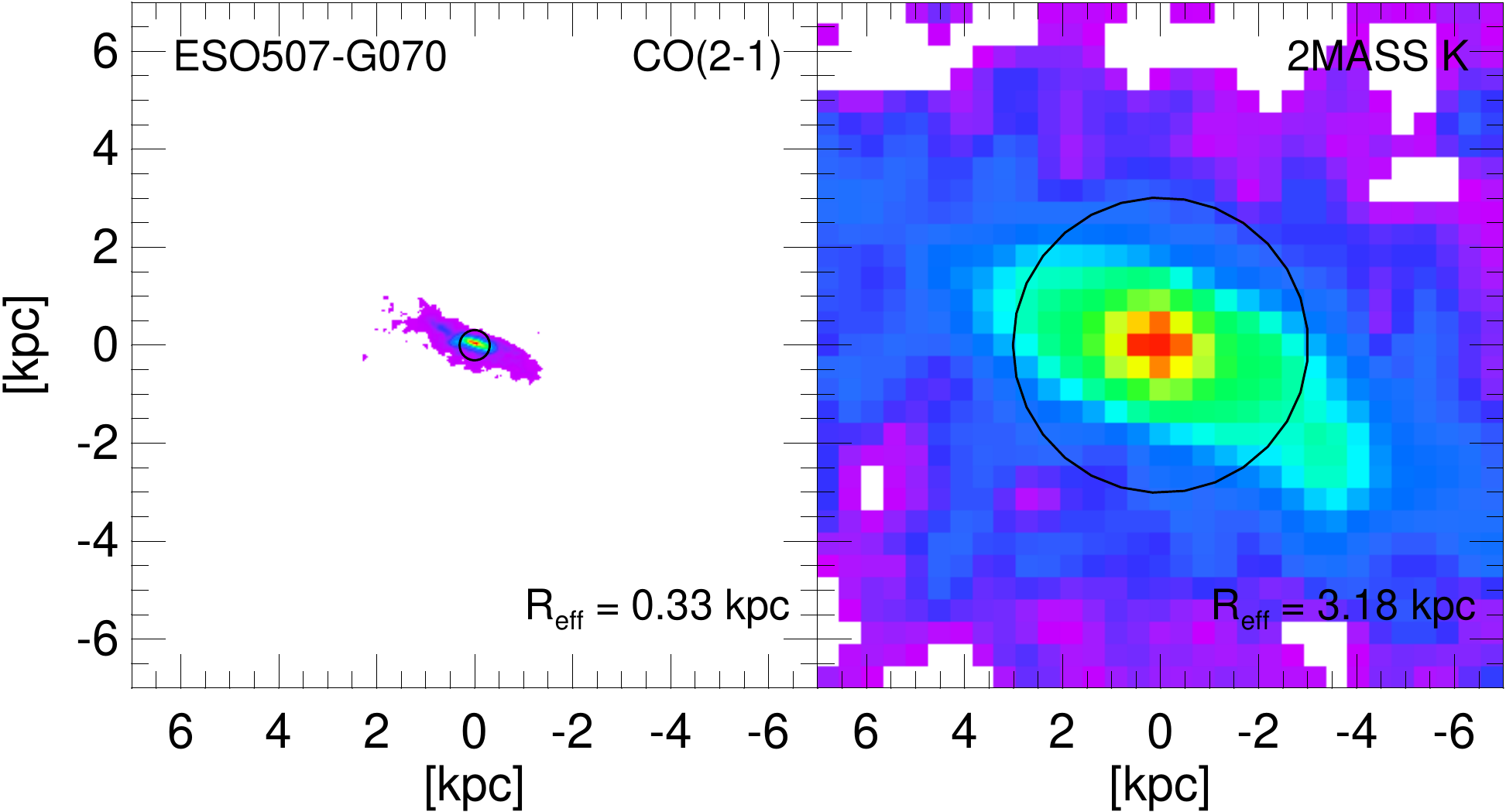}
\includegraphics[width=0.45\textwidth]{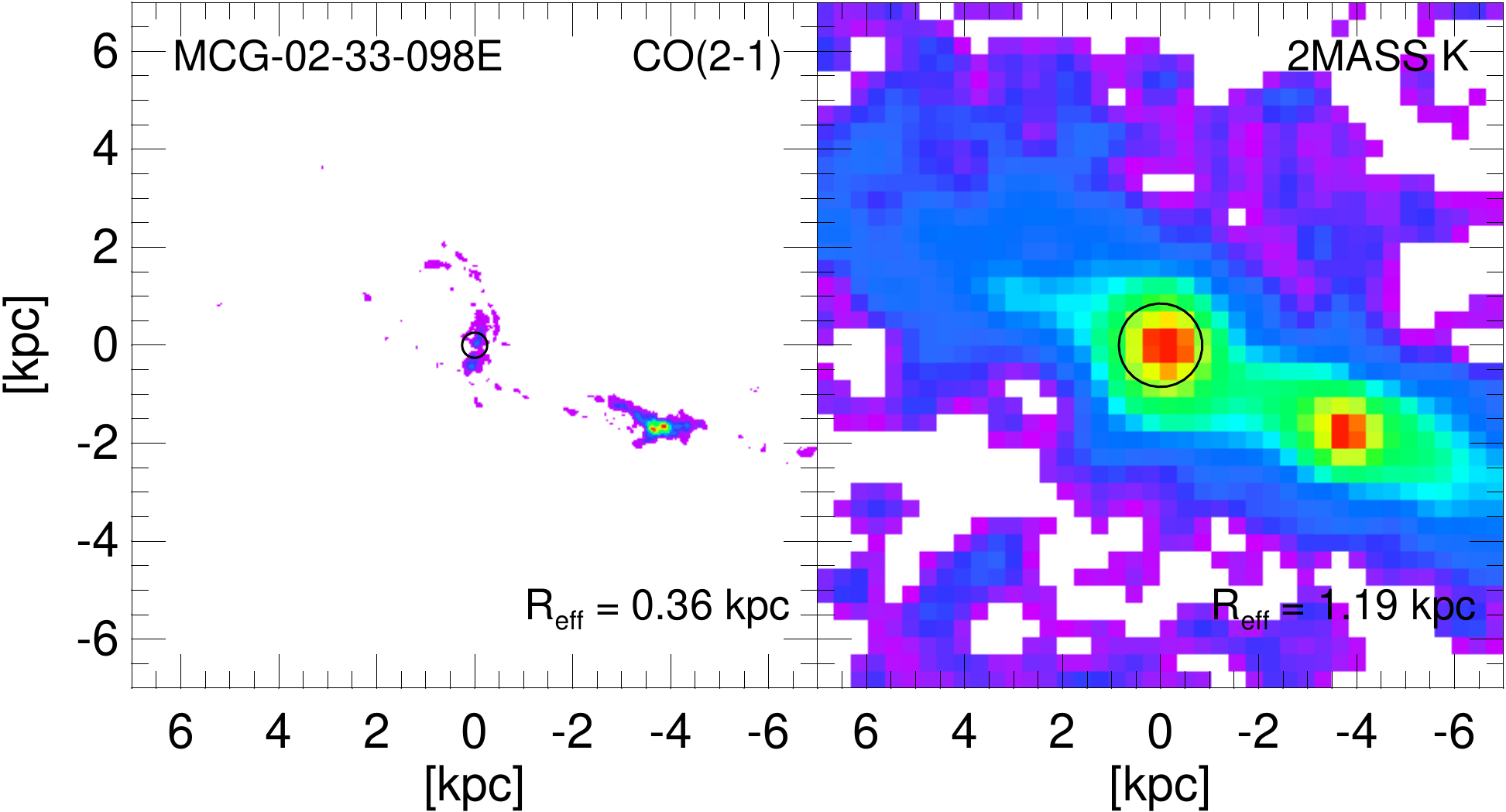}
\vskip3mm
\includegraphics[width=0.45\textwidth]{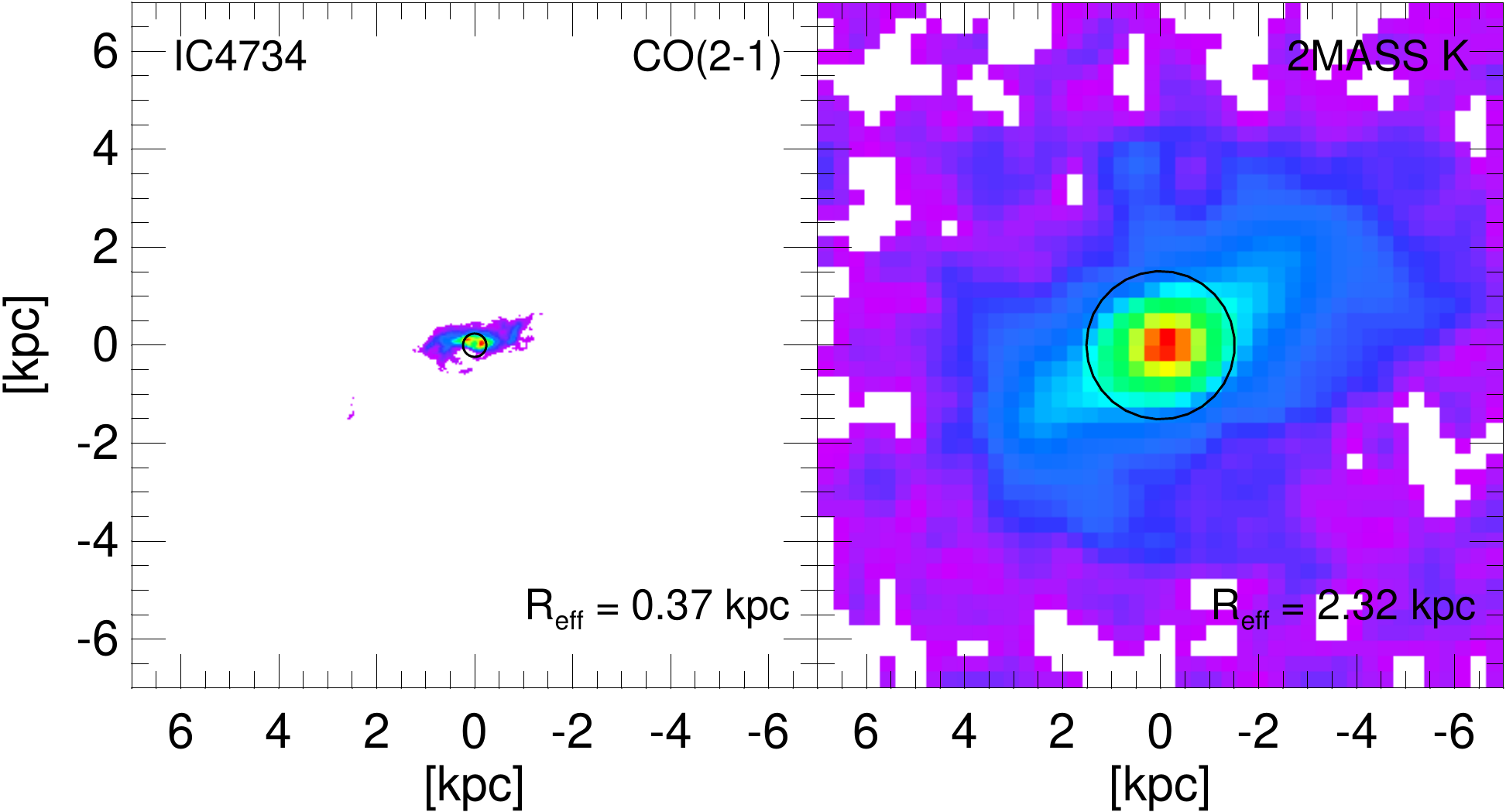}
\includegraphics[width=0.45\textwidth]{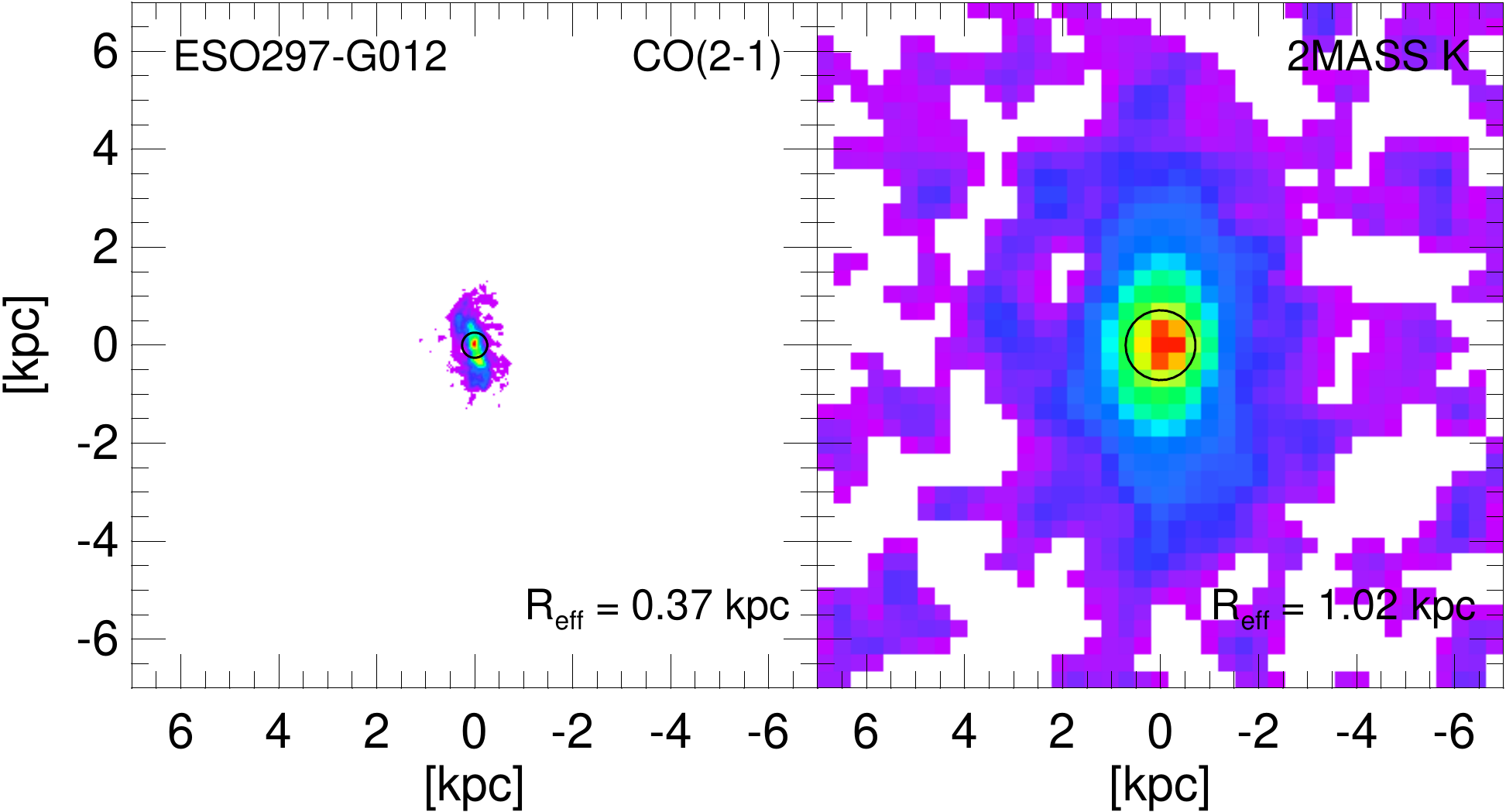}
\vskip3mm
\includegraphics[width=0.45\textwidth]{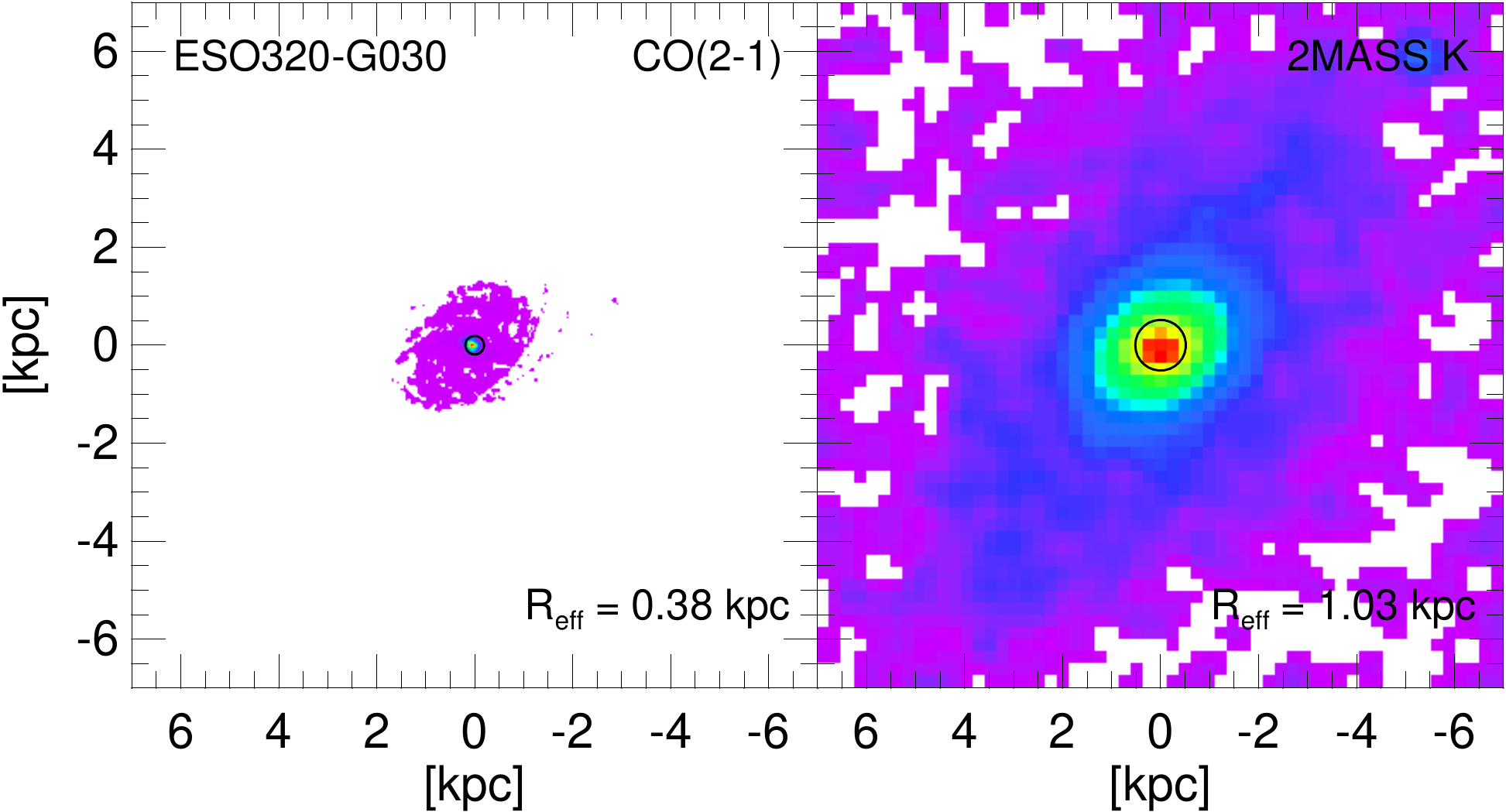}
\includegraphics[width=0.45\textwidth]{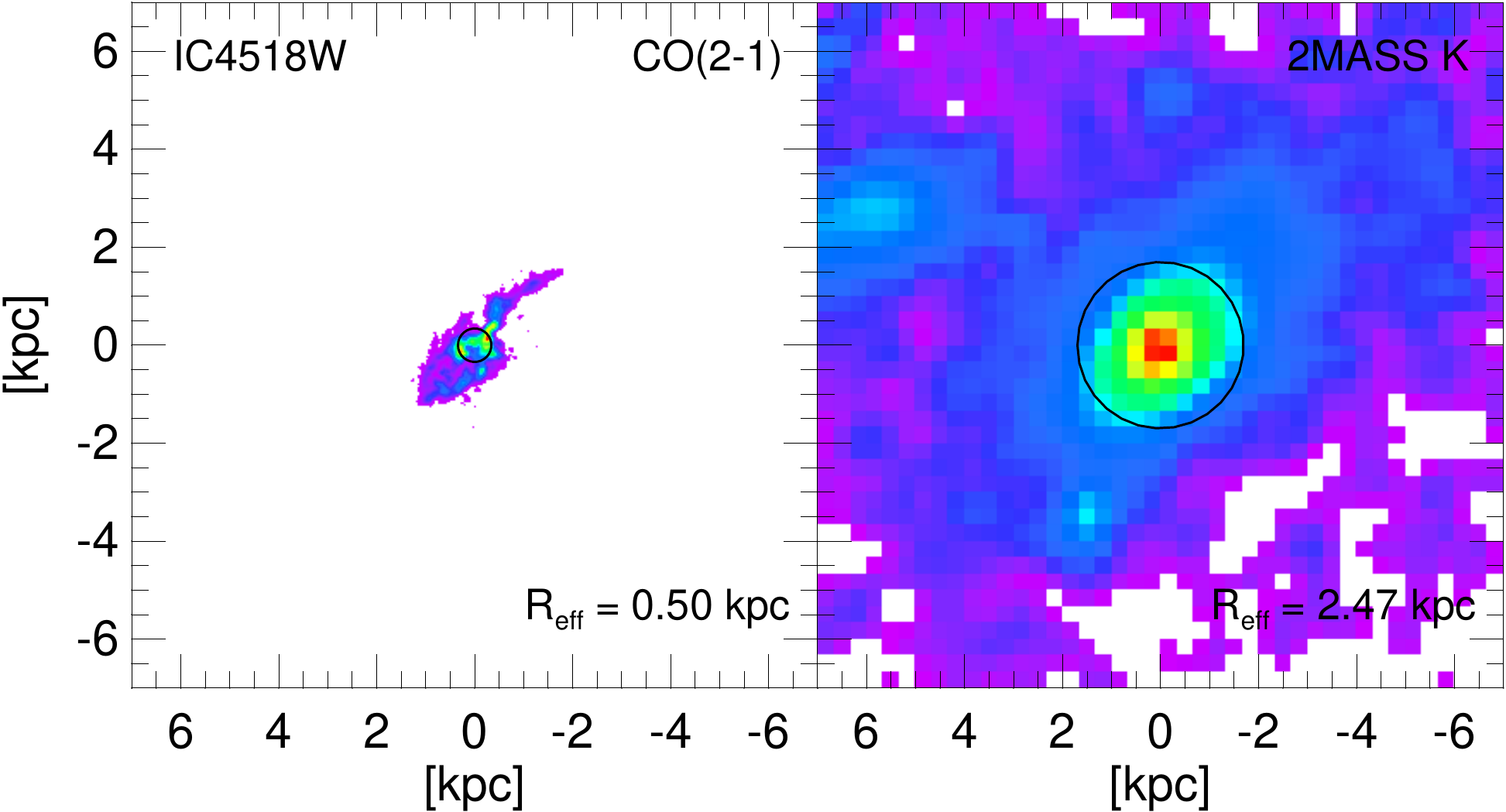}
\caption{{\it From top to bottom, left to right:} CO(2--1) maps (left) and 2MASS images in the K band (right) of the whole sample. The galaxies are ordered according to their increasing R$_{CO}$. The FoV is the same for all the galaxies (14 kpc $\times$ 14 kpc).
The black circle in each map identifies the respective effective radius in the map, whose value is also shown in each panel. North is up and East to the left. 
}
\label{Mix_class}
\end{figure*}

\begin{figure*}
\centering
\includegraphics[width=0.45\textwidth]{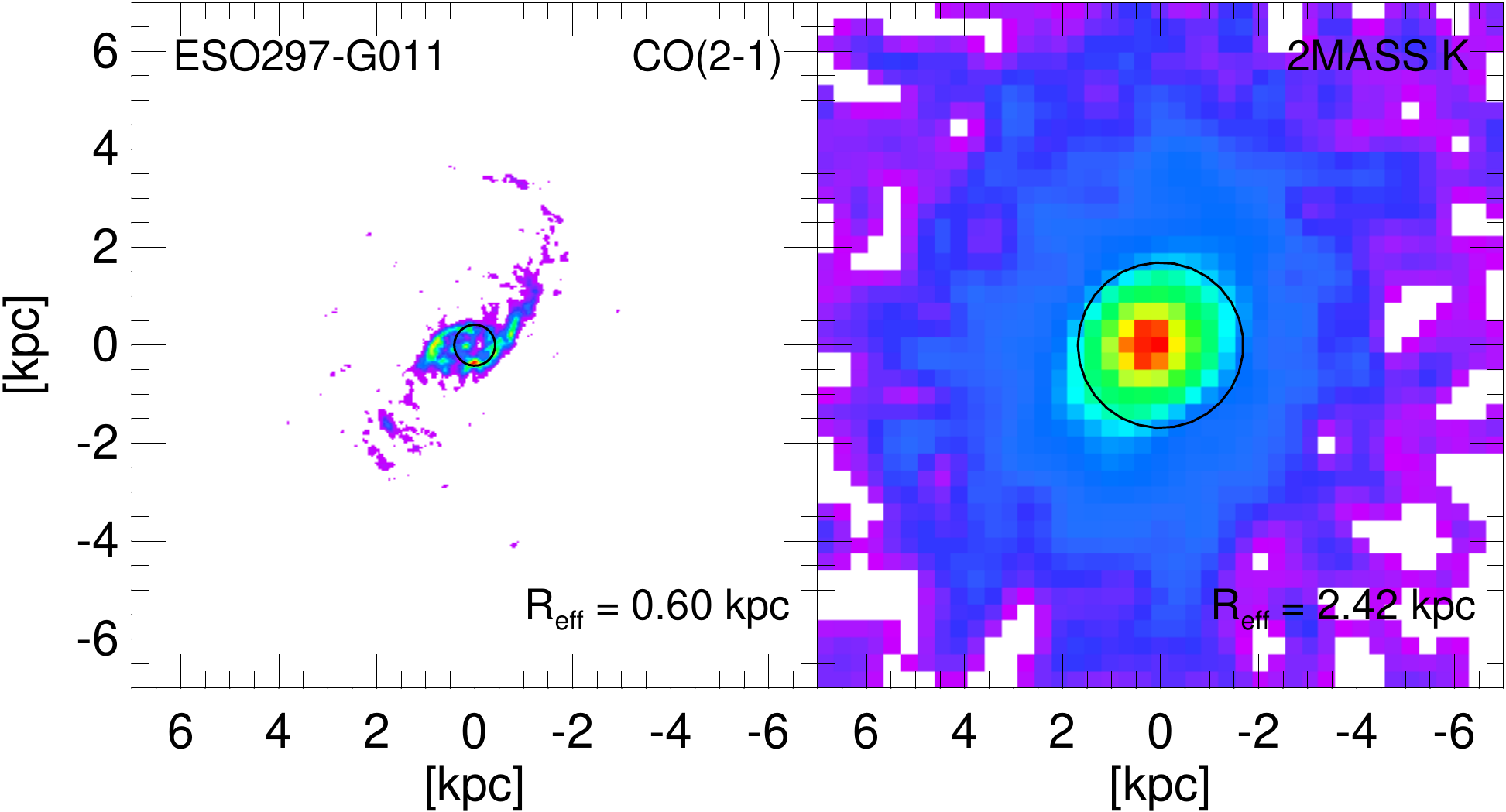}
\includegraphics[width=0.45\textwidth]{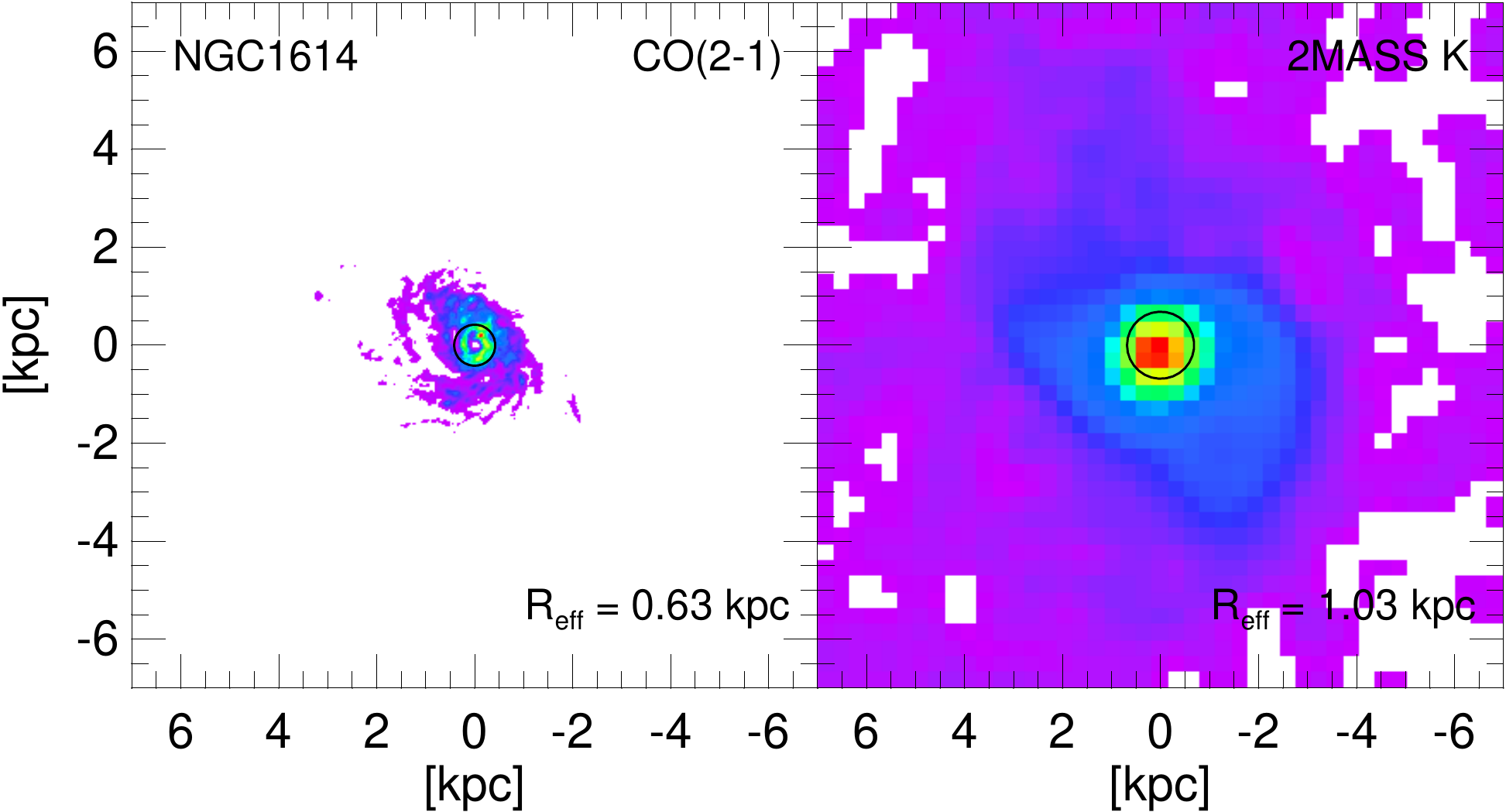}
\vskip3mm
\includegraphics[width=0.45\textwidth]{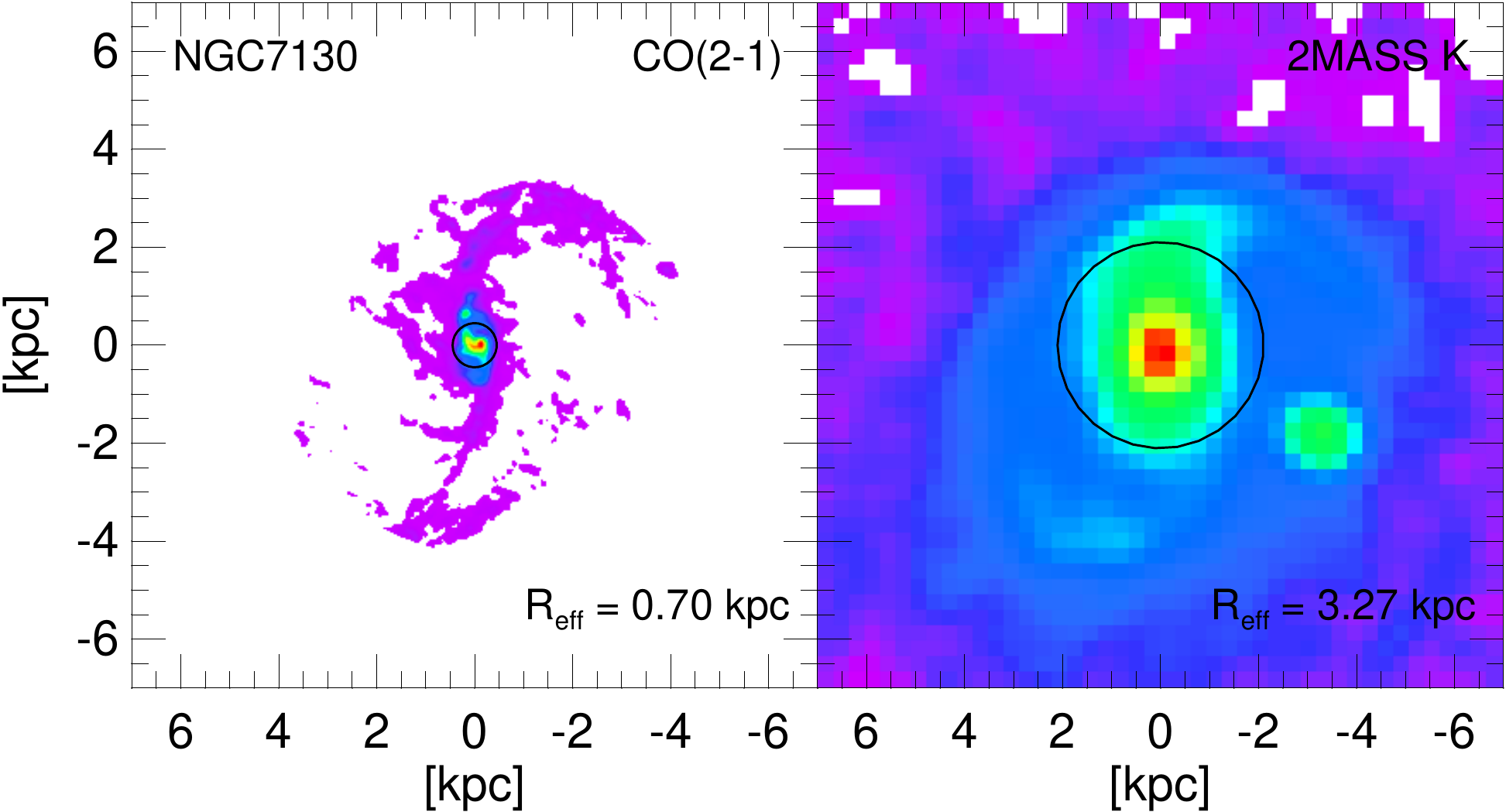}
\includegraphics[width=0.45\textwidth]{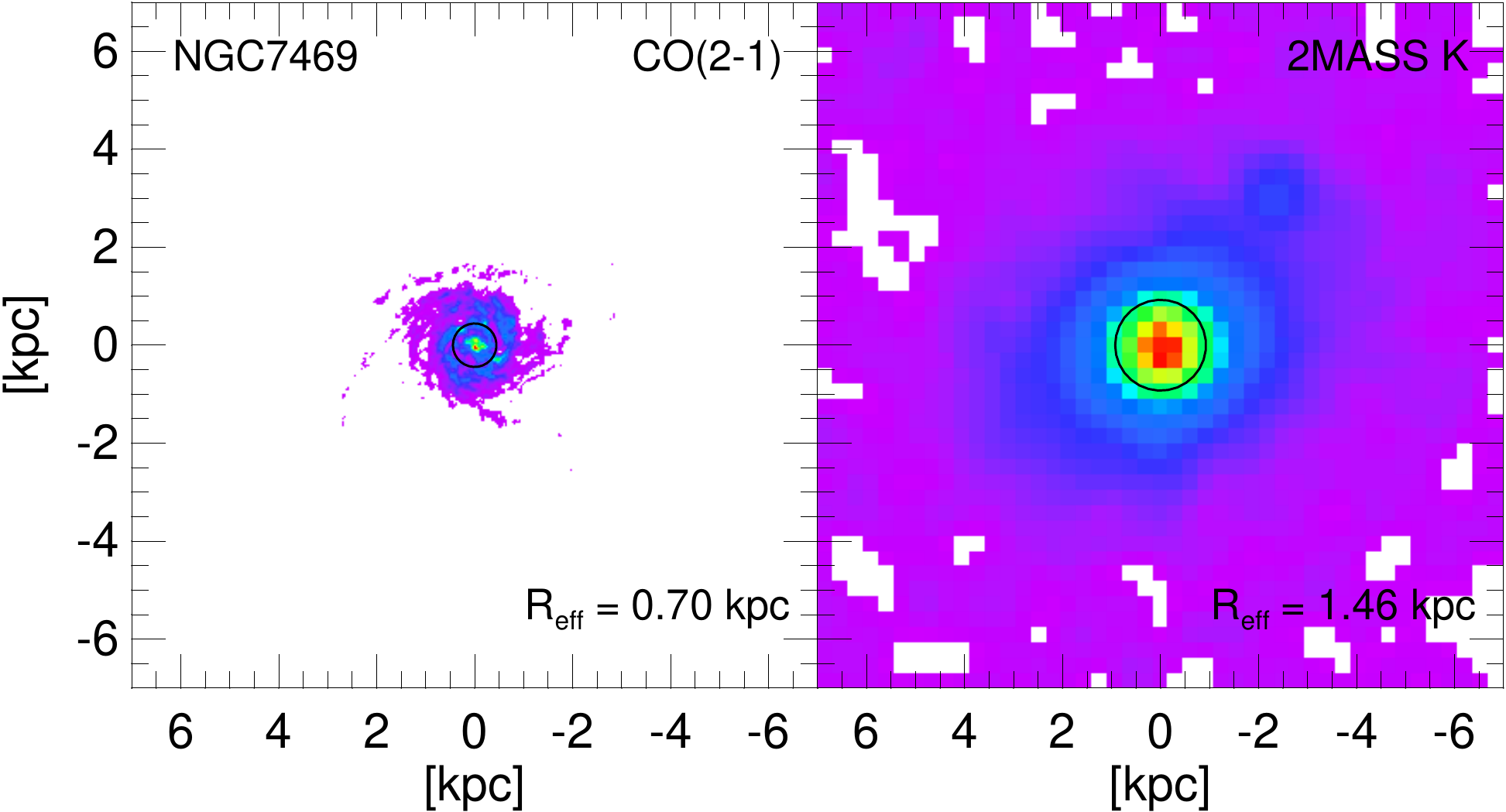}
\vskip3mm
\includegraphics[width=0.45\textwidth]{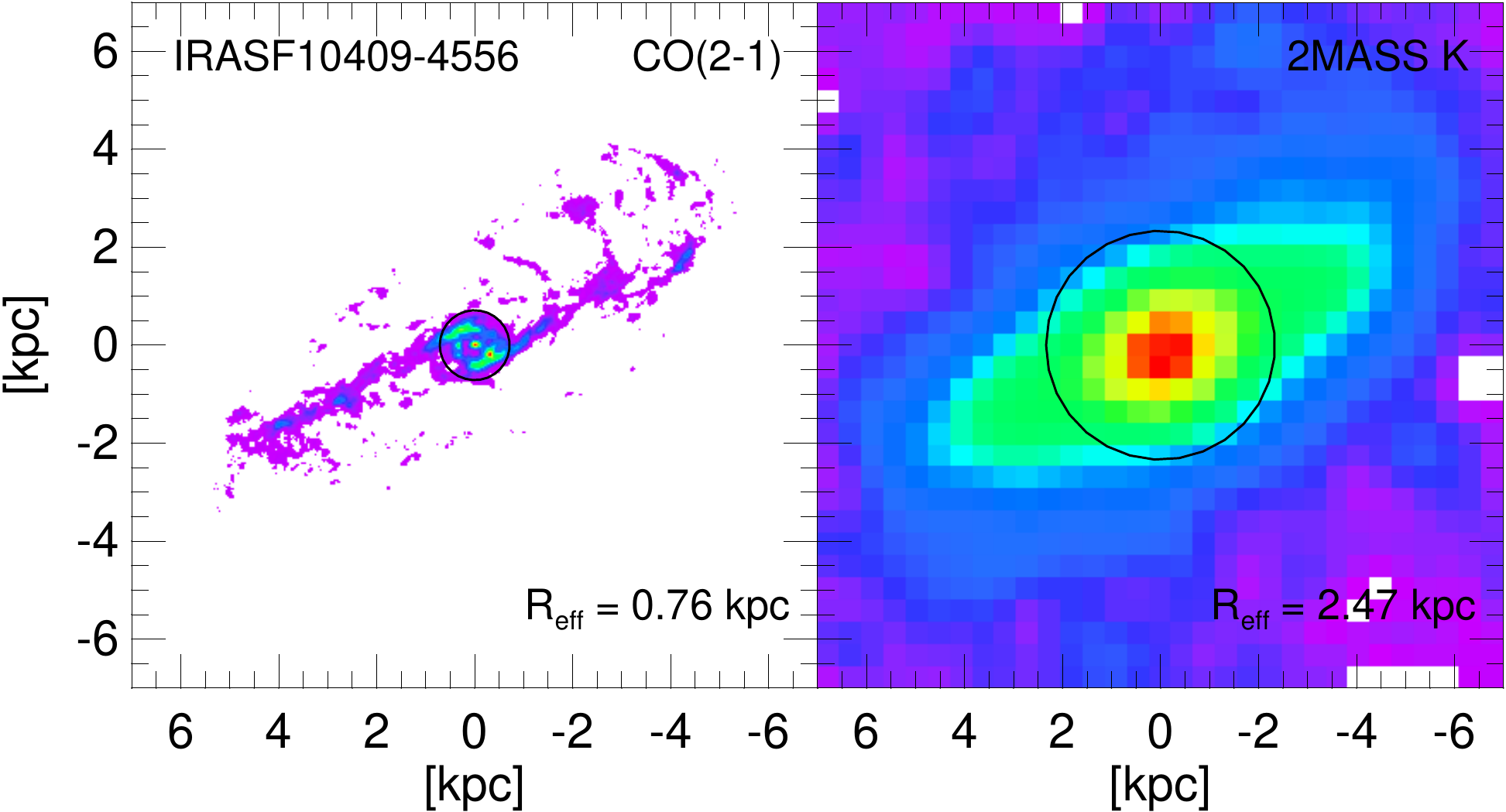}
\includegraphics[width=0.45\textwidth]{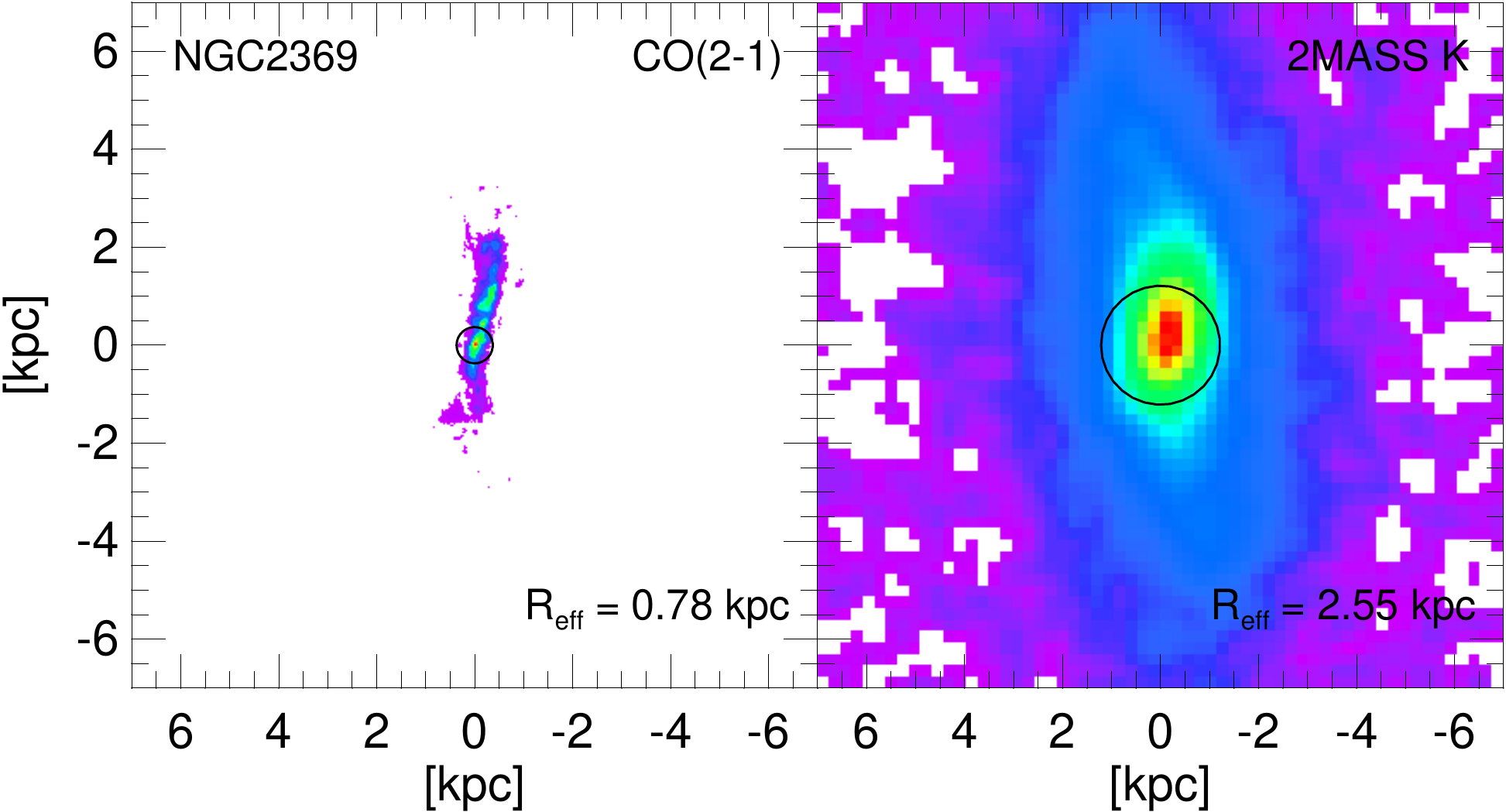}
\vskip3mm
\includegraphics[width=0.45\textwidth]{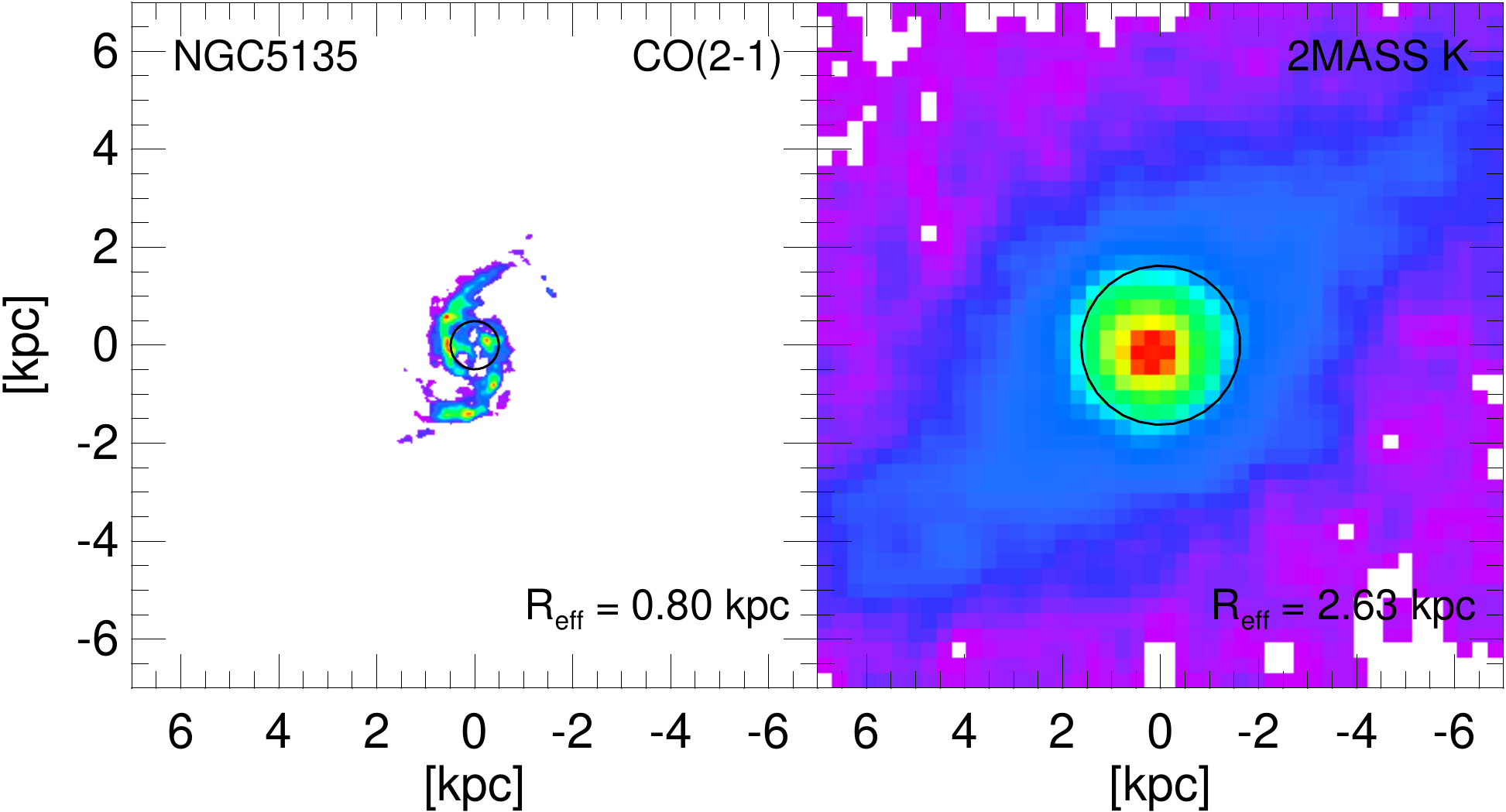}
\includegraphics[width=0.45\textwidth]{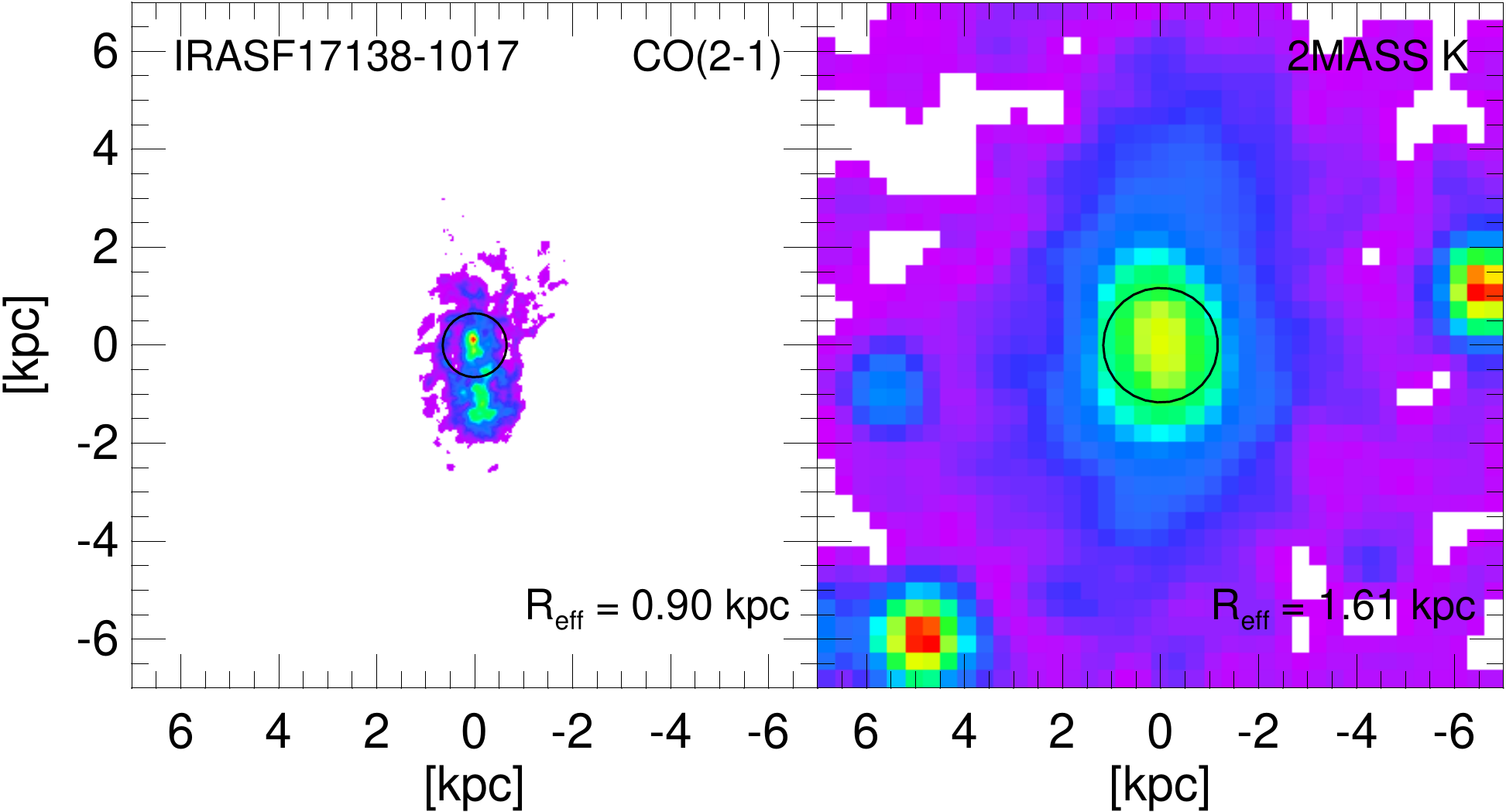}
\vskip3mm
\includegraphics[width=0.45\textwidth]{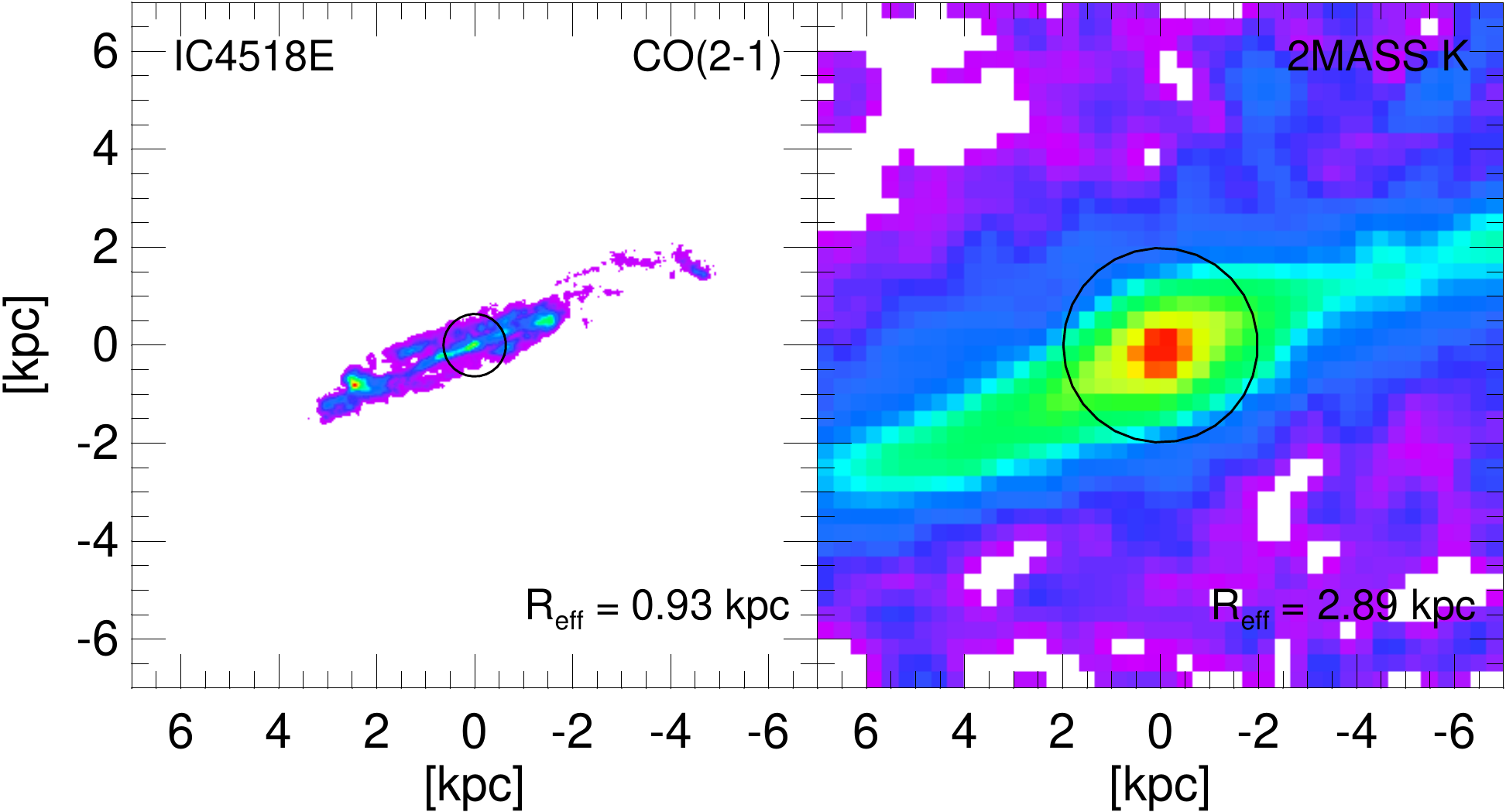}
\includegraphics[width=0.45\textwidth]{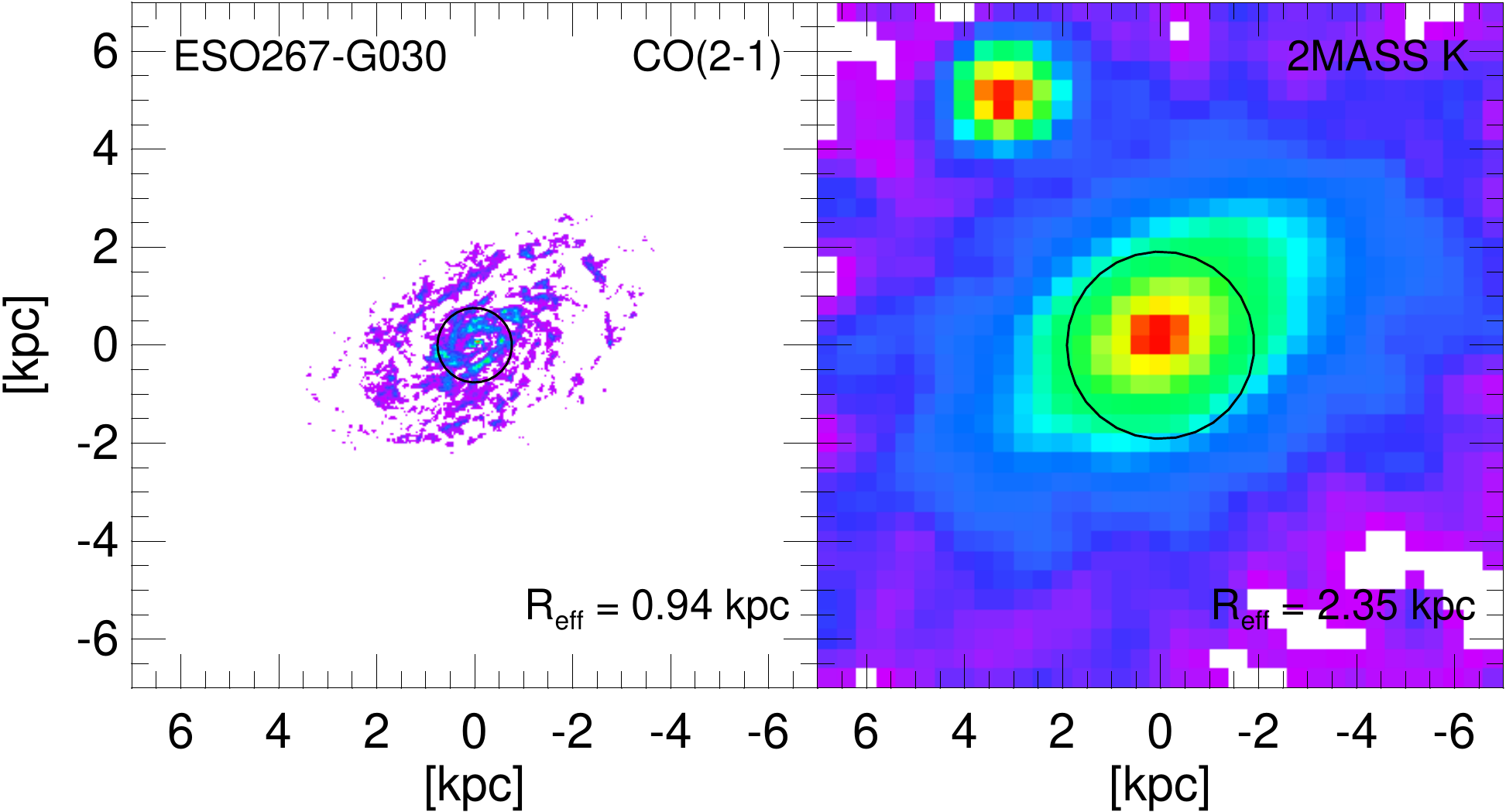}
\caption{Same figure caption as in Fig.~\ref{Mix_class}.}
\label{Mix_class_2}
\end{figure*}

\begin{figure*}
\centering
\includegraphics[width=0.45\textwidth]{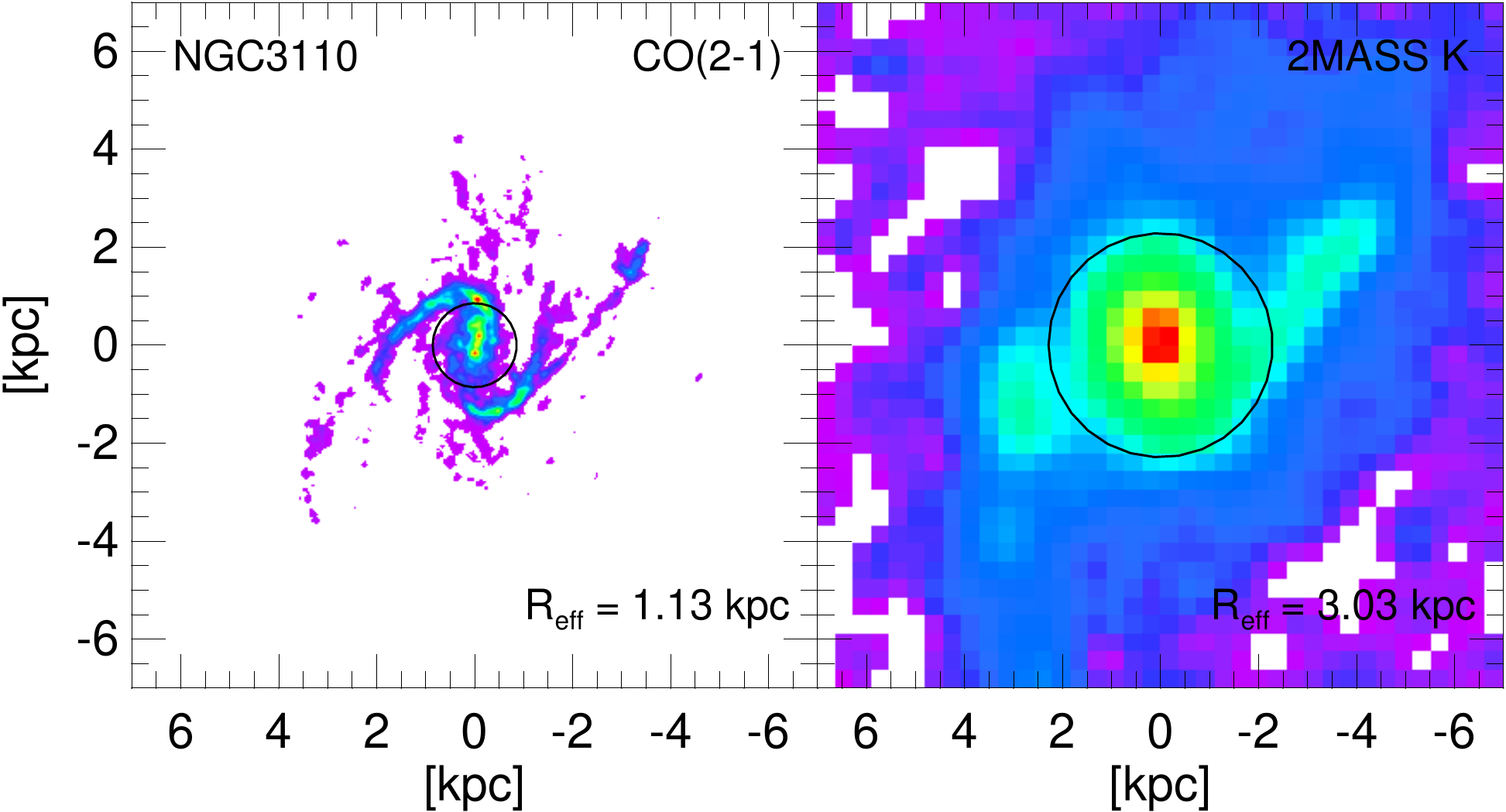}
\includegraphics[width=0.45\textwidth]{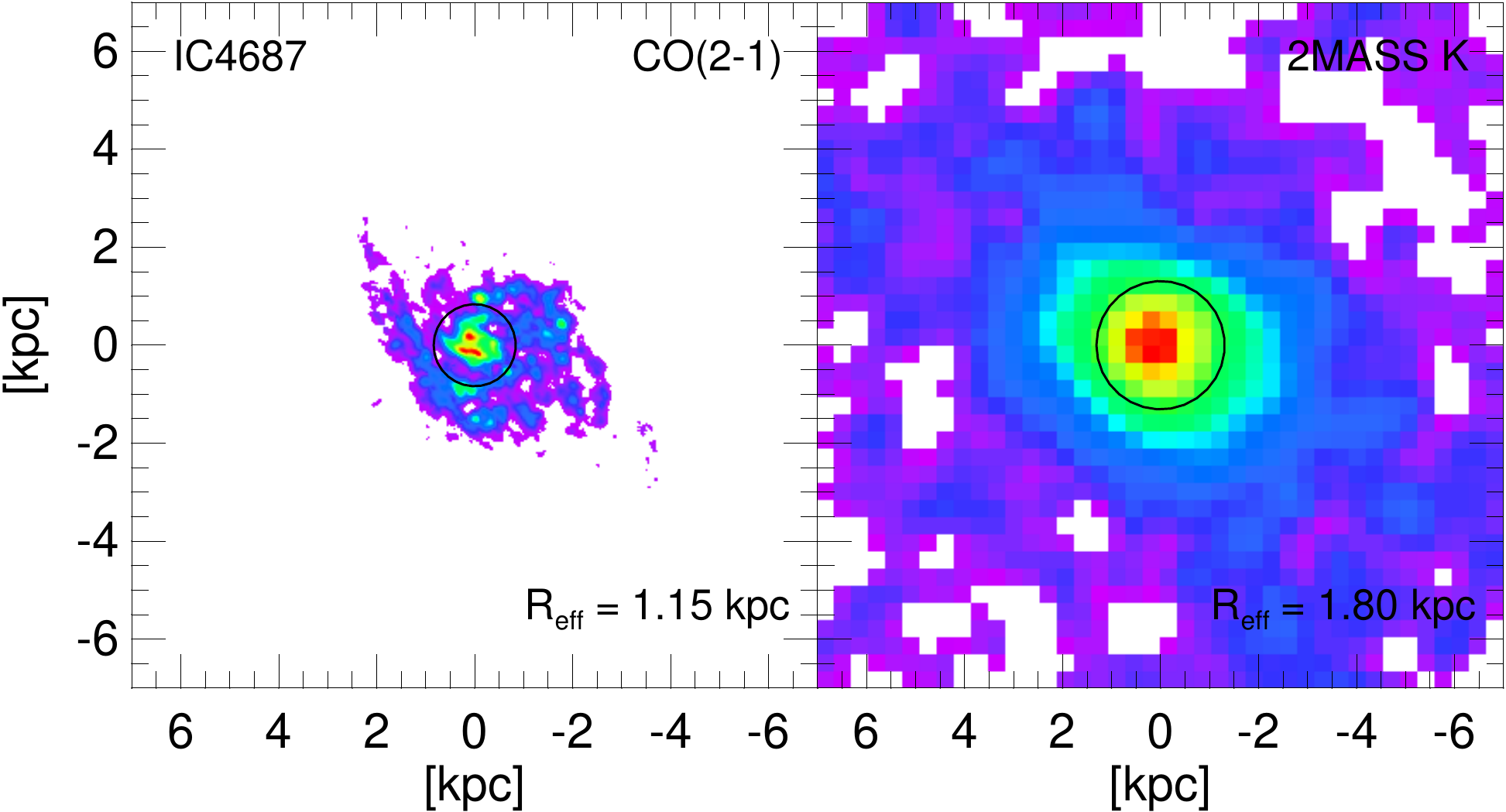}
\vskip3mm
\includegraphics[width=0.45\textwidth]{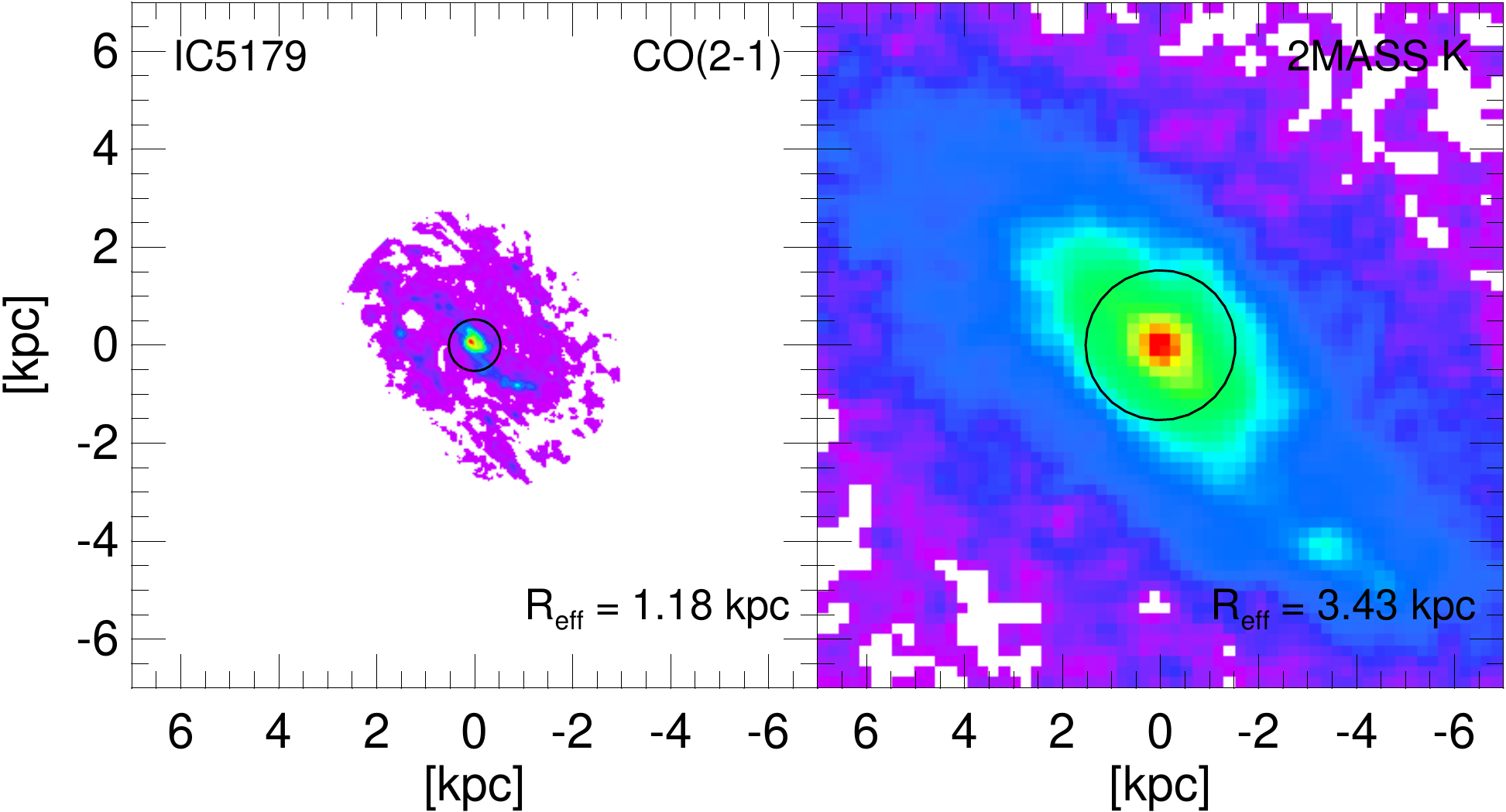}
\includegraphics[width=0.45\textwidth]{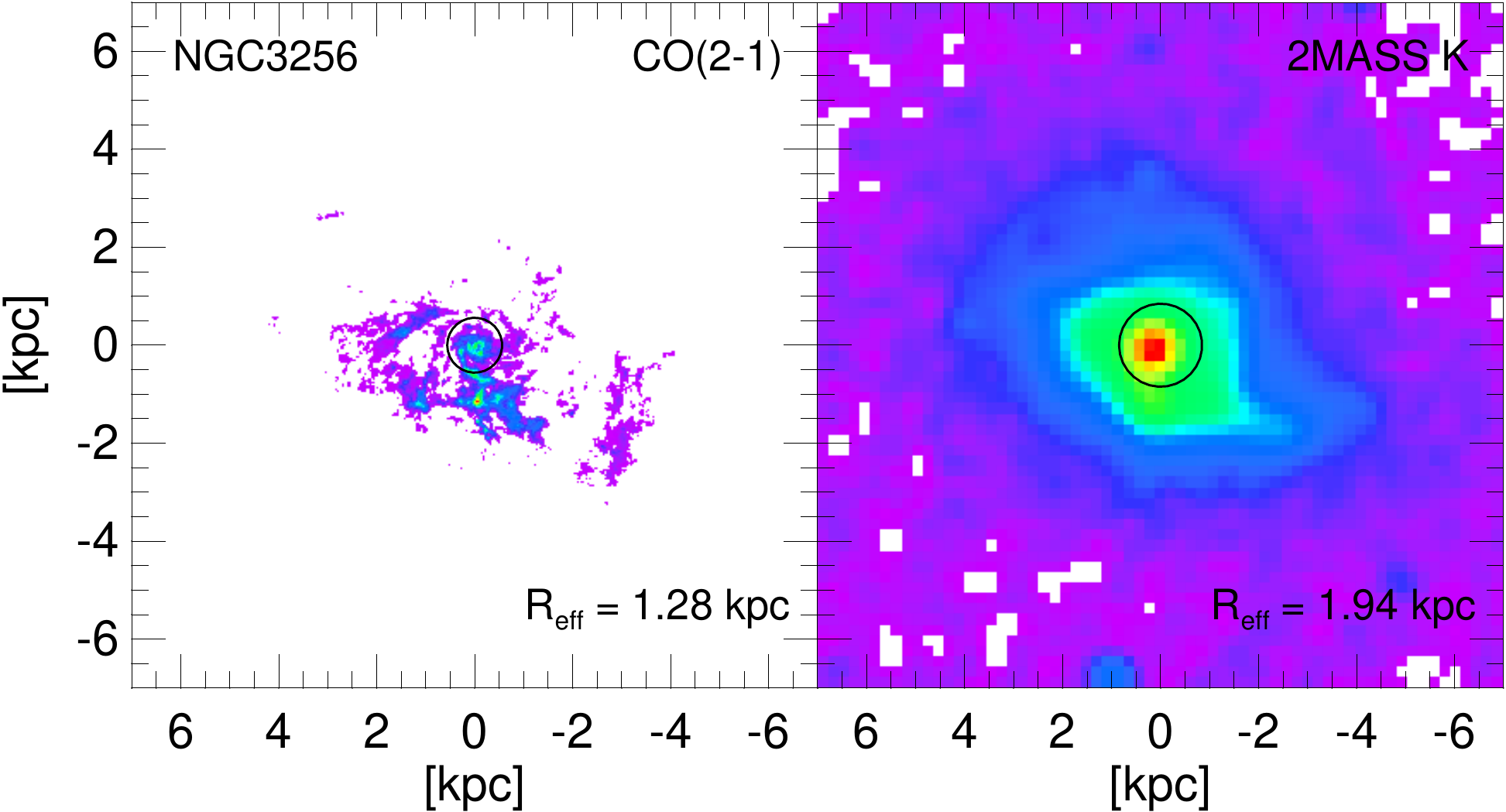}
\caption{Same figure caption as in Fig.~\ref{Mix_class}.}
\label{Mix_class_3}
\end{figure*}

We present the CO(2--1) maps along with the 2MASS K-band maps for the whole sample. The maps are ordered according to the increasing molecular size, R$_{CO}$. We used the same FoV for the whole sample (14$\times$14 kpc$^2$), both in the CO and stellar emissions for a more direct comparison. The effective radii of the molecular and stellar components are highlighted within the figures (Figs.~\ref{Mix_class}, \ref{Mix_class_2}, \ref{Mix_class_3}).

\section{IR SED fit: ALMA flux loss estimation at 1.3 mm}
\label{SEDs_sample}

In this Appendix we present the IR spectral energy distribution (SEDs) for the whole LIRG sample, using {\it Herschel} PACS and SPIRE data (\citealt{Chu17}). The gray-body emission has been derived assuming a $\beta$ parameter of 2 (except for one case, IRAS F17138-1017) and one single gray body temperature, T$_{dust}$. The ALMA continuum fluxes at 1.3 mm are also shown but they have not been included to the fit. In all cases, the ALMA flux densities are below the SED fitting curve. We note that the total continuum flux emission was computed considering $>$5$\sigma$ emission.

The aim of this analysis is to quantify the flux loss derived when using ALMA data at 1.3 mm with respect to the extrapolated flux at 1.3 mm (through the SED fitting analysis). Indeed, at such frequencies ($\sim$230 GHz), in addition to the dust emission we can also have some contribution from non-thermal synchrotron emission and thermal free-free emission (\citealt{Condon16}). 
The ratio between the theoretical flux density and that obtained using ALMA is also highlighted in each fit. The gap between the two values clearly suggests that a considerable part of the 1.3 mm emission ($<$70\%) is not detected by our ALMA observations. The maximum recoverable scale is 3-5 kpc, so extended diffuse and faint emission is unlikely filtered-out, but could be beyond the sensitivity of these data.

\begin{figure*}
\centering
\includegraphics[width=0.3\textwidth]{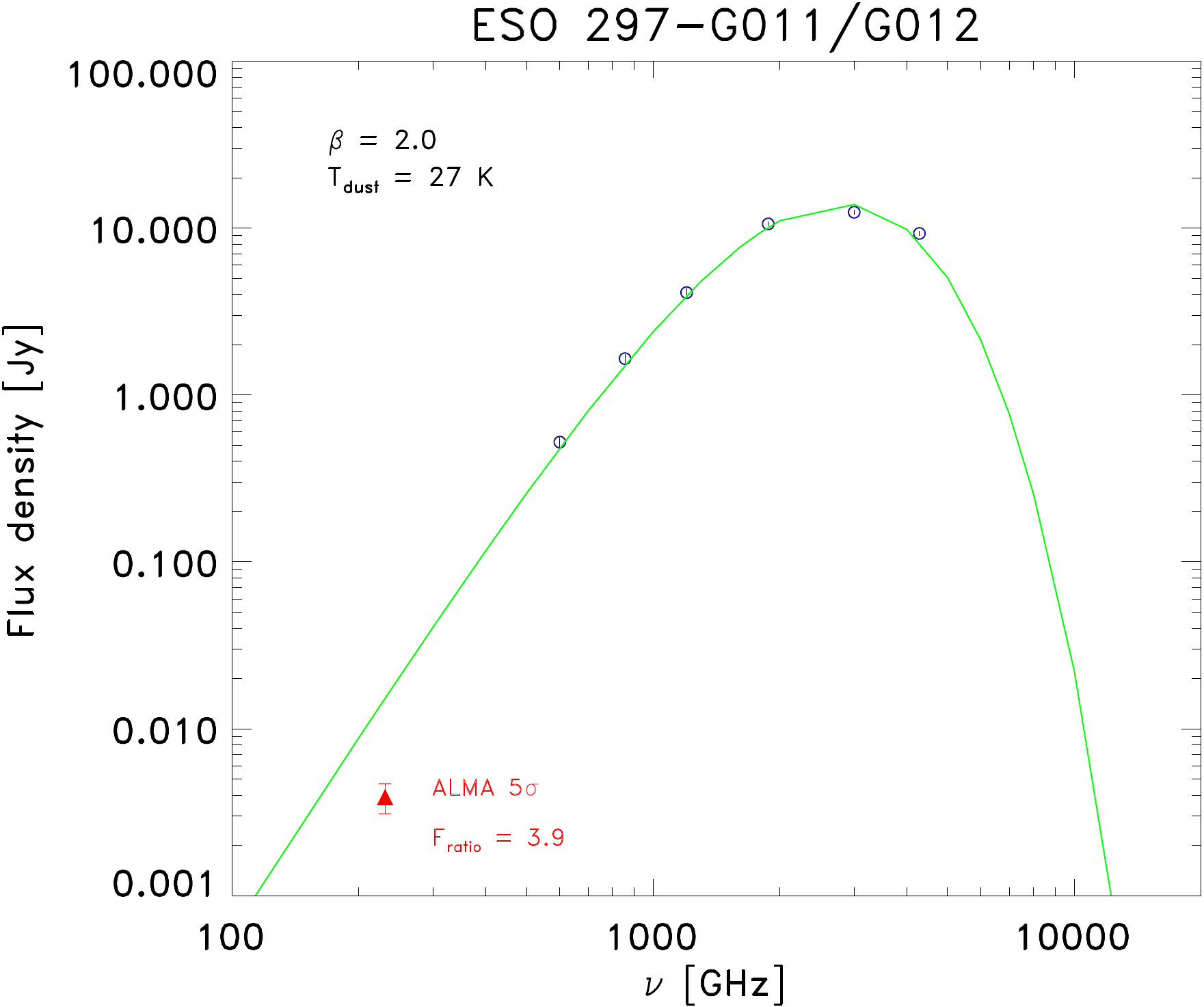}
\hskip2mm\includegraphics[width=0.3\textwidth]{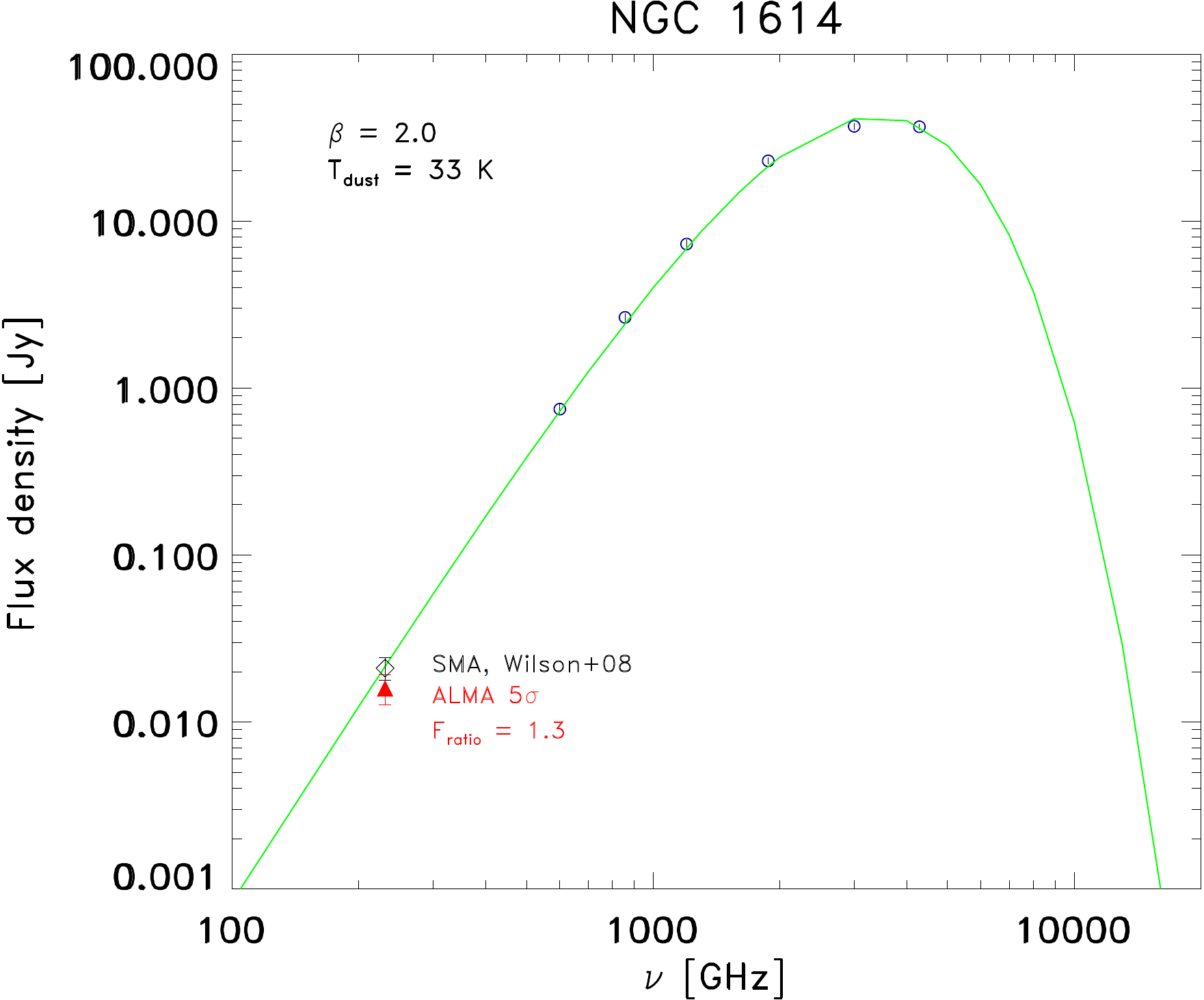}
\hskip2mm\includegraphics[width=0.3\textwidth]{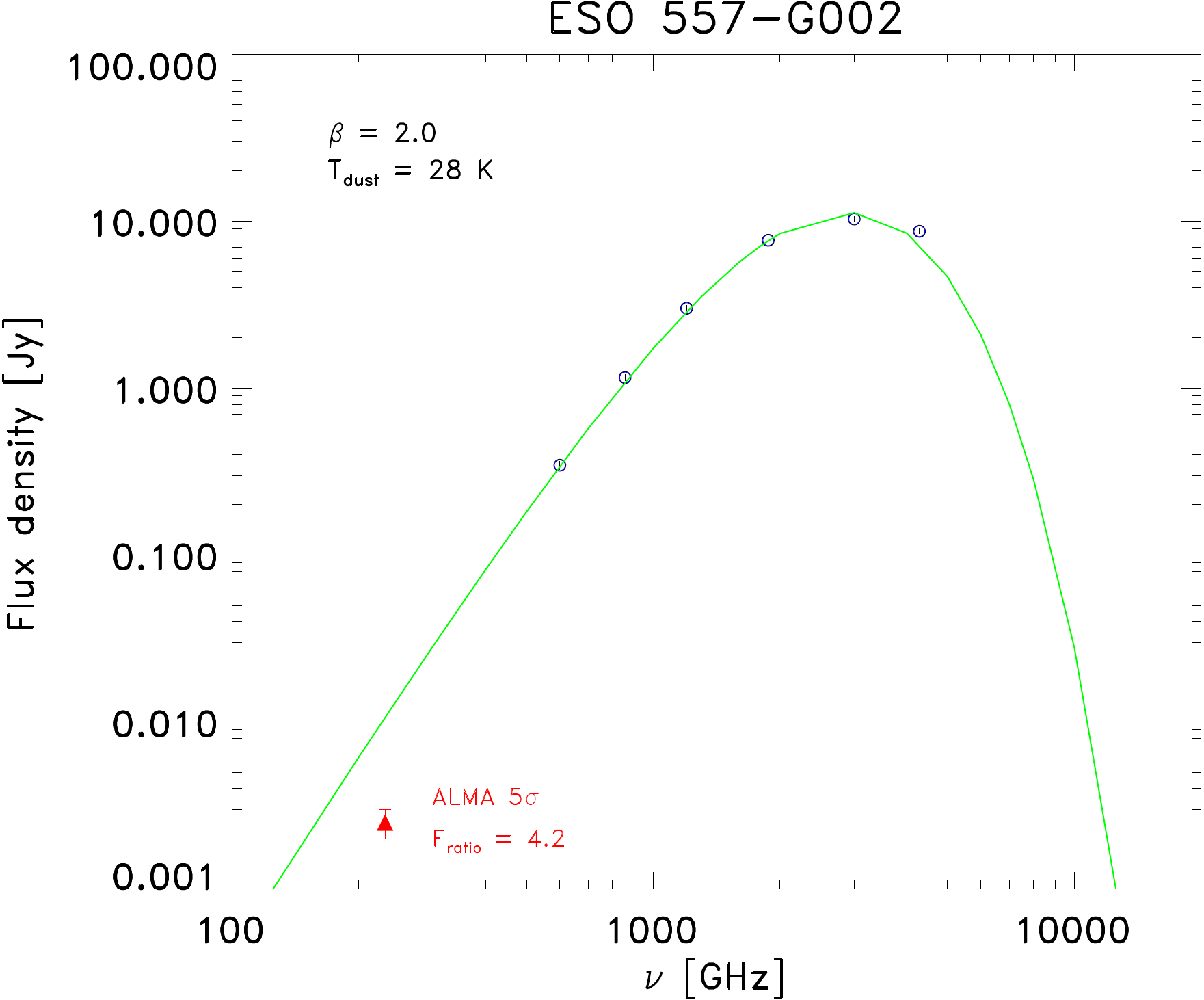}
\vskip3mm
\includegraphics[width=0.3\textwidth]{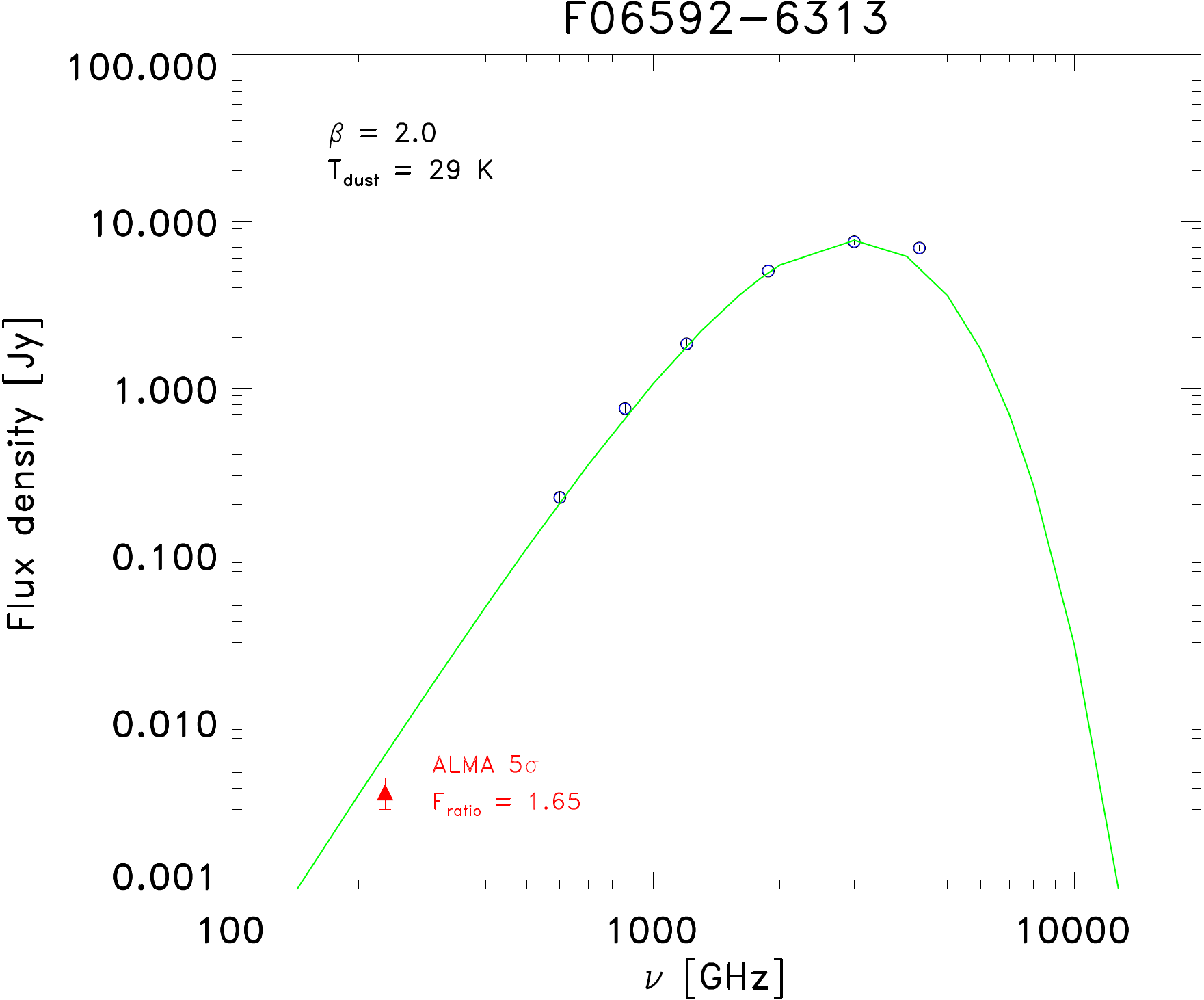}
\hskip2mm\includegraphics[width=0.3\textwidth]{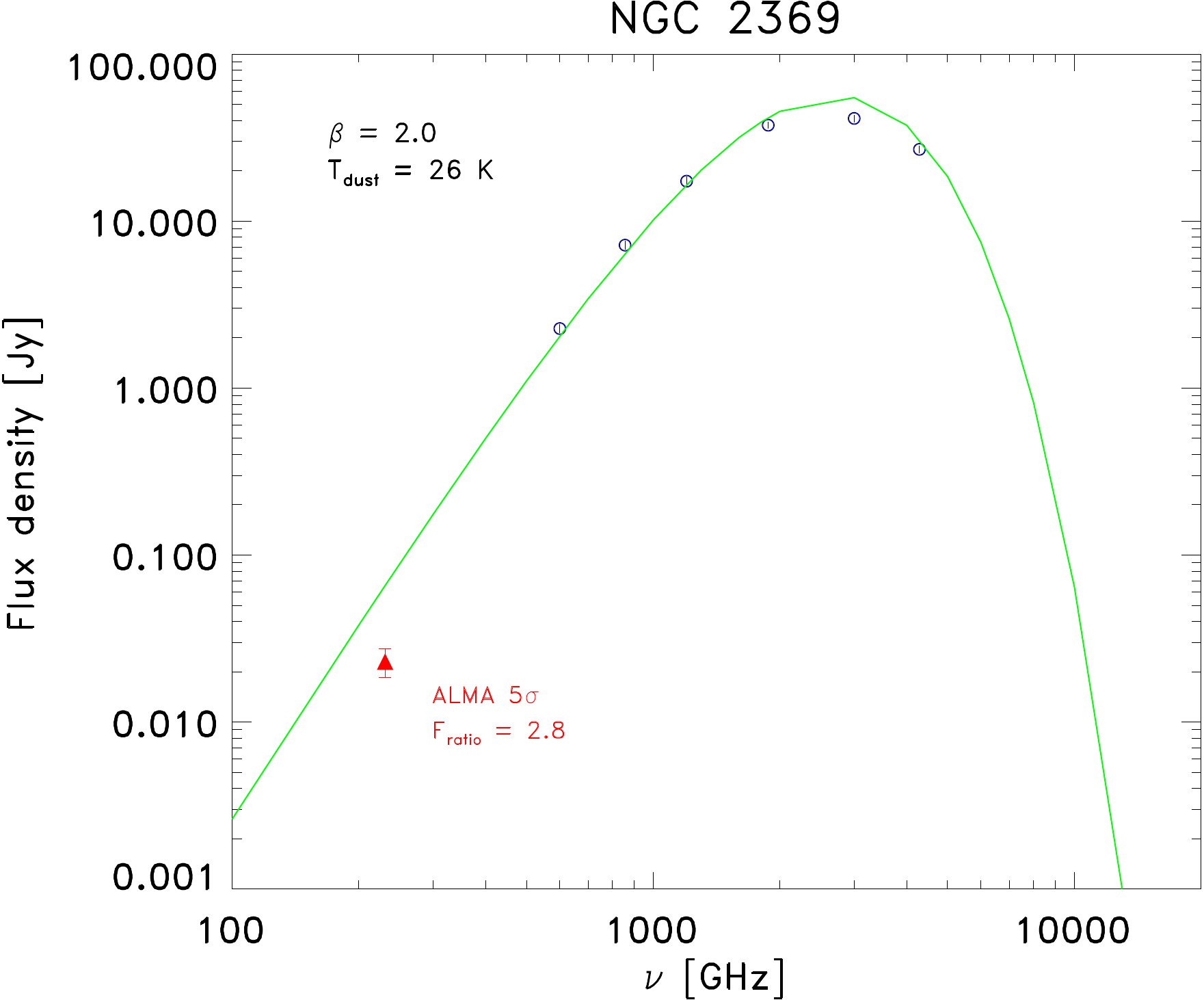}
\hskip2mm\includegraphics[width=0.3\textwidth]{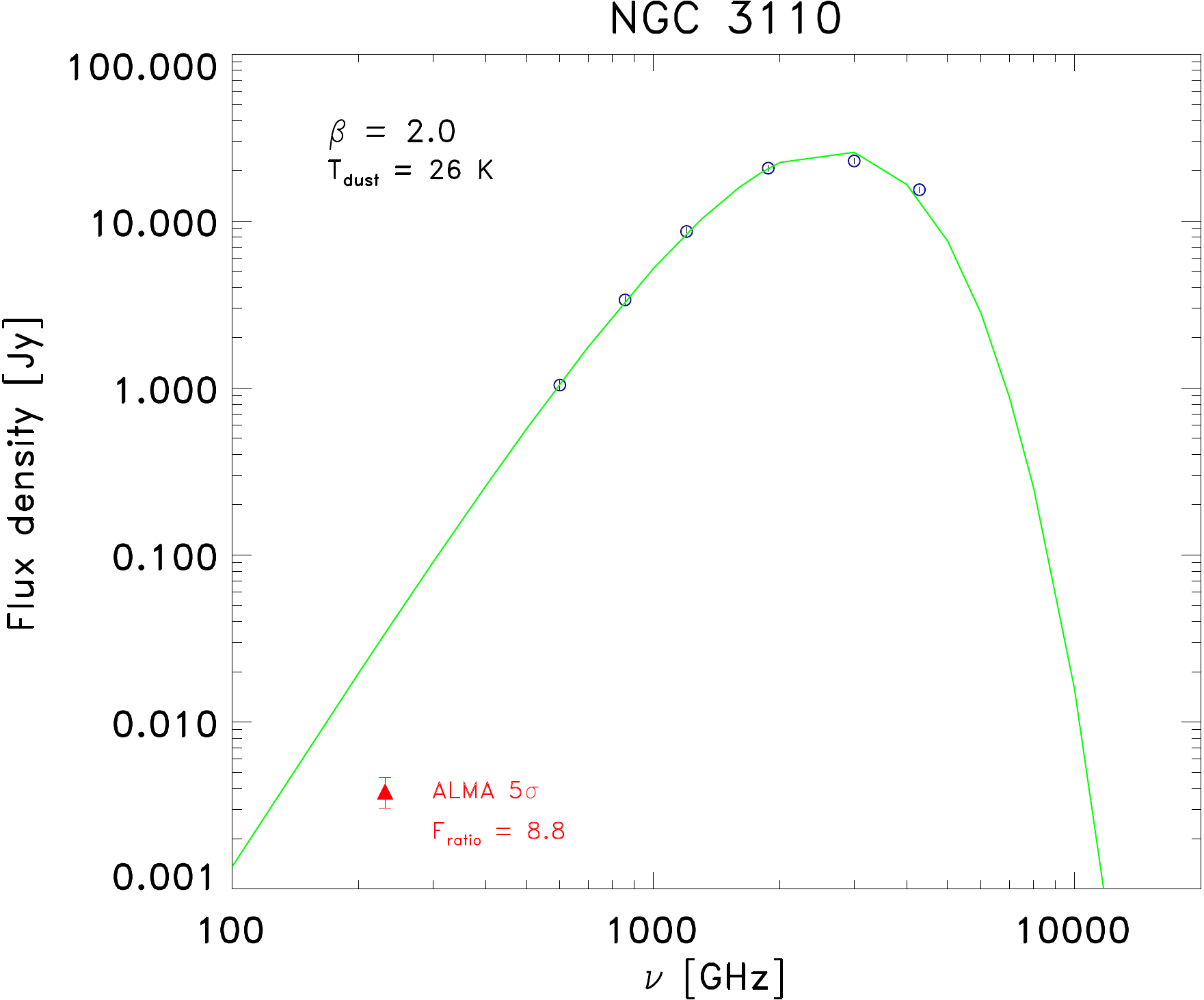}
\caption{Spectral line energy distributions (SEDs) using {\it Herschel} data, assuming $\beta$ = 2.0 and deriving the T$_{dust}$. The ALMA continuum flux at 1.3 mm at 5$\sigma$ is also shown for a direct comparison. The ratio between the flux at 1.3 mm expected by the model and the ALMA value is also highlighted in red (F$_{ratio}$).}
\label{SEDs_mix_13mm}
\end{figure*}

\begin{figure*}
\centering
\includegraphics[width=0.3\textwidth]{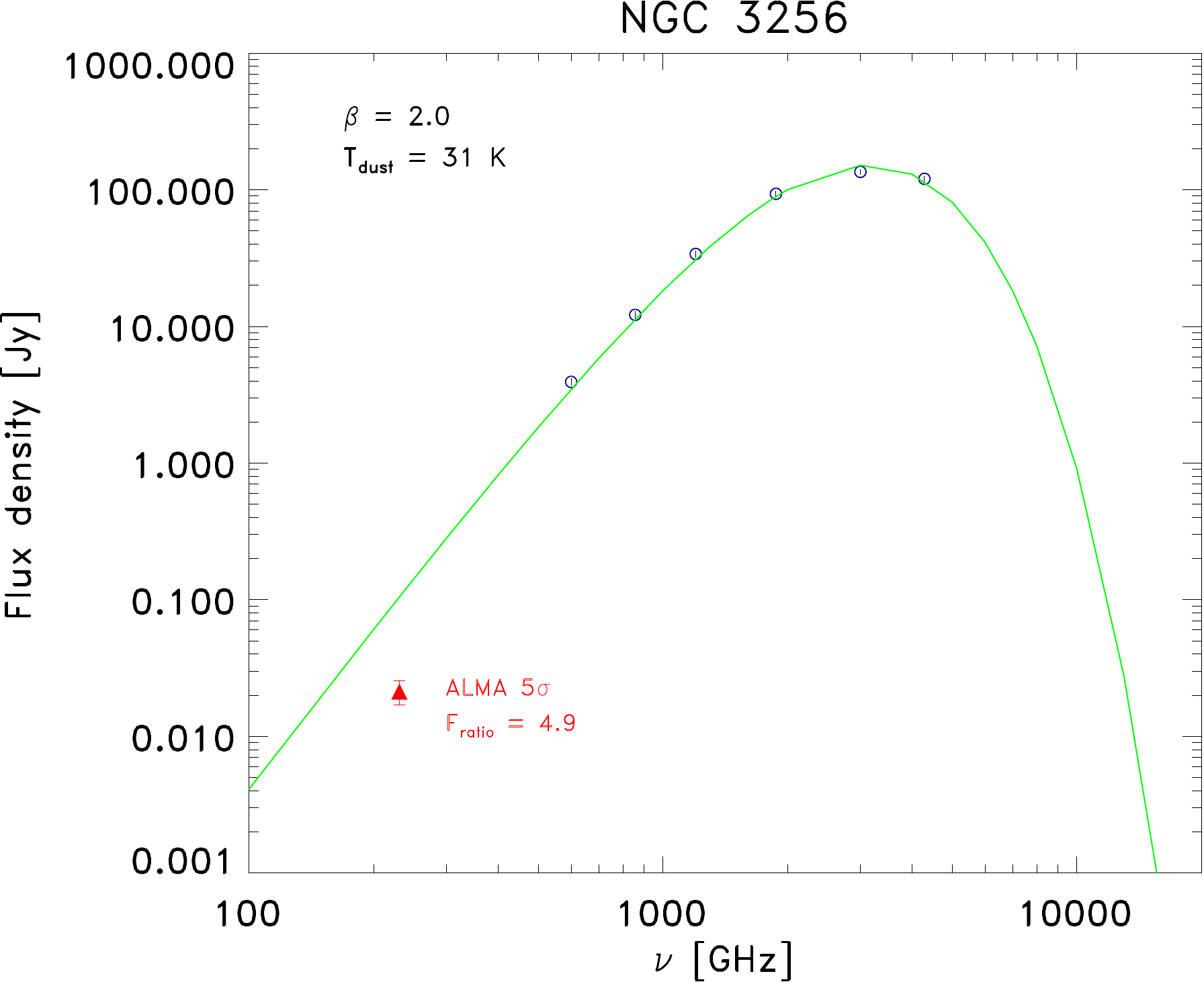}
\hskip2mm\includegraphics[width=0.3\textwidth]{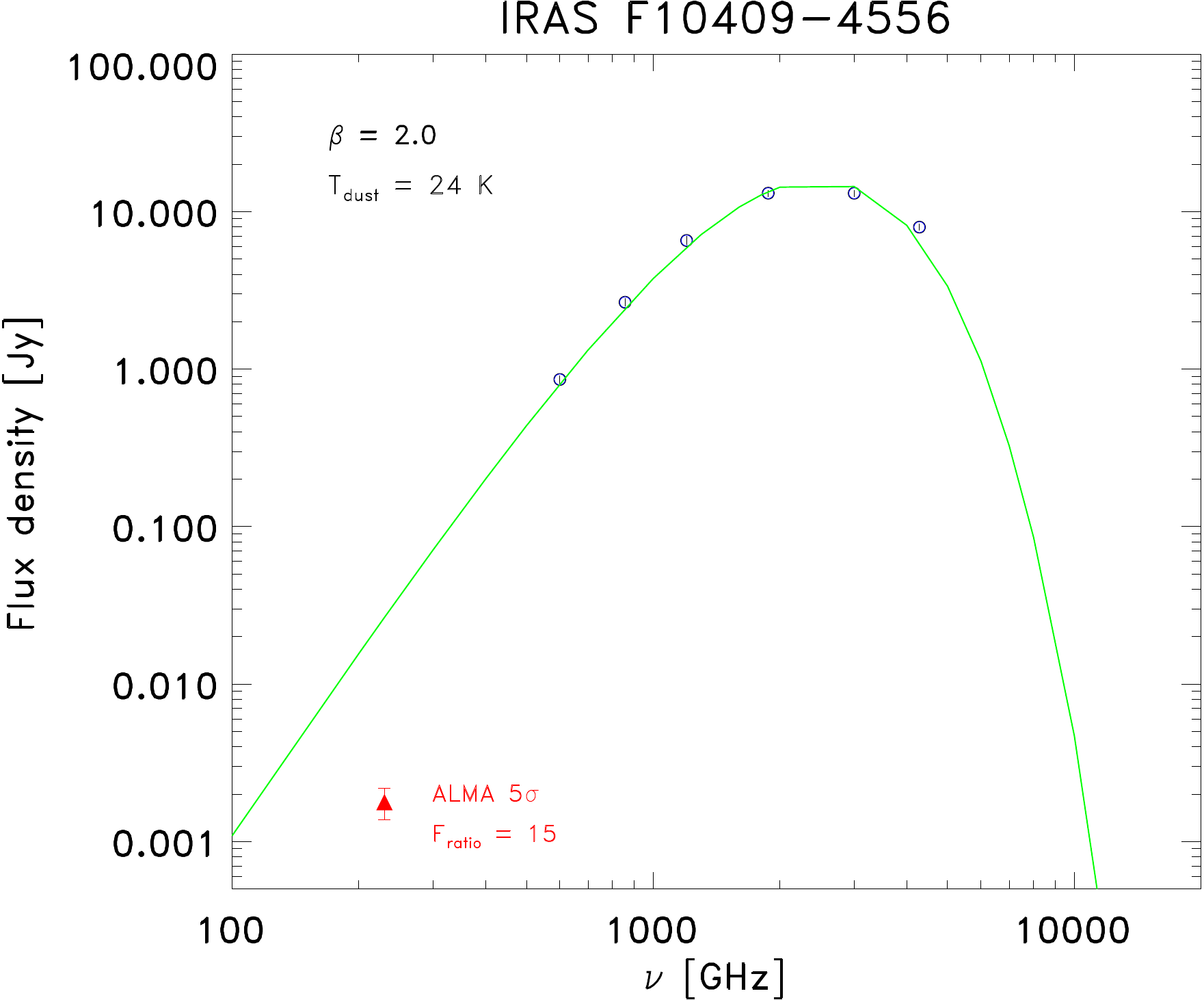}
\hskip2mm\includegraphics[width=0.3\textwidth]{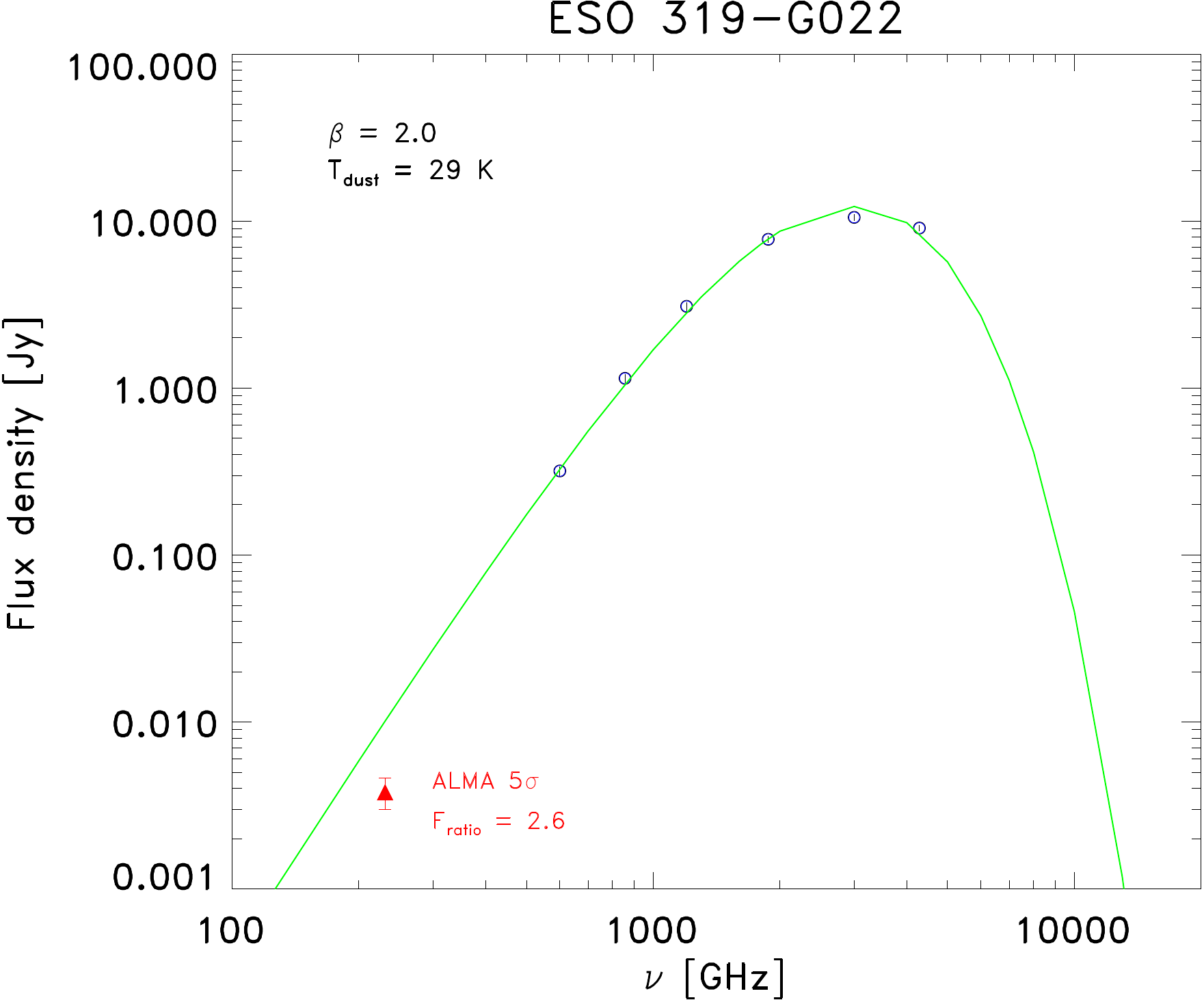}
\vskip3mm
\includegraphics[width=0.3\textwidth]{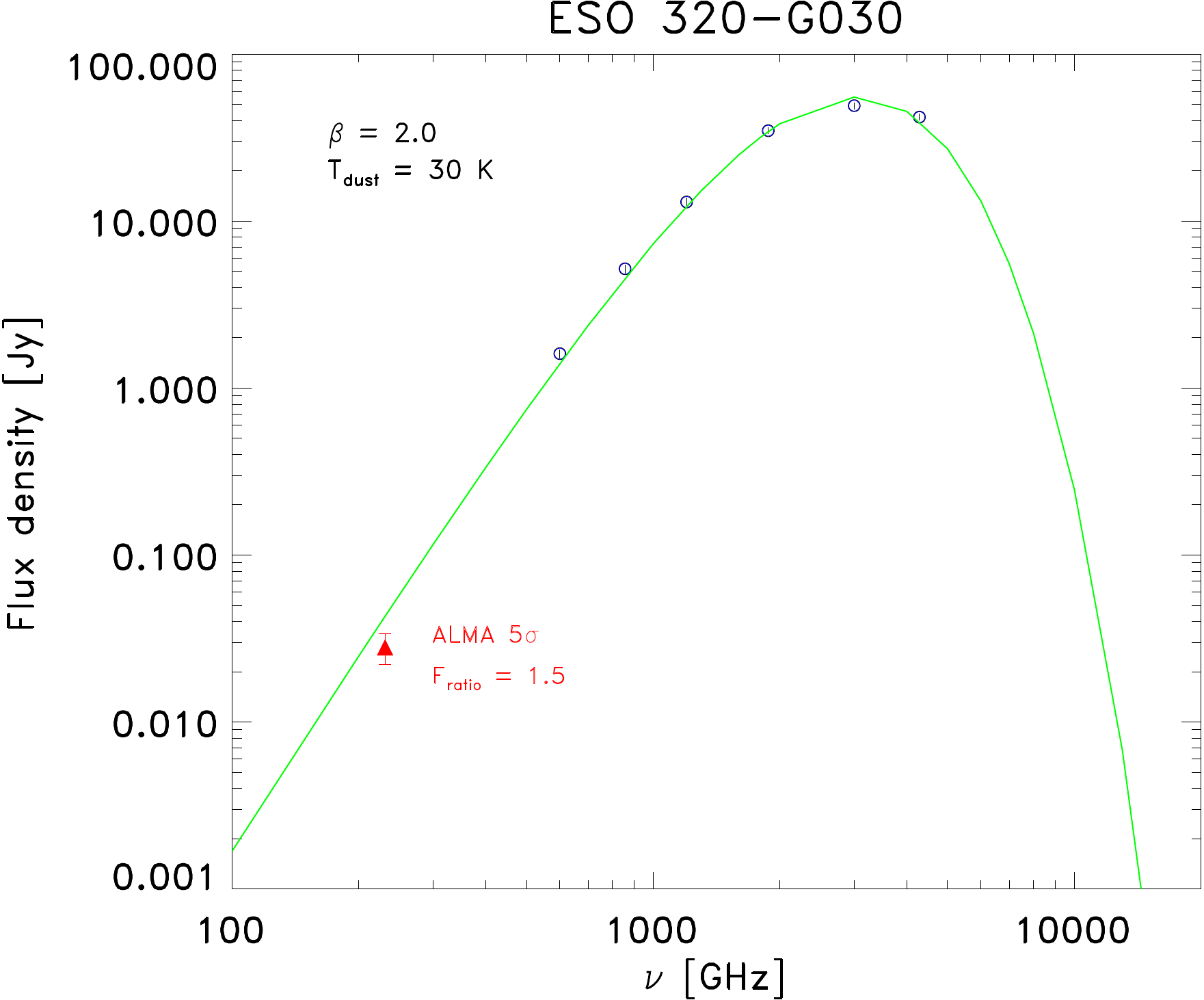}
\hskip2mm\includegraphics[width=0.3\textwidth]{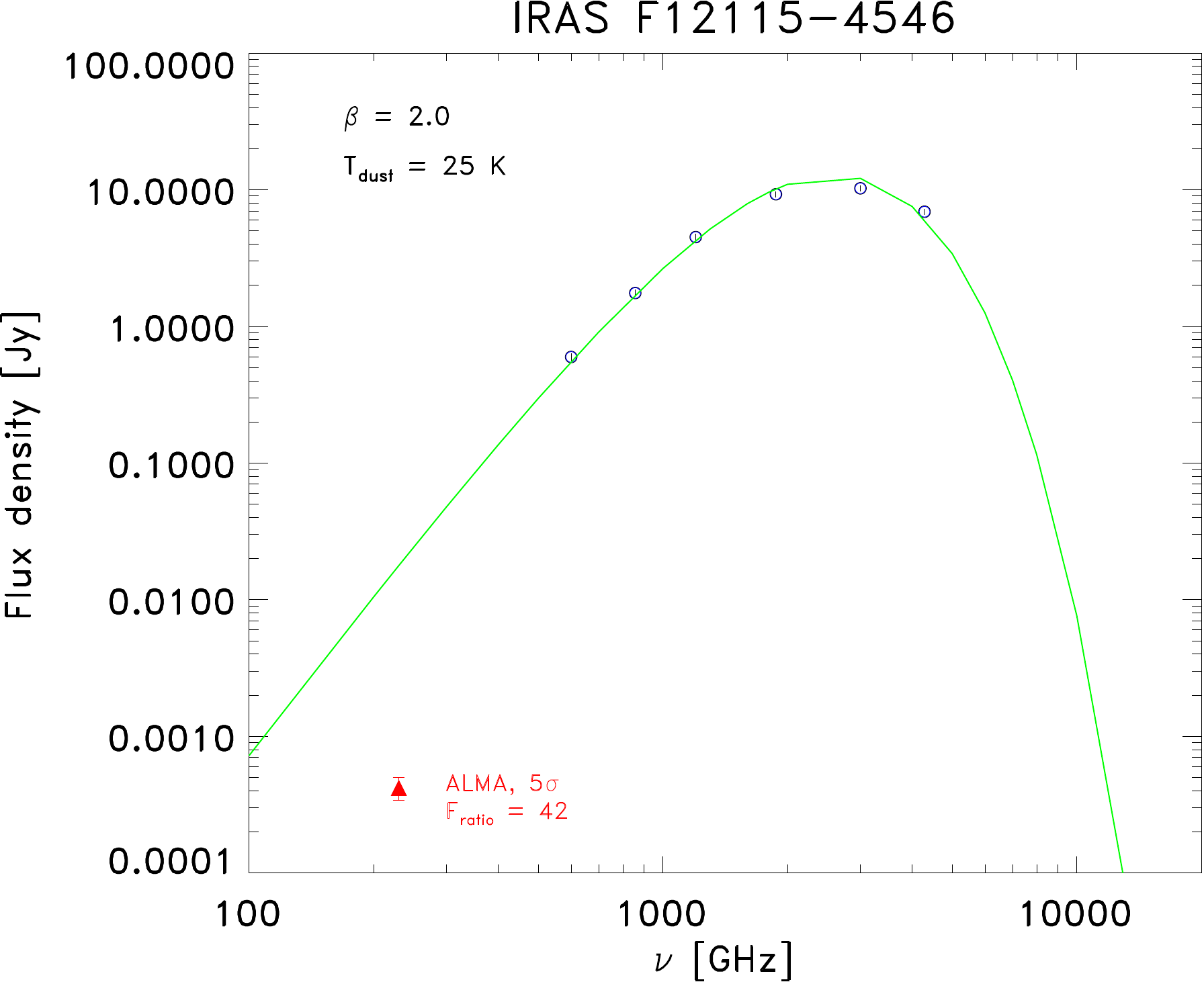}
\hskip2mm\includegraphics[width=0.3\textwidth]{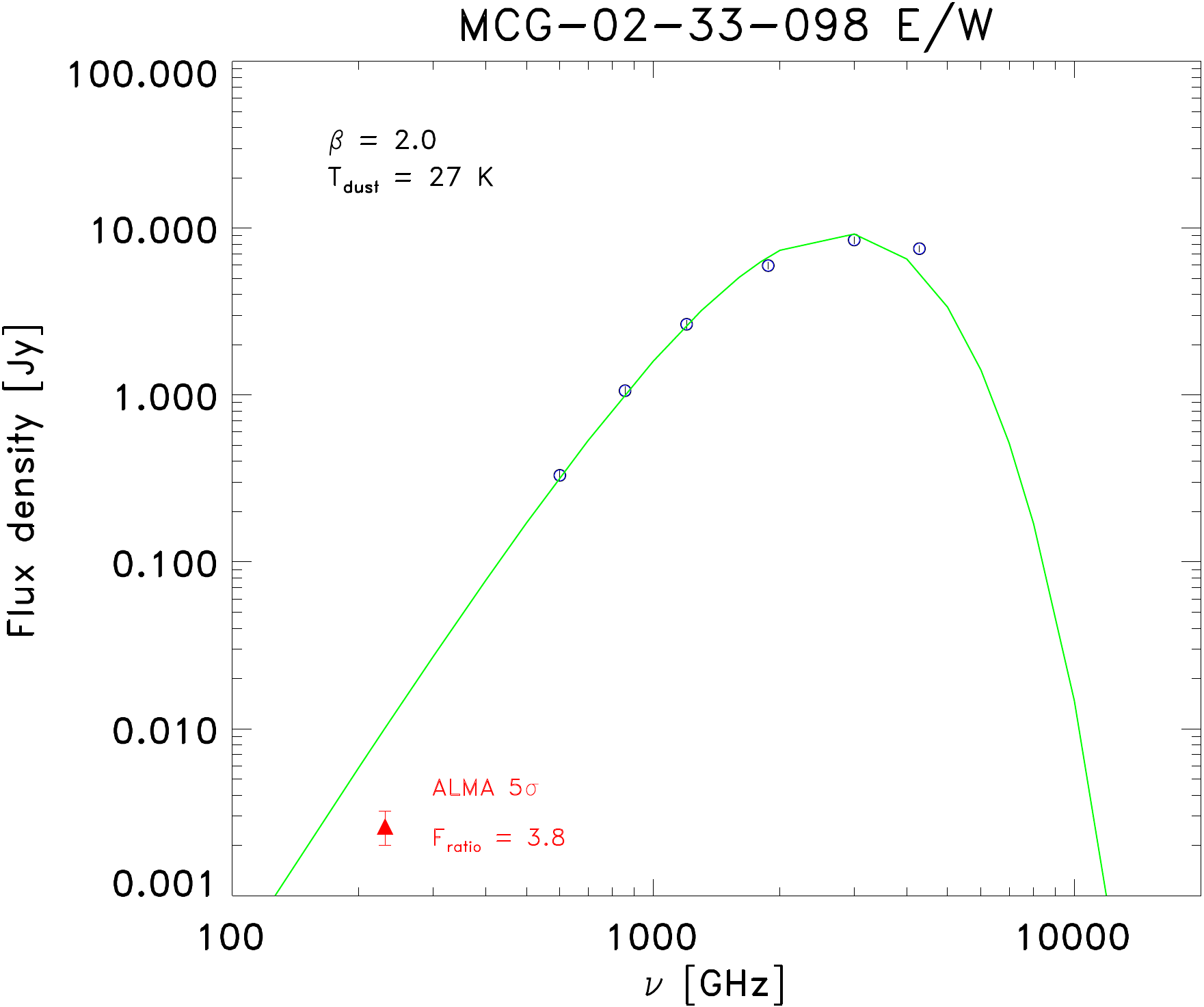}
\vskip3mm
\includegraphics[width=0.3\textwidth]{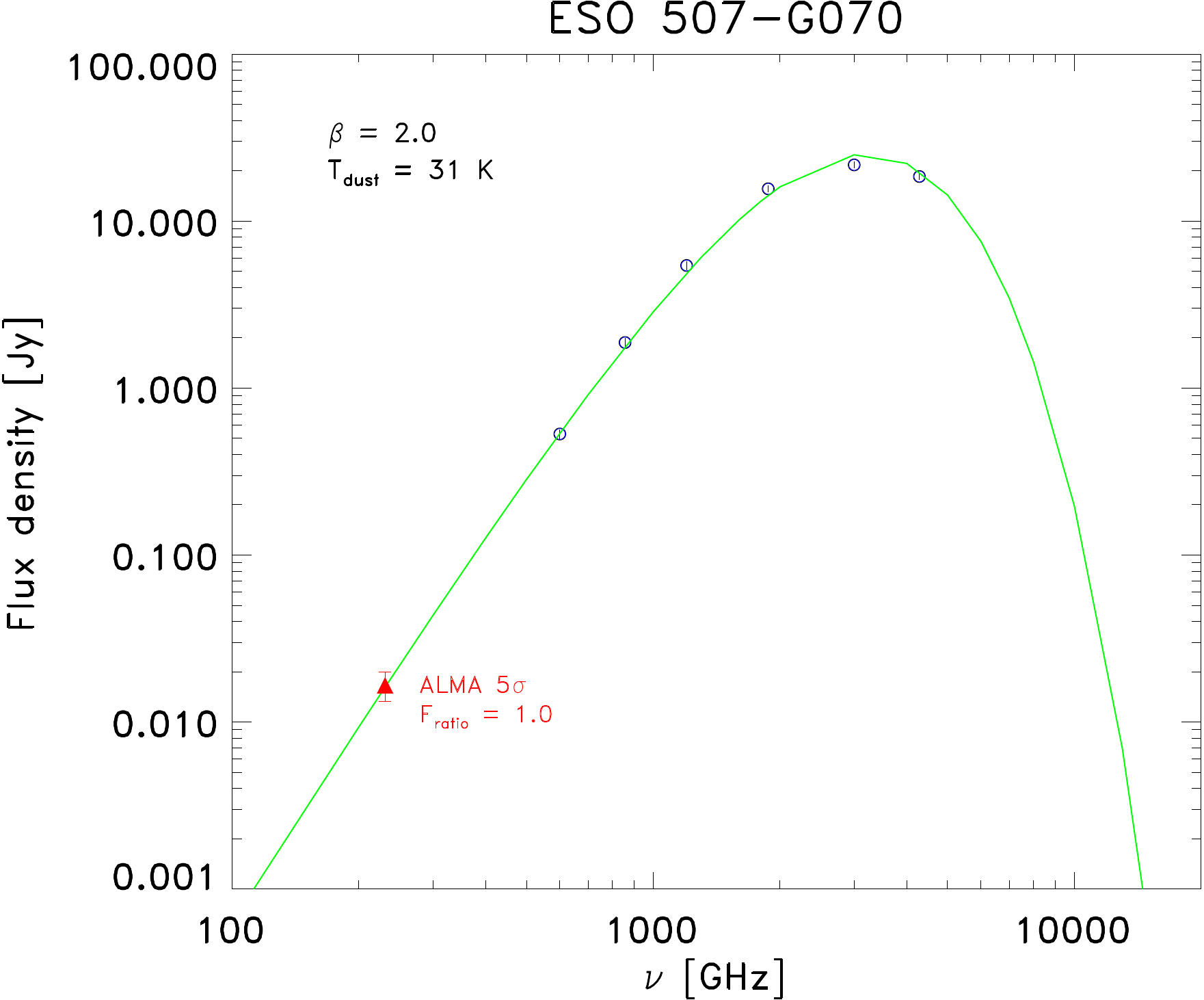}
\hskip2mm\includegraphics[width=0.3\textwidth]{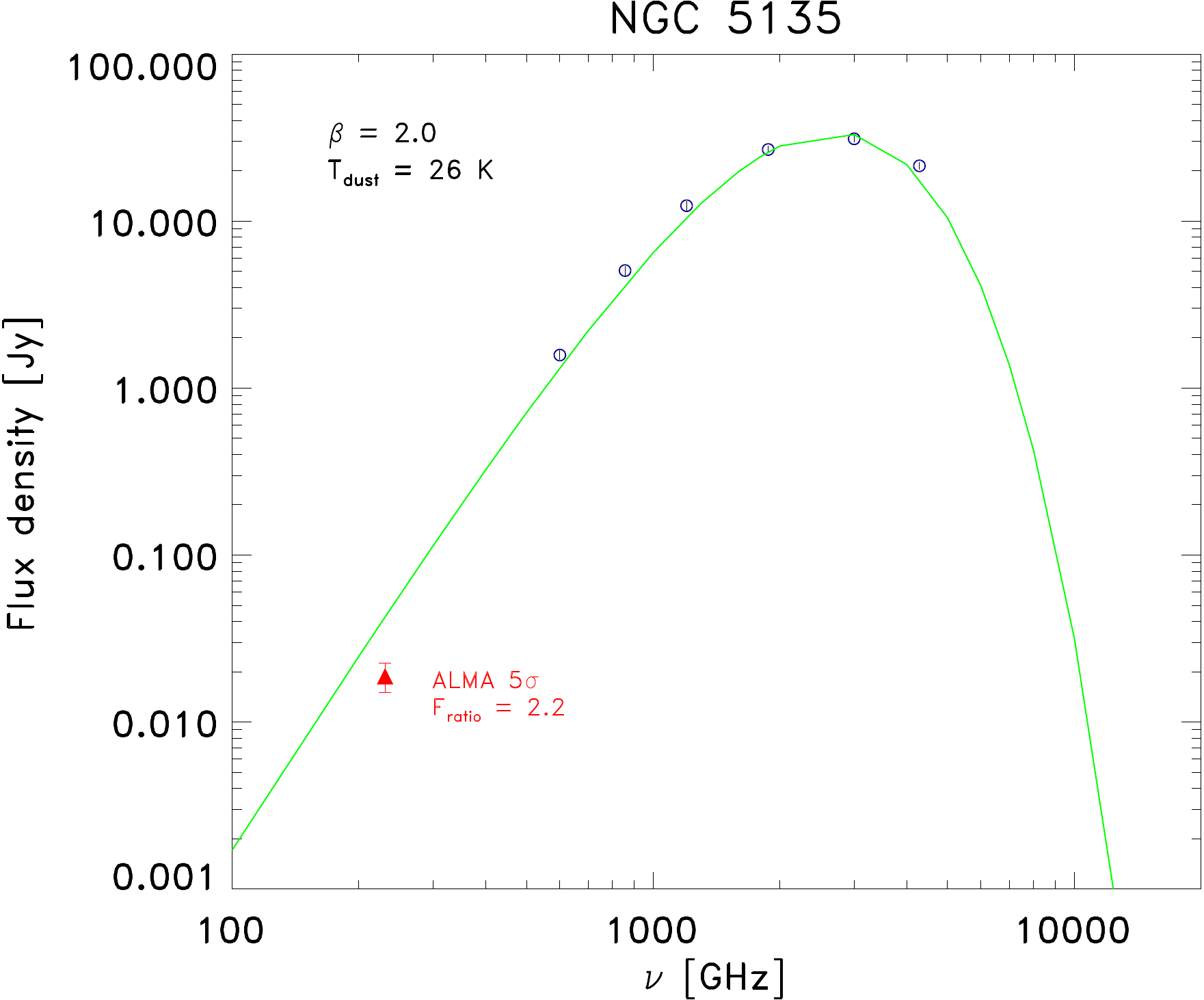}
\hskip2mm\includegraphics[width=0.3\textwidth]{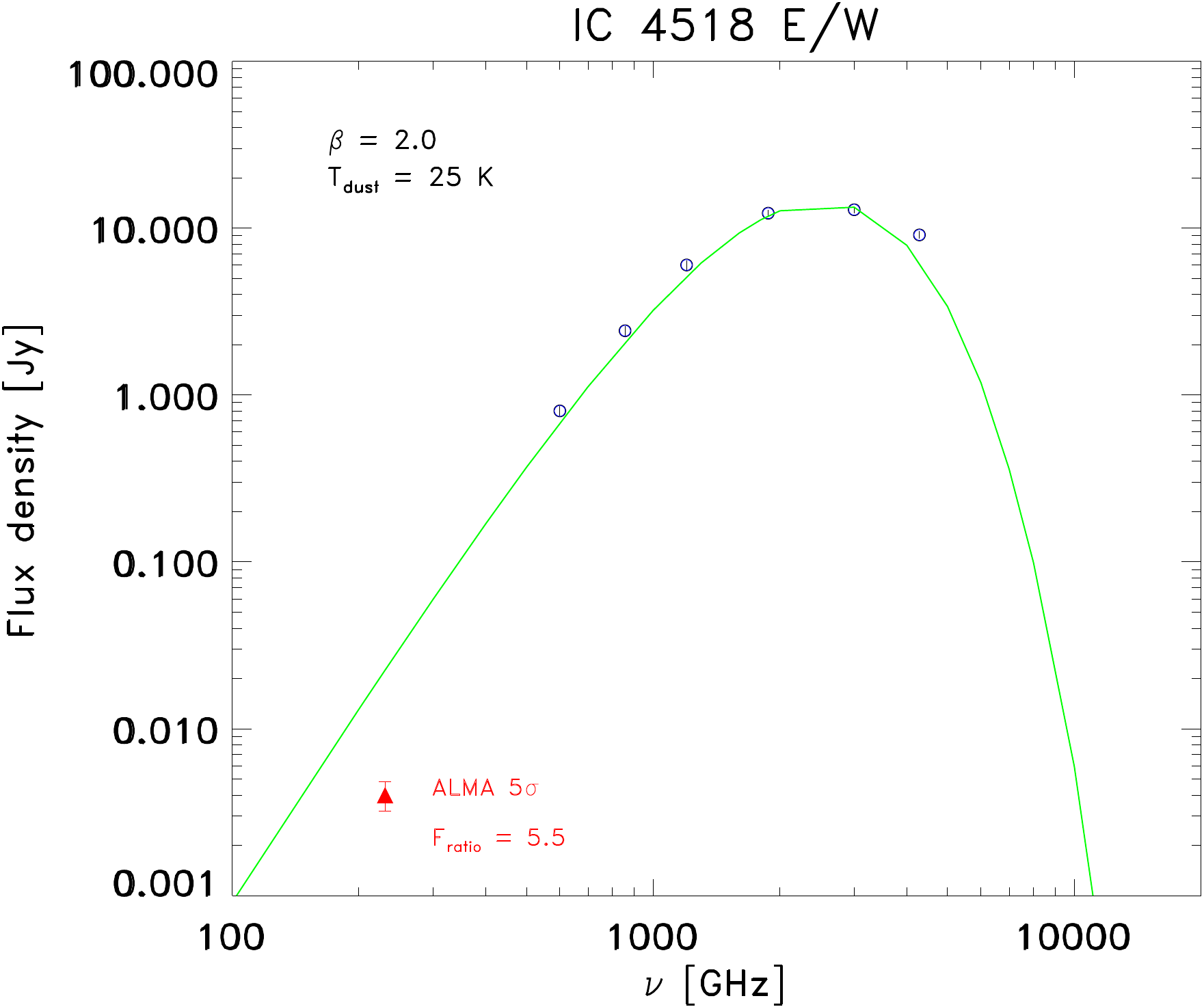}
\vskip3mm
\includegraphics[width=0.3\textwidth]{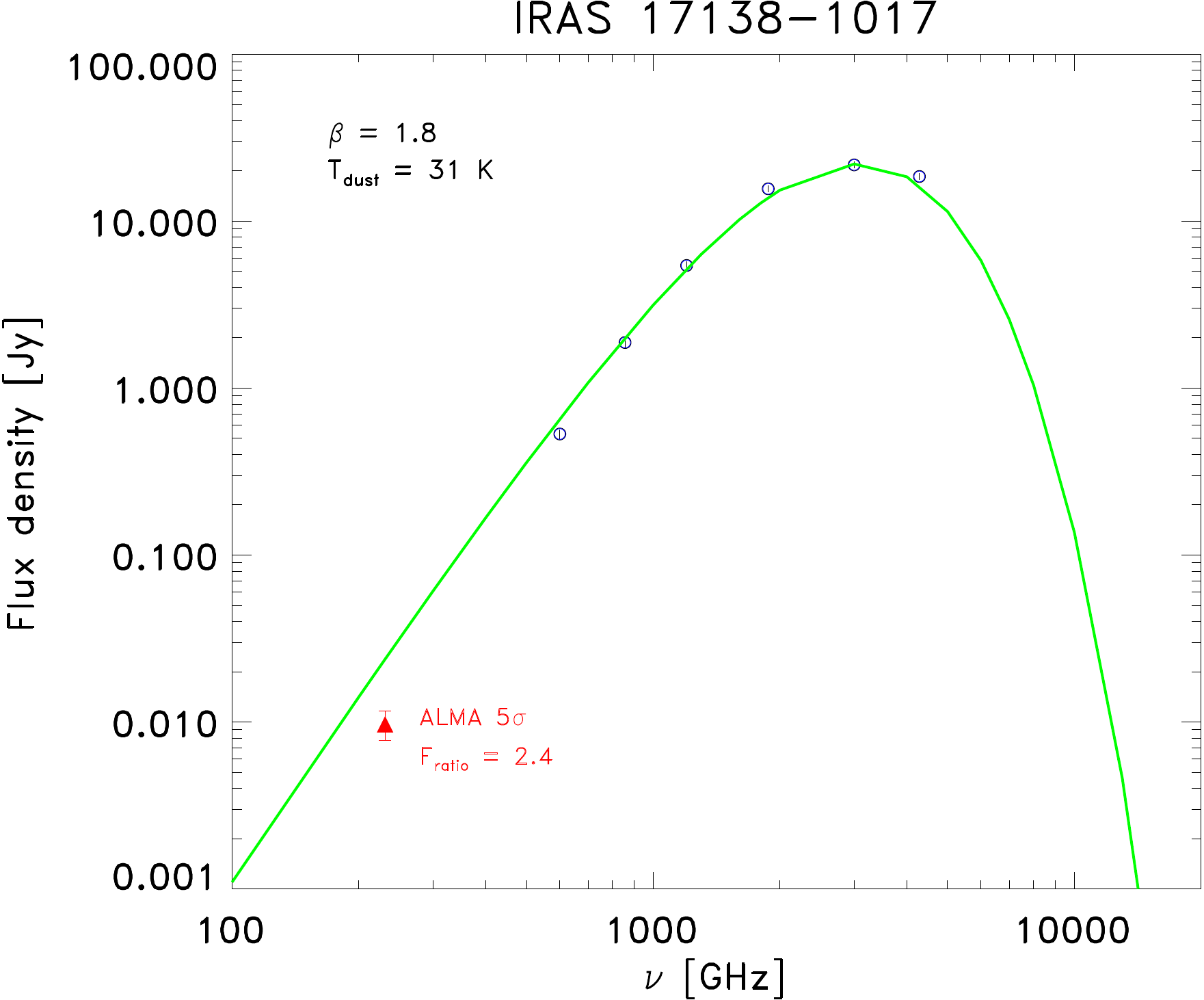}
\hskip2mm\includegraphics[width=0.3\textwidth]{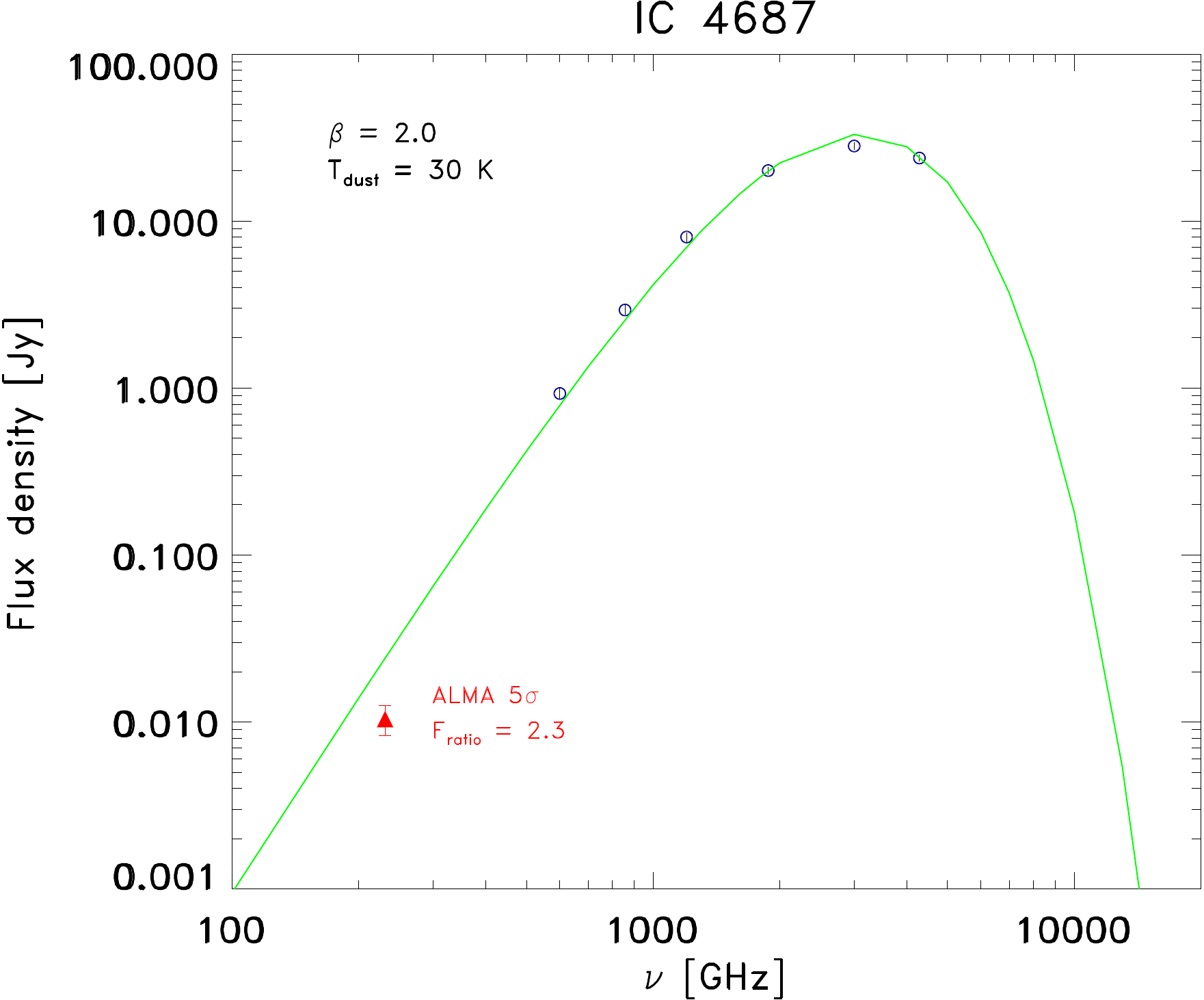}
\hskip2mm\includegraphics[width=0.3\textwidth]{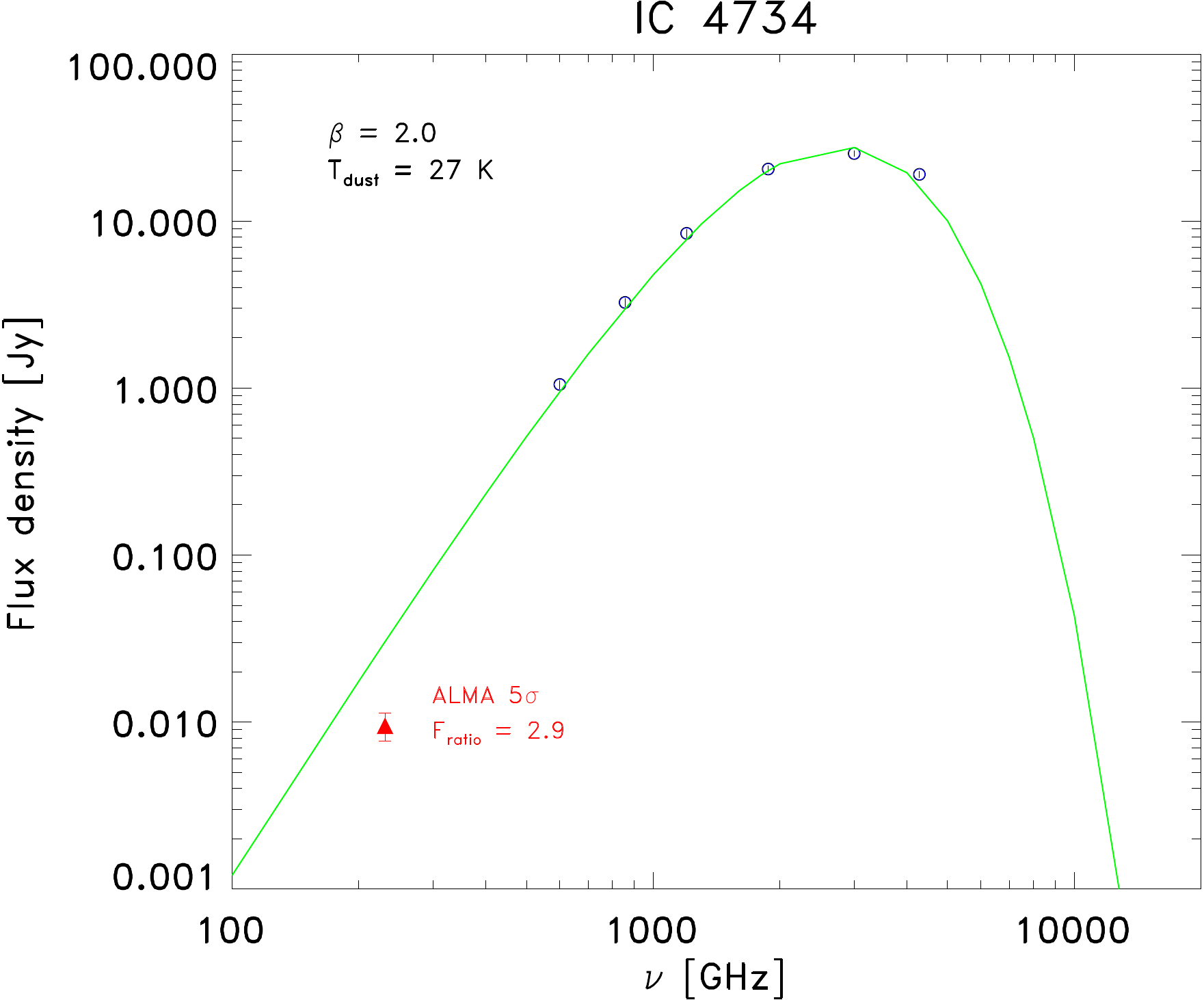}
\vskip3mm
\includegraphics[width=0.3\textwidth]{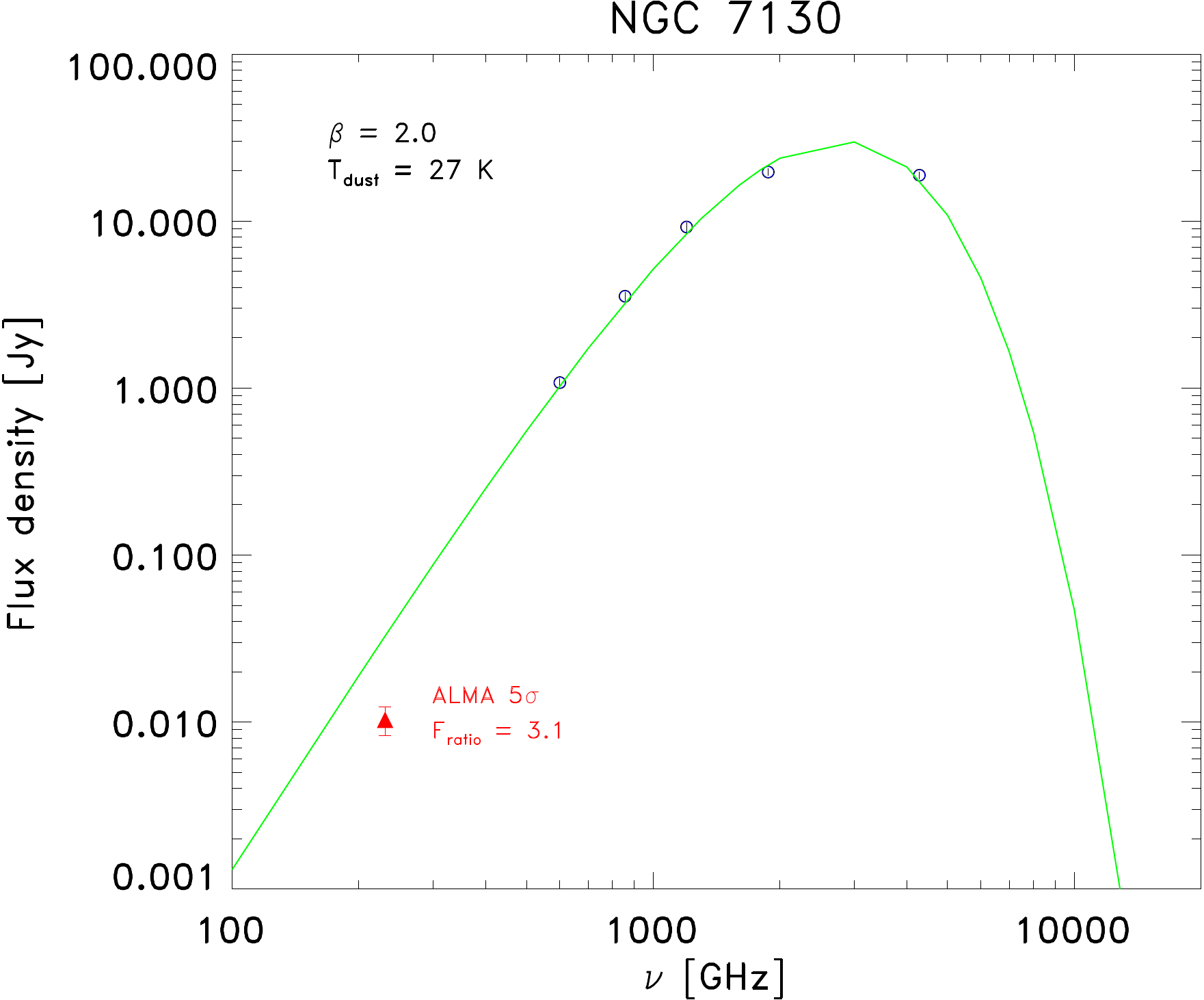}
\hskip2mm\includegraphics[width=0.3\textwidth]{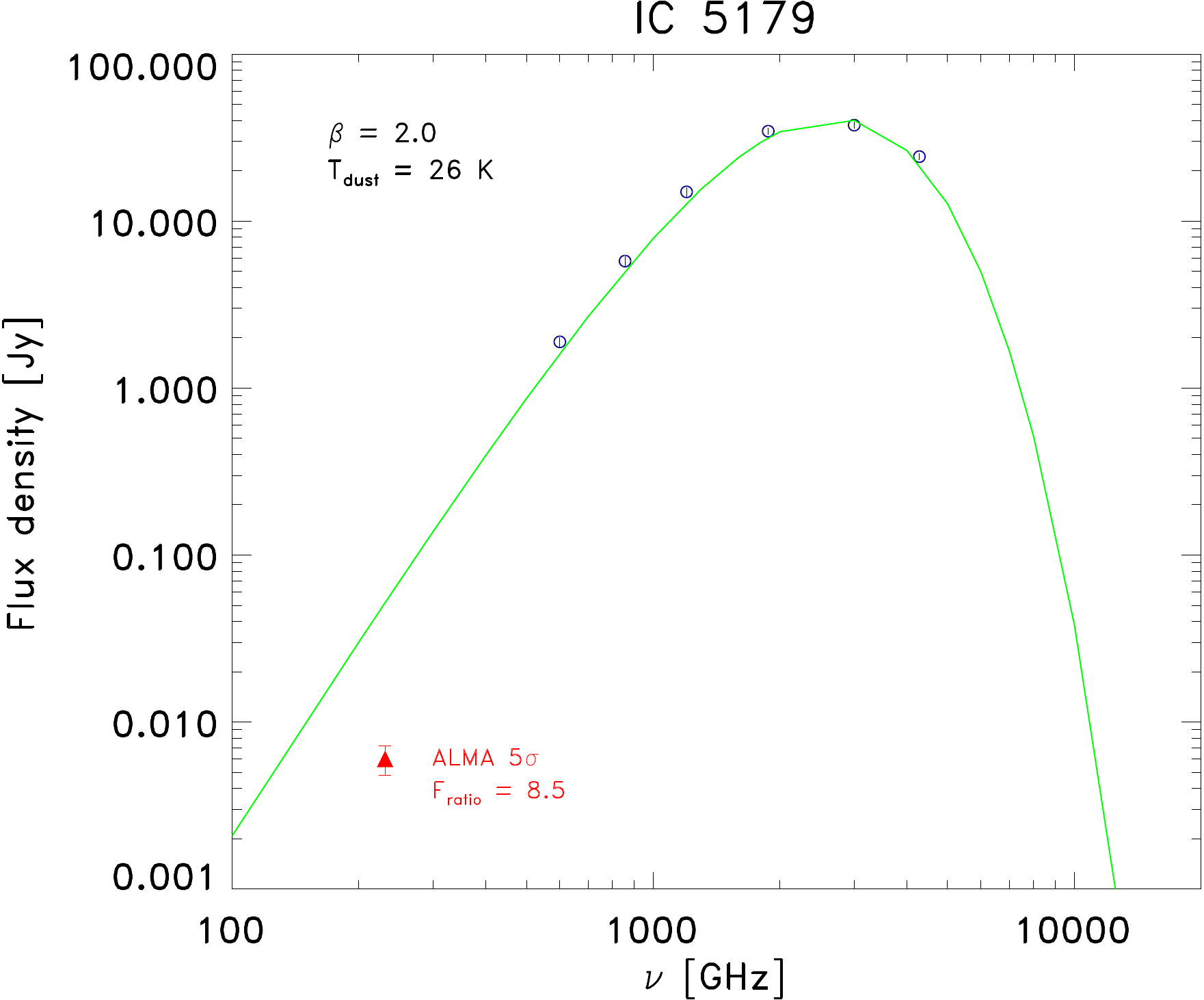}
\hskip2mm\includegraphics[width=0.3\textwidth]{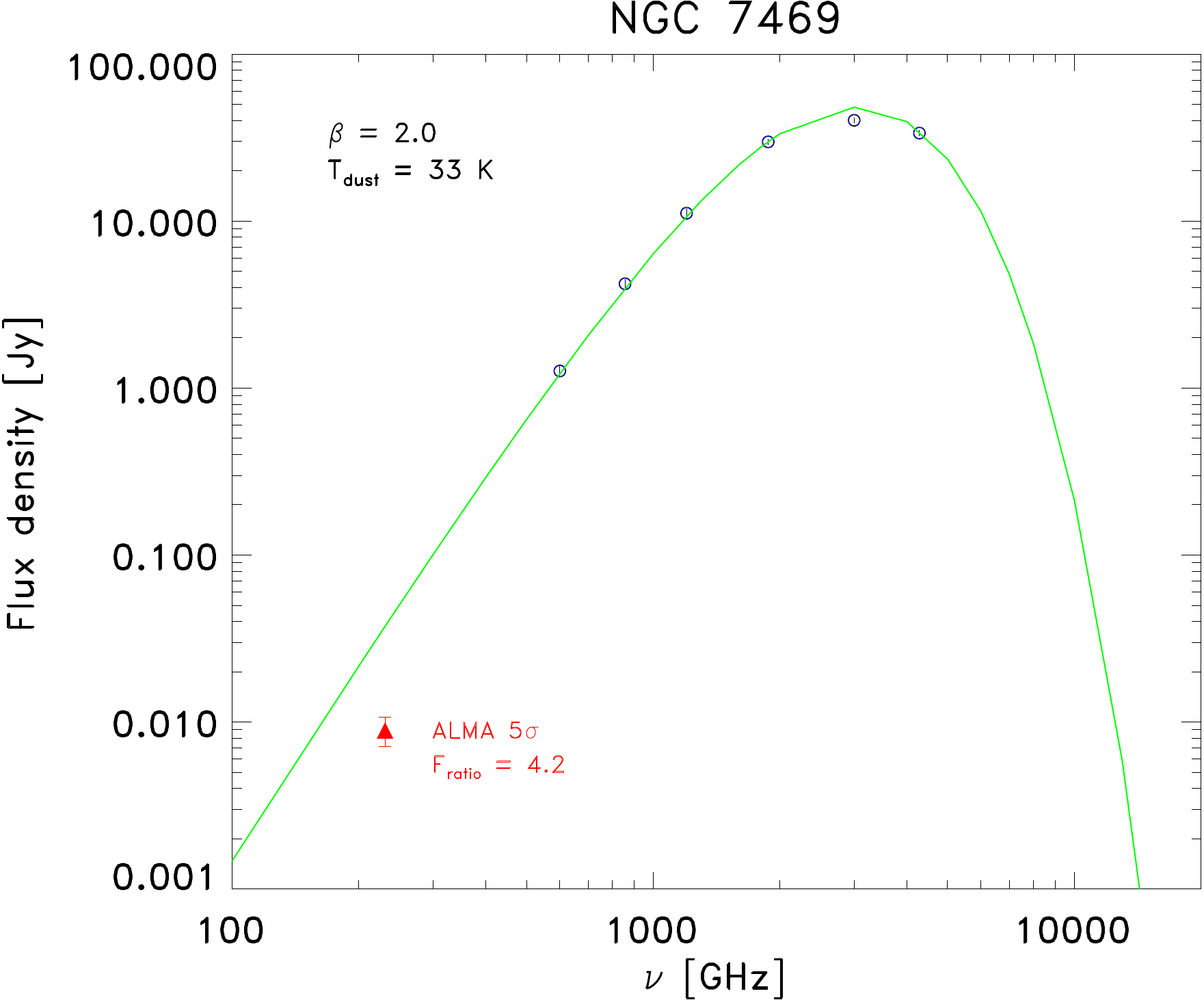}
\caption{Same figure caption as in Fig.~\ref{SEDs_mix_13mm}}
\label{SEDs_mix_13mm_2}
\end{figure*}

\end{appendix}

\end{document}